\documentclass[traditabstract]{aa}

\usepackage{natbib,natbibspacing}
\bibpunct{(}{)}{;}{a}{}{,}
\usepackage{graphicx}
\usepackage{txfonts}
\usepackage{longtable}
\usepackage{rotating}
\usepackage{multirow} 
\usepackage{appendix}
\usepackage{latexsym}
\usepackage{lscape}
\usepackage{afterpage}
\usepackage{caption}

\begin{document}

\title{Neutral gas outflows in nearby [U]LIRGs via optical NaD feature}
\author{S.~Cazzoli\inst{\ref{inst1},\ref{inst2}, \ref{inst3}}\and  S.~Arribas \inst{\ref{inst1},\ref{inst3} } \and R.~Maiolino\inst{\ref{inst2}, \ref{inst4}} \and  L.~Colina\inst{\ref{inst1}, \ref{inst3}}}  
\institute{CSIC - Departamento de Astrofisica-Centro de Astrobiologia (CSIC-INTA),Torrejon de Ardoz, Madrid, Spain \\ \email{scazzoli@cab.inta-csic.es}\label{inst1} \and  Cavendish Laboratory, University of Cambridge 19 J. J. Thomson Avenue, Cambridge CB3 0HE, UK\label{inst2} \and Astro-UAM, UAM, Unidad Asociada CSIC\label{inst3}\and Kavli Institute for Cosmology, University of Cambridge, Madingley Road, Cambridge CB3 0HA, UK\label{inst4}}

\date{Received 19\,July\,2015 / Accepted 15\,February\,2016}

 \abstract
{We studied the properties of the neutral gas in a sample of  38 local luminous and ultra luminous infrared galaxies ([U]LIRGs, 51 individual galaxies at z\,$\leq$\,0.09), which mainly covers the less explored LIRG luminosity range. This study is based on the analysis of the spatially integrated and spatially resolved spectra of the NaD$\lambda$$\lambda$\,5890,5896\,$\AA$ feature obtained with the integral field unit (IFU) of VIMOS at the Very Large Telescope. Analyzing spatially integrated spectra, we find that  the contribution of the stars to the observed NaD equivalent width is small ($<$\,35$\%$) for about half of the sample, and therefore this feature is dominated by inter stellar medium (ISM) absorption. After subtracting the stellar contribution, we find that the pure-ISM integrated spectra generally show blueshifted NaD profiles, indicating neutral gas outflow velocities, V, in the range 65-260 kms$^{\rm-1}$.  Excluding the galaxies with powerful AGNs, V shows a dependency with the star formation rate (SFR) of the type V\,$\propto$\,SFR$^{\rm0.15}$, which is in rather good agreement with previous results. 
The spatially resolved analysis could be performed for 40 galaxies, 22 of which have neutral gas velocity fields dominated by noncircular motions with signatures of cone-like winds. However, a large number of targets (11/40) show disk rotation signatures. Based on a simple model, we found that the wind masses are in the range 0.4\,-\,7.5\,$\times$\,10$^{\rm8}$\,M$_{\rm \sun}$, reaching up to $\sim$\,3$\%$ of the dynamical mass of the host. The mass rates are typically only $\sim$\,0.2-0.4 times the corresponding global SFR indicating that, in general, the mass loss is too small to slow down the star formation significantly. In the majority of cases, the velocity of the outflowing gas is not sufficient to escape the host potential well and, therefore, most of the gas rains back into the galaxy disk. On average V/v$_{\rm esc}$ is higher  in less massive galaxies, confirming that the galaxy mass has a primary role in shaping the recycling of gas and metals. The comparison between the wind power and kinetic power of the starburst associated with SNe indicates that only the starburst could drive the outflows in nearly all the [U]LIRGs galaxies, as the wind power is generally lower than  20$\%$ of the kinetic power supplied by the starburst.  The contribution of an active galactic nuclei (AGN) is, in principle,  significant in two cases.}
\maketitle
\newcommand{\Ha}{H$\alpha$}
\newcommand{\kms}{$\mbox{kms}^{\rm-1}$}
\newcommand{\co}{CO(2-0) }
\newcommand{\ul}{[U]LIRGs }
\newcommand{\li}{LIRGs }
\newcommand{\disp}{$\sigma$ }           
\newcommand{\ampl}{$\Delta$V }
\newcommand{\ebv}{E$_{\rmB-V}$ }
%
\section{Introduction}

Complex feedback phenomena, such as galactic winds (GWs), play a major role in the evolution of  galaxies and  the surrounding intergalactic medium (IGM) in which they are embedded. In particular, GWs self-regulate  the growth of both the stellar and black hole masses in galaxies (e.g., \citealt{Veilleux2005} for a review) and they are also invoked for  the color  transformation of spiral galaxies into passive ellipticals caused by gas exhaustion (e.g., \citealt{Weiner2009}). The GW feedback has been discussed by a number of authors (e.g.,   \citealt{Veilleux2005, Rupke2005c}) and is now part of many galaxy formation models (e.g., \citealt{Dutton2009, Fabian2012, Hopkins2015}). \\
Starburst-driven GWs are generated by the radiation and mechanical energy liberated by massive stars and the sequential explosions of SNe \citep{Chevalier1985, Hopkins2012}; and they are common in a range of galaxy types: dwarfs (e.g., \citealt{Schwartz2004}),  starburst (e.g., \citealt{Chen2010}) and luminous infrared
galaxies  (LIRGs; L$_{\rm IR}$\,=\,L$_{\rm (8-1000 \mu m)}$\,=\,10$^{\rm11}$-$10^{\rm12}$\,L$_{\rm \sun}$) and ultraluminous  infrared galaxies (ULIRGs, L$_{\rm IR}$\,$\geq$\,10$^{\rm12}$;  e.g.,  \citealt{Martin2005,  Rupke2005b, Rupke2005c, Arribas2014}). Active galactic nuclei (AGN) are invoked as the power source for the most extreme (i.e., fastest and powerful) outflows  observed in quasars (e.g., \citealt{Cicone2014, Sturm2011, Villar2014}). \\
Over the past 20 years, the observational multiwavelength data on outflows is steadily increasing both in quality and quantity. Outflows are a multiphase phenomena  and, therefore, their different phases (hot, warm, neutral, molecular) can be studied via different tracers (e.g., FeXXV at 6.7 keV, \Ha, NaD, CO(1-0) at 115.271 GHz) in different wavelength ranges (X-rays, optical, and radio; \citealt{Veilleux2005}).\\
GWs are observed at any redshift. During the peak of star formation (z\,$\sim\,$2-3) warm-ionized outflows are very prominent and ubiquitous (\citealt{Maiolino2012, Cano2012, Newman2012,Forster2014}).\\ 
In the local Universe, ionized  (e.g., \citealt{Arribas2014}), neutral (e.g., \citealt{Rupke2005a},b,c),  and molecular (e.g., \citealt{Cicone2014})  GWs have been found in galaxies undergoing intense and spatially concentrated star formation (i.e., starburst galaxies) in which the star formation rate (SFR) per unit area exceeds $\Sigma_{\rm SFR}$\,$\geq$\,10$^{\rm-1}$ M$_{\rm \sun}$ yr$^{\rm-1}$ kpc$^{\rm-2}$ (\citealt{Heckman2002}).\\
Of particular interest is to investigate the amount of  gas entrained in winds since its evacuation may have a significant  impact on the  gas reservoir available for star formation.  A method to detect the  neutral gas in GWs consists in tracing NaD absorption against regions with a bright stellar continuum.\\
In this context, [U]LIRGs, which are mostly powered by star formation,  are ideal  objects to investigate the neutral phase of SNe-driven outflows although they have a more important AGN contribution with increasing luminosity
(\citealt{Veilleux2009}). In particular, \ul  offer the opportunity to study the feedback phenomenon in environments similar to those observed at high-z, but with a much higher signal-to-noise (S/N) and spatial resolution. Indeed,  local \ul share  some basic structural (\citealt{Arribas2012}) and kinematical (\citealt{Bellocchi2013}) properties with distant star-forming galaxies, although they may differ in terms of gas content (which increases with increasing redshift; \citealt{Tacconi2010}). \\ 
Most of the previous observational works on outflows in local starbursts and [U]LIRGs have been based on long-slit data, which often provide a limited knowledge of their geometry and kinematics.  Integral field spectroscopy (IFS) provides a better characterization of the  morphology and the velocity structure of GWs (e.g., \citealt{Colina1999, Jimenez2007, Westmoquette2011}). \\
In this paper, we use IFS-data taken with the VIMOS at the Very Large Telescope (VLT) to significantly expand in number previous samples of spatially resolved neutral outflows (e.g., \citealt{Rupke2013, Rupke2015}), especially in the less studied LIRG luminosity range. In particular, we analyze the NaD absorption doublet in 30 LIRGs and 8 ULIRGs (51 individual galaxies).  Our main goal is to characterize the kinematics and structure of the neutral outflows  and investigate how they are related to the host properties, for instance,  the infrared luminosity (i.e., star formation), and dynamical mass.  \\
The paper is organized as follows. In Sect.\,2 we briefly describe the sample, observations, data reduction, and line fitting. In Sect.\,3 we present the data analysis, including details about the continuum modeling, and the generation of integrated spectra and spectral maps. In Sect.\,4, the properties of the neutral gas kinematics are commented and discussed with special focus on the characteristics of GWs and their relation with other galaxy properties. Finally, the main results and conclusions are summarized in Sect. 5. In the Appendix, the kinematic maps, comments on the individual objects are presented. Throughout the paper, we assume H$_{\rm 0}$\,=\,70 km\,s$^{\rm-1}$\,Mpc$^{\rm-1}$ and standard $\Omega_{\rm m}$\,=\,0.3,  $\Omega_{\rm \Lambda}$\,=\,0.7 cosmology. 


\section{Sample, observations, data reduction, and line fitting}
\label{SODR}
                \begin{table*}
                        \caption[Sample]{General properties of the \ul sample.}                         
				\begin{center}
                                        \scriptsize{\begin{tabular}{l c c c c c c c c}
                                        \hline \hline
                                        
          ID1                &  ID2           &  z      & log\,(L$_{\rm IR}$/L$_{\rm \sun}$)      & Class     & S.C.          &  C$^{AGN}$    & \multicolumn{2}{c}{Comments}  \\
         IRAS              & Other        &        &       &   &  &                              &       1D &  2D \\
         (1)                  &  (2)            &     (3)   &        (4)  &    (5)  &  (6)      & (7)                            &       (8) & (9)                \\
         \hline                 
        F01159-4443\,(N)   & ESO 244-G012   & 0.0229 & 11.48   & 1      & H                       & AGN $^{a}$                            & S+B           &   GW    \\ 
        F01159-4443\,(S)    &                      & 0.0229 &   $\cdots$        & 1      & H/Sy                  & AGN $^{a}$                            &$\cdots$         &   $\cdots$  \\ 
        F01341-3735\,(N)    & ESO 297-G011   & 0.0173  &  10.99 (11.18) & 1      & H                      & SB                                            & B+B                     & GW \\ 
        F01341-3735\,(S)    & ESO 297-G012   & 0.0173  &  10.72 & 1      & H             & SB                                            &  S                    &  $\cdots$ \\
        F04315-0840           & NGC 1614           & 0.0156  & 11.69  & 2      & H                        & SB                                    & S+B             &  GW \\
        F05189-2524          &                             & 0.0426 & 12.19   & 2      & Sy          & AGN $^{a,\,b,\,e,\,g,\,i}$  &  B+B & GW        \\
        F06035-7102           &                            & 0.0795 & 12.26   & 1      & H                  & AGN $^{a}$                            &  $\cdots$       &   $\cdots$ \\ 
        F06076-2139\,(N)   &                                      & 0.0374 & 11.67   & 1      & $\cdots$           & $\cdots$                              &  B                      &  $\cdots$              \\  
        F06076-2139\,(S)   &                              & 0.0374 &       $\cdots$       & 1      & $\cdots$            & $\cdots$                                      &  B                      & $\cdots$  \\  
        F06206-6315          &                            & 0.0924  & 12.27  & 1      & Sy           & AGN $^{a}$                           &  $\cdots$         &       GW      \\
        F06259-4780\,(N)   & ESO 255-IG 007 & 0.0388 & 11.91   & 1      & H                       & SB                                    & B+B           & GW  \\
        F06259-4780\,(C)   & ESO 255-IG 007 & 0.0388 & $\cdots$          & 1      & $\cdots$             & $\cdots$                              &  S+ R                   &  RD  \\ 
        F06259-4780\,(S)   & ESO 255-IG 007 & 0.0388 & $\cdots$          & 1      & $\cdots$             &$\cdots$                                       &       $\cdots$         &    $\cdots$   \\
        F06295-1735         & ESO 557-G002    & 0.0213 & 11.27  & 0     & H                       &  SB                           &  $\cdots$     & $\cdots$ \\ 
        F06592-6313         &                              & 0.0230 & 11.22  & 0     & H                    & SB                            &   B                   &  GW    \\ 
        F07027-6011\,(N)   & AM 0702-601      & 0.0313 & 11.04 (11.64)& 0      & Sy                       & AGN $^{a}$                            &  B+B            &   GW    \\ 
        F07027-6011\,(S)   & AM 0702-601      & 0.0313 & 11.51 & 0      & $\cdots$        &  $\cdots$                                     &  S+B          &  RD +  GW         \\
        F07160-6215      & NGC 2369          & 0.0108 & 11.16   & 0     & $\cdots$         &  AGN\,(8$\%$) $^{i}$                 &  S                    & GW     \\  
        08355-4944           &                            & 0.0259 & 11.60  & 2      & $\cdots$            &  $\cdots$                                     & $\cdots$        &  $\cdots$      \\
        08424-3130\,(N)    & ESO 432- IG006  & 0.0162 &    $\cdots$ & 1      & $\cdots$          & SB                                    &  $\cdots$     &   $\cdots$   \\ 
        08424-3130\,(S)    & ESO 432- IG006  & 0.0162  &   11.04    & 1      & $\cdots$          & SB                                    &  $\cdots$     & $\cdots$    \\      
        F08520-6850         & ESO 60-IG016     & 0.0463 & 11.83 & 1      & $\cdots$              & AGN\,(9$\%$) $^{e}$           &  $\cdots$     &    $\cdots$   \\ 
        09022-3615      &                               & 0.0596  & 12.32& 2      & $\cdots$       & AGN\,[9$\%$] $^{h}$           &  $\cdots$     &         $\cdots$    \\
        F09437+0317\,(N) & IC 563               & 0.0205  & 10.99 (11.21) & 1(0)   & $\cdots$      &   SB                         & S                     &   RD    \\
        F09437+0317\,(S) & IC 564               & 0.0205  & 10.82 & 1(0)   & $\cdots$      &   SB                                & $\cdots$      &  RD  \\
        F10015-0614          & NGC 3110 & 0.0169  & 11.31        & 0      &  H                   &   AGN $^{i}$                          &  B                    &         GW  \\
        F10038-3338          & IC 2545          & 0.0341 & 11.77        &   2      &  H/L                 & SB                            &  B                    &   GW     \\
        F10257-4339          & NGC3256          & 0.0094  & 11.69        & 2      &  H                   &  SB                                   & B                       & GW \\
        F10409-4556          &  ESO 264-G036   & 0.0210 &  11.26       & 0      &  H/L                   & SB                            & B                     &  RD +  GW       \\
        F10567-4310          & ESO 264-G057    & 0.0172  & 11.07        & 0      &  H                     & SB                            &  B                    &   GW \\
        F11255-4120          &  ESO 319-G022   & 0.0164 & 11.04        & 0      &  H                     & SB                            & B+B           &   GW    \\
        F11506-3851          &  ESO 320-G030   &0.0108  & 11.30        & 0      &  H                     & AGN\,($<$4$\%$)$^{i}$         & R+B           &   GW    \\
        F12043-3140\,(N)  &  ESO 440-IG 058 & 0.0232 &    11.37       & 1      & H/L                      & SB                            &  $\cdots$     &  $\cdots$   \\     
        F12043-3140\,(S)   &                            & 0.0232 &   $\cdots$      & 1      & H               & SB                                    & B                       & RD    \\      
        F12115-4656          &ESO 267-G030    & 0.0185 &   11.11            & 0      &  H                        & AGN $^{f}$                            & B                       & RD   \\       
        12116-5615            &                                  & 0.0271  & 11.61       & 2(0)  &  $\cdots$              & AGN $^{l}$                            &  B+B            &   GW  \\
        F12596-1529          & MCG-02-33-098 & 0.0159 & 11.07        & 1      &  H                       & AGN $^{a}$                            & $\cdots$        &   $\cdots$   \\
        F13001-2339          & ESO 507-G070   & 0.0217 & 11.48        & 2(0/1) & L                     & SB                                            & $\cdots$        &  GW        \\
        F13229-2934          & NGC 5135    & 0.0137 & 11.29        & 0      & Sy                 & AGN\,(1.4$\%$) $^{a,\,i}$     &  B                    &  GW   \\    
        F14544-4255\,(E)  & IC 4518             & 0.0157 & 10.80 (11.11) & 1      &                      & $\cdots$                              & $\cdots$        & RD \\
        F14544-4255\,(W) & IC 4518              & 0.0157& 10.80 & 1      & Sy                    & AGN\,(6$\%$) $^{i}$           & $\cdots$      &   $\cdots$ \\
        F17138-1017          &                           & 0.0173 & 11.41        & 2(0)   & H                     & AGN $^{d}$                            &  $\cdots$       & $\cdots$    \\
        F18093-5744\,(N)  & IC 4687               & 0.0173 &  11.47 (11.57)& 1      & H                      & AGN\,(5$\%$) $^{i}$           &   B                   &  $\cdots$   \\
        F18093-5744\,(C)  & IC 4686               & 0.0173 &  10.87 & 1      & H                  & $\cdots$                             & $\cdots$      &   $\cdots$     \\
        F18093-5744\,(S)  & IC 4689               & 0.0173 &    $\cdots$      & 1      & H                               &   $\cdots$                            &  S                      & RD  \\
        F21130-4446         &                           & 0.0926 &12.09       & 2      & H                               &  SB $^{d}$                            & $\cdots$        & $\cdots$ \\
        F21453-3511         & NGC 7130           & 0.0162  & 11.41        & 2      & L/Sy                        & AGN\,(1.5$\%$) $^{a,\,i}$     &  $\cdots$       &  GW     \\
        F22132-3705         & IC 5179            & 0.0114  & 11.22        & 0      & H                           & AGN\,($<$3$\%$) $^{i}$        &  S                      &  RD     \\
        F22491-1808         &                           & 0.0778  &12.17              & 1      & H                               &  AGN\,($<$4$\%$) $^{e,i}$     &  $\cdots$       &  $\cdots$         \\
        F23128-5919\,(N) & AM 2312-591     & 0.0446  & $\cdots$               & 1      & H/L/Sy                  &   $\cdots$                            &  $\cdots$      & $\cdots$      \\
        F23128-5919\,(S)  & AM 2312-591     & 0.0446  & 12.06         & 1      & H/L/Sy                   & AGN $^{a,\,c}$                        &  $\cdots$       &   GW   \\
        \hline
                \hline
                                        \end{tabular}
                                        \label{Tab_sample}}
                                        \end{center} 
        \tablefoot{\scriptsize{Column~(1): Object designation  in the Infrared Astronomical Satellite (IRAS) Point Source Catalogue (PSC)  and in the IRAS Faint Source Catalogue (FSC; with prefix \lq F\rq). Column~(2): Other name. Column~(3): Redshift from the NASA Extragalactic Database (NED). Column~(4): Infrared luminosity (L$_{\rm IR}$\,=\,L$_{\rm (8-1000\,\mu m)}$) in units of solar bolometric luminosity  calculated with the flux in the four IRAS bands as given in \citet{Sanders2003} when available. Otherwise, the standard prescription in \citet{Sanders1996} with the values in the IRAS-FSC and IRAS-PSC catalogues was used. For some targets (22\,$\%$, i.e.,11/51), new updated L$_{\rm IR}$ values can be found in the \cite{Surace2004}.  However, for sake of homogeneity,  preferred to use the values described above for the whole sample. For those systems for which it was possible  to measure the  L$_{\rm IR}$ of the individual galaxies, that of the system is indicated in parenthesis. Column~(5): Morphology class defined as follows: 0 identifies isolated rotating disks, 1 interacting systems, and 2 mergers. We refer to \citet{Rodriguez2011} and \citet{Bellocchi2013} for further details about this classification. Column~(6): S.C. stands for (nuclear) spectroscopic classification of the ionization type.  H defines HII galaxy,  L,  LINER, and  Sy  stands for Seyfert (see \cite{Rodriguez2011}). Column~(7): AGN and  SB indicate, respectively,  evidence of an AGN (from the infrared, optical, and X-ray observations) and starburst galaxies. When possible we also give an estimate of the AGN contribution to the 24\,$\mu$m and bolometric luminosity (round and square parenthesis, respectively). Columns (8) and (9): Our classification scheme, according to the integrated spectra (i.e., 1D, column~(8)) and the maps (i.e., 2D, column~(9)).  If multiple classifications are quoted for the same object, each classification refers to different kinematic components. The letter S refers to a systemic component, while  R  and  B  indicate, respectively, the presence of redshifted (i.e., V\,$>$\,60\,kms$^{\rm-1}$) and blueshifted (i.e., V\,$<$\,-60\,kms$^{\rm-1}$) components (with respect to the stars).  GW stands for galaxy in which footprint of a galactic winds was found via NaD (see Sect.\,\ref{detectionrate}) and   RD indicates the presence of  a disk of neutral gas (see Sect.\,\ref{disks}). \\
\textit{References.} $^{a}$\,\cite{Arribas2014}; $^{b}$\,\cite{Dadina2007}; $^{c}$\,\cite{Dixon2011}; $^{d}$\,\cite{Farrah2003}; $^{e}$\,\cite{Iwasawa2011};  $^{f}$\,\cite{JimenezBailon2007}; $^{g}$\,\cite{Nardini2009}; $^{h}$\,\cite{Nardini2010}; $^{i}$\,\cite{Pereira2010}; $^{l}$\,\cite{Valiante2009}.}}
                        \end{table*}
                        
\subsection{Sample and observations}
 \label{SO}
The sample contains 51 individual galaxies mainly drawn from the IRAS Revised Bright Galaxy Sample (RBGS; \citealt{Sanders2003}), which have been observed via IFS by \cite{Arribas2008} (see Table~\ref{Tab_sample}). This sample contains a large number (i.e., 43/51) of sources in the less-studied LIRG luminosity range (L$_{\rm IR}$\,=\,2.9 $\times$ 10$^{\rm11}$ L$_{\rm \sun}$, on average), while a smaller number of objects (i.e., 8/51) are classified as ULIRGs with  L$_{\rm IR}$ = 1.6 $\times$ 10$^{\rm12}$ L$_{\rm \sun}$ on average.  The mean redshift of the LIRGs and ULIRGs subsamples are 0.024 and 0.069, respectively (see Table~\ref{Tab_sample}). \\
The sample is not complete either in luminosity or in distance (\citealt{Rodriguez2011, Bellocchi2013}), however, it is representative of the [U]LIRGs population. Indeed, this sample covers a variety of morphological types along the merging process (i.e., rotating disks, interacting systems, and mergers) and  encompasses objects with different  nuclear ionization types (i.e., HII, Seyfert and LINER). See Table~\ref{Tab_sample}, for details. \\
Nearly half of the objects in the sample (i.e., 22/51) show an excess of 24 $\mu$m, hard X-ray emissions, or they have optical line ratios indicating the presence of AGNs (Table~\ref{Tab_sample}). However, the AGN contribution to the infrared luminosity or to the ionized gas kinematics is substantial in the cases of the ULIRGs IRAS F05189-2524, F23128-5919\,(S) and for the LIRG F07027-6011\,(N)  (Table~\ref{Tab_sample}).\\
Throughout the paper, we made use of the SFR derived by \citet{Rodriguez2011} inferred from L$_{\rm IR}$, following Kennicutt (1998; with a Salpeter initial mass function); we also made use of the dynamical masses (M$_{\rm dyn}$) and angle of inclination of the galaxies derived by \citet{Bellocchi2013}.\\
The observations were carried out with the integral field unit (IFU) of VIMOS at the 8.2-meter telescope  of the  ESO-VLT (\citealt{LeFevre2003}).  In brief, the IFS data of all the targets were acquired using the high resolution orange grating, which provide an intermediate spectral resolution  of 3470 (dispersion of 0.62 $\AA$\,pix$^{\rm-1}$) in a wavelength region from 5250  to 7400\,$\AA$. The field of view (FoV) in this configuration is 27$\arcmin$$\arcmin$\,$\times$\,27$\arcmin$$\arcmin$ with a spaxel scale of 0.67$\arcmin$$\arcmin$ per fiber. The 1600 spectra obtained are organized in a 40\,$\times$\,40 fiber array, which constitute one single pointing. A square of 4 pointing dithered pattern for each target was used, providing an effective FoV of  29.5$\arcmin$$\arcmin$\,$\times$\,29.5$\arcmin$$\arcmin$. For more details about the observations, see \citet{Arribas2008}.\\
We excluded from analysis the galaxies 08355-4944 and F21130-4443 from the  \ul sample described above    since they  show very weak NaD absorption preventing any robust study. In addition,  nine objects (i.e., F06035-7102; F06076-2139\,(N, S); F06259-4780\,(S); F06295-1735; F08520-6850; F12596-1529; F17138-1017; F22491-1808) have a  large percentage of low S/N spectra in individual spaxels and were not suitable for the spatially resolved analysis. In summary, of the total sample of 51 individual objects observed, the integrated  and spatially resolved properties have been studied for 49  and 40  galaxies, respectively.

\subsection{Data reduction \label{DR}}
        
We reduced VIMOS raw  data with the pipeline provided by ESO via  ESOREX (version 3.6.1 and 3.6.5), which allows us to perform the following steps: sky and bias subtraction, flat-field correction, spectra tracing and extraction, correction of fiber and pixel transmission,  flux and wavelength calibration flux. We also used a set of customized IDL and IRAF scripts. Once the  four  quadrants were  individually reduced for the four dithered positions, they were combined into a single data-cube  made of 44\,$\times$\,44 spaxels (i.e., 1936 spectra). We checked the width of the instrumental profile and wavelength calibration  using the [O\,I]\,$\lambda$6300.3\,$\AA$ sky line. The average values for the full width half maximum (FWHM) and central wavelength for the whole sample were (6300.29 $\pm$ 0.07)\,$\AA$ and (1.80 $\pm$ 0.07)\,$\AA$, respectively. For each spectrum, we corrected for the effect of instrumental dispersion  by subtracting it in quadrature from the observed line dispersion, i.e., $\sigma_{\rm line}$=\,$\sqrt{\sigma_{\rm obs}^{2}\,+\,\sigma_{\rm INS}^{2}}$. A more detailed description of the data reduction process is given in \citet{MonrealIbero2010}. 

\subsection{Line fitting \label{LF}}

                \begin{figure}
                           \centering
                           \includegraphics[width=.45\textwidth]{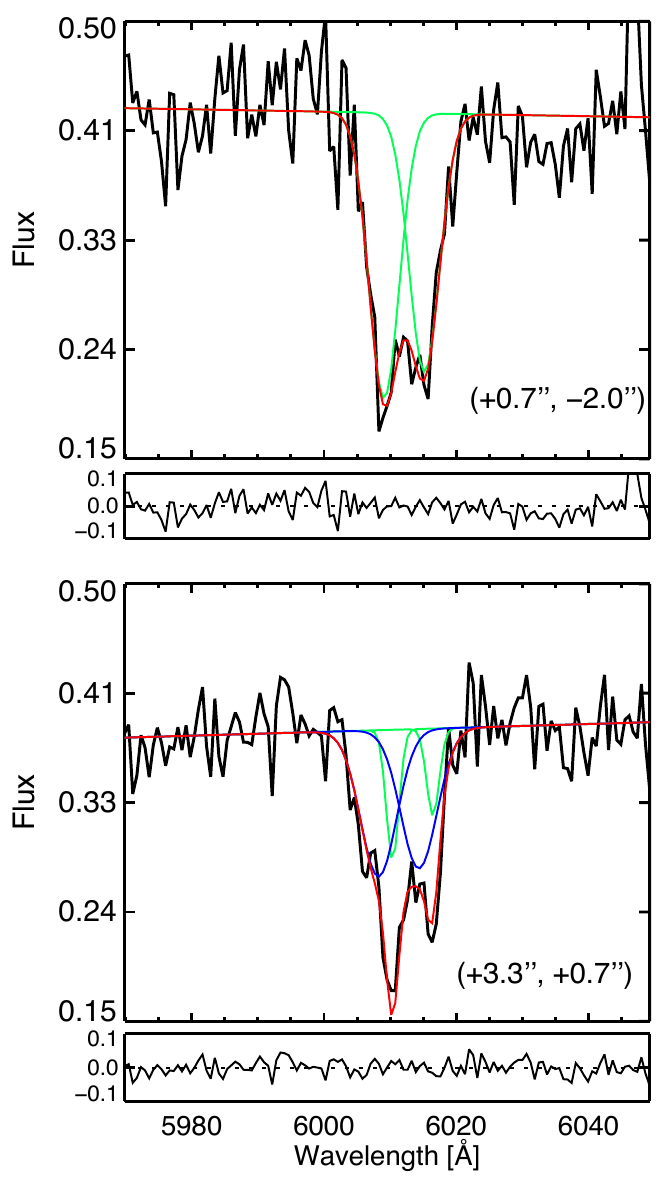}  
\caption{Examples of one (top) and two (bottom) component(s) Gaussian fits to the NaD line profiles in two different spaxels in IRAS F10409-4556. Below, the residuals (i.e., data\,-\,model) are presented. When fitting two components the blueshifted component is shown in blue (bottom panel only). The red curve shows the total contribution coming from the NaD Gaussian fit. In bracket, the spaxels coordinates indicated as distance from the nucleus (i.e., map center, Fig.\,\ref{Panel_F10409}).} 
                           \label{Spaxels}               
                \end{figure}
                
The observed NaD-absorption features are modeled both in the IFS and integrated spectra with two Gaussian profiles (i.e., single kinematic component) as in \citet{Cazzoli2014} (also \citealt{Davis2012}). This approach has the limitation that it is only correct  in the cases of either uniform covering factor and low optical depth or in the case of covering factor that varies with velocity as a Gaussian (independently to the optical depth). Absorption line fitting methods that consider both optical depth and covering factor are presented in  \citet{Rupke2005a}, and references therein. Although  the observed NaD-absorption features show generally well-resolved  lines  and not saturated profiles (i.e., flat-bottomed), we stress that the validity of our fitting approach is limited to the cases mentioned above. As in many cases the line ratio within the two doublet lines indicates an optically thick regime, it is implicity assumed that the covering factor varies with velocity as a Gaussian in these cases.\\
We found some spectra that have strongly asymmetric profiles and their modeling requires two Gaussian pairs (i.e., a single component fitting is certainly an oversimplification). After visual inspecting all the spectra, we decided to apply a two-components line fitting for the spectra of four objects (F07027-6011\,(S), F10409-4556, F11506-3851, and F12116-5615) similarly as in Cazzoli et al.\,(2014). In a few other cases, there is evidence for the presence of multiple kinematic components in individual spectra (e.g., F04315-0840, F10257-4310, F13229-2934). However, the limited S/N does not allow us a robust kinematic decomposition into two different components. Therefore, a single kinematic component  fitting  of the NaD absorption doublet was preferred in these cases. Figure \ref{Spaxels} shows examples of the two different Gaussian fits for the galaxy IRAS F10409-4556.  The existence of two kinematical components is expected as the NaD absorption feature may take part in the galaxy ordered rotation and be entrained in winds with blueshifted velocities. Therefore, we identify these components according mainly to their spatial distribution and velocities.  \\
We also fit the NaD line profile in  the integrated spectra (after the subtraction of the stellar contribution, Sect.~\ref{IntegratedSpectra}) with one or two components following the approach described above.      
The majority (21/28) of these integrated spectra  show a prominent nebular emission line HeI\,$\lambda$5876\,$\AA$, which was included in the line fitting (i.e., modeled selectively with one or two Gaussian components and then subtracted). This strategy has also been applied in a few cases for the spatially resolved data.    


\section{Data analysis \label{DataAnalysis}}

\subsection{Integrated Spectra: Generation and modeling of the continuum and line profile \label{IntegratedSpectra}}

\begin{figure*}
 \vspace{5.0em} 
  \centering                             
\includegraphics[width=.995\textwidth]{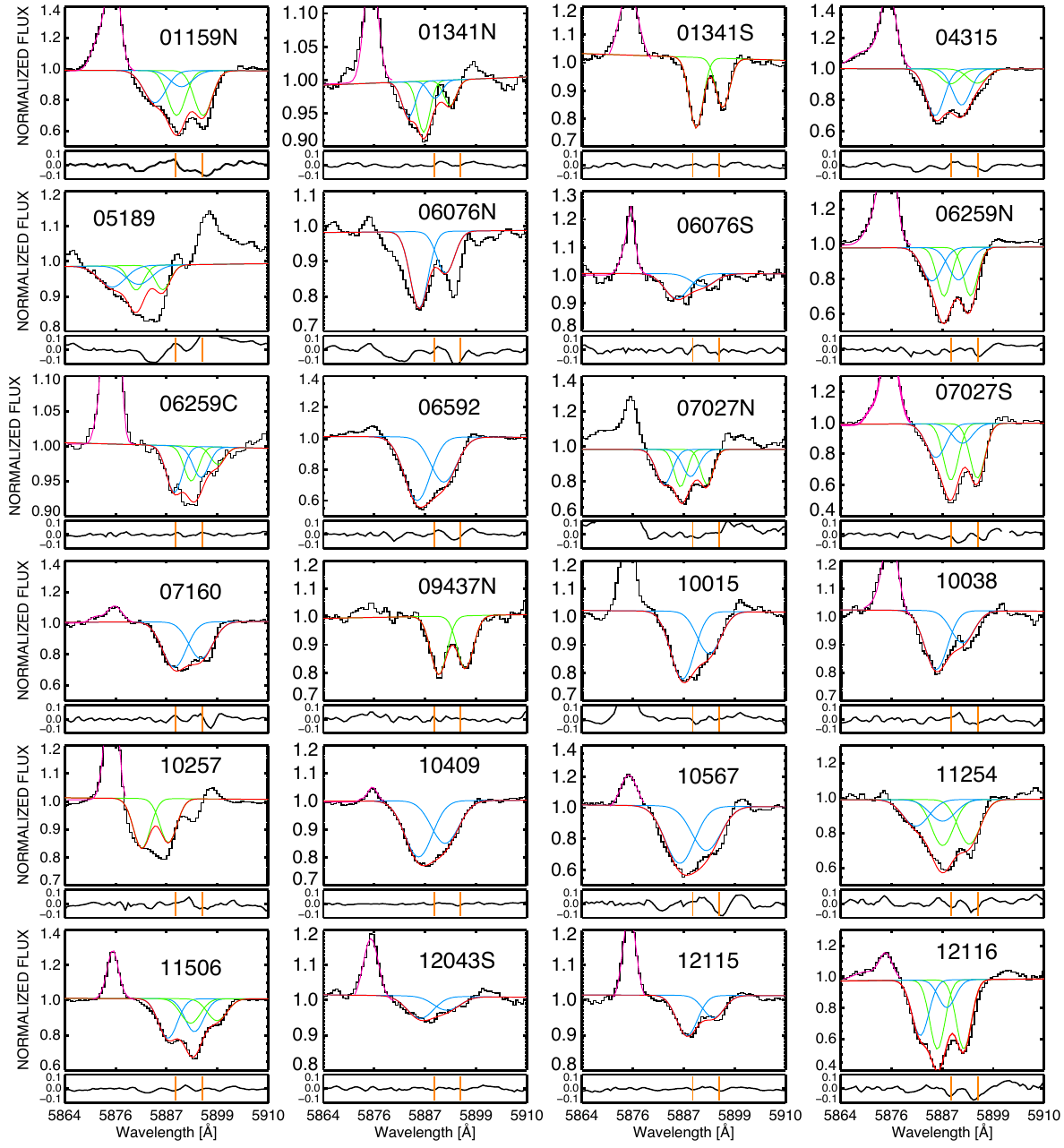}
 \caption{Normalized integrated spectra of the NaD absorption line profile after the subtraction of the stellar contribution in galaxies with medium and high quality stellar modeling of  the integrated spectra (Sect.\,\ref{IntegratedSpectra}). Each  spectrum shown covers a rest-frame range of $\sim$~46~\AA \ (i.e, $\sim$2300 kms$^{\rm-1}$). For each source, in the upper panels, the different kinematic components, are shown in blue and green, i.e., blueshifted and systemic or redshifted component, according to our classification (Table~\ref{Tab_sample}). The red curve shows the total contribution coming from the NaD Gaussian fit.  In pink, we indicate the model of the HeI emission line. This line was successfully modeled with one (e.g., F10257-4339) or two components (e.g., F04315-0840) in  21 cases. In two cases, F07027-6011\,(N) and F10015-0614,  the line modeling of the (prominent) HeI emission cannot  reproduce the observed line profile well. Therefore, in these cases, the HeI is simply excluded when fitting the NaD absorption. In the lower panels, the residuals (i.e., data - model) are presented with the rest-frame NaD wavelengths indicated in orange. The galaxy IDs follow that of Table~\ref{Tab_sample}, although here they are shortened for better visualization.}
  \label{LF_int}        
 \end{figure*}

From the lFS data cubes, we generated a spatially integrated spectrum per galaxy via the S/N optimization method proposed by  \citet{Rosales2012}, using the \textit{pingsoft} tool (\citealt{Rosales2011}). Before integration, the spectra were corrected from the general stellar velocity pattern. For this step, we used the narrow component of the ionized gas velocity field, which describes  the systemic behavior (\citealt{Bellocchi2013}). This method significantly improves the modeling of weak features of the stellar continuum compared to other techniques (e.g., S/N cutoff; see \citealt{Rosales2012} for a detailed discussion). \\
We applied a penalized PiXel fitting analysis (\textit{pPXF}; \citealt{Cappellari2004})  for the recovery of the shape of the stellar continuum. As in Cazzoli et al. (2014), we used the Indo-U.S. stellar library (\citealt{Valdes2004}) to produce a model of the stellar spectra that matches the observed  line-free continuum (any interstellar medium (ISM) features, including  NaD, are ignored in the fitting). The result of this approach is a model that in general reproduces  the  continuum shape  well (the residuals are typically $<$10$\%$)  except few cases. The integrated and model spectra are shown for each galaxy in Appendix\,\ref{comments_maps}. \\
We checked all the stellar continuum modeling by visual inspection of the residuals especially in the line-free continuum regions and nearby the NaD feature (e.g.,  we checked if any potential NaD-HeI blending may affect the continuum model). We then flagged each object with the parameter Q, indicating the quality of stellar modeling (Table~\ref{Tab_stcont}). From the total sample of 49 objects, 31 were flagged with good (Q\,=\,2) or very good (Q\,=\,3) modeling. For these objects, we generated a  pure ISM-spectra by subtracting from the observed NaD profile the model of the stellar spectra obtained via the pPXF method. This model was also used to fix the stellar (systemic) zero velocity of the spectra. 
Then, we applied the Gaussian fitting described in Sect.\,\ref{LF} to the spectra  to derive the neutral gas kinematics. Fig.\,\ref{LF_int} shows the purely ISM NaD absorption and its modeling, while the kinematics properties are summarized in Table~\ref{Tab_stcont} for each galaxy. For only 3 out of 31 (i.e., F01159-4443\,(N), F21453-3511, F23128-5919\,(N)) was not possible to model the purely ISM NaD line profile since the stellar subtraction resulted in an almost undetectable neutral gas NaD doublet (the stellar contamination is $>$\,95$\%$, Table~\ref{Tab_stcont}).  \\
In  $\sim$61$\%$ (i.e., 17/28) of the cases a single kinematic component  (a Gaussian pair) already gives  a good fit, suggesting that if a second component exists in these galaxies, it is weak. Only 11 out of 28 require two kinematic components; we find that a two-Gaussian component model per line led to a remarkably good fit of the NaD absorption, significantly reducing the residuals with respect to one-Gaussian fits (Fig.\,\ref{LF_int}). \\
We tested the feasibility of the stellar continuum modeling on a spaxel-by-spaxel basis. Unfortunately, the spectra in individual spaxels, in general, lack the S/N required making  the determination of the stellar and ISM contributions in 2D highly inaccurate. Therefore,  we used a different (and more simplistic) approach to evaluate the  stellar contamination in the NaD maps for the 2D analysis, as discussed in Sect.~\ref{Origin}.

        \subsection{Two-dimensional maps \label{maps}}

The 2D analysis is based on the spectral maps (i.e., velocity field (V), velocity dispersion  ($\sigma$), and equivalent width (EW)) of the different kinematic components, which were  generated  after the line fitting procedure (Sect.\,\ref{LF}). To produce the maps themselves, we used a set of IDL procedures  (i.e., \textit{jmaplot}) developed by \citet{Maiz2004}. The maps are shown, for each galaxy (for a total of 40 objects),  in Appendix\,\ref{comments_maps}. As a reference we also show the continuum image and H$\alpha$ map of the systemic (narrow) component generated from the VIMOS-IFU data cube (\citealt{Bellocchi2013}). The integrated spectrum (Sect.~\ref{IntegratedSpectra}) is also shown for each object  . The maps of the LIRG F11506-3851 are already discussed in detail in \citet{Cazzoli2014} and, therefore, they are not included in Appendix\,\ref{comments_maps}, but the results are considered in the overall statistics and figures. 


\section{Results and discussion \label{Results}}

\subsection{Contamination from stellar NaD absorption \label{Origin}}

In order to measure the stellar contribution to the NaD doublet (i.e.,  stellar contamination), we computed the ratio between the  EW(NaD) in the stellar model obtained with pPXF (Sect.~\ref{IntegratedSpectra}) and the total EW(NaD) observed in the integrated spectra. This ratio is listed as percentages for each individual galaxy in Table~\ref{Tab_stcont} and a histogram of its distribution is presented in Fig.~\ref{stellarfrac1}.\\
We divided the sample into three groups:  i) objects for which the NaD absorption is dominated by the ISM  (i.e.,  stellar contribution $<$\,35$\%$), ii), medium-contaminated  objects (i.e., stellar contribution between 35$\%$ and 65$\%$), and stellar dominated objects (i.e., stellar contribution $>$\,65$\%$).  Considering only the 31 objects with good stellar continuum modeling (Fig.~\ref{stellarfrac1}),  the stellar contribution is dominant for 23$\%$ of the cases,  while the NaD absorption is dominated by the ISM for
 42$\%$ of the cases. \\
As mentioned above, the stellar continuum fitting method could not be applied  on a spaxel-by-spaxel basis. Therefore, to evaluate the origin of the NaD  absorption in the maps, we  compare the values of the sodium equivalent width, EW(NaD), with that  of  a purely stellar NaD doublet. In fact, we consider that the NaD feature is mainly originated in the ISM when EW(NaD)\,$>$\,1.3\,$\AA$. We established this (conservative) threshold  considering the EW(NaD) of the stellar models obtained with pPXF, which are shown in Fig.~\ref{stellarfrac4} as a function of the stellar NaD fraction.  In fact, independent of the fraction of NaD originated in the stars, the EW of the stellar models is  on average 0.9\,$\AA$. Therefore, our choice of EW(NaD)\,$>$1.3\,$\AA$ to identify a NaD feature dominated by the ISM is rather conservative as it is above the stellar contribution in  90$\%$ of the cases (or 100$\%$ within 1 $\sigma$). Therefore, the generally strong NaD absorption (with EW(NaD)\,$>$1.3\,$\AA$) seen in many regions of the  [U]LIRGs (Appendix\,\ref{comments_maps}) can be robustly associated with the cold neutral ISM. As for comparison, starburst galaxies with EW(NaD)\,$>$\,0.8 $\AA$ are considered strong ISM-NaD absorbers by \citet{Chen2010}.          
\begin{figure*}
\centering                         
 \includegraphics[width=.995\textwidth]{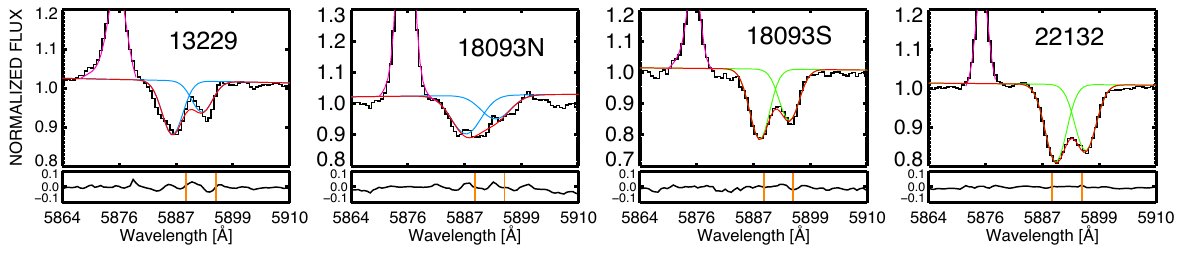} 
 \captionsetup{labelformat=empty}{Fig.\,\ref{LF_int}\,--\,Continued.}                             
\end{figure*}

\subsection{Galactic winds in one dimension (integrated spectra) \label{int_kin}}

When identifying GWs in the integrated ISM spectra (Sect.~\ref{Origin}) only components with central line velocities blueshifted with respect to the systemic by more than 60 kms$^{\rm-1}$  are considered to be outflowing (Colum 9 in Table\,\ref{Tab_stcont}). This cutoff, which is on the order of two spectral elements of $\sim$~0.6\,\AA \ each, is similar to those assumed in previous studies (e.g., 50 kms$^{\rm-1}$ in \citealt{Rupke2005b}). Table~\ref{Tab_sample} lists the classification based on this criterion for each individual object in the survey.\\
Neutral gaseous outflows are found in 19 out of 28\footnote{We excluded  three galaxies (F05189-2524, F06076-2139\,(N) and F10257-4339) from the current 1D-analysis, since modeling of the NaD absorption line profile leaves strong residual (Table\,\ref{Tab_stcont}).} [U]LIRGs. The incidence of winds  in our local [U]LIRGs, i.e., 68$\%$, is  similar to that found in other samples of infrared-luminous galaxies (e.g., \citealt{Heckman2000, Rupke2005b}). In five objects the NaD absorption is found at systemic velocity (i.e., -60 kms$^{\rm-1}$\,$<$\,V\,$<$\,+\,60 kms$^{\rm-1}$),  while in one case (F06259-4780\,(C)) is redshifted.\\ 
We found that, generally, the outflows have velocities, V,  in the range from  65 kms$^{\rm-1}$ to 260 kms$^{\rm-1}$ (on average $\sim$\,165\,\kms). 
In Fig.\,\ref{V_SFR}  (left) we show the neutral gas outflow velocities  obtained from the spatially integrated analysis as a function of the SFR.  A regression of the type V\,$\propto$\,SFR$^{\rm n}$ to this mainly LIRGs sample excluding objects with evidence of strong AGNs yields  \textit{n}\,=\,0.15\,$\pm$\,0.06. This dependency is in rather good agreement with the results of \citet{Rupke2005c} (\textit{n}\,=\,0.21\,$\pm$\,0.04) for their sample of (mostly) ULIRGs. These results are also fairly consistent with those of \citet{Martin2005} (\textit{n}\,=\,0.35), who considered a sample of ULIRGs and three dwarf galaxies of low SFRs. As shown in Figure 5, the different LIRG and ULIRG samples complement  each other well, sampling the SFR rather homogeneously over nearly 2 orders of magnitude. The general trend defined with these 1D (i.e., integrated and long-slit) data is consistent with an index \textit{n} in the range 0.1-0.2. The effects of inclination and the presence of AGNs make it difficult to constrain this value further with those samples. \\
The comparison of the kinematic properties of the neutral outflows with their ionized counterparts  (\citealt{Arribas2014}) for the same objects, is shown in Fig.\,\ref{kin_gw_1D_comparison}.  \\
On the one hand, we find that the outflow velocities (measured at the center of the line) are significantly higher for neutral than for ionized outflows in all cases. On the other hand, the neutral gas velocities seem on average slightly smaller than the commonly used ionized gas maximum outflow velocities, which are defined as V$_{\rm max}$\,=\,$|$-$\Delta$V$|$\,+\,FWHM/2 (\citealt{Westmoquette2011, Genzel2011}). \\
Our measurements seem to contradict previous single aperture studies for which ionized and neutral winds are correlated  in Seyferts but not in starburst \citep{Rupke2005b}. On a spatially resolved basis, only weak evidence of  a correlation between the velocities of the neutral and ionized gas outflow phases (within a given galaxy) are found by \cite{Rupke2013} for a sample of  six nearby merger [U]LIRGs (including three obscured QSOs). However, all these findings are not fully comparable as they differ in terms of type of object and ionized gas tracers. For the velocity dispersions, both neutral and ionized outflows show similar median values of 115 and 138 \kms, respectively, if the strongest AGNs are excluded. \\
\\
\\

\begin{figure}
\includegraphics[width=.49\textwidth]{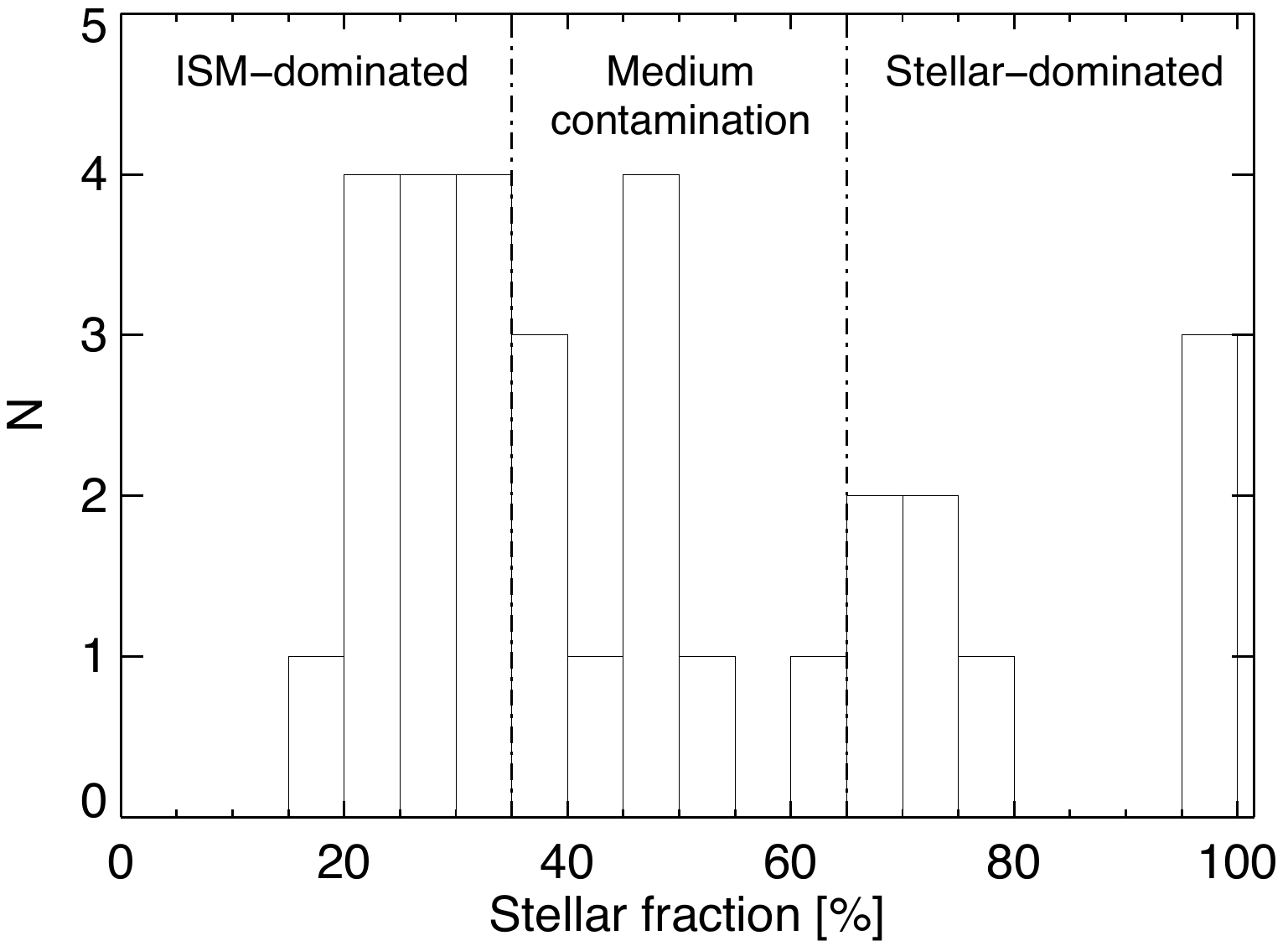}  
\caption{Distribution of the  stellar contribution to the observed integrated NaD  (Table~\ref{Tab_stcont}) for the 31 objects with a good stellar continuum modeling (Sect.\,\ref{IntegratedSpectra}). Vertical lines follow  the adopted definition for interstellar dominated, medium-contaminated, and stellar dominated objects; see Sect.~\ref{Origin} for details.}            
\label{stellarfrac1}             
\end{figure}

\begin{figure}
 \includegraphics[width=.49\textwidth]{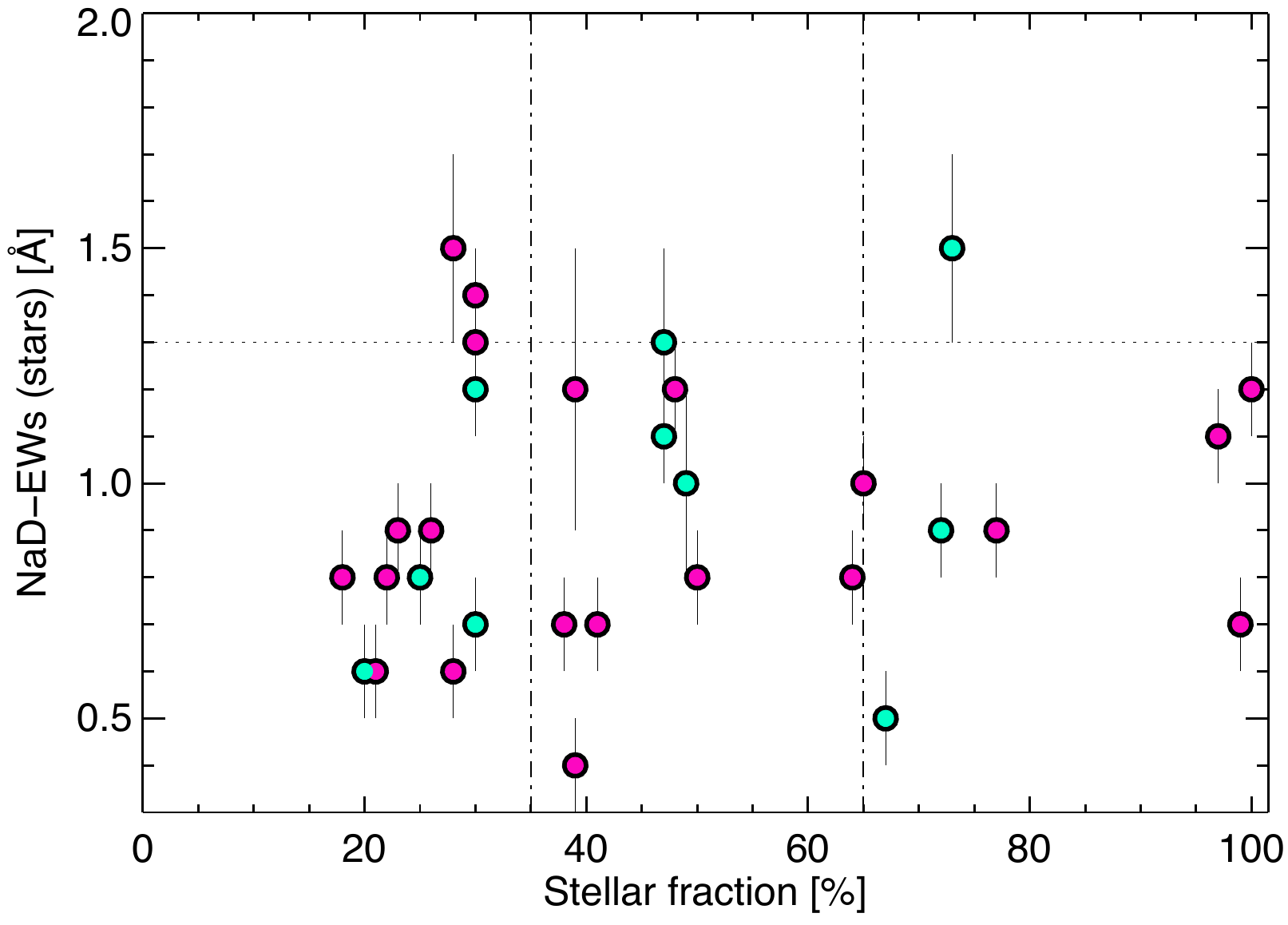}  
\caption{NaD equivalent widths, estimated in the best-fit stellar spectrum, obtained via pPXF (Sect.\,\ref{IntegratedSpectra})  versus the integrated stellar contribution (Table~\ref{Tab_stcont}). Horizontal line indicates our conservative choice of EW(NaD) (i.e., 1.3\,$\AA$) for identifying the NaD originated in the ISM in the maps (Sect.\,\ref{Origin}). Vertical lines follows the adopted definition for interstellar dominated, medium-contaminated,  and stellar-dominated  objects (Sect.~\ref{Origin}). Symbols are color-coded accordingly to the quality of the stellar modeling on the integrated spectrum (e.g., Q, Table~\ref{Tab_stcont}). Specifically, light blue and pink symbols indicate good  and very good quality stellar modeling, Q\,=\,2 and  Q\,=\,3, respectively.} 
\label{stellarfrac4}             
\end{figure}

\begin{table*}
\caption[stellacont]{Results from the integrated spectra analysis. \label{Tab_stcont}}                          
\begin{center}
\scriptsize{\begin{tabular}{l c c c c c c c c c}
                                        \hline \hline                                   
          ID1       &      EW$_{\rm stars}^{\rm ppxf-fit}$   & IS   &   Q      & V$^{\rm B}$         & $\sigma$$^{\rm B}$    & V$^{\rm other}$     & $\sigma$$^{\rm other}$  &                     V$^{\rm adopted}$  & $\sigma$$^{\rm adopted}$ \\ 
 IRAS     &      $\AA$       & $\%$                    & & kms$^{\rm-1}$ &  kms$^{\rm-1}$          &  kms$^{\rm-1}$     &    kms$^{\rm-1}$          &  kms$^{\rm-1}$             &     kms$^{\rm-1}$     \\
         (1)         &   (2)               & (3)        & (4)    &     (5)         &         (6) &       (7)           &         (8)  &         (9)  &         (10)        \\
         \hline                 
        F01159-4443\,(N)        &  1.4 $\pm$ 0.1        &   30          &2  &   -237 $\pm$ 13      &  127 $\pm$ 18         & 13   $\pm$ 3          & 102 $\pm$  16     &     -237 $\pm$ 13   &  127 $\pm$ 18 \\ 
        F01159-4443\,(S)        &  1.2 $\pm$ 0.1        & 100           & 2    &      $\cdots$     & $\cdots$             &   $\cdots$            &  $\cdots$               &$\cdots$ &$\cdots$ \\
        F01341-3735\,(N)        & 0.9 $\pm$ 0.1 &  72           & 3    & -278 $\pm$ 8      &  71 $\pm$ 3    &  -105 $\pm$  4     &   76 $\pm$ 6           & -192 $\pm$ 6  & 74 $\pm$ 5\\          
        F01341-3735\,(S)        &  1.0 $\pm$ 0.2        &  49           & 2   &    $\cdots$               &  $\cdots$             &      41 $\pm$ 4                & 80 $\pm$  8            &      41 $\pm$ 4        & 80 $\pm$  8  \\ 
         F04315-0840                    & 0.6 $\pm$ 0.1 &  20           & 3   &  -186  $\pm$  21  & 104  $\pm$ 11 &     -8  $\pm$ 5               & 109 $\pm$  17    &  -186  $\pm$  21     & 104  $\pm$ 11 \\       
        F05189-2524             &  0.7 $\pm$ 0.1        &  38           & 2   &   -725 $\pm$  26  &  150  $\pm$  13       &     -451  $\pm$ 11    & 87 $\pm$   7        & $\oslash$ & $\oslash$ \\  
        F06035-7102                     &  (0.6  $\pm$ 0.1)     &  (77) & 1   &      $\cdots$     & $\cdots$      &   $\cdots$            &  $\cdots$         &$\cdots$ &$\cdots$\\        
        F06076-2139\,(N)                &  0.7   $\pm$ 0.1      &   30          & 3   &    -172 $\pm$ 29  &  102 $\pm$ 9  &      $\cdots$         & $\cdots$         &  $\oslash$ & $\oslash$ \\  
        F06076-2139\,(S)                &   0.8   $\pm$ 0.1     &  64           & 2  &  -176  $\pm$ 16    &  139 $\pm$ 14 &      $\cdots$         & $\cdots$         &    -176  $\pm$ 16      &  139 $\pm$ 14    \\ 
        F06206-6315                     &  (1.9 $\pm$ 2.3)      & (100)         & 1  &      $\cdots$      & $\cdots$      &   $\cdots$            &  $\cdots$         & $\cdots$&$\cdots$\\                     
        F06259-4780\,(N)                &  0.9  $\pm$ 0.1       &  23           & 2  & -234        $\pm$ 9        &  117  $\pm$ 2 &       -97  $\pm$ 3            &91  $\pm$ 8        &  -166 $\pm$ 6 & 104 $\pm$ 6\\         
        F06259-4780\,(C)                &  0.4  $\pm$ 0.1       &   39  & 2  &  -3 $\pm$ 4                &   92 $\pm$ 14 &    181 $\pm$15                &  81 $\pm$ 24    & 93  $\pm$ 38 & 87 $\pm$10 \\                  
        F06259-4780\,(S)                & (1.0  $\pm$  0.3)  &   (72)   &  1  &      $\cdots$     & $\cdots$      &   $\cdots$            &  $\cdots$          &$\cdots$ & $\cdots$\\                          
        F06295-1735                     &  (1.4  $\pm$ 0.7)     &  (100)         & 1   &      $\cdots$   & $\cdots$      &   $\cdots$            &  $\cdots$       &$\cdots$ &$\cdots$ \\ 
        F06592-6313                &   0.8 $\pm$ 0.1    &   25  & 3   &   -193 $\pm$ 11  &  148 $\pm$ 13 &       $\cdots$                & $\cdots$         &   -193 $\pm$ 11 & 148 $\pm$ 13 \\          
        F07027-6011\,(N)                &   0.6 $\pm$ 0.1       &   21  & 2   &  -328 $\pm$ 18       &  92 $\pm$ 10 &   -140 $\pm$  4             &   71 $\pm$ 4     &  -234 $\pm$ 12 & 87 $\pm$ 7\\           
        F07027-6011\,(S)                &   0.9 $\pm$ 0.1       &   26          &  2  &  -202 $\pm$  18   &  87 $\pm$  4    &  -31  $\pm$ 4  &   112    $\pm$ 10               & -202 $\pm$  18   &  87 $\pm$  4 \\           
        F07160-6215               & 1.2 $\pm$ 0.3       &  39           &   2 &   $\cdots$                &  $\cdots$             &    -13   $\pm$          2      &  132 $\pm$ 12         & -13   $\pm$ 2 &  132 $\pm$ 12   \\ 
        08424-3130\,(N)                 &  (1.0 $\pm$  0.1)     &  (42)          &   1   &   $\cdots$            &  $\cdots$             &    $\cdots$           &$\cdots$                & $\cdots$ & $\cdots$ \\                 08424-3130\,(S)                 &  (1.1 $\pm$ 0.1)      &  (34)          &   1   &       $\cdots$        & $\cdots$              &  $\cdots$                     &  $\cdots$               &$\cdots$ &$\cdots$     \\ 
        F08520-6850                     &   (1.0 $\pm$ 0.1)     &   (100)         &   1   &      $\cdots$         & $\cdots$      &   $\cdots$            &  $\cdots$       & $\cdots$ & $\cdots$ \\ 
        09022-3615                &   (1.2 $\pm$ 0.4)   &   (100)       &   1 &      $\cdots$     & $\cdots$      &   $\cdots$            &  $\cdots$         & $\cdots$ & $\cdots$   \\ 
        F09437+0317\,(N)        &   1.3 $\pm$ 0.2       &         30    & 2   &   $\cdots$                &  $\cdots$             &       54  $\pm$       5         &  93 $\pm$ 14      &   54  $\pm$ 5 &  93 $\pm$ 14   \\ 
        F09437+0317\,(S)          &  (1.7 $\pm$ 0.2)&   (100)           & 1 &      $\cdots$               & $\cdots$      &   $\cdots$            &  $\cdots$       & $\cdots$  & $\cdots$\\  
        F10015-0614                &   1.2 $\pm$ 0.1    & 48            &   2 &   -116 $\pm$ 5    & 135 $\pm$ 4   &      $\cdots$         & $\cdots$         & -116 $\pm$ 5  & 135 $\pm$ 4   \\
        F10038-3338                     & 0.8 $\pm$ 0.1         &         50     &   2 &   -175  $\pm$ 5 &  131 $\pm$ 4  &      $\cdots$         & $\cdots$         &   -175  $\pm$ 5 & 131 $\pm$ 4                \\ 
        F10257-4339                &    0.6 $\pm$ 0.1 & 28      &  2 &  -386 $\pm$ 11        &  97 $\pm$ 10  &      $\cdots$         & $\cdots$      &   $\oslash$ & $\oslash$ \\  
        F10409-4556                &   1.3 $\pm$ 0.2 &  47      &   3   &  -180 $\pm$ 8   &  150  $\pm$ 6 &      $\cdots$         & $\cdots$      & -180 $\pm$ 8 & 150  $\pm$ 6     \\
        F10567-4310                &    1.5 $\pm$ 0.2    &  28          &  2  &  -146 $\pm$ 6     & 160  $\pm$ 4  &      $\cdots$         & $\cdots$         &   -146 $\pm$ 6 &  160  $\pm$ 4 \\        
        F11255-4120                &    0.8 $\pm$ 0.1    &  18          & 2   & -405  $\pm$ 21    &  148 $\pm$ 26 &   -99  $\pm$11                & 127 $\pm$ 15     & -252  $\pm$ 21 & 138 $\pm$ 21\\               
        F11506-3851                &  1.2 $\pm$ 0.1     &         30   &   3 &   -88 $\pm$ 11    &  108 $\pm$ 8  &   173 $\pm$ 18                & 113 $\pm$  14    &  -88 $\pm$ 11        &  108 $\pm$ 8 \\ 
        F12043-3140\,(N)          & (0.7 $\pm$ 0.1)     & (79)          & 1  &      $\cdots$      & $\cdots$              &   $\cdots$            &  $\cdots$        & $\cdots$ &$\cdots$   \\        
       F12043-3140\,(S)           &  0.9 $\pm$ 0.1      &         77    &    2 &  -152 $\pm$ 16   &  162  $\pm$ 13        &       $\cdots$                & $\cdots$                &  -152 $\pm$ 16        &  162  $\pm$ 13 \\                              
        F12115-4656               &   1.5 $\pm$ 0.2     &         73    &    3 & -65  $\pm$         8     &  119 $\pm$ 6  &       $\cdots$                & $\cdots$                &  -65  $\pm$     8     &  119 $\pm$ 6 \\ 
        12116-5615                      &      0.8 $\pm$ 0.1&  22       &  2  &  -371 $\pm$ 21    &  98 $\pm$ 18  &   -156    $\pm$ 11    & 96  $\pm$ 16       &  -264 $\pm$ 16 & 97 $\pm$ 8\\ 
        F12596-1529                     &  (1.5 $\pm$ 0.2)      &  (100)         & 1   &      $\cdots$   & $\cdots$      &   $\cdots$            &  $\cdots$       &$\cdots$ &$\cdots$     \\ 
        F13001-2339              &  (1.2 $\pm$ 1.1)     & (63)          &  1  &       $\cdots$    & $\cdots$              &   $\cdots$                    &  $\cdots$               &$\cdots$ & $\cdots$    \\
        F13229-2934               &  0.5 $\pm$ 0.1      &  67   &  3  &   -140 $\pm$ 4   &  106 $\pm$ 3  &       $\cdots$                & $\cdots$         &   -140 $\pm$ 4        &  106 $\pm$ 3 \\ 
        F14544-4255\,(E)        &  (0.9 $\pm$ 0.1)      &   (41)        & 1  &      $\cdots$      & $\cdots$      &   $\cdots$            &  $\cdots$         & $\cdots$ & $\cdots$\\ 
        F14544-4255\,(W)         & (1.9 $\pm$ 0.4)      & (100)         & 1  &      $\cdots$      & $\cdots$      &   $\cdots$            &  $\cdots$         &$\cdots$ &$\cdots$ \\ 
        F17138-1017                &  (1.2 $\pm$ 0.5)   &   (80)        & 1   &      $\cdots$     & $\cdots$      &   $\cdots$            &  $\cdots$         &$\cdots$& $\cdots$\\ 
        F18093-5744\,(N)        &   1.0 $\pm$ 0.1       &  65   &  2    & -90  $\pm$ 14   & 152  $\pm$  11   &       $\cdots$             & $\cdots$         &   -90  $\pm$ 14         & 152  $\pm$  11  \\ 
        F18093-5744\,(C)                &    (0.5 $\pm$ 0.1)&  (56)     &  1      &    $\cdots$   & $\cdots$      &   $\cdots$            &  $\cdots$         & $\cdots$ &$\cdots$ \\ 
        F18093-5744\,(S)                &  0.7  $\pm$   0.1   &  41     &  2  &$\cdots$           &  $\cdots$             &    -41 $\pm$  4       &  103 $\pm$   13  &       -41 $\pm$      4       &  103 $\pm$   13 \\ 
        F21453-3511              & 1.1   $\pm$   0.1    & 97            &  2  &     $\cdots$              & $\cdots$              &   $\cdots$                    &  $\cdots$               &   $\cdots$      & $\cdots$ \\ 
        F22132-3705             & 1.1   $\pm$   0.1 & 47        &   3 &   $\cdots$               & $\cdots$              &    36  $\pm$  3               & 109 $\pm$ 12     &  36  $\pm$  3                & 109 $\pm$ 12 \\
        F22491-1808                     &  (0.7  $\pm$   0.1)  &   (90)    & 1  &      $\cdots$  & $\cdots$      &   $\cdots$            &  $\cdots$         &$\cdots$ &$\cdots$ \\ 
        F23128-5919\,(N)        &   0.8 $\pm$ 0.1       & 99    &   2 &       $\cdots$           & $\cdots$      &   $\cdots$                    &  $\cdots$               & $\cdots$ & $\cdots$   \\ 
        F23128-5919\,(S)        &   (0.7 $\pm$ 0.1)   &  (98)     &   1 &     $\cdots$            & $\cdots$               &   $\cdots$           &  $\cdots$               &$\cdots$ & $\cdots$ \\    
        \hline 
                                        \hline
                                        \end{tabular}
                                        }
                                        \end{center} 
\tablefoot{\scriptsize{Column\,(1): IRAS name.  Column\,(2): NaD Equivalent width as measured in the model stellar spectra (output of the pPXF routine). Column\,(3): The estimated percentage contribution to the NaD  by old stars (Sect.~\ref{IntegratedSpectra}).  In brackets values obtained from poor fits (i.e., Q\,=\,1, in column (4)). Column\,(4): Index of the quality of the stellar modeling of the integrated spectra (after visually inspecting the results of the pPXF analysis). It is defined as follows: 1 doubtful results, 2 and 3  medium and highly reliable results, respectively. We assign Q\,=\,1 to both galaxies in the system 08424-3130, since only a part of the nuclear region of both galaxies is covered by our VIMOS FoV (Fig.\,\ref{Panel_F08424}). In particular, the area covered by the NaD absorption (and H$\alpha$, \cite{Bellocchi2013}) for the Northern galaxy is very small, so that its correction for the rotation pattern (Sect.\,\ref{IntegratedSpectra}) is rather inaccurate. Columns\,(5-8): Velocity and velocity dispersion of the different components derived for  the NaD line profiles seen in the  decontaminated spectra (i.e., observed - stellar model). The different components are identified according to their velocities. Specifically, the superindex   B stands for a blueshifted component (according to the criterion in Sect.\,\ref{int_kin}), while \lq other\rq \ indicates a second kinematic component.  Such a component is typically narrower with respect to the blueshifted component, and it is found at systemic velocity (i.e., -60\,$<$\,V\,$<$\,60 \kms) or redshifted (V\,$>$\,60 \kms) with respect to the stars.  When two blueshifted components are found for the same object, the component with the lowest velocity is in the column dedicated to the \lq other\rq -component. Columns\,(9-10): Adopted values of velocity and velocity dispersion for  the 1D-analysis (Sect.\,\ref{int_kin}). When two blueshifted components are found, their average values is considered. The symbol $\oslash$ indicates the galaxies excluded for the 1D analysis since the Gaussian modeling of the NaD absorption line profile leaves strong residual. }}
                        \end{table*}

\subsection{Galactic winds in two dimensions: Identification and detection rate  \label{detectionrate}}

We carrid out the search of GWs in 2D  by inspecting the NaD spectral maps for regions with, simultaneously:
        \begin{enumerate}
        \item  significant blueshifted velocities ($\gtrsim$\,60 \kms) with respect to the systemic, which cannot be explained  by rotation. For this identification, we consider  the velocity field of the narrow component of H$\alpha$ as reference, for which Bellocchi et al. (2013) have shown  follows  systemic behavior (Appendix A);
        \item  EW(NaD)\,$>$\,1.3\,$\AA$, to guarantee that the NaD feature is dominated by ISM absorption;
and        \item  a relatively broad  kinematic component ($\sigma$\,$>$\,90 kms$^{\rm-1}$).
        \end{enumerate} 
This procedure identifies  22 objects out of 40 (55$\%$) with outflows (see Table~\ref{Tab_sample}), implying a detection rate that is slightly lower than that obtained from the analysis of the integrated spectra (i.e., 71$\%$). This is likely a consequence of adopting a stricter criterion for the identification of outflows in the spectral maps.  However, the agreement in the identification of a GW and/or rotation between these two procedures is high.   From the 25 objects for which it was possible to make a kinematical classification via both,  spatially resolved and spatially integrated spectra, 18  have outflow signatures and four rotation or systemic signatures, according to the two methods. Therefore, we find inconsistency  only in three cases. In two cases (F12115-4656 and F12043-3140\,(S), Figures \ref{Panel_F12115} and \ref{Panel_F12043}), this is because the putative outflows only cover a small faint region of the maps and, as a result, are undetected in the integrated spectra. In the other case (F07160-6215, Fig.\,\ref{Panel_F07160}), the neutral gas velocity field has a complex structure and the wind partially overlaps the approaching side of the rotation pattern, making the 2D identification difficult.\\
Previous works have shown that the ability of detecting neutral GWs with absorption tracers depends on the galaxy inclination angle. Specifically, \citet{Heckman2000} found  a probability of $\sim$\,70\,$\%$ of detecting outflowing gas in absorption  in starburst galaxies with an inclination less than 60$\degr$. Similar trends are also seen in  more recent studies of the relation between EWs(NaD) and the galaxy inclination angle (e.g., \citealt{Chen2010}). As seen in  Fig.~\ref{distr_multi2}, in our sample GWs are detected in galaxies observed in a wide range of inclination angles (i.e., 10$\degr$-80$\degr$). Although we also find a large percentage of GWs ($\sim$\,78\,$\%$) detected in galaxies at low inclination angles ($<$60$\degr$), this percentage is similar to the total number of galaxies with low inclination whether a GW is detected or not. \\
In addition to the 22 objects with neutral gas  GW detection, 10 targets show spider-like NaD velocity field in at least one kinematic component. This indicates disk rotation and this subsample is discussed in Sect.~\ref{disks}.  Nine sources (eight LIRGs and one ULIRG) lack any clear outflows or rotation signatures either because of a lack of blueshifted velocities or because the putative disk kinematic is very irregular; these sources are excluded for the following discussion. The maps and  integrated spectra of these nine sources, however, are shown in Appendix\,\ref{comments_maps}. Table~\ref{Tab_sample} (Column 9) summarizes the different kinematic patterns found in the maps.

\subsection{Spatially resolved neutral GWs properties: Two-dimensional kinematic and geometry \label{2Dprop}}
\begin{figure*}
\centering              
 \includegraphics[width=1. \textwidth]{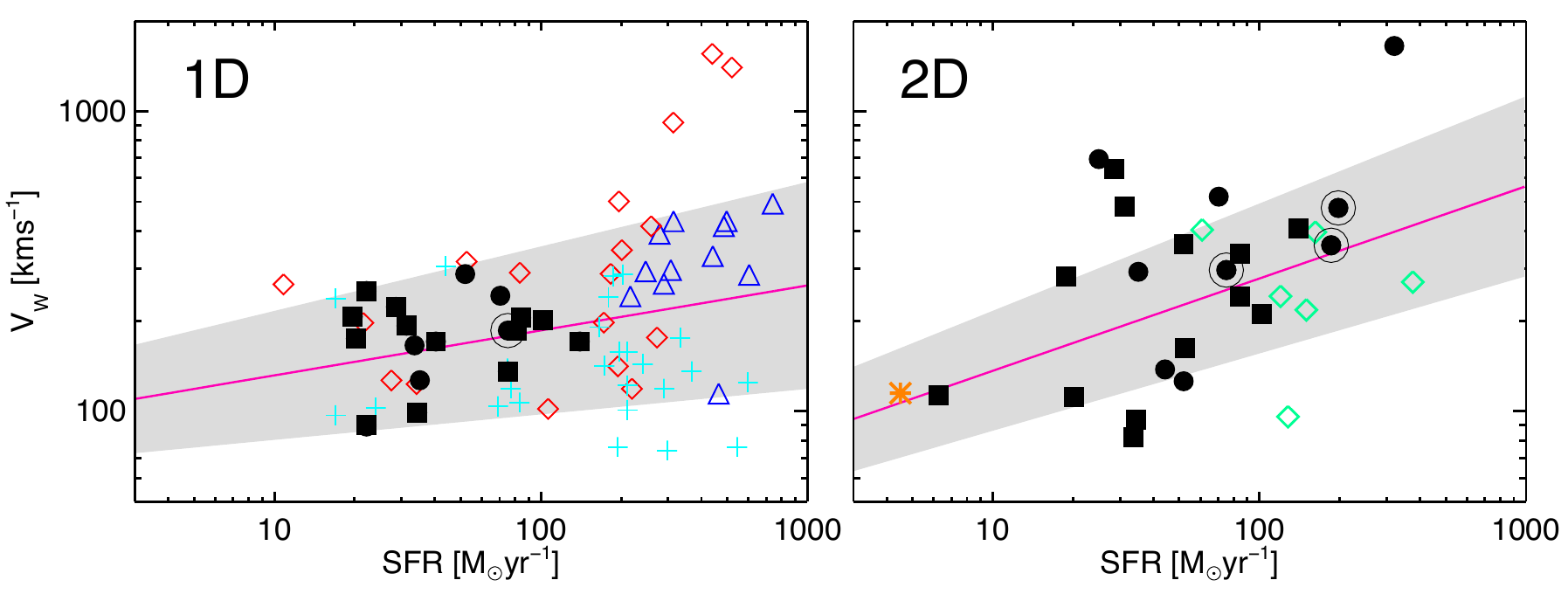}  
\caption{Velocity relative to systemic of the outflowing component  vs. SFR of the host galaxy for samples analyzed on the basis of 1D (i.e., integrated and long-slit, left) and 2D (IFS, right) data. The present sample is represented with black filled symbols in both panels. We identify pure starbursts and AGN hosts with squares and circles, respectively,  while very strong AGN are indicated with an additional circle. In the left panel, other [U]LIRGs samples from 
\citet{Rupke2002}, \citet{Rupke2005b}, and  \citet{Martin2005}  are indicated with red diamonds,  light blue crosses, and blue triangles, respectively. In the right panel, green diamonds indicate the IFS-based major-merger ULIRGs results by Rupke and Veilleux (2013) and the orange asterisk shows the result for the wind in M100 obtained by Jimen\'ez-Vicente et al.\,(2007). The pink lines, in both panels,  represent the trends of the type V\,$\propto$\,SFR$^{n}$ found for our samples (excluding the strongest AGNs).  See text for details. The shaded gray bands indicate uncertainties.}           
\label{V_SFR}    
\end{figure*} 
\begin{figure*}
\centering
\includegraphics[width=.99\textwidth]{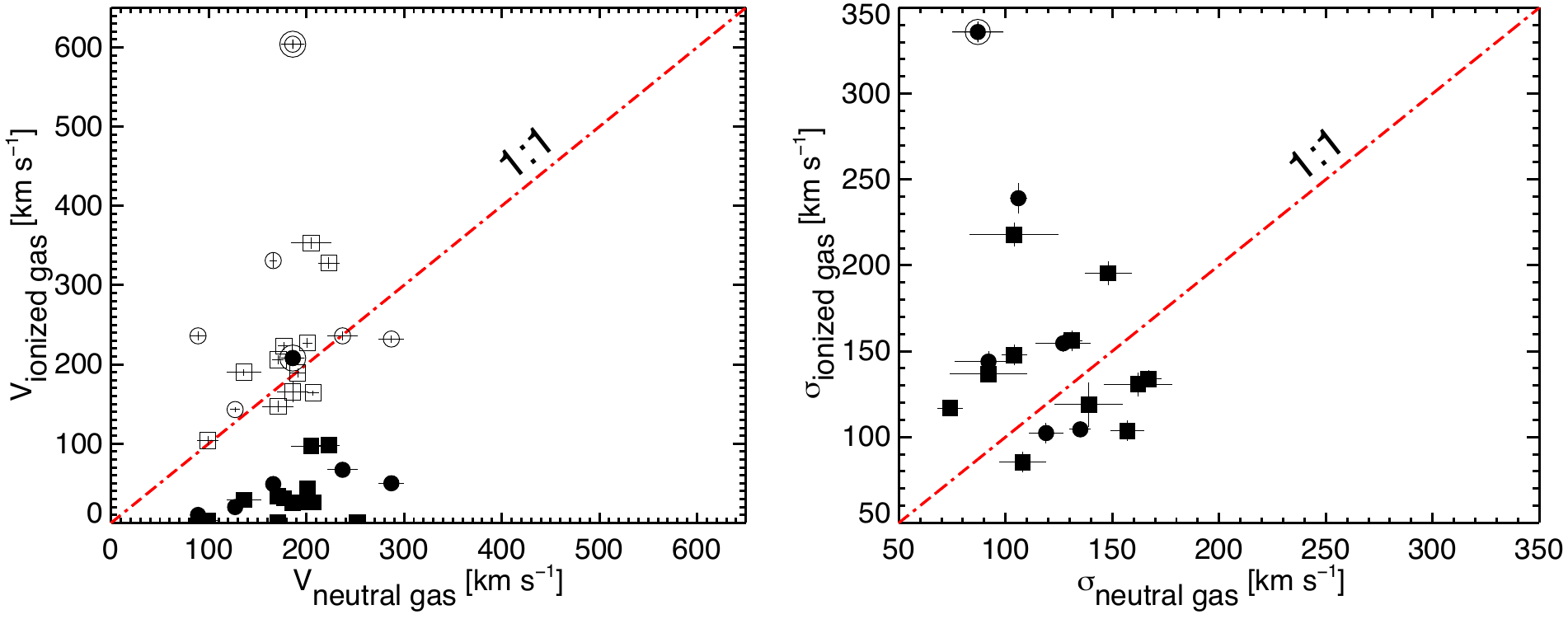}  
\caption{Comparison of neutral and ionized outflows velocities (left) and velocity dispersions (right), as derived from the (1D) integrated spectra. Results for the ionized outflows come from Arribas et al. (2014). The symbols used are the same as in Fig.\,\ref{V_SFR}. Specifically, filled squares and circles are for starbursts and  for AGN hosts, respectively, while very strong AGN are indicated with an additional second circle. In the left panel, solid symbols indicate velocities measured at the center of the line, while the empty symbols indicate the ionized gas velocities, V$_{\rm max}$, from Arribas et al.\,(2014). The dot-dashed line indicates the 1:1 relation in both panels.}          
\label{kin_gw_1D_comparison}             
\end{figure*}

\begin{figure}
\centering
 \includegraphics[width=.49\textwidth]{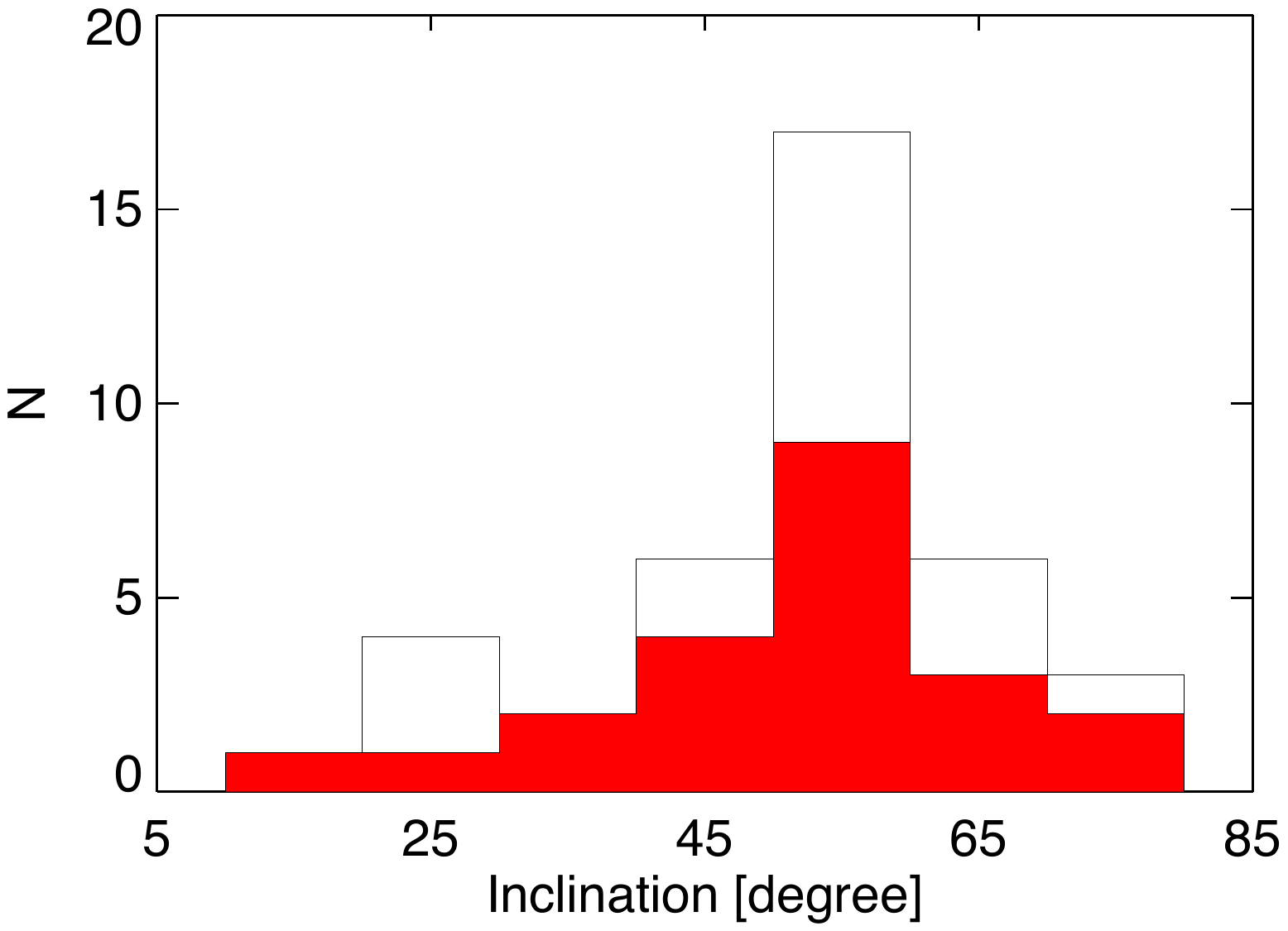}  
\caption{Distribution of the galaxy inclination angles for the 40 [U]LIRGs with reliable NaD-IFS detection (Sect.\,\ref{SO}).  Objects for which a  GW was detected are shown with a red-filled histogram (Sect.\,\ref{detectionrate}). }              
\label{distr_multi2}
\end{figure}
\begin{figure}
 \centering
\includegraphics[width=.492\textwidth]{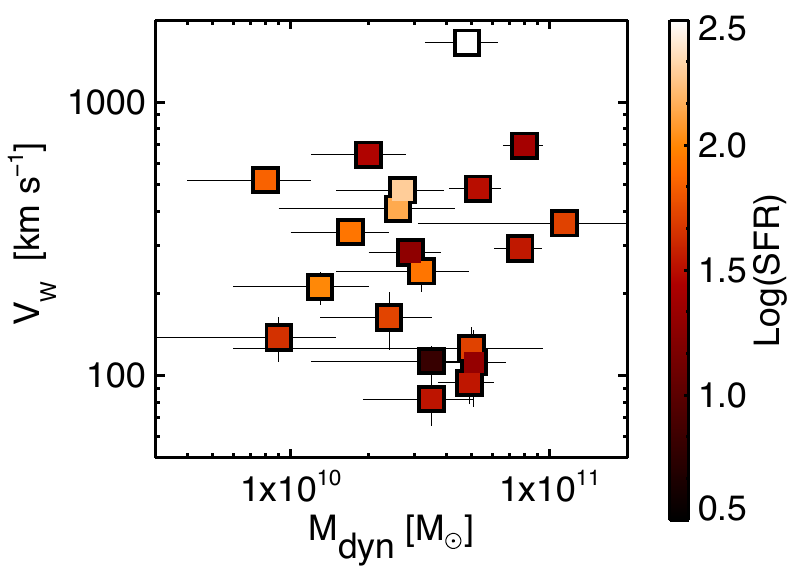}  
\caption{Wind velocity versus galaxy's dynamical mass. Symbols are color coded by the logarithm of the SFR.  Vertical error bars are $1-\sigma$ standard deviations inferred from the wind velocity fields. Points for F05189-2524 and F07027-6011\,(N) are not shown, since there is no  reliable estimation of their dynamical mass (\citealt{Bellocchi2013}).}
 \label{Fig_Kin_mass}            
\end{figure}
\begin{figure}
 \centering
 \includegraphics[width=.49\textwidth]{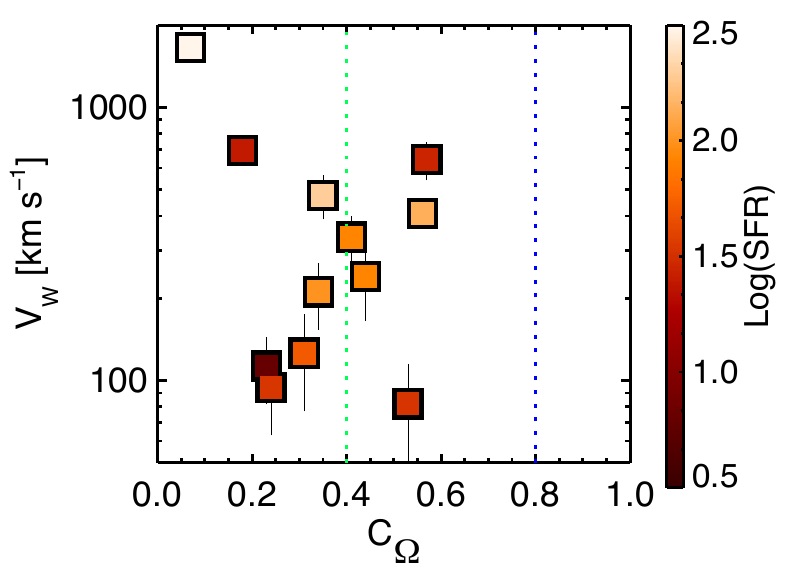}  
 \caption{Wind velocity  plotted against the  3D wind opening angle (C$_{\rm \Omega}$) color coded by the logarithm of the SFR. Vertical green and blue dotted lines represent typical values of C$_{\rm \Omega}$ found in LIRGs and ULIRGS, respectively (\citealt{Rupke2005c}). Uncertainties for velocities are the 1-$\sigma$ standard deviations 
(inferred from the wind velocity fields) while, for C$_{\rm \Omega}$, assuming a 20$\%$ error, they are typically $\sim$\,0.1.}
 \label{omega}           
\end{figure}

We derived the (typical) wind velocity  (V$_{\rm w}$)  as the deprojected median value\footnote{In this work,  we consider that the GWs are  perpendicular to the disk. Therefore,  we deproject  using the inclination of the galaxy (as listed in \citealt{Bellocchi2013}).} over the region identified as GW from the spectral maps.  The inclination-corrected outflow velocities  of the neutral gas entrained in GWs are in the range $\sim$\,80-700\,\kms, which is similar to the velocities found previously for  starburst driven winds in local star-forming galaxies and [U]LIRGs (\citealt{Heckman2000, Martin2005, Rupke2005c,Chen2010}).  An  exception to the general behavior is the outflow in the  ULIRG F06206-6315  with velocities that exceed 1000\,\kms \ , which host an AGN (Table\,\ref{Tab_sample}). \\
The outflow velocity inferred from the maps also has a clear correlation with the global SFR (Fig 5, right). In particular, on the one hand, V\,$\propto$\,SFR$^{n}$ with  \textit{n}\,=\,0.40\,$\pm$\,0.07 for all cases, and \textit{n}\,=\,0.30\,$\pm$\,0.05  excluding the objects with an AGN. This is in rather good agreement with previous results, including those obtained from the integrated spectra (see  Fig.\,\ref{V_SFR}, left). On the other hand, the wind velocity does not seem to correlate with the dynamical mass of its host  (Fig.\,\ref{Fig_Kin_mass}). \\Results from a previous  IFS survey of local major-merger [U]LIRGs (\citealt{Rupke2013}) indicate that in such systems the typical neutral gas outflow velocity dispersion is high,  up to 1000\,\kms (FWHM) in the case of MRK231. In our sample of (mainly) LIRGs, the typical velocity dispersions of the observed winds are lower ($\sigma$\,$\sim$\,95-190\,\kms; FWHM\,$\sim$\,230-460\,\kms).  However, these values are significantly higher than the thermal velocity dispersion of the warm neutral gas (i.e., 8 km$^{\rm-1}$, \citealt{CalduPrimo2013}). This indicates that a wide range in neutral gas velocity is integrated along the line of sight or that the winds are turbulent and associated with shocks (as seen in the ULIRG F10565+2448 by \citealt{Shih2010}). In  the standard GW scenario, (e.g., \citealt{Heckman2000}) broadening effects are consistent with outflowing gas in interaction (e.g., via shocks) with the surrounding material and with the presence of turbulent mixing layer on cloud surface. \\
In our sample, outflows are in many cases consistent with being along the minor axis of the ionized gas rotation (Appendix A). Thanks to the present IFS data, we are also able to infer the morphology outflow in about half of the sample  (see figures in Appendix A). These outflows appear to be extra-planar, conical (the projected area is triangular in shape), and extended on kpc scales, which is consistent with the expectations from the standard GW model \citep{Heckman2000}.   With the present data set, it is unpractical to develop a customized and more detailed model invoking superbubbles or different geometries  (which is beyond the scope of this paper). We, therefore, assume a simple outflow model where the GWs are emerging perpendicular to the disk in all cases. \\
The measured outflow extension values (Table\,\ref{Table_GWs}) are in good agreement with those reported previously for neutral gas GWs in [U]LIRGs (e.g., \citealt{Veilleux2005}). However, because of the lack of a bright continuum at large radii, the outer regions of the outflows may not have been detected with absorption line techniques (e.g., via NaD), and thus the quoted extensions should be considered  lower limits. \\
In those outflows with a cone-like morphology, we identify  the cone apex (near the galaxy nucleus) and boundaries in the spectral maps, and then we measure the wind 3D opening angles (i.e., C$_{\rm \Omega}$; see Table\,\ref{Table_GWs}). We find  C$_{\rm \Omega}$ values are in the range between 0.1 and 0.6,  in agreement with those (indirectly) estimated from the wind detection rate by \citet{Rupke2005c} and \citet{Veilleux2005} in local starburst and LIRGs (i.e., C$_{\rm \Omega}$\,$\sim$\,0.4).  These are, however, lower than those inferred by these authors in local ULIRGs (C$_{\rm \Omega}$\,$\sim$\,0.8).  Our results do not support a correlation between C$_{\rm \Omega}$ and SFR  (or the V$_{\rm w}$/$L_{IR}$, Fig.\,\ref{omega}), although we note that the uncertainties associated with the determination of  C$_{\rm \Omega}$  are large.\\
While in half of the cases it is not possible to infer the morphology of the outflows, it is relatively straightforward to estimate the wind projected area for all 22 GWs detected  (Table\,\ref{Table_GWs}). On average,  the projected area of the neutral winds is 5 kpc$^{\rm2}$. A spatially resolved broad H$\alpha$ emission is seen in many objects of our sample (\citealt{Bellocchi2013}), but its presence shows no obvious correlation with the observed neutral GWs. Specifically,  the neutral outflows are more extended with respect to the H$\alpha$ broad component  in 15 of the cases. Considering these cases, the areas covered by the outflow and the broad H$\alpha$ component generally only overlap partially (12 cases). 

\begin{landscape}
 \begin{table}
                        \caption[Sample]{Kinematics, geometry, and feedback properties of detected neutral GWs in the sample of [U]LIRGs.}
                                \begin{center}
                                        {\begin{tabular}{lcccccccccccc}
                                        \hline
                                        \hline 
        ID1                                      &           R     &    Area &V$_{\rm w}$     & $\sigma$$_{\rm w}$ & C$_{\rm \Omega}$ & N$_{\rm H}$    & M$_{\rm w}$         & $\dot{M_{\rm w}}$ &  $\eta$   & E$_{\rm w}$   & P$_{\rm w}$    \\
         IRAS                                  &         kpc    &  kpc$^{\rm2}$ & kms$^{\rm-1}$      &    kms$^{\rm-1}$         &   &  10$^{\rm21}$ cm$^{\rm-2}$ &    10$^{8}$  M$_{\rm \sun}$  &    M$_{\rm \sun}$yr$^{\rm-1}$ &       & 10$^{56}$\,erg$^{\rm-1}$  &   10$^{\rm41}$\,erg$^{\rm-1}$\,s$^{\rm-1}$      \\
         (1)                                      &       (2)        &        (3)              &          (4)          &  (5)    &      (6)        &   (7)         &             (8)  & (9)        &  (10)  & (11)  & (12)   \\
         \hline   
F01159-4443\,(N)                        &  2.3       & 1.6             &  126 $\pm$ 50 &    197 $\pm$ 50   &  0.3           &  9.5 $\pm$ 1.1  &  7.4 $\pm$ 0.7   &    21.2 $\pm$ 4.2   &  0.4 $\pm$ 0.2 &  11.1 $\pm$ 2.7 &  9.8 $\pm$ 1.0    \\
F01341-3735\,(N)                        &  3.0         & 1.8    &  113  $\pm$  31 &   159 $\pm$  19  &  0.2          &  2.8 $\pm$ 0.3   &  2.7 $\pm$ 0.3   &      7.3 $\pm$  1.9  &   1.2 $\pm$ 0.2 &  2.4 $\pm$  0.5 &   2.1 $\pm$  0.4 \\  
F04315-0840                             &  2.3         & 2.5    &  335  $\pm$  64 &   150 $\pm$  44  &  0.4          &  4.3 $\pm$ 0.8   &  1.0 $\pm$ 0.3   &    22.1 $\pm$ 5.8   &  0.3 $\pm$  0.1 &  2.1 $\pm$ 0.8  &  15.4 $\pm$ 3.8   \\
F05189-2524                                     &  $\cdots$ &  13.3     & 358 $\pm$ 115  &  192 $\pm$ 97    & $\cdots$ &  3.2 $\pm$ 0.3   &  (0.6 $\pm$ 0.1) &   (7.9 $\pm$ 0.8)   &  (0.05 $\pm$ 0.01) &  (1.5 $\pm$ 0.2)  &   (6.4 $\pm$ 1.4)   \\
F06206-6315                             & 10.1        &  19.9   & 1654  $\pm$  168 &  203 $\pm$  79  &  0.1        &  3.8 $\pm$ 0.7   &  4.5 $\pm$ 1.0   & 113.5 $\pm$ 27.7 &  0.4 $\pm$ 0.1 &  147.2 $\pm$ 43.3  &  1260 $\pm$ 120  \\    
F06259-4780\,(N)                        & 3.1        & 9.8              &  408  $\pm$ 45 &    121 $\pm$ 32  &  0.6             &  4.5 $\pm$ 0.4   &  2.8 $\pm$ 0.5  & 47.8  $\pm$ 8.8    &  0.3 $\pm$ 0.1  &  6.8 $\pm$  1.8 & 35.8  $\pm$ 6.3  \\    
F06592-6313                             &  2.2       & 4.4      & 643   $\pm$  103 & 157 $\pm$  53  &  0.6       &  5.6 $\pm$ 0.6    &  2.1 $\pm$ 0.6      &  70.7 $\pm$  17.9 &  2.5  $\pm$  0.5 &  11.9 $\pm$  4.2 & 131.2  $\pm$ 33.8    \\
F07027-6011\,(N)                        &  $\cdots$ &  7.2      & 296 $\pm$  59  &   166 $\pm$ 33  & $\cdots$    &  5.1 $\pm$ 0.6   &  (1.0 $\pm$ 0.1) &   (9.4 $\pm$ 1.3)  &  (0.1 $\pm$ 0.01)  &  (1.5 $\pm$ 0.1)  &   (4.7 $\pm$ 1.6)   \\
F07027-6011\,(S) $^{\ddagger}$            &  $\cdots$ & 4.0     & 163  $\pm$ 77  &    140 $\pm$ 46  & $\cdots$   &  4.6 $\pm$ 1.1   &  (0.9 $\pm$ 0.2) &   (5.4 $\pm$ 1.2)  &  (0.1 $\pm$ 0.01)  &  (0.8 $\pm$ 0.1)  &   (1.5 $\pm$ 1.0)   \\
F07160-6215                             &  3.0                 & 1.1    &  694  $\pm$  55 &   114 $\pm$  50  &  0.2          &  3.6 $\pm$ 0.8   &  0.6 $\pm$  0.2   &  23.0  $\pm$ 6.2  &  0.9 $\pm$ 0.2 & 3.6  $\pm$ 1.0  &  45.5 $\pm$ 12.1  \\
F10015-0614                             &  $\cdots$ &  3.2      & 292 $\pm$ 29  &    106 $\pm$  33 & $\cdots$    &  6.1 $\pm$ 1.1   &  (1.1 $\pm$ 0.1) &   (11.1 $\pm$ 1.6)  &  (0.3 $\pm$ 0.06)  &  (1.3 $\pm$ 0.1)  &   (3.9 $\pm$ 0.8)     \\
F10038-3338                             &  4.0        & 5.8             & 211  $\pm$ 58 &    145 $\pm$ 27   &  0.3            &  4.2 $\pm$ 0.8   &  2.4 $\pm$ 0.6   &   17.1 $\pm$  5.5 &  0.2 $\pm$ 0.04 &  3.0 $\pm$ 1.3 & 7.2  $\pm$  1.8 \\
F10257-4339                             &  2.3              & 3.1                &  241 $\pm$  76 &  163 $\pm$  52  &   0.4        &  2.4 $\pm$ 0.5   &  1.3 $\pm$ 0.6       &   26.2 $\pm$ 8.9  &  0.3 $\pm$ 0.06  & 2.0  $\pm$ 0.9 &  18.6 $\pm$  6.1 \\
F10409-4556 $^{\ddagger}$               &  $\cdots$   & 10.9     & 483  $\pm$ 70  &  174 $\pm$  63 & $\cdots$    &  3.9 $\pm$ 0.5   &  (0.7 $\pm$ 0.1) &   (12.6 $\pm$ 1.0)  &  (0.4 $\pm$ 0.1)  &  (2.5 $\pm$ 0.1)  &   (13.3 $\pm$ 1.4)     \\
F10567-4310                             &  $\cdots$   &  2.1     & 112  $\pm$ 70 &   119 $\pm$ 40   & $\cdots$   &  7.2 $\pm$ 1.1   &  (1.3 $\pm$ 0.2) &   (5.8 $\pm$ 1.8)  &  (0.3 $\pm$ 0.06)  &  (0.8 $\pm$ 0.1)  &   (1.0 $\pm$ 0.3)   \\
F11255-4120                             &  $\cdots$   &   1.8    &  282 $\pm$  70 &  203 $\pm$  68  & $\cdots$  &  5.6 $\pm$ 0.9   &  (1.1 $\pm$ 0.2 ) &  (9.6 $\pm$ 1.4)  &  (0.5 $\pm$ 0.1)  &  (1.7 $\pm$ 0.1)  &   (4.9 $\pm$ 2.1)   \\
F11506-3851 $^{\ddagger}$                   &  1.8            &  0.9    &  98   $\pm$  31  &   98  $\pm$ 20  & 0.2           &  5.0 $\pm$ 0.4    &  2.5  $\pm$ 0.3     &  23.6  $\pm$ 5.1 &  0.9 $\pm$  0.1&  0.4 $\pm$ 0.1 &  1.4 $\pm$ 0.1  \\
12116-5615 $^{\ddagger}$                &  $\cdots$   & 3.6     &  520 $\pm$ 114 & 115 $\pm$  58 & $\cdots$  &  4.0 $\pm$ 1.3    &  (0.7 $\pm$ 0.2)   &   (11.6 $\pm$ 1.0)  &  (0.2 $\pm$ 0.03)  &  (1.8 $\pm$ 0.1)  &   (8.7 $\pm$ 0.6)   \\
F13001-2339                             &  $\cdots$   & 4.8     &  361 $\pm$ 46   &  120 $\pm$  65 & $\cdots$   &  3.5 $\pm$ 0.6     &  (0.6 $\pm$ 0.1) &  (7.8 $\pm$ 0.9)  &  (0.2 $\pm$ 0.03)  &  (1.1 $\pm$ 0.2)  &   (4.0 $\pm$ 0.6)   \\
F13229-2934                             & 1.4                    & 1.1  &  82  $\pm$ 33  &    139 $\pm$ 25   & 0.5          &  2.7 $\pm$ 0.2    & 0.4  $\pm$ 0.1    &   2.6 $\pm$ 0.7  &  0.1   $\pm$   0.01  &  0.3 $\pm$ 0.1 & 0.6  $\pm$ 0.3  \\
F21453-3511                             &  $\cdots$   &  5.6    &  138  $\pm$  51  &   146 $\pm$  70 &  $\cdots$ &  2.6 $\pm$ 0.6    &  (0.5 $\pm$ 0.1) &  2.6 $\pm$ 0.7  &  (0.05 $\pm$ 0.01)  &  (0.4 $\pm$ 0.1)  &  (0.7 $\pm$0.3)   \\
F23128-5919                             &  3.4                   & 7.3  & 477 $\pm$  88 &    142 $\pm$ 57 & 0.4              &  1.7 $\pm$ 0.8    & 0.6 $\pm$ 0.1    &   12.9 $\pm$ 7.4  &   0.1   $\pm$   0.01  &  1.8 $\pm$ 0.4  & 13.5 $\pm$ 6.8    \\
         \hline
        \hline
                        \end{tabular}
\label{Table_GWs}}
\end{center}  
\tablefoot{Column\,(1): IRAS name. Column\,(2): Observed  extent of the neutral winds corrected for the inclination using the inclination values of Bellocchi et al.\,(2013). Column\,(3): Area covered by the GW. Column\,(4):  Inclination-corrected median velocities.  Column\,(5): Median velocity dispersion over the GW area. Column\,(6): 3D Wind opening angle. Column\,(7): Average wind column density. Column\,(8): Wind mass in the neutral phase. Column\,(9): Mass outflow rate.  Column\,(10):  Outflow loading factor defined as the mass outflow rate normalized to the corresponding SFR. Column\,(11): Wind energy.  Column\,(12):  Wind power. The symbol:  $^{\ddagger}$, denotes the galaxies for which we performed two-component NaD modeling (Sect.~\ref{LF}). In brackets are values obtained for those GWs for which the morphology is not  constrained well (Sect.~\ref{2Dprop}).}
                        \end{table}                     
\end{landscape}

\subsection{Galactic winds feedback in 2D  \label{feedback}}
While it is relatively straightforward to demonstrate that a GW is present, it is more difficult to robustly calculate the rates at which mass and energy are  transported out by the wind (i.e., its feedback effects).  We  adopted a  free wind (FW) model to quantify the neutral wind feedback in the form of  outflowing mass, outflow mass rate, and loading factor. Details of the model are given in  \citet{Heckman2000},  \citet{Rupke2002}, and  \citet{Rupke2005c}.
\begin{figure*}
\centering
 \includegraphics[width=1.\textwidth]{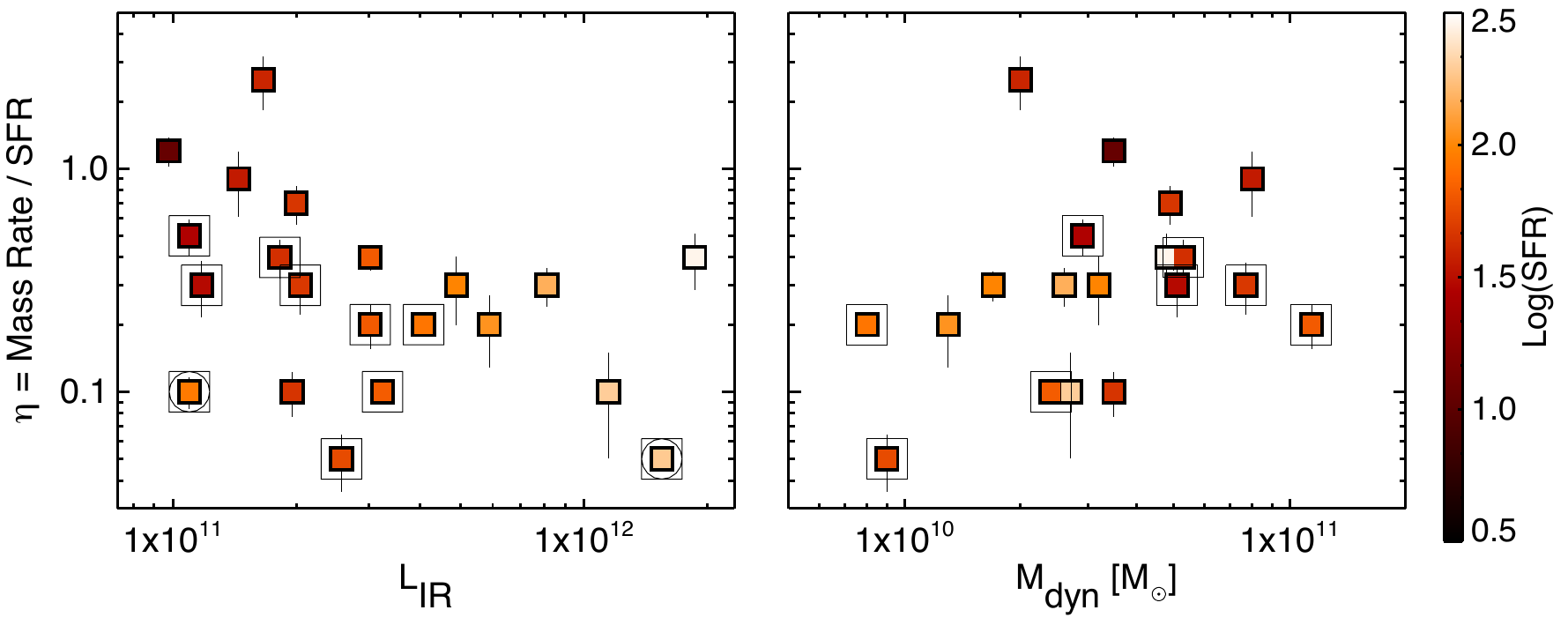}  
\caption{Wind loading factor plotted against the infrared luminosity (left, Table\,1) and dynamical mass  (right) color coded by the logarithm of the SFR. In both panels, those galaxies for which we were unable to infer  the wind morphology in
detail are indicated with an additional  square. In the left panel, the  galaxies that lack of a reliable estimation of the dynamical  mass  (F05189-2524 and F07027-6011\,(N); \citealt{Bellocchi2013}) are shown with circles.}
\label{Fig_feedback}   
\end{figure*}
\noindent Briefly, this model consists of a starburst  surrounded by thin shells of a free-flowing wind, a shocked wind (i.e., GW ionized phase), and  entraining clouds of neutral ISM.  One drawback of this model is that a thin shell could be broken leading to the formation of filaments. Indeed, a starburst-driven wind, in its expansion through the ISM, entrains clouds that are denser than the ambient ISM. A dense cloud embedded in a subsonic flow experiences pressure differences along the surface. This may lead to Rayleigh-Taylor instabilities and the fragmentation of clouds with the consequent formation of filaments that are not accounted for by the FW\ model (\citealt{Heckman2000}). Although this scenario is possible, the spatial resolution of the VIMOS data does not allow us to probe possible filaments and, therefore, the shell is a natural and simple interpretation of our observations of GWs.  Other approaches are, however, possible (e.g., single radius FWs or superbubbles; \citealt{Rupke2013}).\\
To estimate how much of the neutral gas is brought out into the outflow, knowledge of the column density of the wind  (N$_{\rm H}$) is required.  To obtain spatially resolved column densities, we used a method that relates N$_{\rm H}$ to the strength of the NaD absorption (i.e., its equivalent width) via  reddening (E$_{\rm B-V}$). Specifically, first, \citet{Turatto2003} via a light-curve fitting of low resolution spectroscopy of SNe found that the EW of the NaD  correlates  with the reddening well (see also \citealt{Veilleux1995}). Second,  \citet{Bohlin1978} found that the E$_{\rm B-V}$ follows a well-defined linear relation with N$_{\rm H}$, analyzing  a large sample of galactic stars. Combining these relations\footnote{For the present work, we considered the average relation within the two extreme relationships found by Turatto et al.\,(2003).  In this work, the relation for heavily reddened objects, used in  \citet{Cazzoli2014}, encompasses only one data point for EW(NaD)\,$>$\,1.2\,\AA.}, we found the following dependence of N$_{\rm H}$, with EW(NaD): 
 \begin{center}
\begin{equation} 
N_{\rm H}\,=\,\frac{-0.025\,+\,0.335\,\times\,EW(NaD)}{0.2\,\times\,10^{-21}} \ cm^{-2}.
\label{eq_nh}
\end{equation}
  \end{center}
The median column density  in our sample is 4.2\,$\times$\,10$^{\rm21}$\,cm$^{\rm-2}$ (see Table\,\ref{Table_GWs}). This approach, which has also been  followed in \citet{Davis2012} and \citet{Cazzoli2014}, is applied here on a spaxel-by-spaxel basis. However, this ignores issues such as the spatial variation of the ionization state of NaD \citep{Murray2007}, metal depletion on dust grains, line saturation effects (e.g., the square-root part of the curve of growth), and variations in gas-to-dust ratio (\citealt{Wilson2008}). \\
Alternatively, column densities may be computed following \citet{Hamann1997}. By using spatially resolved and spatially integrated data, we found values in rather good agreement with the current estimate and previous works (e.g., 1.6\,$\times$\,10$^{\rm21}$\,cm$^{\rm-2}$ by \citealt{Rupke2005c}). Specifically, we found average values of 4.6\,$\times$\,10$^{\rm21}$\,cm$^{\rm-2}$  and  2.0\,$\times$\,10$^{\rm21}$\,cm$^{\rm-2}$  for 2D and 1D cases, respectively.\\
Additionally, we also estimated N$_{\rm H}$ with the EW-ratios (R$_{\rm NaD}$) of the (1D) integrated spectra (after removing the stellar contribution) via the sodium curve of growth (\citealt{Draine2011}). For this, we only considered optically thin (R$_{\rm NaD}$\,$\ge$\,1.5, \citealt{Rupke2005a}) outflowing components (14/24 cases), obtaining a median  value of 1.1\,$\times$\,10$^{\rm21}$\,cm$^{\rm-2}$. This is  smaller  than the results obtained from Eq.\,\ref{eq_nh} quoted above and the typical value found by \citet{Rupke2005c}. \\
For the present paper, we consider the values obtained from Eq.\,\ref{eq_nh} which are in good agreement with those values from the method described in \citet{Hamann1997}. A more detailed estimation of N$_{\rm H}$ (e.g., from more complex line-profile modeling as in \citealt{Rupke2005a}) is beyond the aim of this work.\\
Our IFS observations probe the wind shape in half of the cases (as discussed in Sect.\,\ref{2Dprop}) allowing us to constrain the radial extent and variation of the velocity of the outflow with position. We customized the FW model, allowing each spatial element of the wind to have its own velocity  and distance from the wind's origin. We used the column density and velocity measured in each spatial resolution element (\textit{k}) to derive the wind mass (M$_{\rm w}$) and outflow rate ($\dot{M_{\rm w}}$), following \citet{Rupke2005c},  as 
\begin{equation} \label{masswind}
\begin{array}{c}
M_{\rm w}\,=\,5.6 \times 10^{8} \sum \limits_{\rm k=1}^N \left(\frac{C_{\rm \Omega,k}}{0.4} C_{\rm f}\right) \left(\frac{R_{\rm w,k}^{2}}{100 \, kpc^{2}}\right)   \left(\frac{N_{\rm H,k}}{10^{21} cm^{-2}}\right) \ M_{\rm \sun}.
\end{array}
\end{equation}  
\begin{equation} \label{massdotwind}
\begin{array}{rl}
\dot{M}_{\rm w}\,=\,11.5 \times \sum \limits_{\rm k=1}^N \left(\frac{C_{\rm \Omega,k}}{0.4} C_{\rm f}\right) \left(\frac{R_{\rm w,k}}{10  \, kpc}\right) \times \\
\left(\frac{N_{\rm H,k}}{10^{21} cm^{-2}}\right) \left(\frac{V_{\rm w,k}}{200 \ kms^{-1}}\right) \ M_{\rm \sun}yr^{-1}.
\end{array}
\end{equation}
These equations describe the  mass and mass outflow rate for a GW flowing into a solid angle  C$_{\rm \Omega,k}$ with a cloud covering factor C$_{\rm f}$.  We consider the observed GWs as a series of thin shells, each one located at radius R$_{\rm w,k}$ with the corresponding  inclination-corrected velocity V$_{\rm w,k}$ and column density (N$_{\rm H,k})$ inferred, as  previously discussed. The  extent \textit{R},  median velocity \textit{V} and average column density $N_{\rm H}$,  as well as the mass  and mass outflow rate, which are derived from our observations assuming the FW model, are listed in Table\,\ref{Table_GWs}. The parameter C$_{\rm f}$ is assumed to be 0.37 as in \citet{Rupke2005c}.\\
As mentioned in the previous section,  we cannot infer the wind morphology in  ten cases; this is mainly because of projection effects (Table\,\ref{Table_GWs}). In these cases, to calculate the wind mass and mass rate,  we assumed a fiducial radius of 3 kpc and a wind opening angle of 0.4, which  are the characteristic values seen for the subsample for which we can  derive the actual wind morphology. We also considered the median values of V, and N$_{\rm H}$ distributions measured over each GW region. \\
The wind mass estimates range between  4\,$\times$\,10$^{\rm7}$\,M$_{\rm \sun}$ to  7.5\,$\times$\,10$^{\rm8}$\,M$_{\rm \sun}$ with a mean value of 1.6\,$\times$\,10$^{\rm8}$\,M$_{\rm \sun}$.  These values are roughly in agreement (lower by a factor of 3 when comparing averages) with those reported by \citet{Rupke2005c} obtained for neutral outflows in their local LIRGs  sample (i.e., 6.3\,$\times$\,10$^{\rm8}$\,M$_{\rm \sun}$ on average).\\
Previous long-slit studies of \citet{Rupke2005b, Rupke2005c} and \citet{Martin2006}, on the basis of the FW wind model, found that the mass of neutral outflows in [U]LIRGs can be up to 10$\%$ of the dynamical mass (estimated via CO measurement). For our sample, the wind mass is typically $\sim$\,3$\%$  of the dynamical mass of the galaxy. These results indicate that neutral outflows in [U]LIRGs may carry away a significant amount of gas mass that would otherwise be available for further star formation (\citealt{Sato2009}).   \\
However, in general, we found  that the mass outflow rate (i.e., $\dot{M_{\rm w}}$) is not larger than the global SFR  (Table\,\ref{Table_GWs}). Specifically,  the outflow loading factor (i.e.,  $\eta$\,=\,$\dot{M_{\rm w}}$/SFR) is $\eta$\,$<$\,1 in nearly all objects (Table\,\ref{Table_GWs}), indicating that the mass-loss rates (in the neutral phase) are relatively unimportant for quenching the SFR. 
However, an estimation of the mass-loss rates, which take into account other wind phases (e.g., molecular) may lead to an increase of the significance of the feedback.\\
These results are in rough agreement with previous measurements of $\eta$ in nearby [U]LIRGs \citep{Rupke2005c} and empirical models  (e.g., \citealt{Zahid2012}) for which $\eta$\,$\leq$\,1.  Values of $\eta$ above the unity, which indicates  a strong mass loading of the outflow by the galaxy ISM, have been found in active galaxies (\citealt{Veilleux2005}). \\
In the left panel of Fig.\,\ref{Fig_feedback}, we show the loading factor as a function of the infrared luminosity, which are in negative correlation. Excluding those galaxies for which we were unable to estimate  the wind morphology in detail,  the Paerson coefficient, r$_{\rm PC}$ hereafter, is $\sim$\,-0.6. A similar trend was found by \citet{Rupke2005c}, although these authors used the K-band magnitude as a tracer of the infrared luminosity. Nevertheless, considering the same subsample, our analysis also supports the absence of any dependence of $\eta$ on dynamical masses of the host for the neutral phase (r$_{\rm PC}$\,$<$\,0.1); this is seen in the right panel of Fig.~\ref{Fig_feedback}, where we present the dynamical mass versus the mass loading factors. This is in contrast to the inverse dependency on $\eta$ and the dynamical mass found for the ionized phase of GWs  in 32 LIRGs without AGNs  \citep{Arribas2014}. We note, however, that the present sample (12 objects) is relatively small and it includes 6 galaxies hosting weak AGNs.\\ 
In summary, the present results indicate that the neutral gas loading factors are small ($<$1) and, therefore, the feedback effects are not expected to be large if we only take the neutral gas phase of the outflows into consideration. For the objects with a well-defined morphology,  the loading factors show a trend with the infrared luminosity, but they seem uncorrelated with the host mass. The relative small number statistics (i.e., 12 objects, including 6 weak AGNs) prevent us from considering the latter trend as a firm conclusion.

\subsection{The gas cycle: IGM and ISM pollution via outflows \label{fountain}}

The GWs that originated in the disk of spiral galaxies, in principle, could have the ability  to eject large percentage of the cold gas reservoir available for star formation into the IGM \citep{Veilleux2005}.  We compare the wind velocity to the  escape velocity of each galaxy to investigate the
fate of the outflowing material. \\
Following a simple recipe  (\citealt{Arribas2014} and reference therein) for a galaxy with dynamical mass (M$_{\rm dyn}$), we estimate the escape velocity (v$_{\rm esc}$) at r\,=\,3\,kpc (considering an isothermal sphere truncated at R$_{\rm max}$ as the gravitational model for the host galaxy) as 
\begin{center}
\begin{equation} 
\label{vescape}
v_{\rm esc}\,=\,\sqrt{ \frac{2 G M_{\rm dyn} \times \left[ 1 + ln \left(\frac{R_{\rm max}}{r}\right) \right] }{3 r}   } \ kms^{-1}, 
\end{equation}
  \end{center}   
where G  is the gravitational constant (i.e., 4.3\,$\times$\,10$^{\rm-3}$ pc\,M$_{ \sun}$\,(\kms)$^{\rm2}$) and R$_{\rm max}$/r\,=\,10. This approach allows us to take into account both  the dependence of the dynamical mass estimation to rotation and dispersion motions and assumes that the halo drag is negligible. We refer to \citet{Bellocchi2013} for a detailed description of the dynamical mass estimation. \\
For the two galaxies (F05189-2524 and F07027-6011\,(N)) for which the dynamical mass estimation was not possible as a result of AGN contamination (\citealt{Bellocchi2013}), we calculate v$_{\rm esc}$, as in \citet{Veilleux2005} as follows:
  \begin{figure*}
 \centering
\includegraphics[width=.99\textwidth]{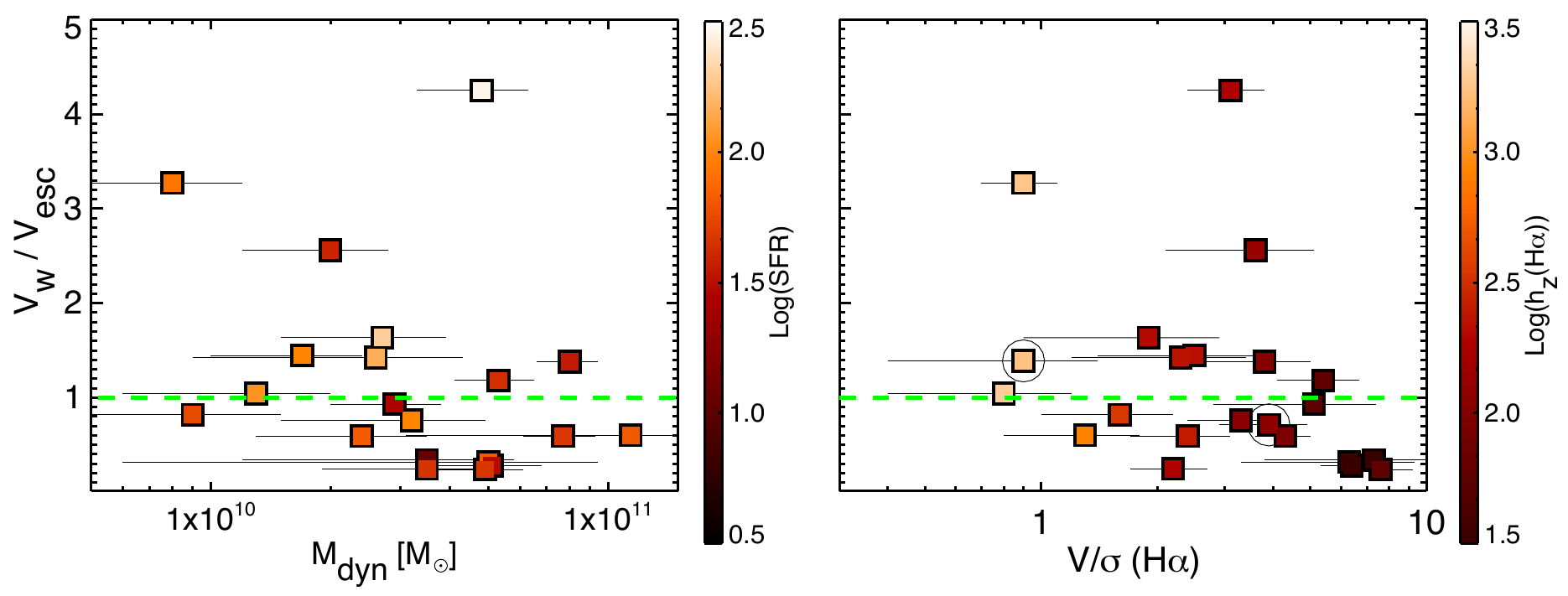}  
  \caption{Ratio of wind velocity  and galaxy escape velocity (Table\,\ref{Table_esc}) plotted against the dynamical mass (left), and the  \Ha \ dynamical ratio from \citet{Bellocchi2013} (right). These plots are color coded by the logarithm of the SFR and disk thickness of the ionized gas disks (Table~\ref{Table_esc}), respectively, right and left. The green horizontal line indicates where V$_{\rm w}$/v$_{\rm esc}$ is 1. In the right panel,   galaxies with no dynamical mass estimation (F05189-2524 and F07027-6011\,(N); \citealt{Bellocchi2013}) are indicated with circles (as Fig.\,\ref{Fig_feedback}). } 
 \label{escape}                  
\end{figure*}   
\begin{center}
\begin{equation} 
\label{vescape}
v_{\rm esc}\,=\,\sqrt{2 \times v_{\rm rot}^{2} \times \left[1 + ln \left(\frac{R_{\rm max}}{r}\right)\right]} \ kms^{-1}. 
\end{equation}
\end{center}  
Similar to the previous case, this simpler approach  assumes a truncated isothermal gravitational potential and no halo drag.  We consider that the  \Ha \ velocity amplitude, which is defined as the half of the observed peak-to-peak velocity corrected for the inclination of the galaxy, (\citealt{Bellocchi2013}) is a good proxy for v$_{\rm rot}$ and R$_{\rm max}$/r\,=\,10 (as in the previous case). \\
Table\,\ref{Table_esc} lists the escape velocities and outflow velocity in units of the escape velocity i.e., V$_{\rm w}$/v$_{\rm esc}$.  This ratio ranges from  0.2 to  4.2 (Fig.\,\ref{escape}) with a median (average) value of 0.9 (1.1).  \\
There are ten objects  for which V$_{\rm w}$/v$_{\rm esc}$\,$>$\,1 and, therefore, a significant amount of their outflowing gas could pollute the IGM (Fig.\,\ref{escape}, right). The most extreme case is F06206-6315, where an AGN is likely playing a role in boosting the velocity of the outflowing gas to exceed the escape velocity  (see also Sect.\ref{windorigin}). Excluding this object, the sample indicates a (weak) correlation in the sense that  V$_{\rm w}$/v$_{\rm esc}$ is higher for the less massive galaxies as found for the ionized outflows \citep{Arribas2014}. This result is consistent with the prediction that the less massive galaxies have the most severe impact on the IGM pollution.\\
In the majority of the cases, we find that the median wind velocities do not exceed the escape velocities (i.e., V$_{\rm w}$/v$_{\rm esc}$$<$\,1, Table\,\ref{Table_esc}), indicating that most of the cold outflowing material that we probe with our NaD measurement is likely falling back to the disk.  This circulation of the gas could in principle favor the redistribution of metals, modifying the abundance gradients in the galaxy disk (\citealt{Spitoni2010}). \\
This fountain scenario is consistent with the presence of clouds of neutral gas, either above the main disk in disk-like layers (in slow rotation) or sparsely distributed (i.e., raining back showing, in projection, an irregular velocity field). In the latter case,  the neutral gas is still raining back in the form of high velocity clouds (HVCs; \citealt{Spitoni2013}), which are gravitationally bound to the host galaxy, but not virialized. \\
However, V$_{\rm w}$ refers here to the median velocity in the outflow and, therefore, even for values of V$_{\rm w}$/v$_{\rm esc}$$<$\,1, part of the gas entrained in the wind could in principle escape.  \\
The efficiency of the observed winds of polluting the IGM, is in any case, a lower limit. Specifically, by using absorption lines as tracers of the outflowing gas we are limited to observing gas at a distance smaller than the projected size of the stellar disk. Tenuous and hot material (in absence of radiative cooling) or molecular gas, may be  found at larger radii (up to 10 Kpc in M82, \citealt{Lehnert1999}) and is  more likely to escape. \\
In Fig. 11 we consider the ratio V$_{\rm w}$/v$_{\rm esc}$ as a function of dynamical mass and dynamical ratio of the ionized disk(i.e., V/$\sigma$,
from \citealt{Bellocchi2013}) in order to study the factors controlling the recycling of gas and metals. As expected, gas and metals are more likely to escape to the IGM  (higher V$_{\rm w}$/v$_{\rm esc}$) in galaxies with small potential wells. \\
In addition, we note that there is a  tendency for V$_{\rm w}$/v$_{\rm esc}$ to increase in turbulent  and thicker ionized gas disk  (i.e., higher disk height and lower dynamical ratio, Fig.~\ref{escape}, right). In such disks, the gas turbulence may help the wind to escape, thereby enhancing its vertical growth. This is in agreement with a scenario in which the wind fluid follows the path of least resistance \citep{Cooper2008}.

 \subsection{Wind engine and energetics \label{windorigin}}
                                              
The outflow kinematic properties are more extreme in galaxies that host a powerful AGN (\citealt{Veilleux2005}). Therefore, the outflow velocity can be used as a discriminant of starburst-driven versus AGN-driven winds as suggested by Rupke, Veilleux, $\&$ Sanders (2005) and Rupke $\&$ Veilleux (2013). The influence of AGNs on the  velocity (and power) of outflows has also been seen  in  ionized (\citealt{Arribas2014}) and molecular (\citealt{Cicone2014, Sturm2011}) wind phases.\\
As mentioned in Sec.\,\ref{detectionrate},  the ULIRG F06206-6315 likely hosts an AGN-driven GW, according to the observed outflow kinematics  (Table \ref{Table_GWs}). In addition to this extreme case, there are ten other  objects where  evidence for the presence of an AGN has been found according to X-ray and infrared observations or optical emission line ratios and kinematics (Table~\ref{Tab_sample}). Despite this, from the observed neutral wind kinematics in these objects  (Table\,\ref{Table_GWs}) it is not obvious that the AGN is significantly boosting the outflows; this is probably because the AGN is not powerful enough.\\
To further address the issue about the wind driver, the energy (E$_{\rm w}$) and power (P$_{\rm w}$) of the winds are calculated, using the shell formalism and the FW model outlined in Sect.~\ref{feedback}, as follows:
\begin{center}
\footnotesize{ 
\begin{equation} 
E_{\rm w}\,=\,\sum \limits_{\rm k=1}^N M_{\rm w,k}\, \times \left(\frac{v_{\rm w,k}^{2}}{2}+\frac{3\sigma_{\rm w,k}^{2}}{2}\right) ergs.
\end{equation}}
\end{center} 
\begin{equation} \label{energydot}
\begin{array}{c}
P_{\rm w}\,=\,1.4 \times 10^{41} \sum \limits_{\rm k=1}^N \left(\frac{C_{\rm \Omega,k}}{0.4} C_{\rm f}\right) \left(\frac{R_{\rm w,k}}{10  \, kpc}\right) \left(\frac{N_{\rm H,k}}{10^{21} cm^{-2}}\right) \left(\frac{V_{\rm w,k}}{200  kms^{-1}}\right)  \times 
\\ 
\left(\frac{V_{\rm w,k}}{200  kms^{-1}}\right)^{2} + 1.5 \times \left(\frac{b_{\rm w,k}}{200  kms^{-1}}\right)^{2} ergs s^{-1},
\end{array}
\end{equation}
   \begin{figure}
\centering
\includegraphics[width=.493\textwidth]{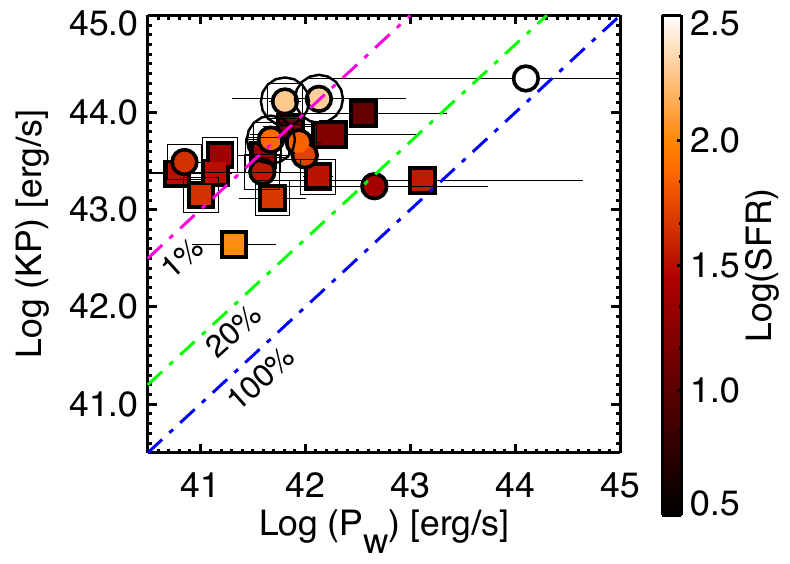} %
\caption{Logarithm of kinetic power of the starburst associated with SNe (KP) as a function of the wind  power (P$_{\rm w}$), color coded by their SFR. Symbols are the same as in Fig.\,\ref{V_SFR}. Specifically, filled squares and circle are for starburst and  for AGN hosts, respectively, while very strong AGN are shown with an additional second circle (as in Fig.\,\ref{Fig_feedback}). In addition,  those galaxies for which we were unable to estimate  the wind morphology in detail are indicated with an additional second square. The pink, green, and blue lines represent the positions for which the power of the wind is equal to the 1$\%$, 20$\%$ and 100$\%$, respectively, of the kinetic power supplied by the starburst.}
 \label{energy_SFR}              
\end{figure} 
where $\sigma$ is the velocity dispersion and b, the Doppler width (defined as: b\,= FWHM/2$\sqrt{ln2}$; \citealt{Rupke2005c}) for each spaxel.  As in Sect.\,\ref{feedback}, for the ten cases for which we cannot estimate the wind morphology, we measured  V,  N$_{\rm H}$, and b on spaxel-by-spaxel basis and used their median values (calculated over the area where the wind is observed). In addition, we assumed a  wind radius and opening angle of 3 kpc  and 0.4, respectively. \\
Excluding the case of F06206-6315, the estimated wind energies are in the range from 3\,$\times$\,10$^{\rm55}$ ergs to 1.2\,$\times$\,10$^{\rm57}$ ergs, while the powers are in the range between 6\,$\times$\,10$^{\rm40}$\,ergs\,s$^{\rm-1}$ and 1.4\,$\times$\,10$^{\rm43}$\,ergs\,s$^{\rm-1}$. The median energy and power are 4\,(2)\,$\times$\,10$^{\rm 56}$  and 14\,(6)\,$\times$\,10$^{\rm 41}$\,ergs\,s$^{\rm-1}$,
respectively, excluding (including) the winds because it was not possible to determine their morphology. \\
For comparison, \citet{Rupke2005c} found that the typical wind energy and power in local LIRGs are 5\,$\times$\,10$^{\rm56}$ ergs and 4\,$\times$\,10$^{\rm41}$\,ergs\,s$^{\rm-1}$, respectively. The average wind energy in the present  work is generally consistent within errors with that of the above mentioned work, although our estimation of the wind power is  generally larger (by a factor of about three).\\
To investigate the wind origin, in Fig.~\ref{energy_SFR}, we  compared the wind power of the neutral outflows and the kinetic power of the starburst associated with SNe (i.e., KP). We calculated the latter with the recipe of \citet{Veilleux2005}: KP\,$\sim$\,7\,$\times$\,10$^{41}$\,SFR/M$_{\rm \sun}$yr$^{\rm-1}$. In nearly all of the cases, the  wind power is within 1$\%$ and 20$\%$ of the kinetic power supplied by the starburst (about $\sim$\,5$\%$ on average). This is in agreement with a scenario in which these winds are originated in dense core of powerful nuclear starbursts \citep{Veilleux2005}. However, our result is in partial contrast with hydrodynamical simulations, which often assume wind  thermalization efficiencies in the range 10-100$\%$ (e.g., \citealt{Strickland2000}) and direct measurements of the thermalization efficiency. For example, for the nearby starburst M82, \citet{Strickland2009} found that medium-to-high thermalization efficiencies ($>$\,30$\%$) are required in hydrodynamical models to match the set of observational constraints derived from hard X-ray observations. \\
We see only two cases,  for which an unlikely thermalization efficiency of $\sim$\,90-100$\%$ could be required. For the ULIRG F06206-6315, the presence of an AGN  suggests that its outflow is rather driven by the energy liberated by the accretion of gas into the black hole. For the LIRG F06592-6313, an AGN-driven outflow could be a reasonable possibility, although there is not evidence for this outflow at optical wavelengths. However, we note that the large uncertainties associated with P$_{\rm w}$ also make these cases consistent with lower thermalization efficiencies. For the other galaxies, the AGN contribution is in principle not required to explain the observed energetics.  

\begin{table*}
                        \caption[Sample]{Galaxy properties: escape velocities and disks thickness. \label{Table_esc}}
                                \begin{center}
                                        \tiny{\begin{tabular}{l c c c c }                                        \hline
                                        \hline
                                        
          ID1 &         v$_{\rm esc}$   & V$_{\rm w}$/v$_{\rm esc}$  &  h$_{z}$(H$\alpha$)   &  h$_{z}$(NaD)    \\
         IRAS&       \kms        && pc & pc                 \\
         (1)    &       (2)          &        (3)       &          (4)  &(5)               \\         
         \hline 
         F01159-4443\,(N)               &  397  & 0.3    & 35 $\pm$ 7                   & $\cdots$\\              
         F01341-3735\,(N)                       &  332  & 0.3    &30 $\pm$ 5                       & $\cdots$\\    
         F04315-0840                      &  232        & 1.4    &218 $\pm$ 44                      & $\cdots$\\    
         F05189-2524                      &  258  & 1.4          &1757 $\pm$ 352                     & $\cdots$\\
         F06206-6315                    &  389  & 4.2   &187 $\pm$  37                          & $\cdots$ \\              
         F06259-4780\,(N)                &  286 & 1.4    &227 $\pm$ 45                          & $\cdots$\\      
         F06259-4780\,(C)                &  $\cdots$ & $\cdots$ &       246$\pm$  49     &   2759 $\pm$  552 \\ 
         F06592-6313                       &  251       & 2.6 & 135 $\pm$ 27                      & $\cdots$ \\ 
         F07027-6011\,(N)                &  416 & 0.7   &       110 $\pm$ 22              & $\cdots$ \\  
         F07027-6011\,(S)  $^{\ddagger}$ &  275 & 0.6   &248 $\pm$ 50                   & 2431 $\pm$  486  \\
         F07160-6215                       &  502      & 1.4 &  100 $\pm$ 20                              & $\cdots$\\
         F09437+0317\,(N)               & $\cdots$ &$\cdots$&   111 $\pm$22             &  713   $\pm$ 143 \\ 
         F09437+0317\,(S)               & $\cdots$ &$\cdots$&    46 $\pm$ 7               & 1879  $\pm$ 376 \\ 
         F10015-0614                      &  492        & 0.6 & 87 $\pm$ 17                      & $\cdots$\\
         F10038-3338                      &  203        & 1.0 & 1969 $\pm$ 394                     & $\cdots$\\ 
         F10257-4339                      &  318        & 0.8 & 109 $\pm$ 22                      & $\cdots$\\ 
         F10409-4556  $^{\ddagger}$ &  409 & 1.2 & 57 $\pm$ 11                  &  1390 $\pm$  278 \\ 
         F10567-4310                       &  401       & 0.3 & 32 $\pm$ 6                               & $\cdots$\\
         F11255-4120                       &  303       & 0.9 & 56 $\pm$ 11                              &  $\cdots$\\ 
         F11506-3851 $^{\ddagger}$  &  393 & 0.2 &35 $\pm$ 6                            &  1541 $\pm$ 308\\ 
         F12043-3140\,(S)               & $\cdots$ &$\cdots$  &  347 $\pm$ 69      &  1315 $\pm$  263 \\ 
         F12115-4656            & $\cdots$ &$\cdots$  & 40 $\pm$ 8              & 1845 $\pm$ 369  \\ 
         12116-5615 $^{\ddagger}$       & 159   & 3.2 &          1779 $\pm$ 355                             & $\cdots$\\ 
         F13001-2339                      & 600         & 0.6 &  828 $\pm$  166                            & $\cdots$\\ 
         F13229-2934                       & 332        & 0.2 &    191 $\pm$ 38                      & $\cdots$\\
         F14544-4255\,(E)               & $\cdots$ &  $\cdots$ &        142  $\pm$ 28       &1287  $\pm$  257 \\ 
         F18093-5744\,(S)               & $\cdots$ &  $\cdots$ &        122  $\pm$ 24       & 1261  $\pm$ 252  \\ 
          F21453-3511                   &  169   & 0.8  &350 $\pm$ 70                   & $\cdots$ \\
          F22132-3705                   &  $\cdots$     &$\cdots$ &     43 $\pm$ 9         &   922 $\pm$ 184\\                
        F23128-5919                              &  292  & 1.6 &        200 $\pm$ 40                & $\cdots$\\
         \hline
         \hline
                                        \end{tabular}
                                        }
                                        \end{center} 
        \tablefoot{Column\,(1): ID. Column\,(2): Escape velocity for those galaxies hosting a GW (Table\,1) as derived in Sect.\,\ref{fountain}. Column\,(3): The ratio between the inclination-corrected wind median velocities (Table\,\ref{Table_GWs}) and the escape velocity.  Columns (4) and (5): Vertical ionized and neutral gas disk heights traced via H$\alpha$ (narrow component, \citealt{Bellocchi2013}) and NaD, respectively (Sect.\,\ref{disks}).  As in Table~\ref{Table_GWs}, the symbol:  $^{\ddagger}$, marks the galaxies for which a two-components NaD modeling have been done (Sect.~\ref{LF}).   We assumed a conservative 20$\%$ systematic error for all the h$_{\rm z}$ values.} 
                                \end{table*}

\subsection{Thick disks in slow rotation \label{disks}}

The velocity fields of the neutral gas observed in 11 [U]LIRGs have the spider diagram pattern characteristic of a rotating disk (Figures in Appendix A; Cazzoli et al.\,2014 for the case of the LIRG F11506-3851). In  three galaxies (F07027-6011\,(S), F10409-4556, and F11506-3851), the rotation is  detected in one kinematic component, but an additional outflowing component is also detected (Figures \ref{Panel_F07027S} and \ref{Panel_F10409}, and \citealt{Cazzoli2014}). However, while in case of an ideal rotating disk the velocity dispersion map should be centrally peaked, the observed neutral gas velocity dispersion maps are generally rather irregular.\\
The observed NaD feature in these 11 galaxies shows a wide range of stellar contribution (Table\,\ref{Tab_stcont}), though in most of the objects (8/11) the stellar absorption is not dominant (i.e., $<$\,50$\%$). For the three galaxies where the NaD absorption is dominated by the stellar contribution (F09437+0317\,(S), F12043-3140\,(S) and F12115-4656),  the rotation curves may be  difficult to interpret in terms of neutral gas motions. However, it is important to consider that the stellar contribution is computed in the integrated light. Therefore, in some cases even if the stellar absorption in the integrated spectrum is substantial, the 2D distribution may help to identify regions where the NaD absorption is dominated by neutral gas or the stellar component, as in  F11056-3851 \citep{Cazzoli2014}.  Without the knowledge of the stellar properties and kinematics in these galaxies, we cannot investigate this further. \\
From our spectral maps, we extracted the velocity and velocity dispersion values for both  the neutral and ionized (i.e., H$\alpha$ narrow disk-like component) ISM phases in a $\sim$\,1\,$\farcs$0 pseudo-slit along the NaD major kinematic axis. The corresponding position-velocity diagrams (PV diagrams) are plotted in the upper panels of Fig.~\ref{DPV}, and show a variety of  shapes for the rotation curves of both ionized and neutral gas components  in the inner regions  (typically within 3 to 8 kpc). Similarly, in the lower panels of that figure, the velocity dispersion radial profiles are shown.\\
We found that for 9 out of 11 galaxies, the rotation axes of the neutral and ionized gas disks are fairly aligned, with offsets smaller than 15$\degr$ (Table\,\ref{Table_disks}), although in two cases (F10409-4556 and F09437+0317\,(S)) the kinematic axes of the neutral gas disks are poorly constrained. The neutral gas is found in slower  rotation than the ionized gas in the  majority of the cases (i.e., 8/11), while the neutral gas seems to corotate with the ionized gas in only two cases. In addition, in F06259-4780\,(C) the neutral gas disk is observed in counter-rotation with respect to the ionized gas disk. In this context, it is interesting to note that a counter-rotating neutral gas component was  also found in the inner few kpc of  M82 (Westmoquette et al.\,2012).\\
The radial profiles of the velocity dispersion are either flat (e.g., F18093-5744\,(S)) or with considerably large deviations from what is expected for a thin rotating disk at large radii. These deviations  are particularly evident in the case of F22132-3705 (with values in the range of 110-200 \kms) .\\
In Fig.~\ref{comparison_vsigma}, we show a comparison of the ionized and neutral V/$\sigma$ ratios of the ISM disks. We found that the neutral disks are dynamically hotter  (i.e., low V/$\sigma$ values, Table\,\ref{Table_disks}) than the ionized disks and,  therefore, they are likely to be thicker (i.e.,  larger h$_{\rm z}$).  This confirms that the ionized gas resides in regions of high density close to the   innermost regions of the disk, while the neutral gas is located further out.\\ 
Therefore, to estimate the h$_{\rm z}$ of the disks we follow two different approaches, which are outlined in  \citet{Cresci2009} and Binney $\&$ Tremaine (2008). Specifically, for the ionized disks we used the approximation proper for thin disks,  for which h$_{\rm z}$ can be computed as

\begin{center}
        \begin{equation} 
        h_{\rm z} \sim \frac{\sigma^{2}\times R}{\Delta V^{2}}.
        \end{equation}
        \label{eq_hz}
\end{center}   
For the neutral disks, however, we used an approximation more suitable for thick disks that is written as 
\begin{center}
        \begin{equation} 
        h_{\rm z} \sim \frac{\sigma \times R}{\Delta V}.
        \end{equation}
        \label{eq_hz}
\end{center}   
In both cases,  we considered as $\Delta$V the (inclination-corrected) semiamplitude of the velocity field, as $\sigma$ the mean velocity dispersion across the galaxy disk (excluding the nuclear regions) and a nominal radius (R)  of 2\,kpc. We chose this radius since it is the typical distance at which the rotation curve starts to flatten (Fig.\,\ref{DPV}). The results are  summarized in Table~\ref{Table_esc}. We find that the neutral gas disks are thicker by a factor up to 46 ($\sim$\,15 on average) than the ionized gas disks. \\ 
When the same approach (formula)  is used for the neutral  and ionized disks, the results indicate that the neutral disks are still  significantly thicker (i.e., by a factor 8 on average) than the ionized disks.

\begin{table*}
                        \caption[Sample]{Kinematics and dynamical support of thick neutral gas disks.}
                                \begin{center}
                                        \begin{tabular}{l c c c c c c c}
                                        \hline
                                        \hline
                                        
          ID1&          \ampl        & V$_{\rm shear}$ & \disp   &  $\Delta$PA &V/\disp     \\
         IRAS         &    \kms          &  \kms& \kms         &   $\degr$ &              \\
         (1)  &  (2)    &      (3)          &      (4)        &  (5) & (6)            \\
         \hline          

        06259-4780\,(C)                 &  78  $\pm$ 32  & 203 $\pm$ 105  & 95   $\pm$ 6    & 205 & 2.7 $\pm$  1.4      \\
F07027-6011\,(S) $^{\ddagger}$          &  87  $\pm$ 14  & 108  $\pm$ 51   &  81 $\pm$ 2     & 10 & 3.7 $\pm$ 1.1        \\
        F09437+0317\,(N)                & 221 $\pm$ 35  & 227 $\pm$ 34    & 73 $\pm$ 2     &  5 & 3.6  $\pm$ 0.6      \\
        F09437+0317\,(S)                &  98  $\pm$ 28  & 150 $\pm$ 29    & 82 $\pm$ 1     & 0 & 2.9  $\pm$ 0.5       \\
F10409-4556  $^{\ddagger}$              & 166 $\pm$ 22  & 168 $\pm$ 33   &  107 $\pm$ 1    & 0 &1.8  $\pm$ 0.3    \\
F11506-3851 $^{\ddagger}$               &   83 $\pm$ 12  &       61 $\pm$  22      &  83 $\pm$  12  & 6    & 0.8  $\pm$ 0.4      \\
         F12043-3140\,(S)               & 166 $\pm$ 22  & 130 $\pm$  36   & 86 $\pm$ 2     & 20 & 2.6   $\pm$ 0.6     \\  
         F12115-4656                            &162  $\pm$ 26  & 177 $\pm$  41   &  100 $\pm$ 1  & 10  & 4.0  $\pm$ 0.7     \\
         F14544-4255\,(E)               & 143 $\pm$ 36  & 178 $\pm$  55   & 90 $\pm$ 2     & 10 & 2.1 $\pm$  0.6      \\
         F18093-5744(S)                 & 149 $\pm$ 43  & 122 $\pm$ 31    & 75 $\pm$ 1     &  5 & 2.6 $\pm$ 0.6     \\
         F22132-3705                            & 212 $\pm$ 36  &  119 $\pm$  28   &   81 $\pm$  3  & 5 & 2.1       $\pm$  0.5          \\
                                        \hline
                                                                                \hline
                                        \end{tabular}
                                        \label{Table_disks}
                                        \end{center} 
        \tablefoot{Column\,(1): IRAS name. Column\,(2):  NaD velocity amplitude defined as  half of the observed  peak-to-peak velocity (i.e., half the difference between the maximum and minimum values is considered without applying the inclination correction) measured in the PV diagrams. Column\,(3): NaD velocity shear defined as  half of the difference between the median of the 5 percentile at each end of the velocity distribution, as   in \cite{Bellocchi2013} (not corrected for the inclination). Column\,(4): Neutral gas mean velocity dispersion.  Column\,(5): Kinematic misalignment between the position angles (i.e., PAs) of the neutral gas rotational major axis and that of ionized gas. The kinematic PAs are derived by inspecting the ionized and neutral velocity maps. Col (6): Neutral gas dynamical ratio between the velocity shear (corrected for the inclination) and mean velocity dispersion values. The symbol:\,$^{\ddagger}$ indicates the galaxies for which we carried out  two-component NaD modeling (Sect.~\ref{LF}).} 
                        \end{table*}


\section{Summary and conclusions}

\label{conclusions}

We have studied the properties of neutral gas outflows in a sample of 51  local \ul  (z\,$\leq$\,0.09) on the basis of VLT/VIMOS IFS of the NaD feature. For the analysis, we  followed two approaches. First, for each galaxy we combined the spectra  in the data cube to obtain a high S/N spatially integrated spectrum. Second,  for a subsample of [U]LIRGs,  we analyzed the NaD spectra on a spaxel-by-spaxel basis to  trace the spatially resolved 2D structure of the neutral gas.  The main conclusions can be summarized as follows:

\begin{enumerate}

\item \textit{Stellar and ISM contributions to the NaD feature.}\,We evaluated the contribution of old stars to the NaD absorption via a stellar continuum modeling (pPXF analysis) of the spatially integrated spectra. The fraction of stellar contribution ranges from about 20$\%$ to 100$\%$, but in nearly half of the cases for which high quality modeling was possible, the NaD is mainly originated in the ISM (i.e., stellar fraction $<$\,35$\%$). This analysis also allows us to derive a threshold of EW(NaD)\,$=$\,1.3\,$\AA$, above which the NaD absorption in [U]LIRGs is likely dominated by the ISM.  Since the results from the stellar continuum modelling on a spaxel-by-spaxel basis could be uncertain, we consider that threshold  to identify  regions dominated by ISM absorption in the maps. Ê\\

\item \textit{Outflow kinematics from the integrated spectra.}\, For the objects with a reliable stellar modeling, we generate a purely ISM NaD absorption spectrum after subtracting the stellar model.  In 22 objects we measure  blueshifted NaD profiles, which indicate typical neutral gas outflow velocities in the range 65-260 kms$^{\rm-1}$.  Excluding the galaxies with powerful AGNs, the neutral outflow velocity shows a dependency with the SFR of the type V\,$\propto$\,SFR$^{0.15}$, which is in fair agreement with previous results. The neutral outflow (central) velocities are significantly higher than those for the ionized gas,  but they are in rather good agreement with the ionized gas maximum outflow velocities that we considered. This suggests that the wind entrains and accelerates the cold ambient gas, which is likely located  at relatively large distances from the regions where the ionized gas resides. The velocity dispersions for both neutral and ionized outflows show similar median values (if the strongest AGNs are excluded). \\

\item \textit{2D Neutral outflows: Detection, morphology,  and kinematics.}\, In the neutral gas velocity fields of 22 out of 40 targets, we found clear signatures of GWs. These neutral winds are conical in shape in 12 out of 22 objects. For the remaining objects, we were not able to constrain the morphology mainly owing to projection effects.  We generally observe collimated outflows with C$_{\Omega}$\,$\sim$\,0.4\,.The inclination-corrected outflow velocities of the neutral gas entrained in GWs is in the range $\sim$\,80-700\,\kms, except for the ULIRG F06206-6315 for which the typical wind velocity exceeds 1000\,\kms. The typical velocity dispersions are in the range $\sim$\,95-190\,\kms \ (i.e., 230-460\,\kms in FWHM),  indicating either that a  wide range of velocities are integrated along the line of sight or that these winds are turbulent. The V-SFR relation inferred for the 2D analysis is similar to that obtained from the integrated spectra. No clear correlation is seem between the outflow velocity and the dynamical mass.  \\

\item \textit{GWs Feedback: Slowed star formation and gas recycling.}\,  Based on a simple FW model,  we found that the wind mass estimates range from  0.4\,$\times$\,10$^{\rm8}$\,M$_{\rm \sun}$ to  7.5\,$\times$\,10$^{\rm8}$\,M$_{\rm \sun}$ (1.6\,$\times$\,10$^{\rm8}$\,M$_{\rm \sun}$, on average), reaching up to $\sim$\,3$\%$ of the dynamical mass of the host. The mass rates are only $\sim$\,0.2-0.4 times the corresponding SFR in most cases indicating that, generally, mass losses are small for slowing down significantly the star formation. The derived mass loading factors ($\eta$\,=\,$\dot{M_{\rm w}}$/SFR) correlate better with the starburst infrared luminosity (r$_{PC}$\,$\sim$\,-0.6) than with the host galaxy mass (r$_{PC}\,$$<$\,0.1). The comparison of the median wind velocity and host escape velocity  indicates that,  in the majority of the cases, most of the outflowing neutral gas rains back into the galaxy disk.  We found that on average V$_{\rm w}$/v$_{\rm esc}$ is higher in less massive galaxies, confirming that the galaxy mass has a primary role in shaping the recycling of gas and metals. We also found a tendency in the sense that V$_{\rm w}$/v$_{\rm esc}$ is higher in more turbulent and thicker ionized gas disks.\\

\item \textit{GWs power source.}\, The comparison between the wind power and kinetic power of the starburst associated with SNe indicates that the starburst  could be the only main driver of the outflows in nearly all the [U]LIRGs galaxies, as wind power is generally lower than 20$\%$ of the kinetic power supplied by the starburst. In the case of  F06206-6315, the outflow is likely  AGN driven, as also indicated by its kinematics. \\
                                
\item \textit{Kinematics of disks.}\, A significant number of [U]LIRGs (11/40) show spider diagrams, such as neutral gas velocity fields, plus rather irregular velocity dispersion maps in one kinematic component. The comparison between the rotation curves of the ionized and neutral disks indicates that the neutral disk lags compared to  ionized disk in the majority of the cases (i.e., 8/11). While in two cases the neutral gas seem to nearly corotate with the ionized gas, we also find a case (IRAS F06259-4780\,(C)) in which the neutral gas disk counter-rotates with respect to the ionized disk. Our kinematics measurements indicate that nearly all the neutral gas disks are dynamically hotter and thicker (by a factor up to 46, 15 on average) with respect to the ionized disks.
\end{enumerate}

\begin{figure*}
 \centering
\includegraphics[trim = .55cm 11.8cm 12.3cm 3.cm, clip=true, width=.3\textwidth]{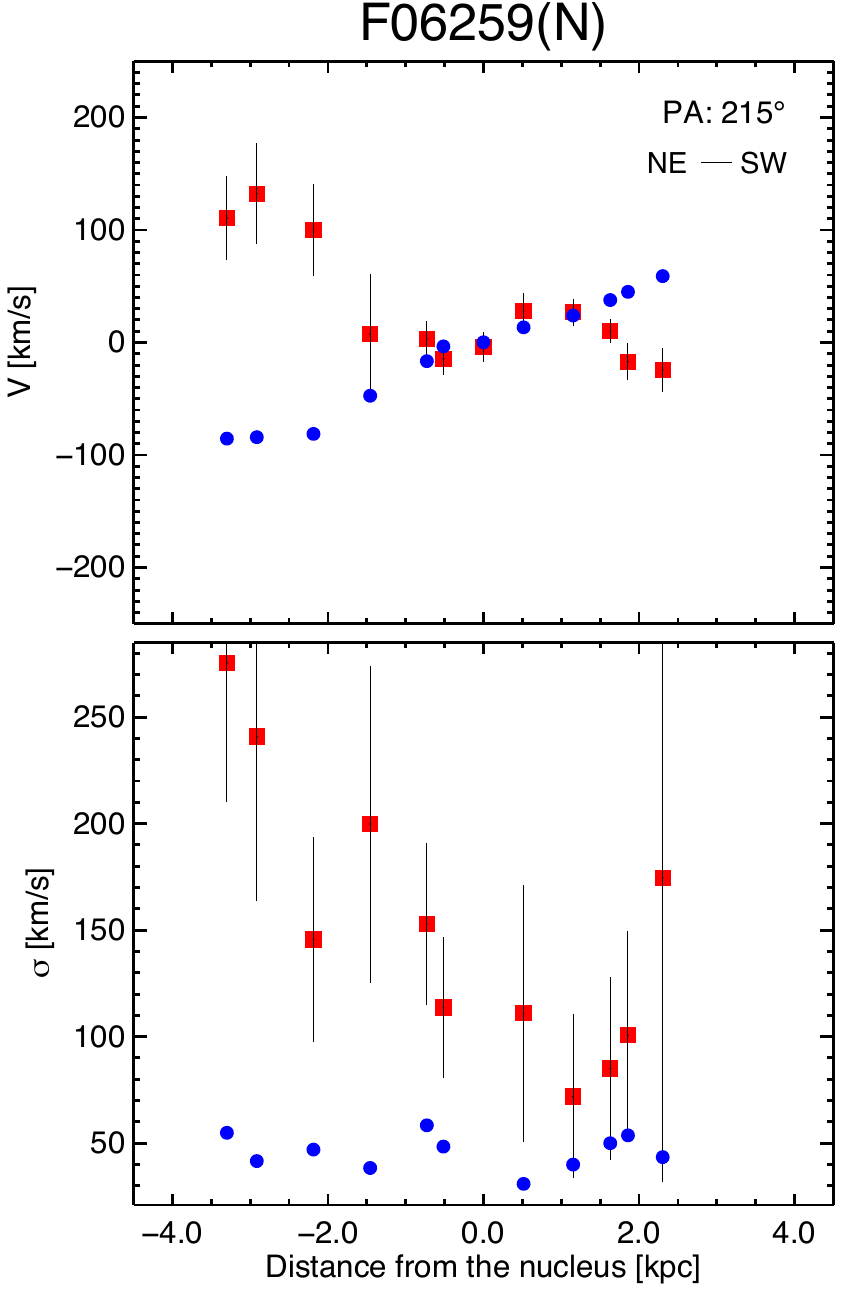} 
\hspace*{.25cm}
\includegraphics[trim = .55cm 11.8cm 12.3cm 2.5cm, clip=true, width=.3\textwidth]{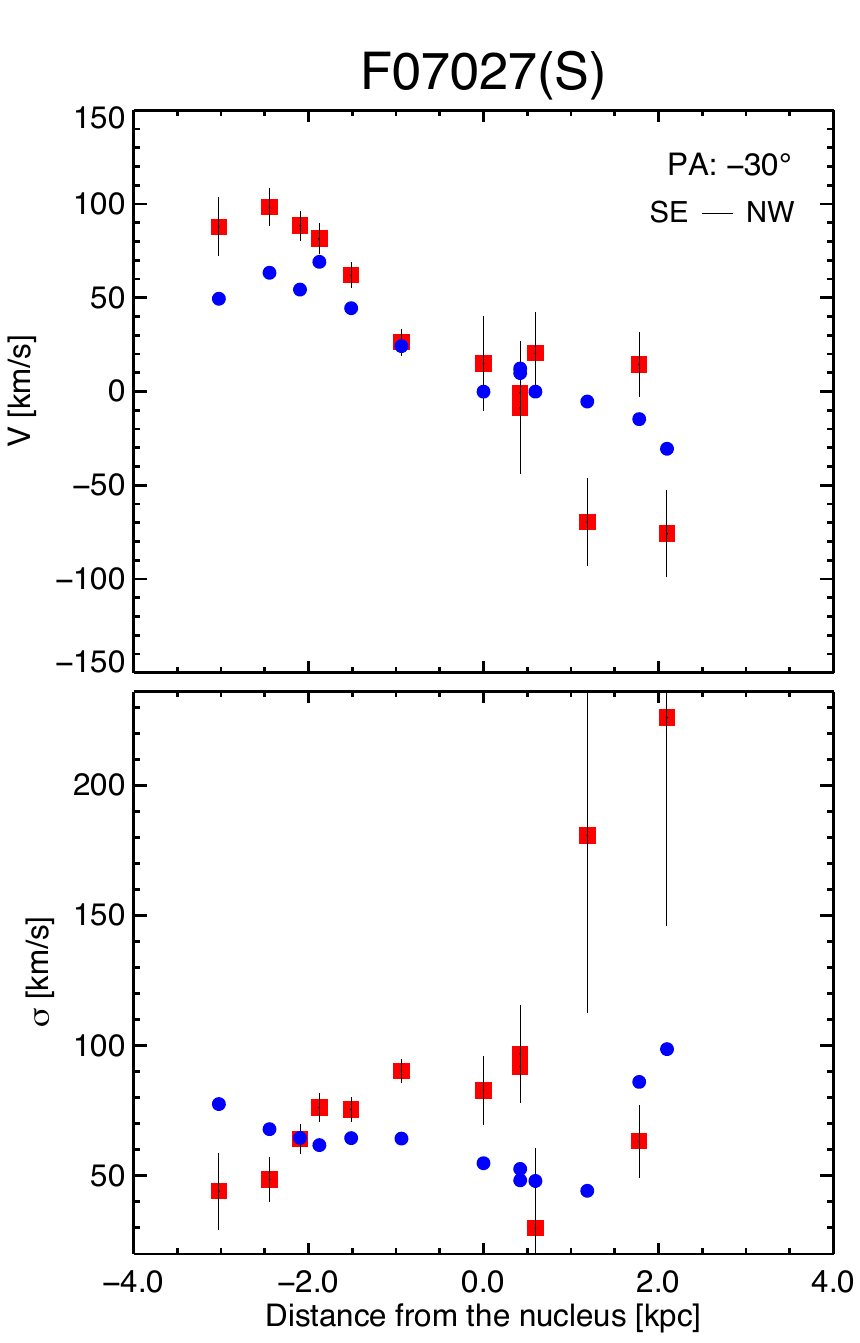}   \\  
\hspace*{.25cm}
 \includegraphics[trim = .55cm 11.8cm 12.3cm 2.5cm, clip=true, width=.3\textwidth]{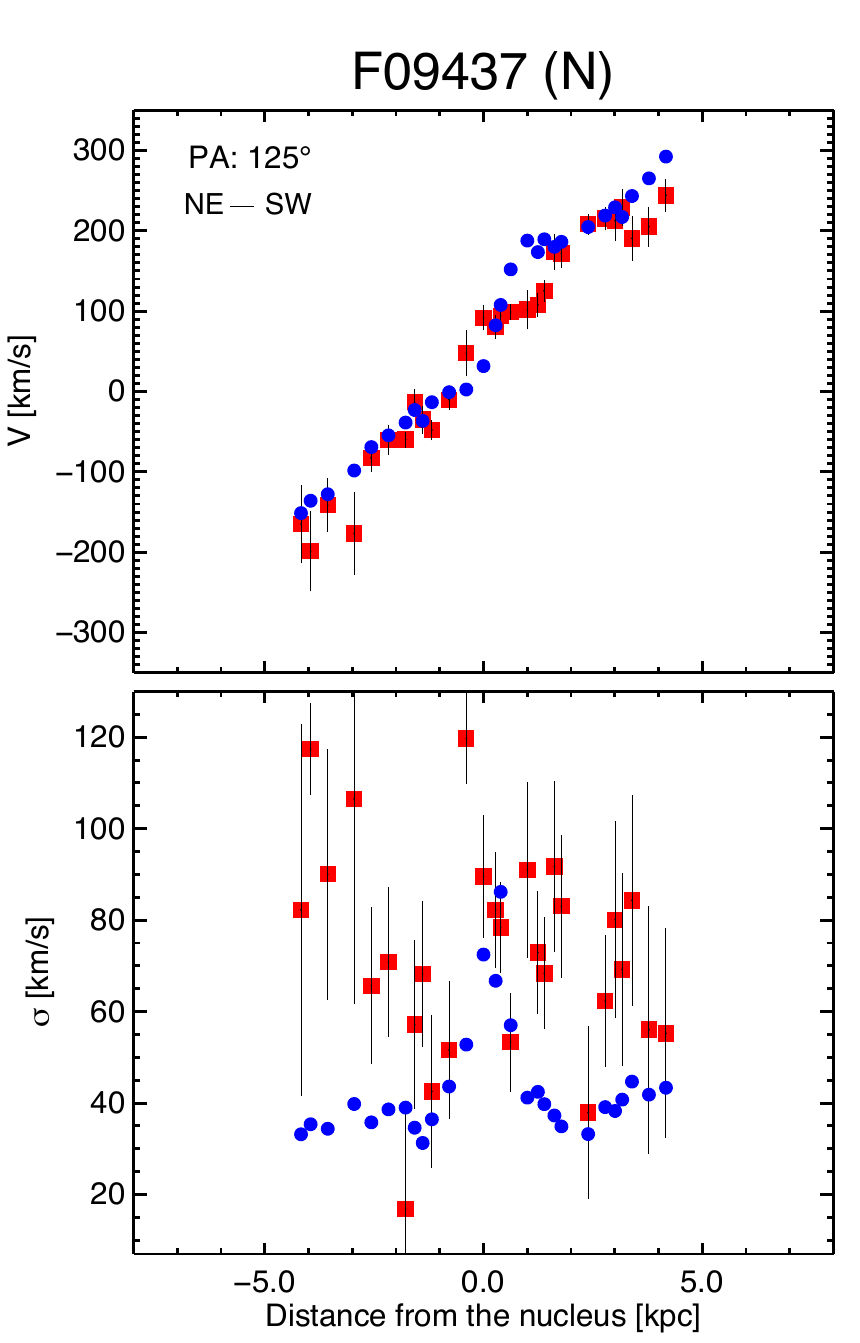} 
\hspace*{.25cm} 
 \includegraphics[trim = .55cm 11.8cm 12.3cm 2.5cm, clip=true, width=.3\textwidth]{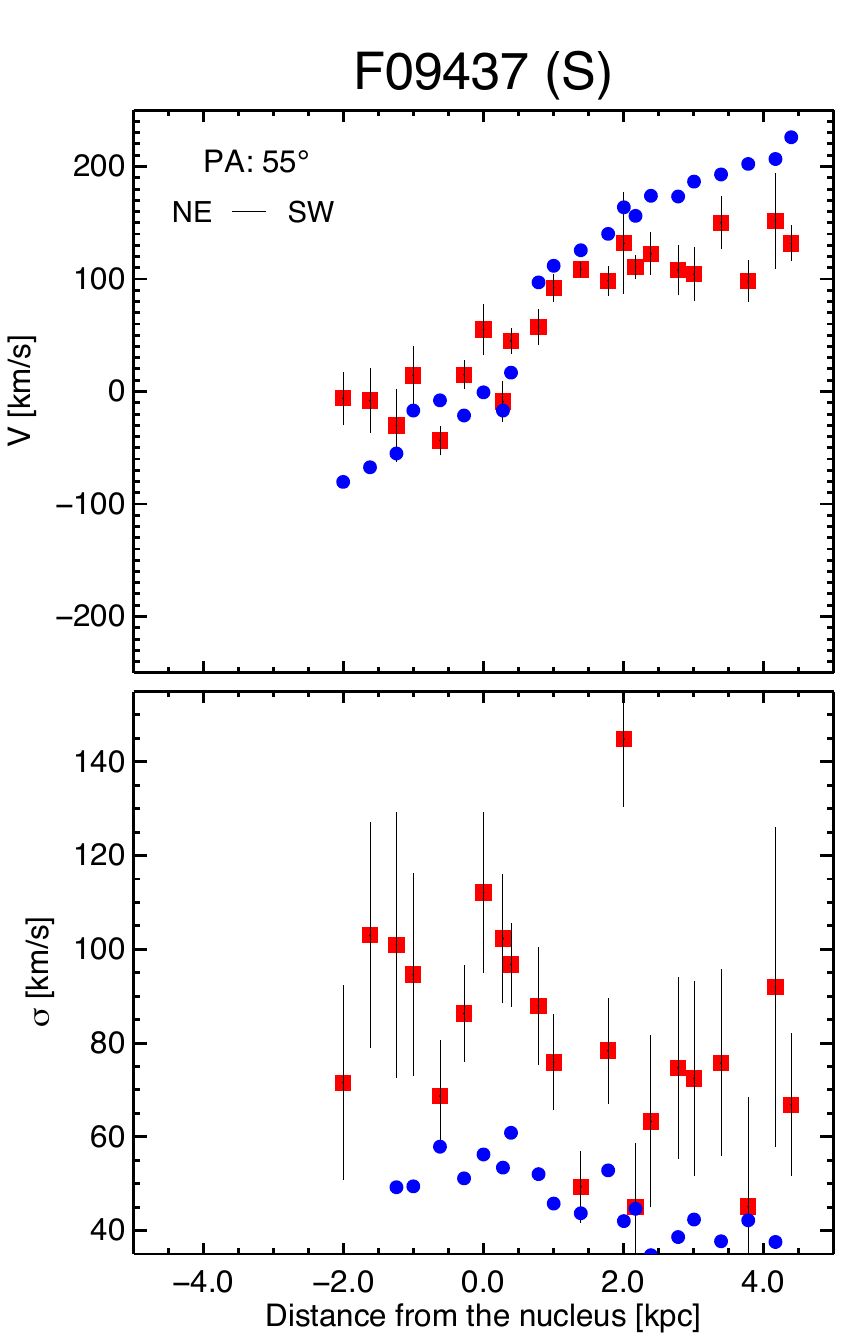}  \\
\hspace*{.25cm}                            
\includegraphics[trim = .55cm 11.8cm 12.3cm 3.cm, clip=true, width=.3\textwidth]{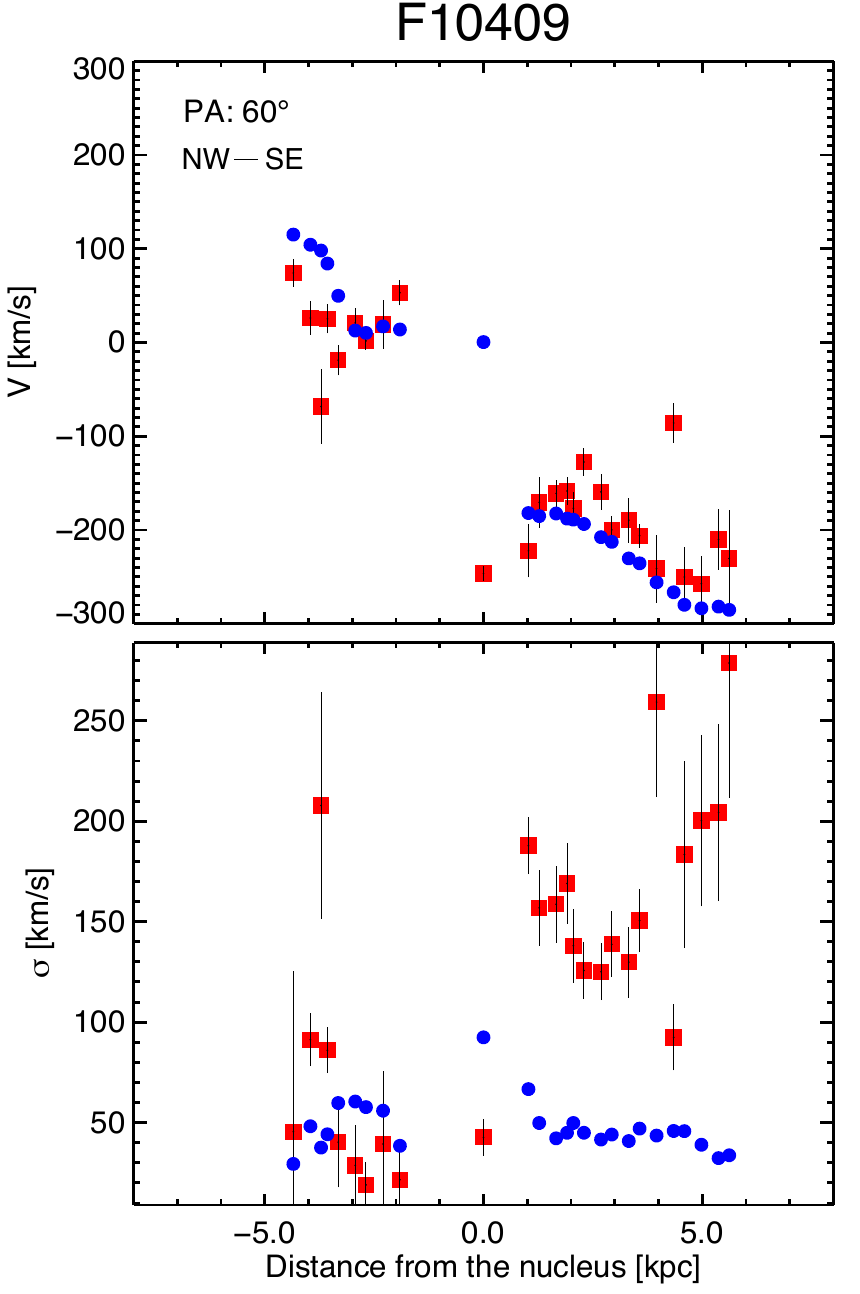} 
\hspace*{.25cm}
\includegraphics[trim = .55cm 11.8cm 12.3cm 2.5cm, clip=true, width=.3\textwidth]{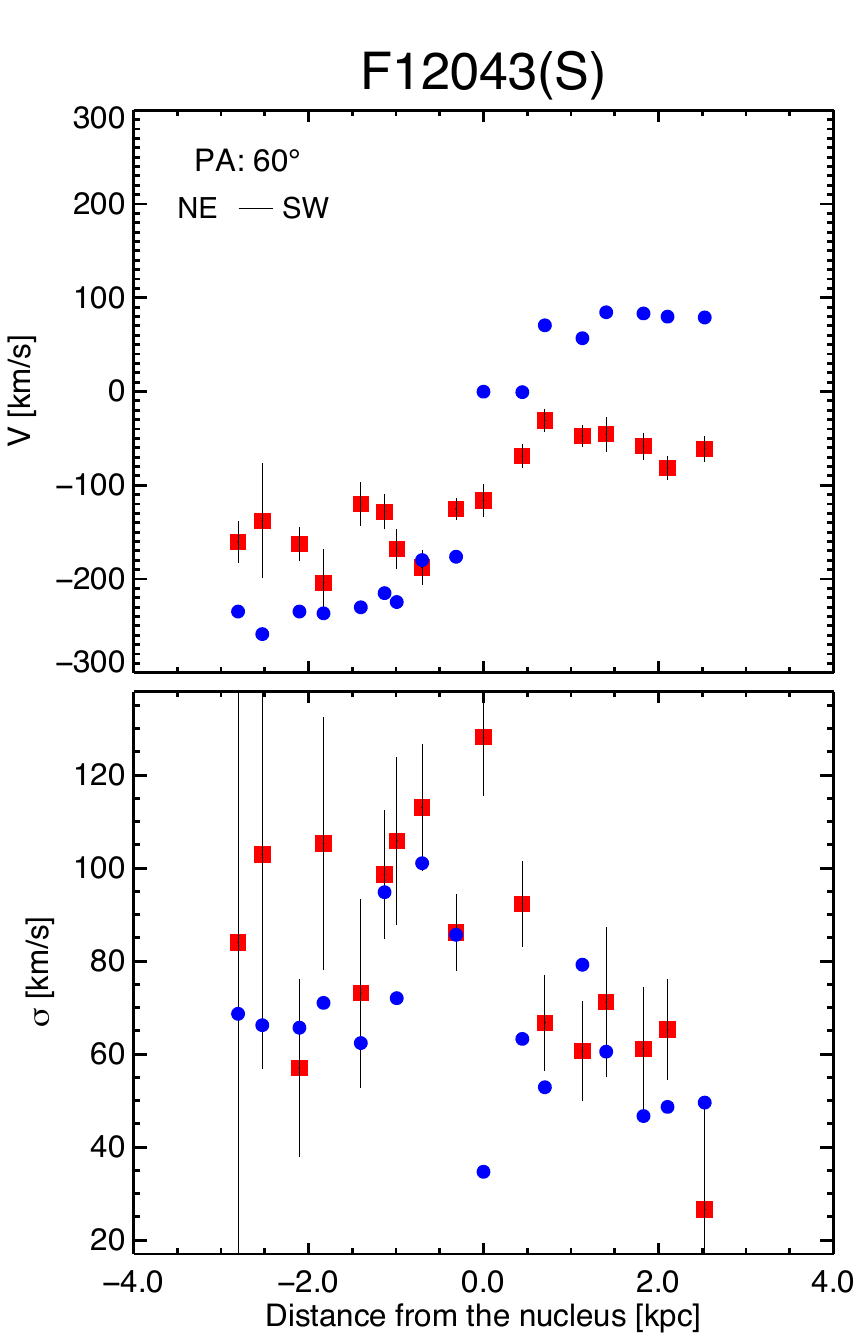}  \\
\captionsetup{labelformat=empty}{Fig.\,\ref{DPV}.} 
\end{figure*}
\begin{figure*}
\centering
 \includegraphics[trim = .55cm 11.8cm 12.3cm 2.5cm, clip=true, width=.3\textwidth]{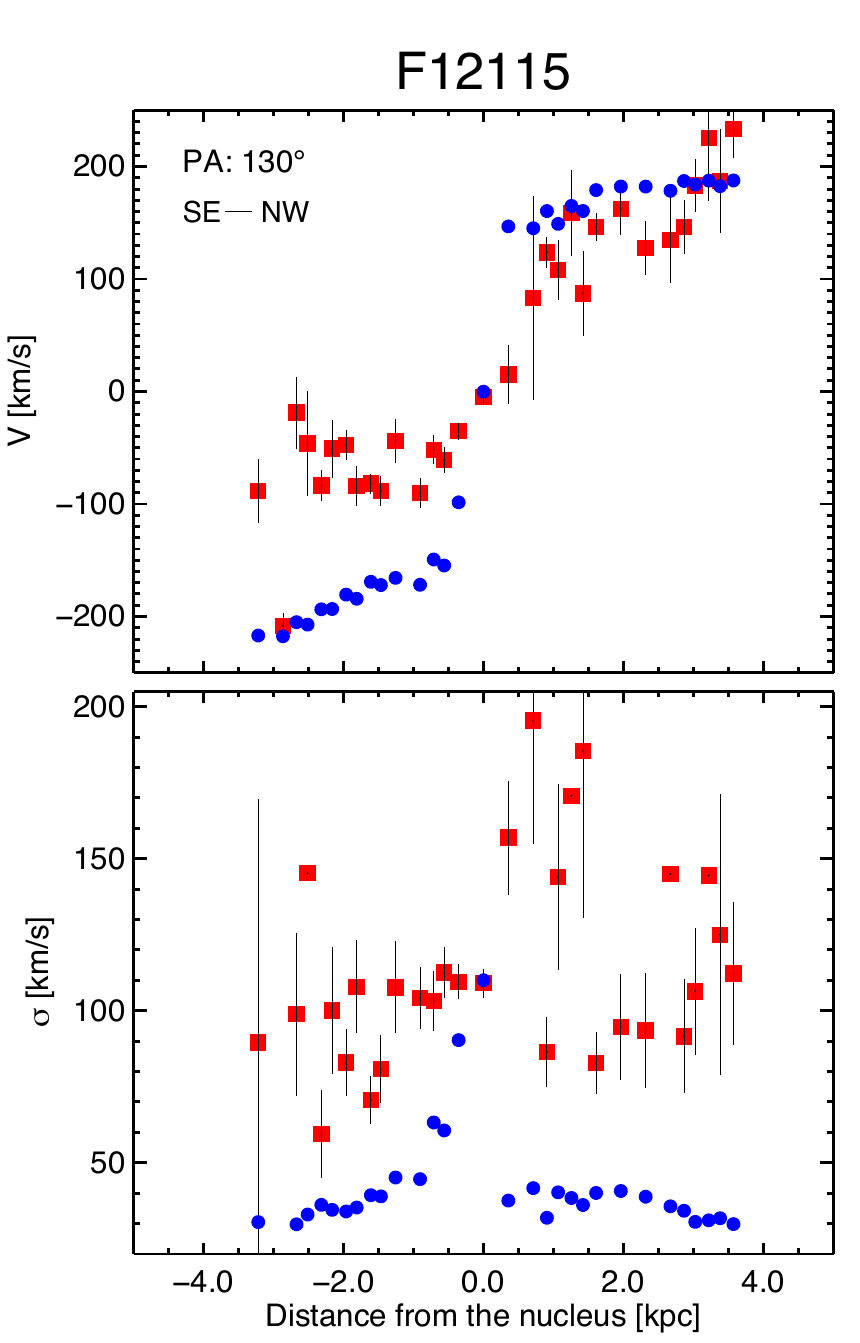}
\hspace*{.25cm}  
\includegraphics[trim = .55cm 11.8cm 12.3cm 2.5cm, clip=true, width=.3\textwidth]{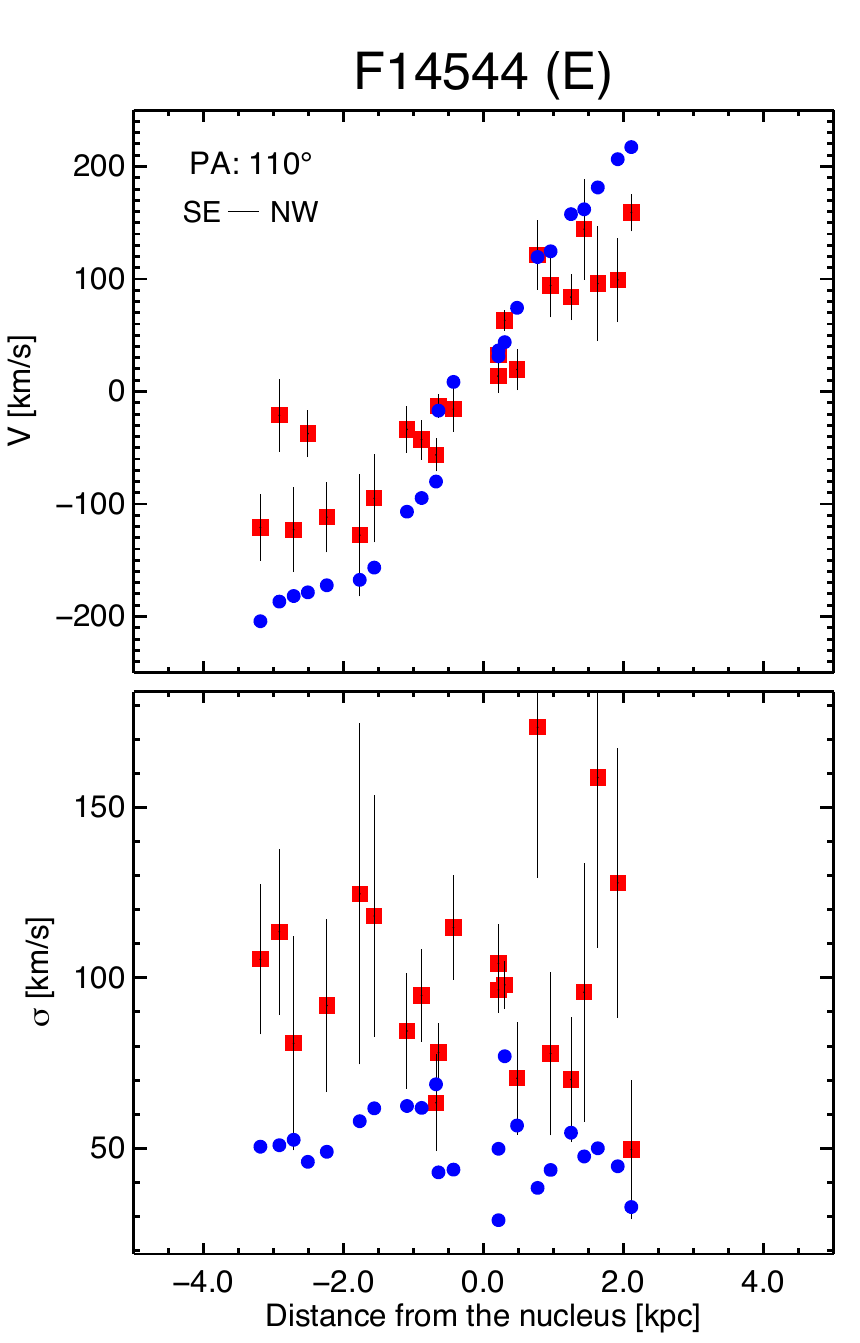}\\
\hspace*{.25cm}
 \includegraphics[trim = .55cm 11.8cm 12.3cm 2.5cm, clip=true, width=.3\textwidth]{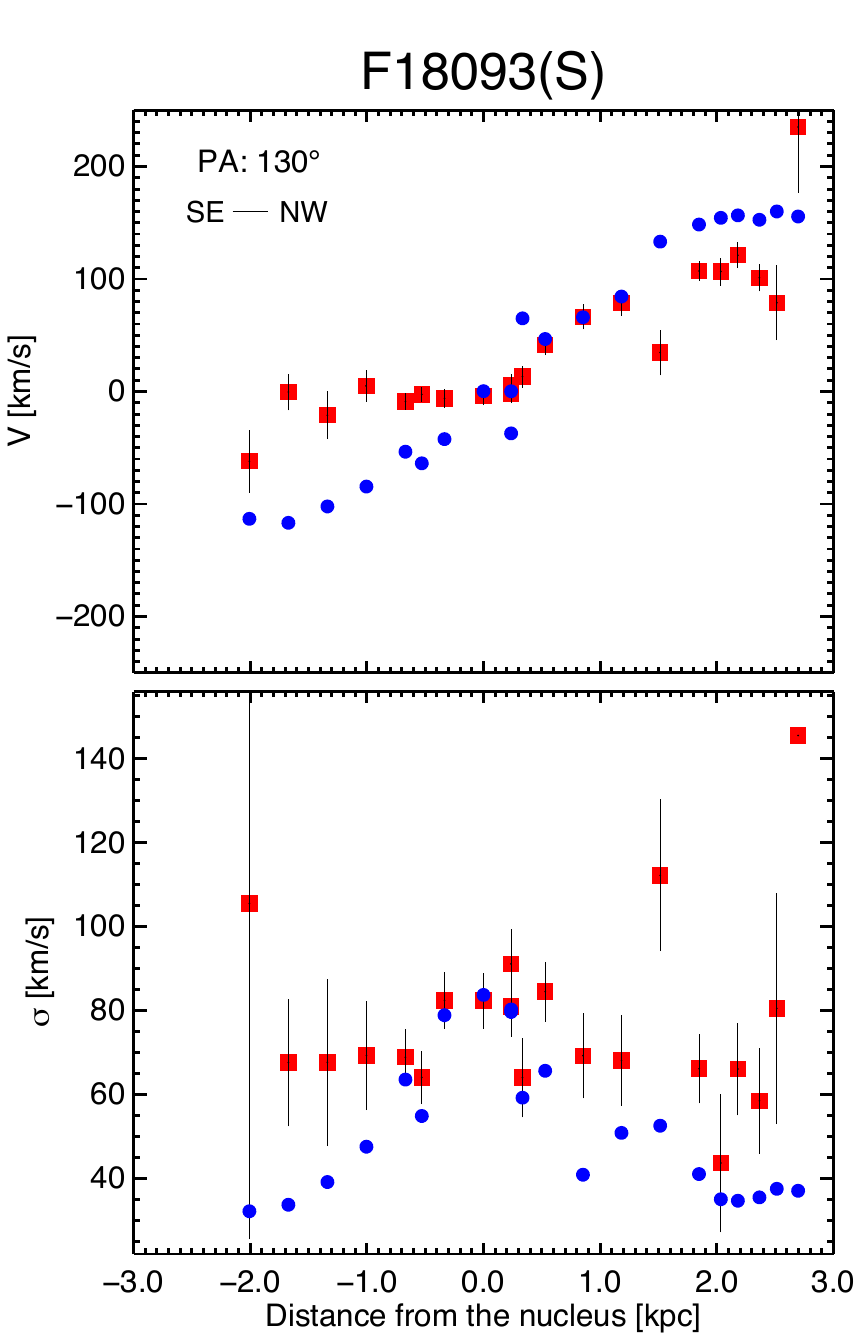}   
 \hspace*{.25cm}                
\includegraphics[trim = .55cm 11.8cm 12.3cm 2.5cm, clip=true, width=.3\textwidth]{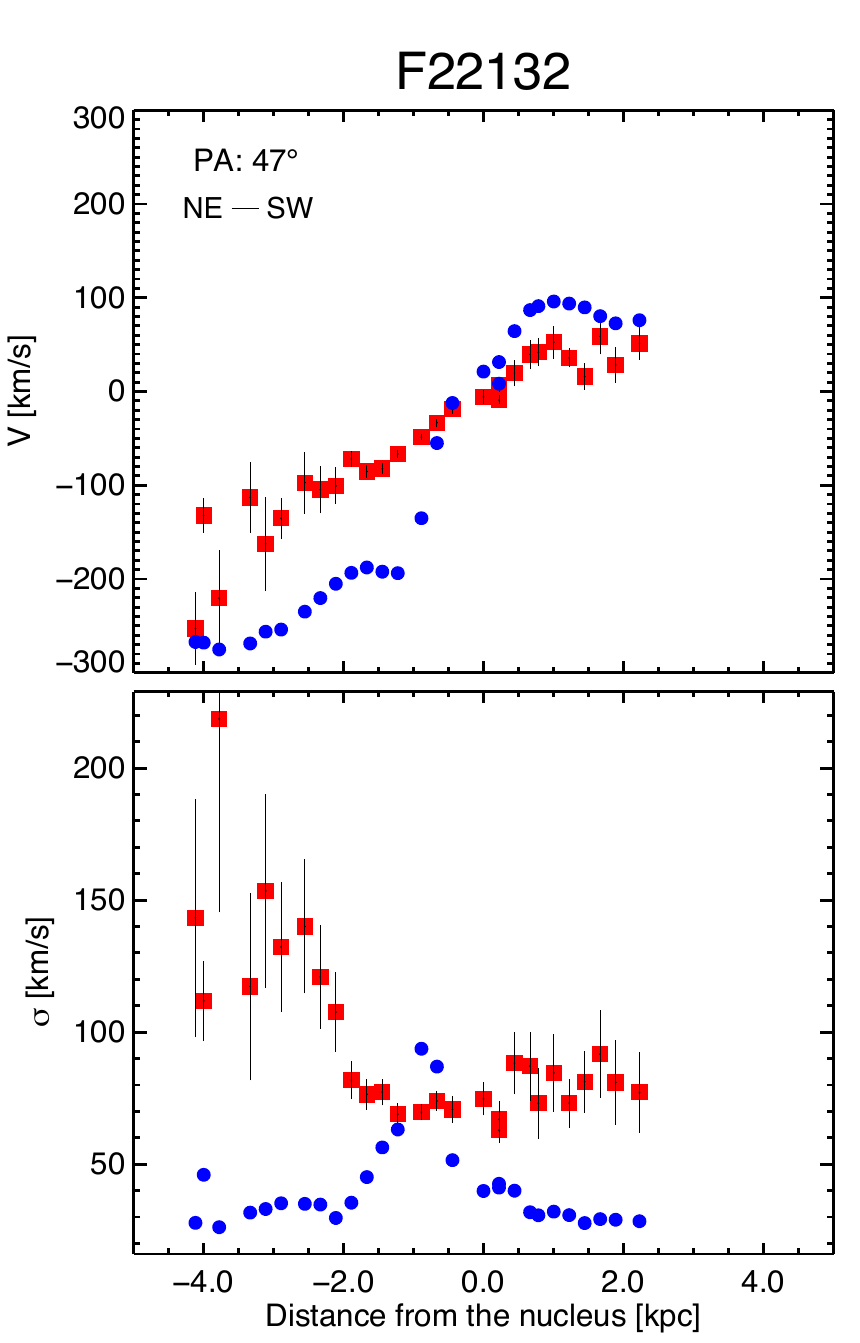}                            
\hspace*{.25cm}
 \caption{For those galaxies classified as disks (Table~\ref{Table_disks}), two panels aligned vertically show the PV curves (top) and the velocity dispersion radial profiles (bottom) along a pseudo-slit aligned according to the major axis of the neutral gas rotation (traced via NaD). In all  panels, the red squares and blue circles indicatethe points for the neutral and ionized gas, respectively, which is traced via the NaD absorption and  H$\alpha$ emission (narrow component). The galaxy IDs follow that of Table\,1, although here they are shortened for a better visualization. The velocity fields of  H$\alpha$ have been taken as reference for an ordinary rotation and typically extend up to a radius larger than the NaD velocity field. For a detailed analysis of the ionized gas kinematics, we refer to \cite{Bellocchi2013}. The LIRG IRAS 11506-3851 is discussed in detail by \cite{Cazzoli2014} thus the correspondent panel is not included here. }
                           \label{DPV}           
                \end{figure*}

\begin{figure}
\centering
\includegraphics[width=.49\textwidth]{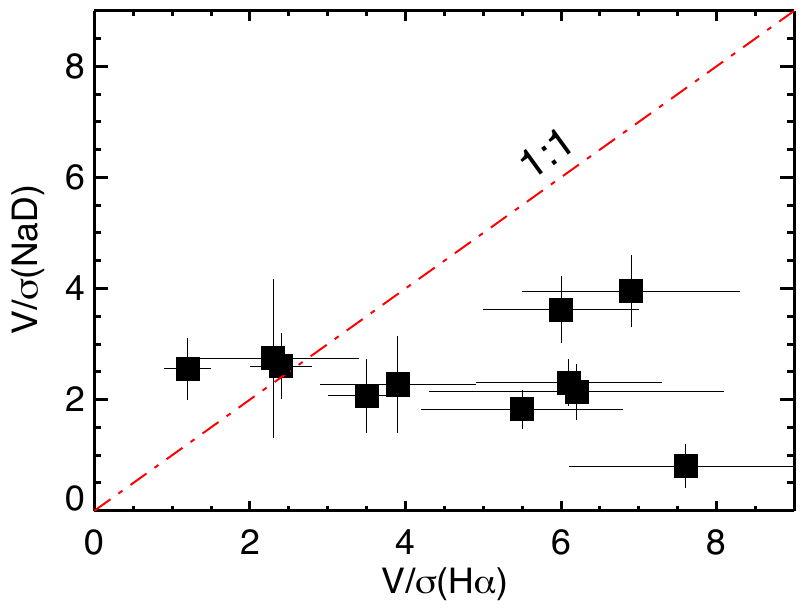}  
\caption{Comparison within the V/$\sigma$ ratios of ionized and neutral  ISM disks (Table\,\ref{Table_disks}). We refer to \cite{Bellocchi2013}  for the V/$\sigma$ values for the ionized gas. The red dash-dotted line represents the 1:1 correlation. Neutral disks are typically thicker and more dispersion dominated with respect to the ionized disks.}
\label{comparison_vsigma}                
\end{figure}

  \begin{acknowledgements}
  
We  thank the referee for useful comments and suggestions. This work was funded by the Marie Curie Initial Training Network ELIXIR of the European Commission under contract PITN- GA-2008-214227 and the  grants AYS2010 and AYA2012  by the Spanish Ministry of Science and Innovation (MICINN).  S.C. gratefully acknowledges the logistic and financial support provided by  the Cavendish Laboratory (Cambridge, UK). We are grateful to E.\,Bellocchi  for kindly providing us the H$\alpha$ maps shown in the Appendix. We also appreciate some information and help provided by M.\,Pereira-Santaella.\\
This work is based on observations carried out at the European Southern Observatory, Paranal (Chile), Programs 076.B-0479(A), 078.B-0072(A), and 081.B- 0108(A). This research has made use of the NASA/IPAC Extragalactic Database (NED), which is operated by the Jet Propulsion Laboratory, California Institute of Technology, under contract with the National Aeronautics and Space Administration.   
   \end{acknowledgements}


\bibliographystyle{aa} 
\bibliography{Bibliography}

  \begin{appendix}
  \section{Comments, maps, and integrated spectra of individual sources}
        \label{comments_maps}

The first part of this appendix is devoted to comment briefly on the spatially resolved and/or the integrated characteristics for each galaxy. In the second part, we present the neutral gas equivalent width (EW), velocity field (V), and velocity dispersion ($\sigma$) maps for those objects for which we were able to obtain spectral maps (Sect.~\ref{SO}). The VIMOS data for the LIRG IRAS 11506-3851 are discussed in detail by \citet{Cazzoli2014} thus the comments and correspondent maps are omitted here. \\

\noindent\textbf{F01159-4443\,(N).} The neutral gas kinematics of this galaxy shows deprojected blueshifited velocities ($\sim$\,130 \kms) that are larger than those of the systemic component of the ionized gas at the same location. Therefore, the observed neutral gas velocities cannot be explained with rotation and are interpreted as outflowing. \\ 
\noindent\textbf{F01159-4443\,(S).} This galaxy is not analyzed in this work either with the spatially integrated spectrum or via spatially resolved IFS of NaD because of  the low quality of  stellar modeling and   lack of any clear  GW or disk signature, respectively. \\ 

\noindent\textbf{F01341-3735\,(N).} We found  a prominent blue wing in the pure-ISM NaD line profile in the spatially integrated
spectrum of this LIRG. In the spectral maps,  we identified an outflowing neutral gas component with high velocity and velocity dispersion.  \\
\noindent\textbf{F01341-3735\,(S).} This galaxy was only studied via the integrated spectrum, where the NaD
doublet was modeled with a single kinematic component found at the systemic velocity.\\
 
\noindent\textbf{F04315-0840.} The neutral gas spatial distribution and kinematics of this post-coalescence late-merger starburst LIRG is very different from that of the ionized gas (traced via the H$\alpha$ narrow component). The neutral gas is mainly seen in the main body of the object, while it is almost absent in the spiral arm. A neutral wind  originates from the nucleus toward the southeast with velocity dispersions values higher than 90 \kms. The blue velocities of the \Ha \ broad component \citep{Bellocchi2013}  spatially overlaps the region where the neutral wind is detected, indicating the presence of a multiphase GW. Interestingly, the velocities of both wind phases are similar at $\sim$\,200-350 \kms.  In the integrated spectrum, we found a blueshifted kinematic component  at V\,$\sim$\,190 \kms \ in agreement with our  spatially resolved results. \\

\noindent\textbf{F05189-2524.} This merger ULIRG has a very compact nucleus and hosts an AGN that powers the high velocity outflow seen via NaD. The morphology of the outflow could not be constrained well  owing to projection effects. A noticeable blue wing is seen in the NaD ISM-absorption line profile in the integrated spectrum, along with the NaD resonant emission line (which is not a common feature of GWs in nearby galaxies) that is studied in detail by \citep{Rupke2015}. \\

\noindent\textbf{F06035-7102.} This galaxy is not analyzed in this work either with the spatially integrated spectrum or via spatially resolved IFS of NaD  because of the low quality of stellar modeling and  lack of any clear  GW or disk signature, respectively.\\

\noindent\textbf{F06076-2139\,(N) and (S).}  We only study these galaxies via their spatially integrated spectra. In both spectra, the NaD line profile were modeled with one broad and blueshifted kinematical component. However, the complex NaD line profile observed in the spectrum of the northern galaxy  likely require more than one (or two) kinematical component(s).\\       

\noindent \textbf{F06206-6315.} This ULIRG has two nuclei, which are clearly seen in the near-IR images \citep{Rodriguez2011} but not in the optical  continuum image. The tidal tail,  starting in the north and bending toward the southeast, is almost undetected via NaD.  While we did not analyze the low S/N integrated spectrum, in 2D, we observed part of the neutral gas as entrained in an AGN-driven outflow. The comparison with the galaxy escape velocity and wind velocities indicates that the neutral gas entrained in the wind is able to escape. A molecular gas outflows, traced by the infrared OH lines, was found by \citet{Spoon2013} at higher observed (terminal) velocity ($\sim$\,800\,\kms). \\

\noindent\textbf{F06259-4780\,(N).} In this galaxy, we detected extended neutral gas that is  outflowing nearly everywhere  in contrast to what is seen for the ionized gas (narrow component). In addition, neutral and ionized ISM phases  also have remarkably different spatial distribution. We identify the GW according to the most blueshifted velocity,  the highest velocity dispersion values, and also its direction. Indeed, the outflow is oriented along the minor axis of the rotation  seen weakly in the ionized velocity field. \\
\noindent\textbf{F06259-4780\,(C).} The 2D-kinematic of the neutral gas in the central galaxy of the system (the southern in Fig.\,\ref{Panel_F06259}) is classified as rotating disk. This gas is observed in counter-rotation with respect to the ionized disk. This makes this object unique in our sample since, in the nearly all  cases, we found slow neutral gas rotation with respect to the ionized gas. \\
\noindent\textbf{F06259-4780\,(S).} The southern galaxy of this system is not analyzed in this work.  Difficulties in modeling the observed weak NaD absorption  in the spatially integrated and spatially resolved data of this galaxy are partially due to its edge-on orientation.\\

\noindent\textbf{F06295-1735.} This LIRG is not analyzed in this work either with the spatially integrated spectrum or via spatially resolved IFS of NaD. The  quality of the stellar modeling is low and  a large number of spectra in individual spaxels were not suitable for the analysis. \\

\noindent\textbf{F06592-6313.} The integrated spectrum of this LIRG shows a strong NaD absorption in which the stellar contribution is rather low (25$\%$). We modeled the purely ISM NaD line profile  with one blueshifted kinematic component. In the spectral maps, toward the south, we also observe strongly blueshifted absorption doublets (with velocities up to 500-600\,\kms), which we interpret as a signature of GW.  The broad component of the \Ha \ emission line (\citealt{Bellocchi2013}), seen with velocities up to 500\,\kms, is rather oriented   with the wind well (PA\,$\sim$\,130$\degr$) and partially overlaps the area covered by the neutral wind. However, the overall morphology of the broad H$\alpha$ component is poorly constrained and it is unclear how it is related to NaD.\\     

\noindent\textbf{F07027-601\,(N).} The velocity map of this galaxy shows outflowing neutral gas with velocities up to $\sim$\,300\,\kms \ extended within 1.5\,kpc of the nucleus. Despite evidence of an AGN, found by Arribas et al.\,(2014) according to its optical spectrum, we found  no strong evidence that the outflow we detect is boosted by the AGN. The morphology of the outflow could not be  constrained well owing to projection effects. Excluding the outflowing gas, the NaD is seen with an irregular spatial distribution, velocity field, and velocity dispersion map. This gas in not-ordered motion could be gas falling back onto the disk in HVCs.\\ 
\noindent\textbf{F07027-601\,(S).} This southern galaxy  is an important  case to study because its NaD IFS spectra, when fitted with multiple components, provide the possibility of disentangling the ordinary rotation and outflow. One kinematic component was found with rather regular rotation pattern, but lagging with respect to the kinematics of the ionized gas (i.e., H$\alpha$-narrow component) with a small positional misalignment. The other (outflowing) kinematic component is evident within  $\sim$\,1.3 kpc of the nucleus toward the northeast. This neutral gas  component is broad ($\sigma$\,$\sim$\,140\,\kms, on average) and has (observed) velocities within 100-300\,\kms. The morphology of the outflow could not be constrained well owing to projection effects.\\  
     
\noindent\textbf{F07160-6215.}  The velocity field of the neutral gas  as a disk-like pattern  in the main body of this dusty edge-on LIRG as seen for the ionized gas (traced via  \Ha\,  and  Br$\gamma$\,$\lambda$2.1655\,$\mu$m; \citealt{Bellocchi2013} and \citealt{Piqueras2012}, respectively). The velocity dispersion is almost flat, excluding the region toward the northeast, where  a broad and blueshifted NaD component was found and interpreted as a signature of  GW.  However, no ionized wind-phase counterpart is seen in  VIMOS and SINFONI IFS data (\citealt{Bellocchi2013}; \citealt{Piqueras2012}). \\   

\noindent\textbf{08355-4944.} This LIRG is not analyzed in this work either with the spatially integrated spectrum or via spatially resolved IFS of NaD. The very weak NaD absorption, seen in both the integrated and the spatially resolved spectra,  prevents any robust study.\\

\noindent\textbf{08424-3130\,(N) and (S).}  These spiral galaxies in interaction are only partially observed in our VIMOS data.  For this reason, we did not study  these galaxies in detail even though in the northern galaxy, we detect neutral gas with blueshifted velocities which are higher than those of the ionized gas (i.e., H$\alpha$-narrow component) at the same location, which can be interpreted as outflowing.\\

\noindent\textbf{F08520-6850.} This LIRG is not analyzed in this work either with the spatially integrated spectrum or via spatially resolved IFS. The  quality of the stellar modeling is low and  a large number of spectra in individual spaxels were not suitable for the analysis.\\

\noindent\textbf{09022-3615.} This ULIRG is not analyzed in this work either with the spatially integrated spectrum or via spatially resolved IFS of NaD. The  quality of the stellar modeling is low and the spectral maps lack of any clear  GW or disk signature. \\

\noindent\noindent\textbf{F09437+0317\,(N).} Two pointings (i.e., northeast (NE) and northwest (NW); \citealt{Rodriguez2011}) sample the northern galaxy. However the NW is not useful because of the low S/N of the spectra of the data cube. In the spectral maps obtained from the NE data cube, we observe that the neutral gas disk kinematic shares the same major kinematic axis with the ionized gas rotation, although the neutral  gas is found in slower rotation compared to the warm ionized gas \citep{Bellocchi2013}. \\
\noindent\noindent\textbf{F09437+0317\,(S).} In the southern galaxy (observed with one VIMOS pointing) we observe a shared major kinematic axis and a high  kinematical coupling  within the neutral and ionized ISM phases. The neutral gas rotation is only slightly slower (of about 40 \kms, on average) than that measured for the ionized gas at the same distance. \\      

\noindent\textbf{F10015-0614.} The NaD-clouds spatial distribution  only partially  reproduce the spiral structure of this LIRG. No hints of rotation pattern are found  in the neutral gas velocity field and velocity dispersion map. On the contrary,   we found blueshifted velocities up to 200 \kms \ along the kinematic minor axis of the ionized gas
rotation. We did not consider as part of the outflow the NaD-clouds toward the north and northeast because of their overlap with the approaching side of the ionized gas rotation pattern. The morphology of the outflow could not be constrained well because of projection effects.\\ 

\noindent\textbf{F10038-3338.} Our data for this LIRG,  show  deep, broad, and blueshifted  NaD line profiles in the integrated spectrum and  spectral maps (toward the northeast). We interpret this feature as a GW.\\      

\noindent\textbf{F10257-4339.} This merger LIRG is the nearest object in our VIMOS-IFS survey. The  neutral outflow is observed out to distances of 1.4 kpc. We did not consider  the NaD-clouds (with negative velocities) toward the north and northwest as part of the outflow, since  we observe an irregular and asymmetric NaD line profile in the IFS spectra. We tested  two-component modeling, but this modeling gives ambiguous results. \\

\noindent\textbf{F10409-4556.} The observed NaD feature  in this LIRG shows generally  complex absorption line profiles, best fitted with two absorption doublets. We considered the most negative velocities as part of the outflow, which have nearly no spatial overlap with the approaching side of the ionized gas rotation. Toward the north, this fast kinematic component associated with a GW dominates the NaD profile. However, the morphology of this outflow could not be  constrained well because of the ambiguity in the identification of the region where it is originated. The outflow is also seen via the integrated spectrum.\\     

\noindent\textbf{F10567-4310.} The NaD absorption seen in the integrated spectrum and spectra of the IFS cube of this LIRG galaxy is significantly deep and mainly  interstellar in origin. Indeed, our stellar continuum modeling indicates a rather low stellar contamination to the NaD doublet (28$\%$). The neutral gas spatial distribution is very compact and the neutral gas velocity field is dominated by a  starburst driven GW with outflowing velocity ranging 100 to 300 \kms.  The morphology of the outflow could not be  constrained well due to projection effects.\\

\noindent\textbf{F11255-4120.} The observed NaD integrated line-profile in this LIRG show the lowest  stellar contamination of the full sample (i.e., 18$\%$). The purely-ISM line profile was modeled with two kinematical components both having blueshifted velocity. Despite of that, the observed NaD line profile in our IFS-data is   modeled  with one kinematic component well.  Our NaD IFS-maps show  evidence of a GW, in the inner region (within R\,$>$\,1.2 kpc), that consist in: large blueshfited velocities (up to 300 \kms) and high turbulence (\disp(NaD) up to 280 \kms). The morphology of the outflow  could not be  constrained well owing to projection effects. Weak evidence for an ionized wind counterpart was found  \citep{Bellocchi2013}. \\ 

\noindent\textbf{F12043-3140\,(N).}  This galaxy is not analyzed in this work. The stellar continuum modeling of the integrated spectrum is rather uncertain. Our spectral maps indicate a generally blueshifted velocity pattern and rather low values for the velocity dispersion and EWs. Therefore, the evidence for claiming a GW detection are weak.\\
\noindent\textbf{F12043-3140\,(S).} The absorption due to stars contributes significantly (77$\%$) to the NaD line profile, so the rotation pattern seen in this VIMOS-spectral map might be tracing the stellar kinematics.  This result is also supported by the low depth of the absorption  found in the EW map. However,  additional observations are needed to confirm this interpretation. The 2D neutral gas disk rotation is slower with respect to that of the ionized gas and the major kinematic axes are not aligned well. The high velocity dispersion values seen toward the north are possibly due to additional turbulence this outer region. \\
Along the southern minor axis there is a region with high blueshifted velocities, and high dispersions that may indicate the presence of an outflow. However, the values of high velocity dispersion are found in just a few spaxels, likely due to low S/N, as the general region has low values. Additionally, in the region of the putative wind, the average value of the residual maps (i.e., V$_{\rm NaD}$-V$_{\rm H\alpha}$) is slightly lower then the limit we consider. Therefore, we did not include this case from the sample with detected outflows.\\ 

\noindent\textbf{F12115-4656.} The velocity field is fairly regular with the kinematic center in positional agreement with the continuum and \Ha \ flux peaks. The neutral gas disk is $\sim$\,45 times thicker than that of the neutral gas. This LIRG was already studied in great detail by \citet{Arribas2008}.  We do not excluded the presence of a GW oriented along the kinematic minor axis, but it is not the main feature either  in  the spectral maps or in  the integrated spectrum (where the stellar fraction to the NaD is rather high, 73\,$\%$) \\

\noindent\textbf{F12116-5615.} The NaD line profile seen in this LIRG (both in the integrated spectrum and in the spectra of individual spaxels)  was fitted with two kinematic components. One kinematical component was found dominated by not-ordered motions and the other by a neutral GW  via IFS. The wind is seen almost face-on preventing us from measuring its extent and opening angle. The neutral gas entrained in the wind  is likely swept out polluting the IGM. The GW is likely multiphase, since the \Ha \ broad component is seen as outflowing \citep{Bellocchi2013}. \\

\noindent\textbf{F12596-1529.} This LIRG is not analyzed in this work either with the spatially integrated spectrum or via spatially resolved IFS of  NaD. The  quality of the stellar modeling is low and  a large number of spectra in individual spaxels were not suitable for the analysis. \\
\noindent\textbf{F13001-2339.} The ionized gas velocity field and velocity dispersion map (traced via the H$\alpha$ narrow component) have an irregular pattern and the neutral gas is seen mainly as outflowing. The outflow spatially overlap the region where the ionized gas velocity field has positive velocities. The morphology of the outflow could not be constrained well because of projection effects and the lack of any rotation pattern of reference.\\

\noindent\textbf{F13229-2934.}  This LIRG  hosts a neutral gas wind detected with both the spatially integrated and spatially resolved spectra. As for the galaxy F04315-0840, an ionized wind counterpart was observed  \citep{Bellocchi2013}. This has  roughly  the same velocities observed in the present neutral gas velocity map. Even though this galaxy hosts an AGN, the comparison of the wind power and kinetic power of the starburst indicate that the multiphase GW is likely driven by SNe. \\  

\noindent \textbf{F14544-4255\,(E).}  We found  that the NaD originated in stars gives a modest contribution (41\,$\%$) to the integrated line profile for this galaxy, but the modeling is rather uncertain. The neutral gas velocity field is disk-like and the kinematic center seems to be in positional agreement with the continuum peak. The rotation seen for the neutral gas lags compared to that of the ionized gas. For both ISM phases we did not observe a central peak in the velocity dispersion maps (see also \citealt{Bellocchi2013}); this may indicate that the contribution of the bulge component is negligible. \\
\noindent \textbf{F14544-4255\,(W).}  The western galaxy has not been classified either in 1D or in 2D. In the spectral maps, the EWs(NaD) are generally low and we found no clear rotation pattern or evidence indicating a GW. \\

\noindent\textbf{F17138-1017.} This LIRG is not analyzed in this work either with the spatially integrated spectrum or via spatially resolved IFS of NaD. The  quality of the stellar modeling is low and  a large number of spectra in individual spaxels were not suitable for the analysis. \\

\noindent \textbf{F18093-5744\,(N).} The  purely ISM NaD profile seen in the integrated spectrum of this galaxy  is modeled with one blueshifted kinematic component well. However, in the spectral maps the putative wind (likely seen toward the northeast) has  both low velocities  and EWs(NaD). This characteristic makes its detection is rather uncertain.\\
\noindent \textbf{F18093-5744\,(C).} This galaxy is not analyzed in this work since it was not possible to make a kinematical classification via both the spatially-resolved and spatially-integrated spectroscopy of the NaD absorption. \\
\noindent \textbf{F18093-5744\,(S).} The spider pattern seen in the neutral gas velocity field of the galaxy is fairly regular. The rotation is seen for a disk three times thicker and slower with respect to that of the ionized gas. The stellar contamination, estimated with our stellar continuum modeling of the integrated spectrum, is modest (41\,$\%$).\\  

\noindent\textbf{F21130-4446.} This ULIRG is not analyzed via NaD in this work. The  NaD absorption is very weak both in the  integrated and  spatially resolved spectra.\\

\noindent\textbf{F21453-3511.} In this galaxy, a GW  at rather slow velocity is seen emerging from the nucleus. The outflow is also seen via a blue wing in the NaD integrated line profile (despite its strong stellar contamination, i.e., 97$\%$). The morphology of the outflow could not be constrained well because of a patchy distribution of the NaD outflowing clouds and projection effects. Only a weak indication of the presence of an ionized wind was found \citep{Bellocchi2013}. Excluding the region where the wind is detected, the neutral gas velocity  field is rather irregular and roughly follows the kinematic of the  \Ha \  disk but with strong deviations (up to 150 \kms).  \\

\noindent\textbf{F22132-3705.} The disk-like velocity field is fairly regular with the kinematic center in positional agreement with the continuum and \Ha \ flux peaks. The neutral gas disk is $\sim$\,20 times thicker than that of the neutral gas. The stellar fraction to the NaD is modest (47\,$\%$).  The  high values found in the velocity dispersion map  toward the northeast (particularly evident in the velocity dispersion radial profile) can be explained by either addition turbulence in these outer regions or GW. The evidence for claiming a GW detection  however is  weak.\\  
\noindent\textbf{F22491-1808.} This ULIRG is not analyzed in this work either with the spatially integrated spectrum or via spatially resolved IFS of NaD. The  quality of the stellar modeling is low and  a large number of spectra in individual spaxels were not suitable for the analysis.\\

\noindent\textbf{IRAS-F23128-5919\,(N) and (S).} Despite the high integrated  stellar contamination, the NaD velocity field of the two galaxies does not show any ordered rotation pattern typically seen, e.g., the stars. Most of the cold gas associated with this system is outflowing reaching velocities up to 500-600\,\kms. The fast neutral outflowing gas has a  conical shape (with the apex in the southern galaxy) as also seen via the Pa$\alpha$ and \Ha \ line profiles  (\citealt{Piqueras2012, Bellocchi2013}). For this system, the capability for the wind of polluting the IGM is doubtful because of the inability to estimate  the contributions to the velocity field of each individual galaxy reliably \citep{Bellocchi2013}, which leads to additional uncertainties for the calculation of v$_{\rm esc}$.


\begin{figure*}
\centering
 \includegraphics[trim = -0.0cm 0.2cm .5cm 13.0cm, clip=true, width=1.\textwidth]{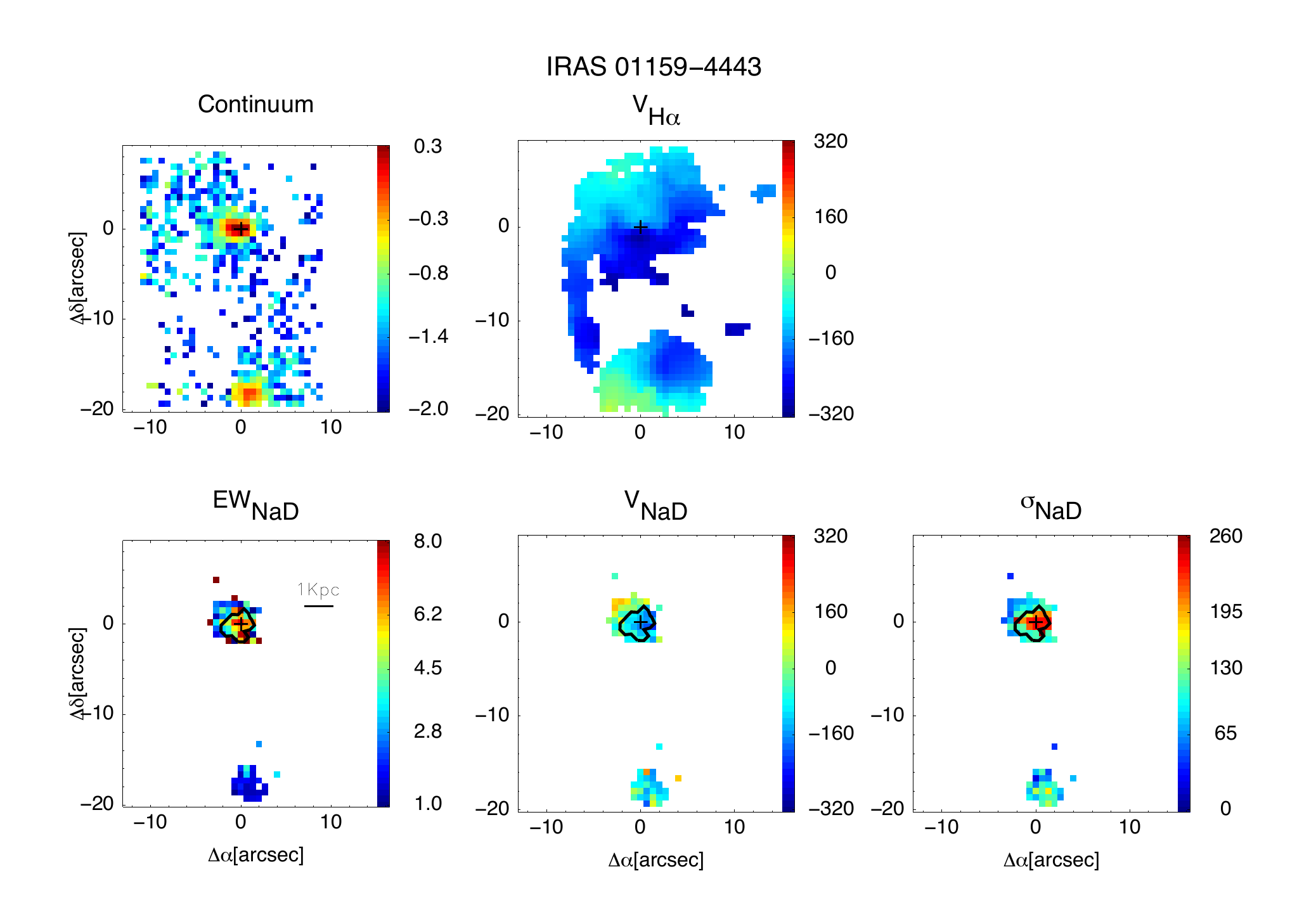}  
\centering
\includegraphics[trim = -0.0cm 5.5cm .5cm 2.0cm, clip=true, width=.95\textwidth]{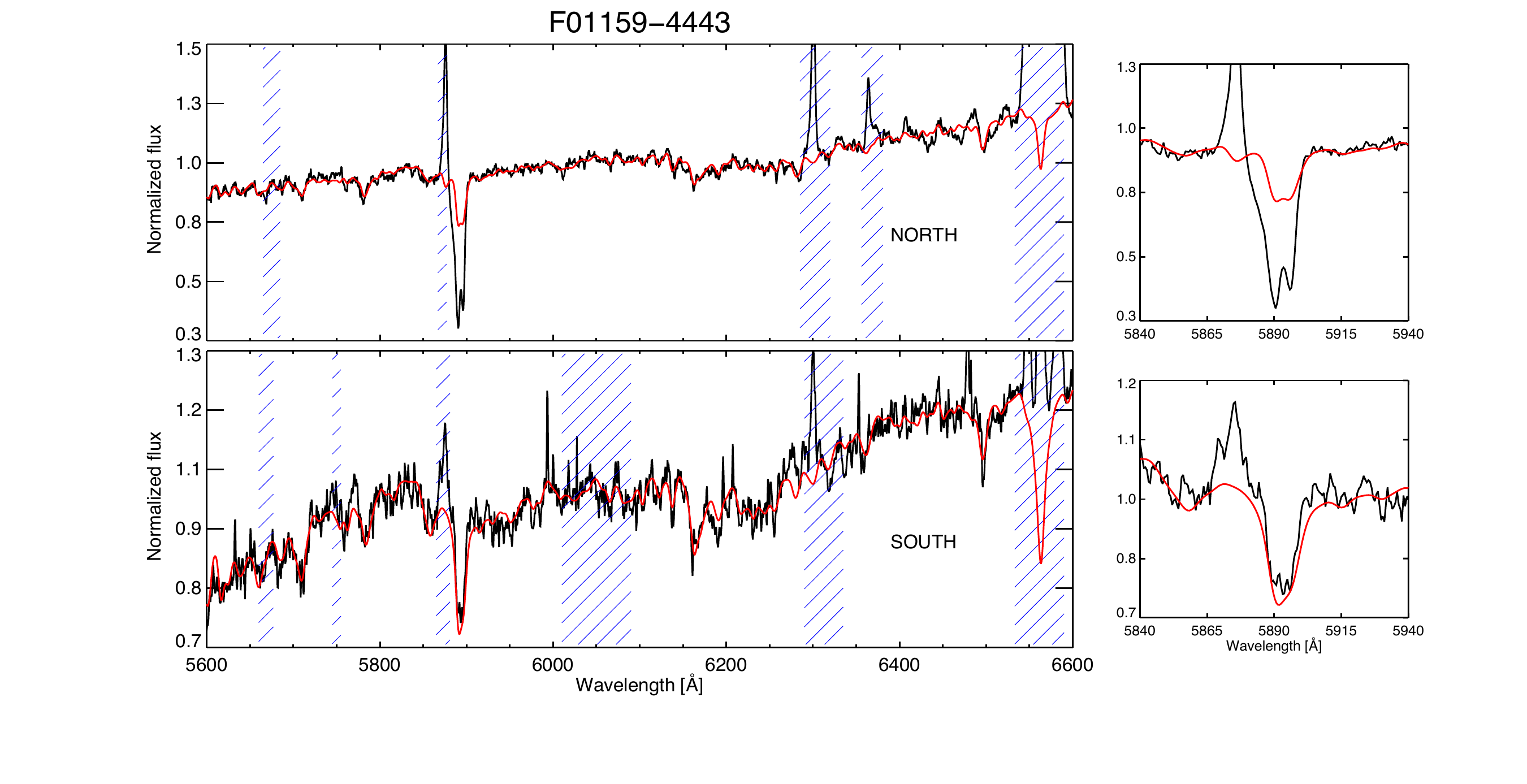}
\vspace{-1.7em}  
\caption{IRAS F01159-4443.  \textit{Top:} The  continuum image obtained from the mean of the line-free continuum nearby the doublet in a 100\,$\AA$\, rest-frame wavelength range in units of  erg\,s$^{\rm-1}$\,cm$^{\rm-2}$\,$\AA$$^{\rm-1}$ after applying a factor of 10$^{\rm-16}$ and the ionized gas velocity field (in  \kms units) traced via the H$\alpha$-narrow component (both included as reference). \textit{Center: }VIMOS observed maps obtained modeling the NaD line profile ($\lambda$$\lambda$\,5890,5896\,$\AA$). From left to right: equivalent width (in $\AA$ units),  velocity, and velocity dispersion (both in \kms units). In all the maps, the brightest spaxel of the VIMOS continuum is indicated with a cross and the orientation of the galaxies is north up, east to the left. The maps are color coded according to their own scale (i.e., range of the velocity, velocity dispersion, and EWs sampled) to facilitate the contrast and to highlight weak features. Black contours (if present) indicate the spaxels in which the neutral gas is identified as entrained in a GW.  \textit{Bottom:} the rest-frame spectra extracted from the original cube via a S/N optimization algorithm (Sect.~\ref{IntegratedSpectra}) for the northern and southern galaxies. The red line indicates the modeled stellar spectrum that matches the observed  continuum, obtained applying the \textit{pPXF} method. The most relevant spectral features blocked  for modeling a line-free continuum are shown in blue.}
\label{Panel_F01159}             
\end{figure*}
\clearpage


\begin{figure*}
\centering
\includegraphics[trim = -0.0cm 0.2cm .5cm 13.0cm, clip=true, width=1.\textwidth]{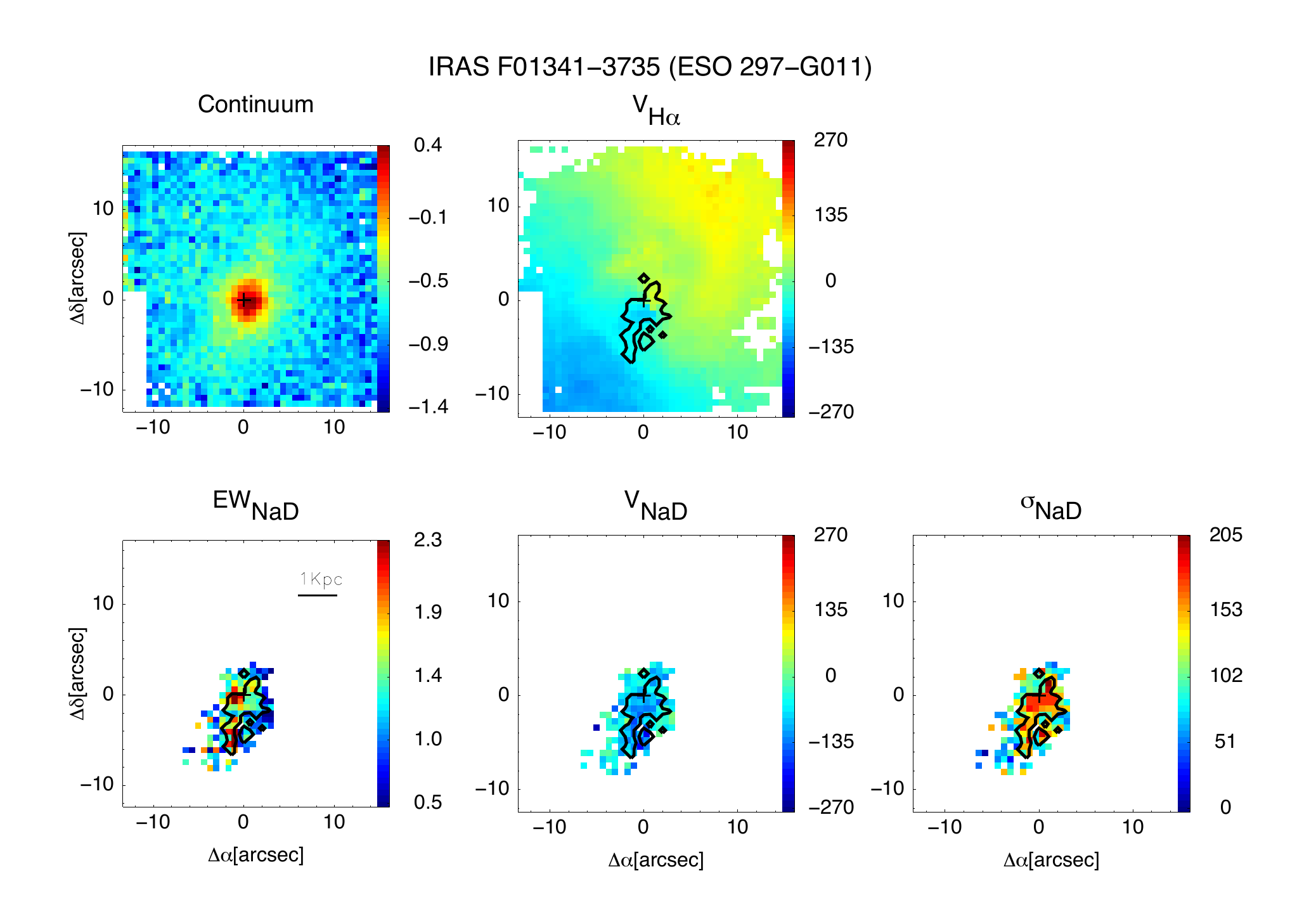}  
\centering
\includegraphics[trim = -0.0cm 8.5cm .5cm 7.0cm, clip=true, width=.8\textwidth]{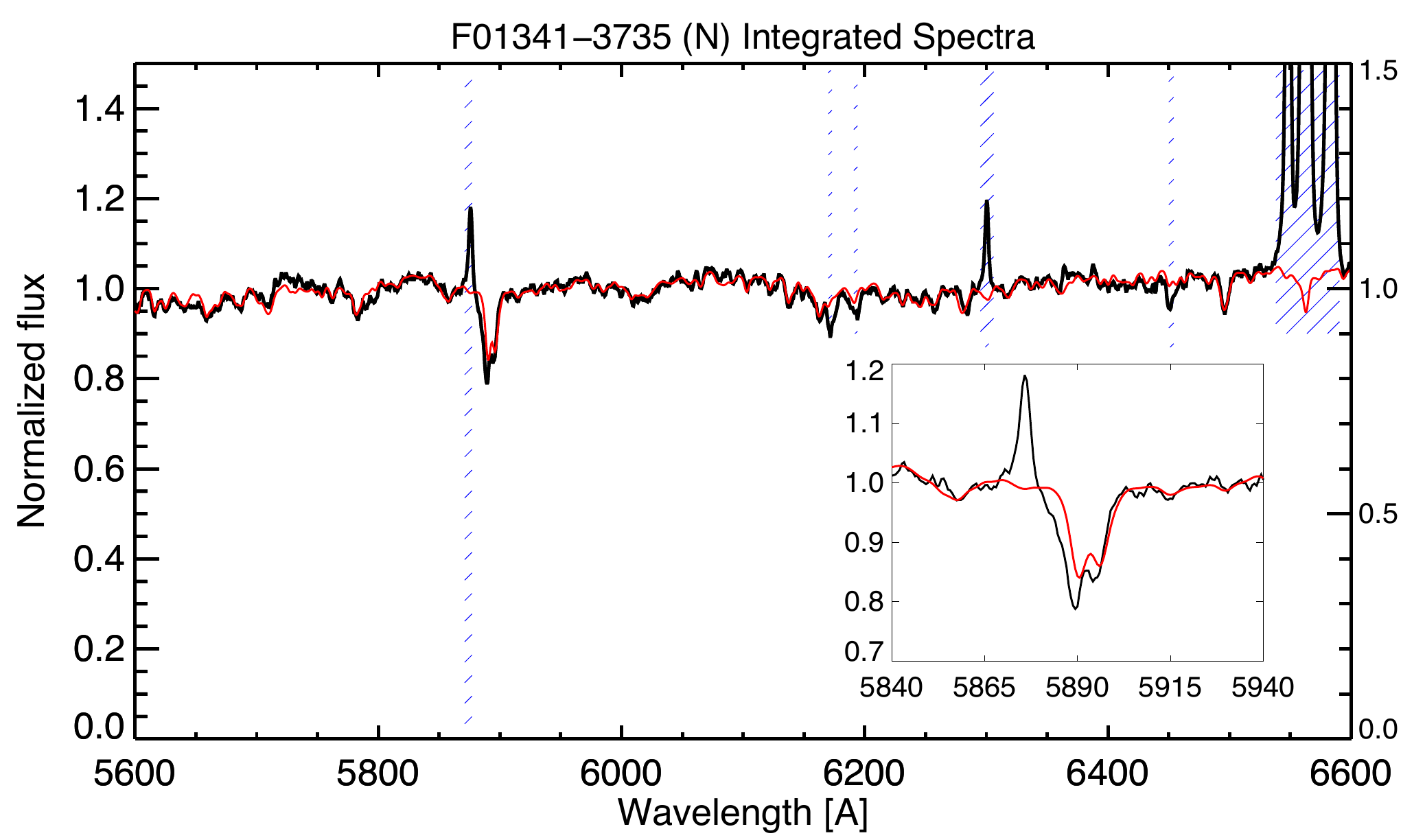} 
 \caption{As Fig.~\ref{Panel_F01159} but for IRAS F01341-3735 (ESO 297-G011).}                      
\label{Panel_01341a}             
\end{figure*}
\clearpage
                            
                    
\begin{figure*}
\centering
\includegraphics[trim = -0.0cm 0.2cm .5cm 13.0cm, clip=true, width=1.\textwidth]{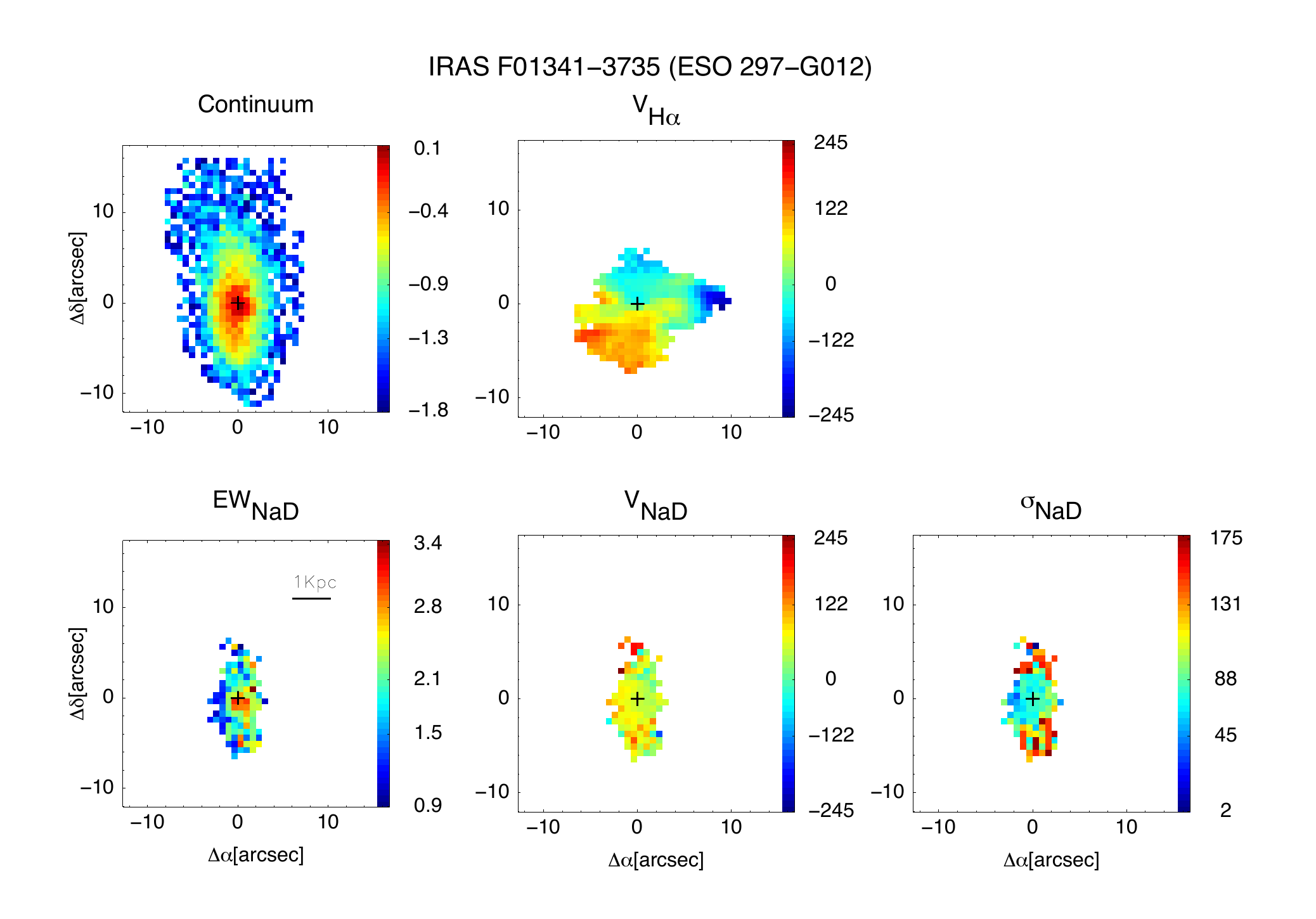}  
 \centering
\includegraphics[trim = -0.0cm 8.5cm .5cm 7.0cm, clip=true, width=.8\textwidth]{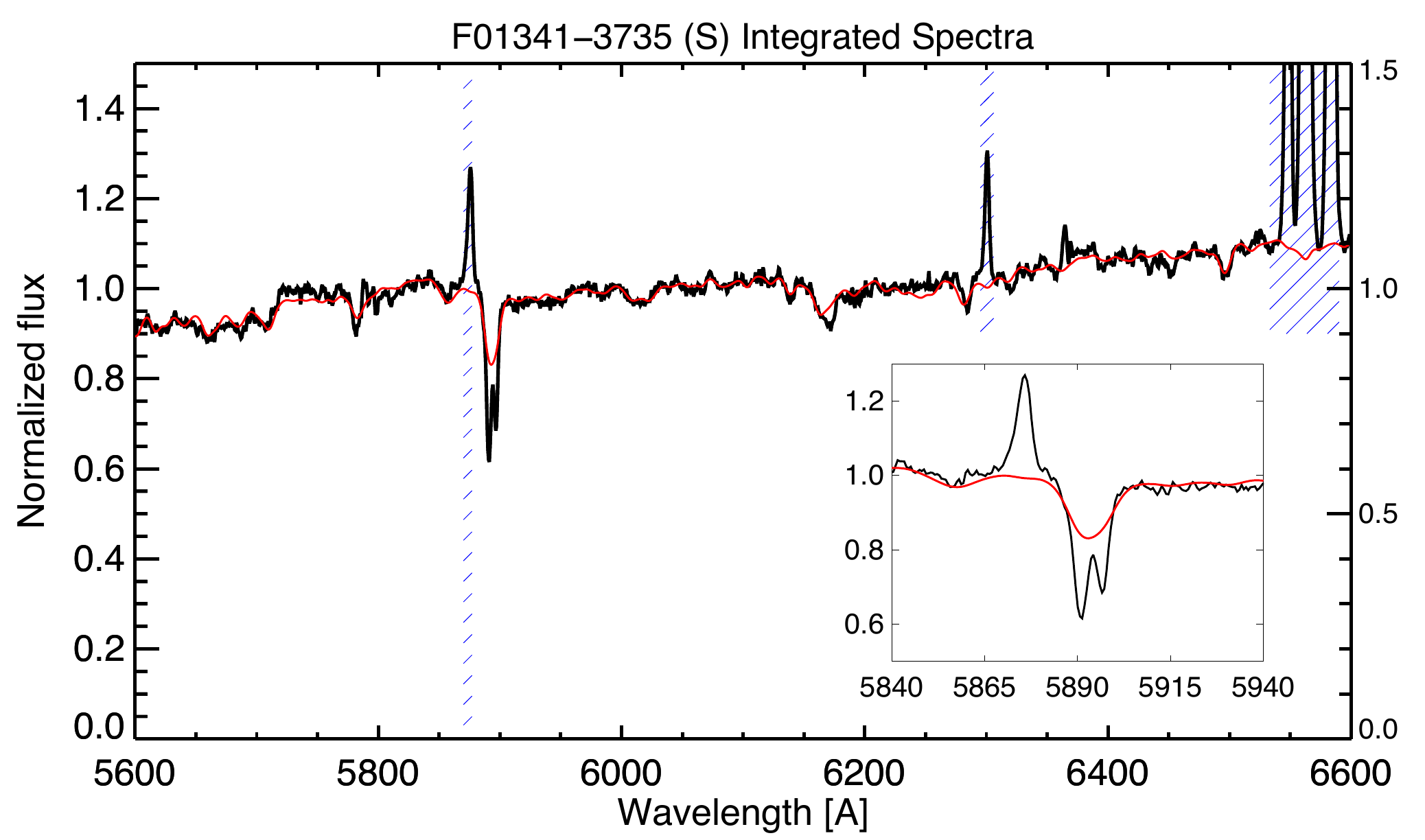} 
\caption{As Fig.~\ref{Panel_F01159} but for IRAS F01341-3735 (ESO 297-G012).}
 \label{Panel_F01341a}           
\end{figure*}
\clearpage   
   
               
\begin{figure*}
\centering
 \includegraphics[trim = -0.0cm 0.2cm .5cm 13.0cm, clip=true, width=1.\textwidth]{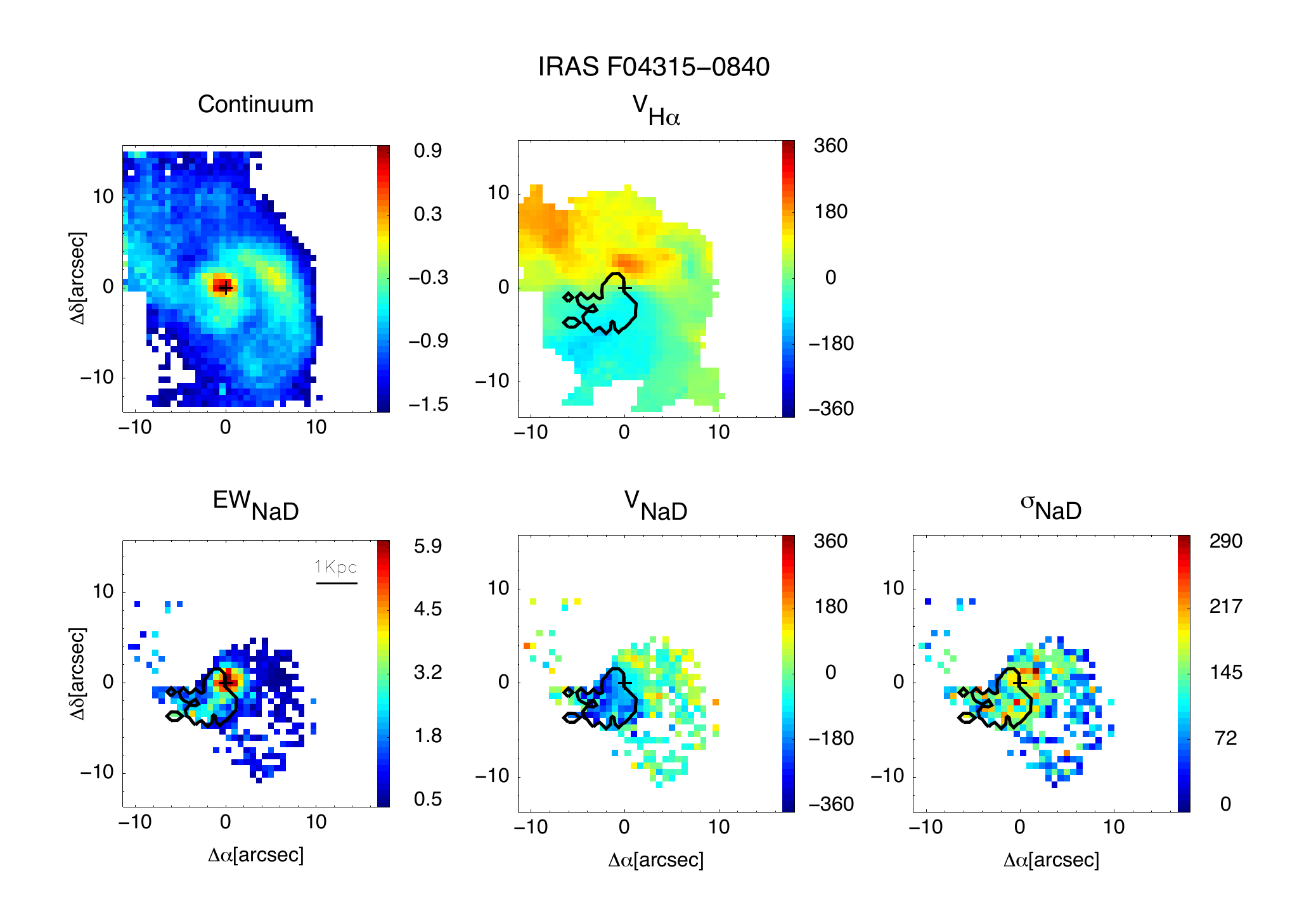}  
\centering
\includegraphics[trim = -0.0cm 8.5cm .5cm 7.0cm, clip=true, width=.8\textwidth]{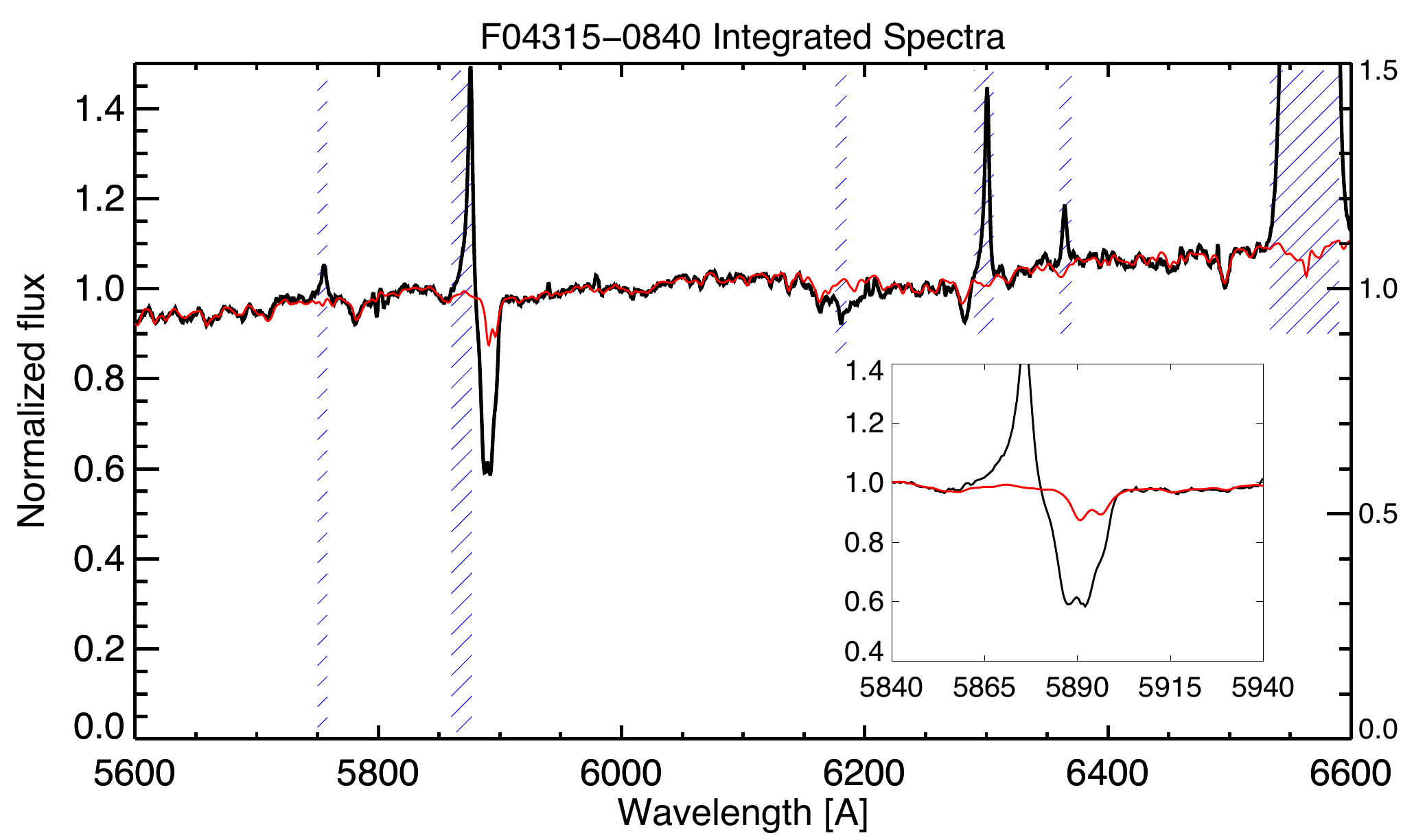}  
\caption{As Fig.~\ref{Panel_F01159} but for IRAS F04315-0840.}
\label{Panel_04315}              
\end{figure*}
\clearpage

        
\begin{figure*}
\centering
 \includegraphics[trim = -0.0cm 0.2cm .5cm 13.0cm, clip=true, width=1.\textwidth]{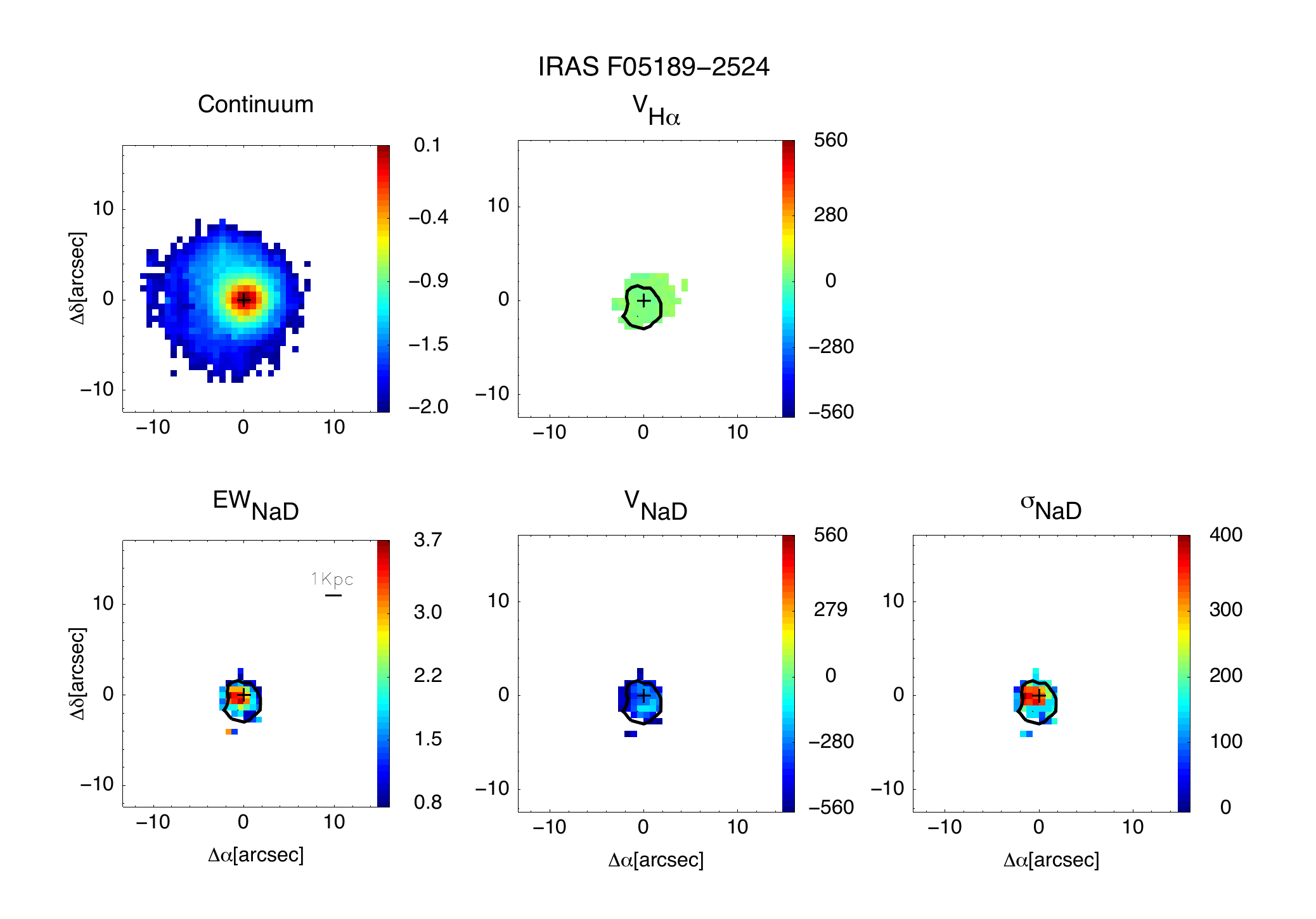}  
\centering
 \includegraphics[trim = -0.0cm 8.5cm .5cm 7.0cm, clip=true, width=.8\textwidth]{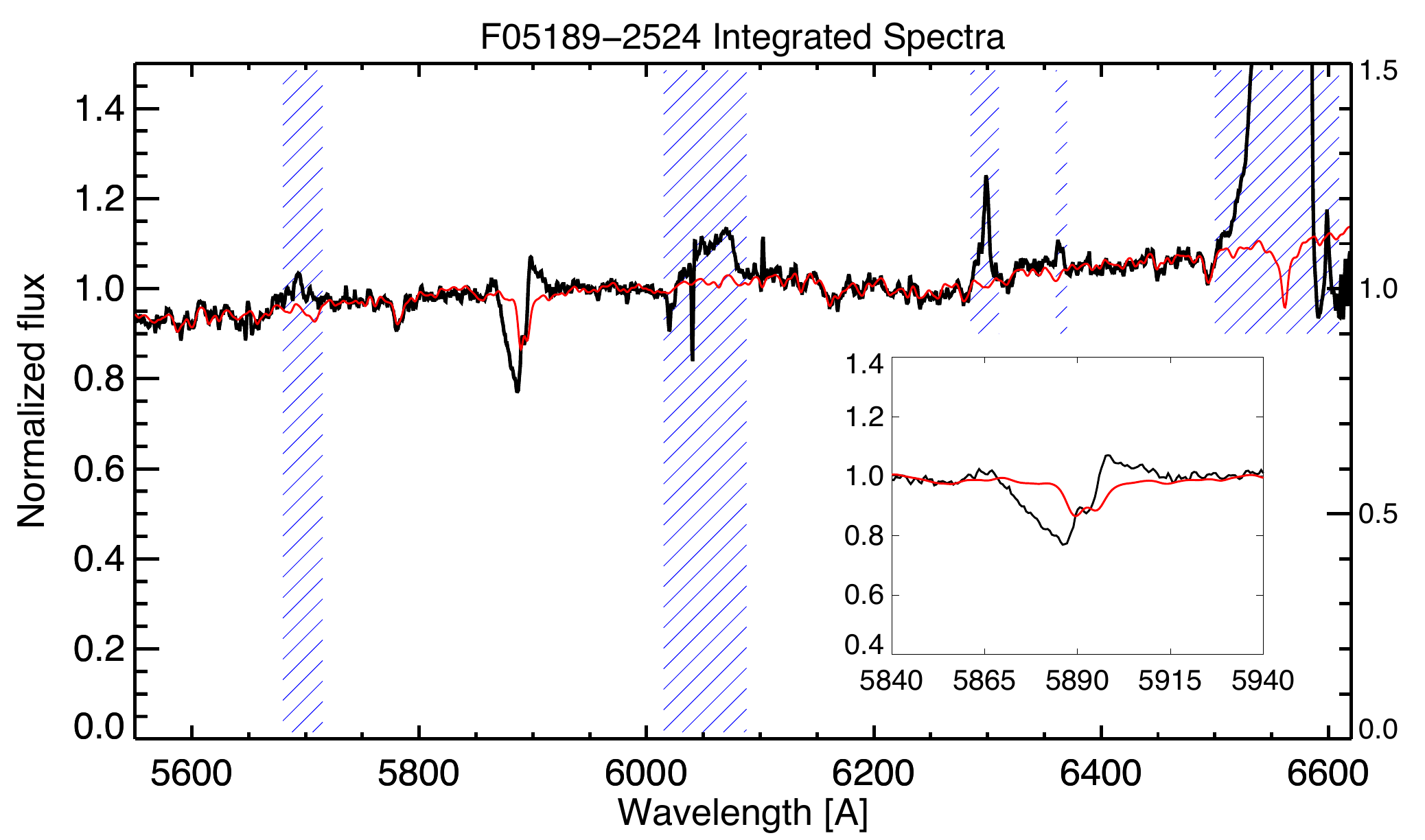}  
\caption{As Fig.~\ref{Panel_F01159} but for IRAS F05189-2524.}
  \label{Panel_F05189}           
\end{figure*}
\clearpage


\begin{figure*}
\centering
 \includegraphics[trim = -0.0cm 8.5cm .5cm 7.0cm, clip=true, width=.72\textwidth]{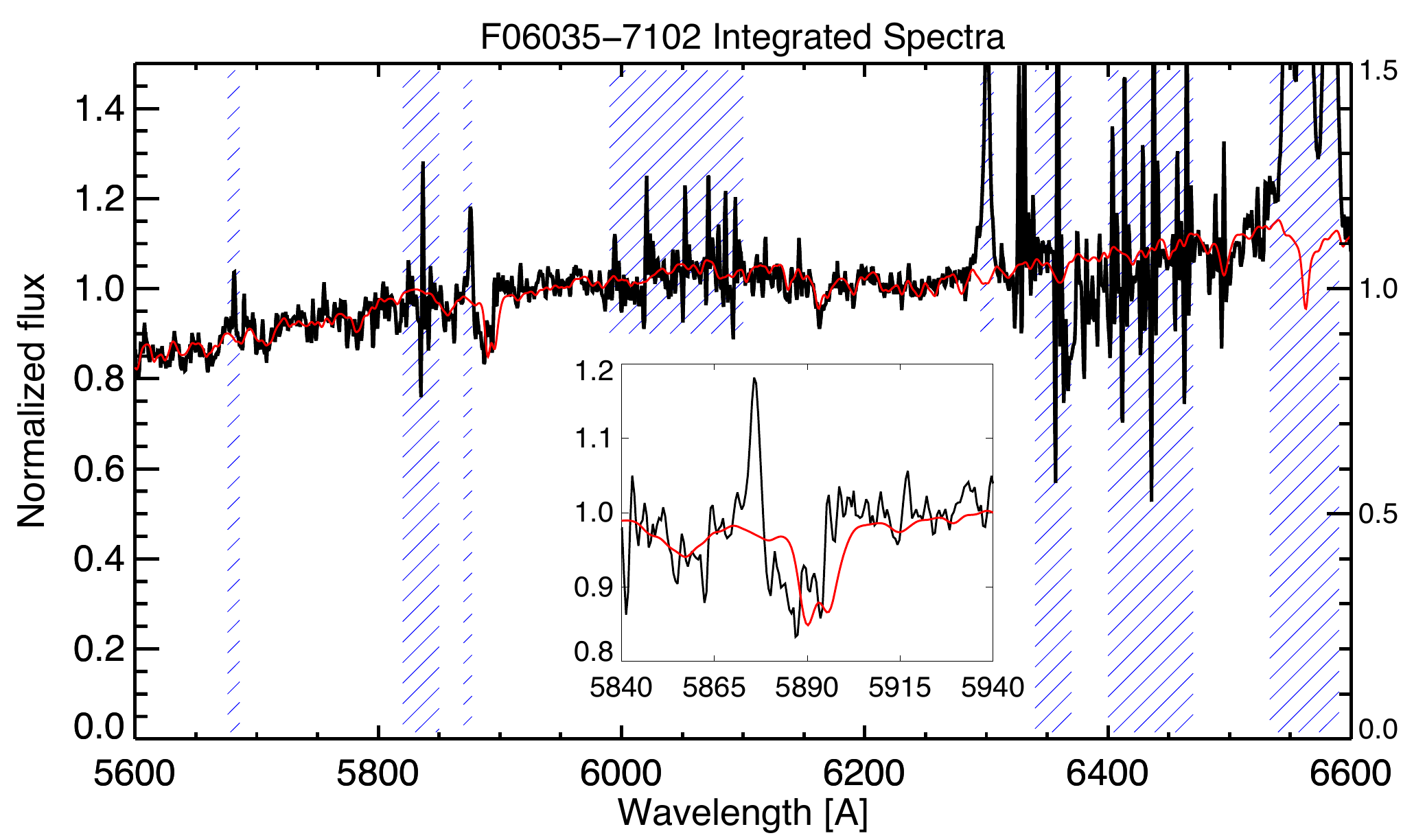}
 \vspace{1.em}  
\centering
 \includegraphics[trim = -0.0cm 8.5cm .5cm 7.0cm, clip=true, width=.72\textwidth]{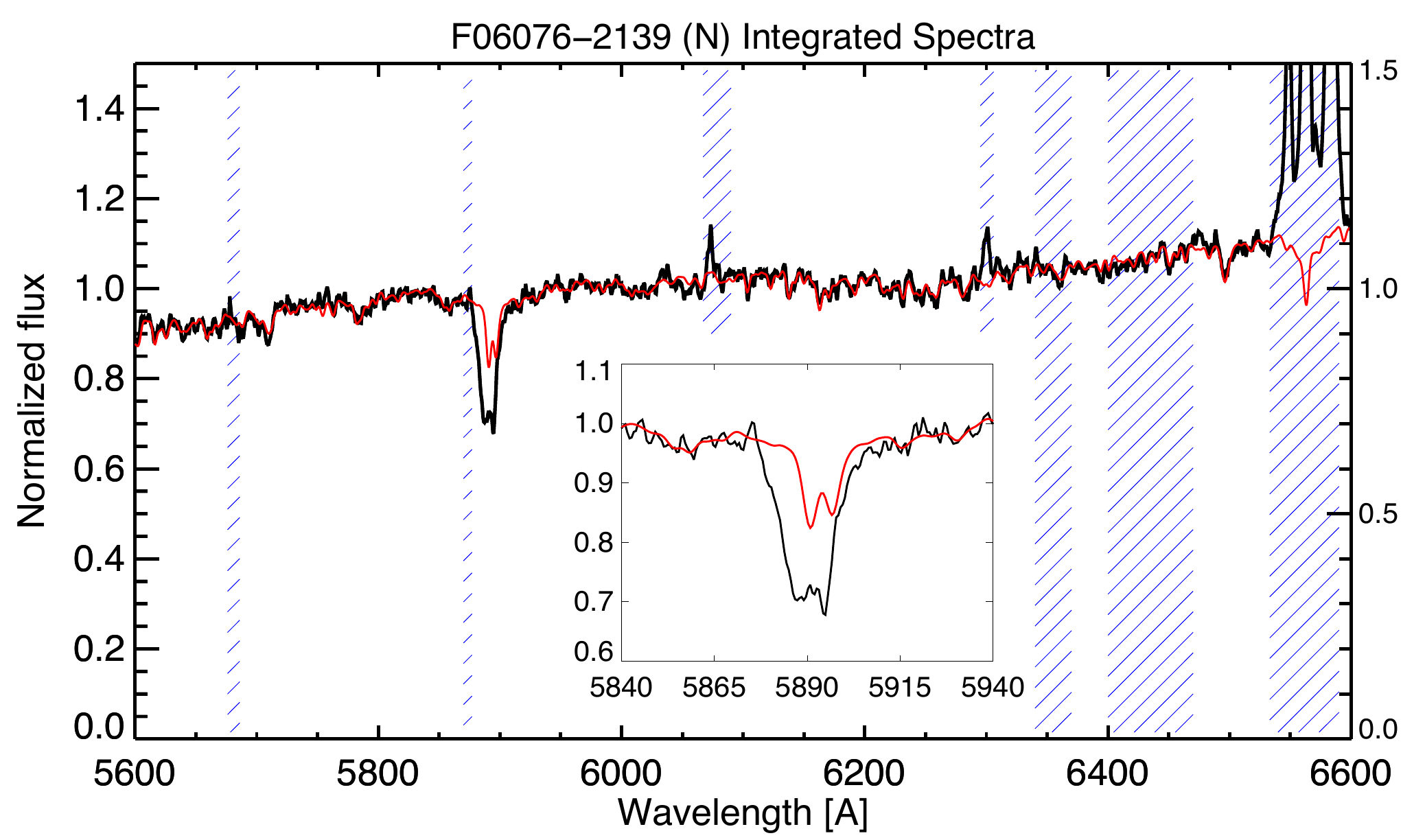}
\vspace{1.em}       
\centering
\includegraphics[trim = -0.0cm 8.5cm .5cm 7.0cm, clip=true, width=.72\textwidth]{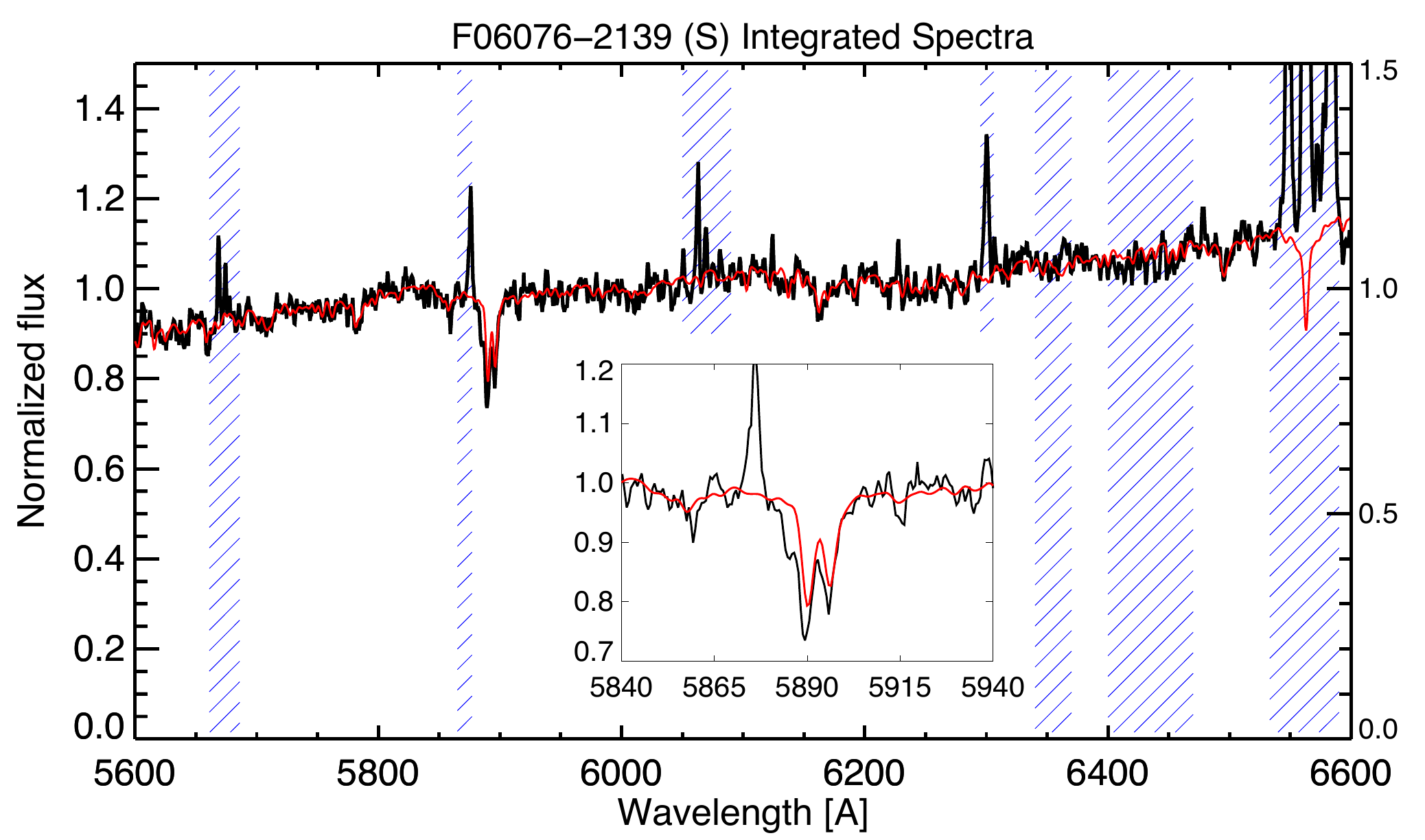}
\caption{As in the lower panel of Fig.~\ref{Panel_F01159} but for IRAS 06035-7102 and IRAS F06076-2139 north and south (from top to bottom).}
\end{figure*}
\clearpage

                
\begin{figure*}
\centering
 \includegraphics[trim = -0.0cm 0.2cm .5cm 13.0cm, clip=true, width=1.\textwidth]{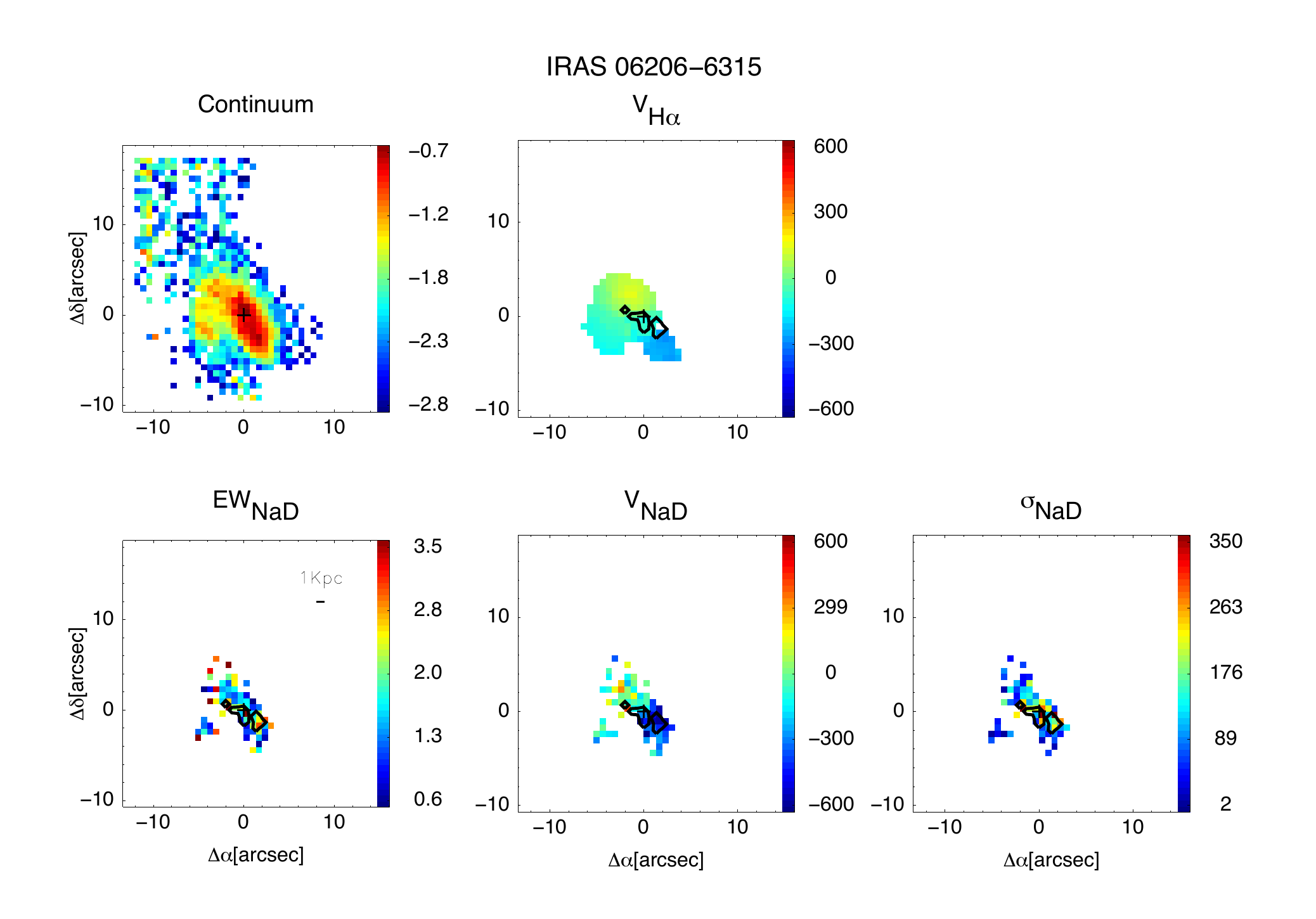}  
\centering
\includegraphics[trim = -0.0cm 8.5cm .5cm 7.0cm, clip=true, width=.8\textwidth]{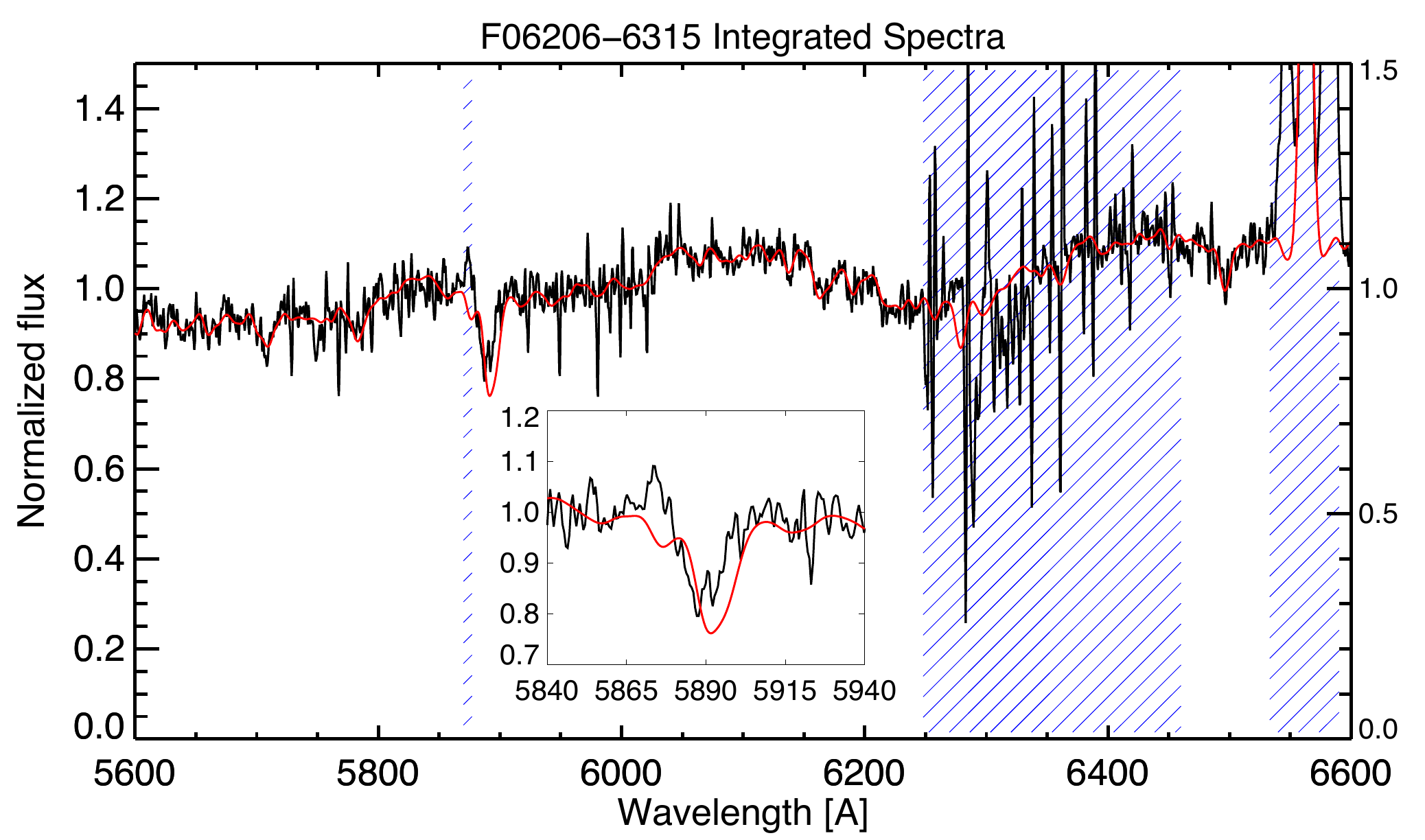}  
\caption{As Fig.~\ref{Panel_F01159} but for IRAS F06206-6315.}
\label{Panel_F06206}               
\end{figure*}
\clearpage

               
\begin{figure*}
\centering
\includegraphics[trim = -0.0cm 0.2cm .5cm 13.0cm, clip=true, width=1.\textwidth]{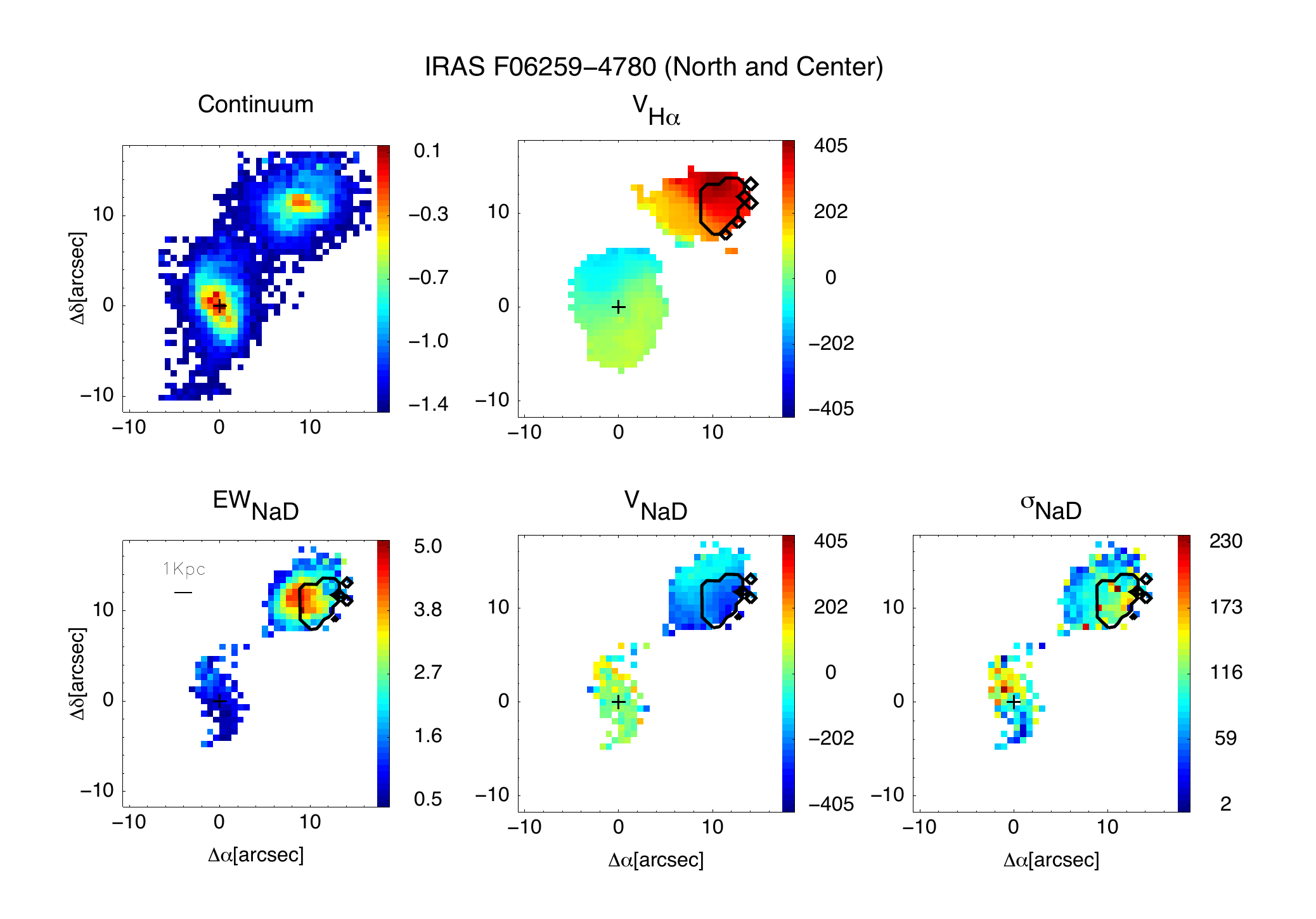}  
\centering
\includegraphics[trim = -0.0cm 5.5cm .5cm 2.0cm, clip=true, width=1.\textwidth]{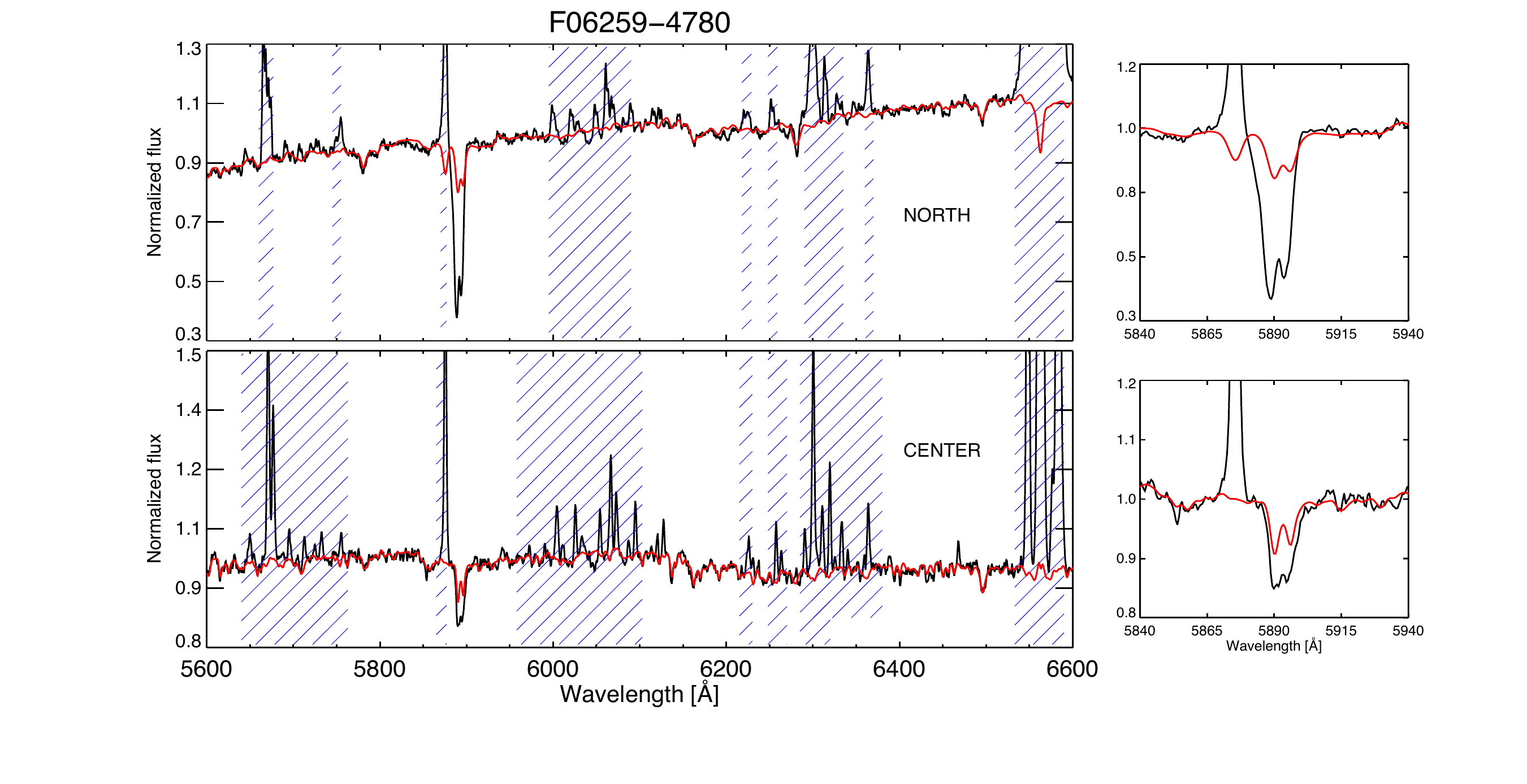}  
\caption{As Fig.~\ref{Panel_F01159} but for IRAS F06259-4780 north and center.}
 \label{Panel_F06259}            
\end{figure*}
\clearpage


\begin{figure*}
\vspace{4.em}   
\centering
\includegraphics[trim = -0.0cm 8.5cm .5cm 7.0cm, clip=true, width=.72\textwidth]{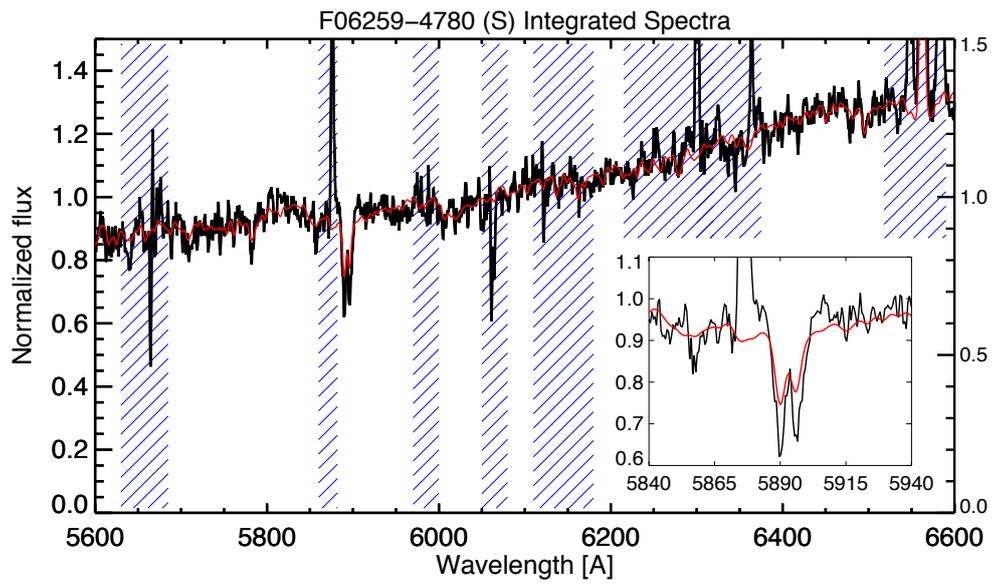} 
\caption{As in the lower panel of Fig.~\ref{Panel_F01159} but for IRAS  F06259-4780\,(S).}
\end{figure*}
        
                                
\begin{figure*}
\vspace{4.em}   
\centering
\includegraphics[trim = -0.0cm 8.5cm .5cm 7.0cm, clip=true, width=.72\textwidth]{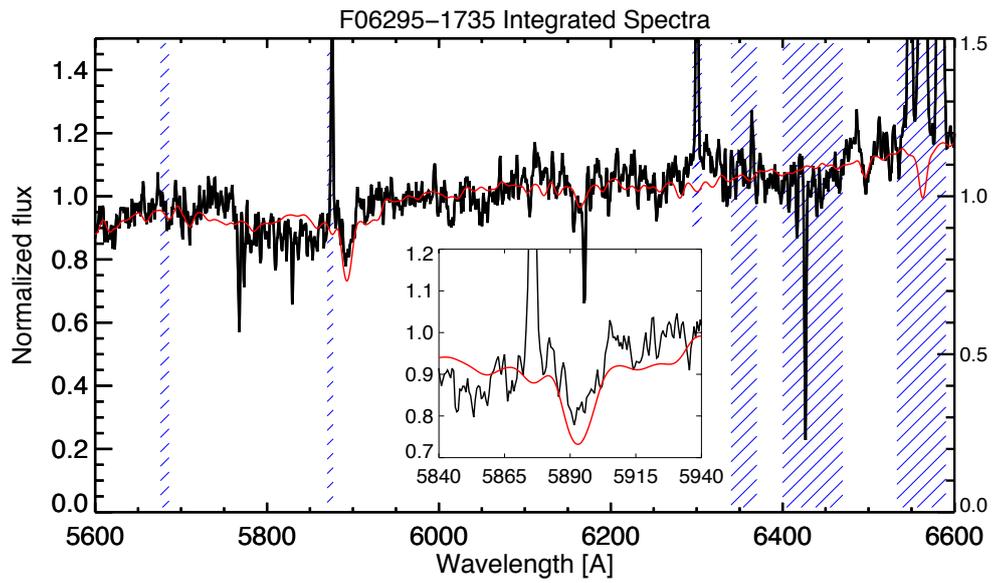} 
\caption{As in the lower panel of Fig.~\ref{Panel_F01159} but for IRAS  06295-1736.}
\end{figure*}
                                

\begin{figure*}
\centering
\includegraphics[trim = -0.0cm 0.2cm .5cm 13.0cm, clip=true, width=1.\textwidth]{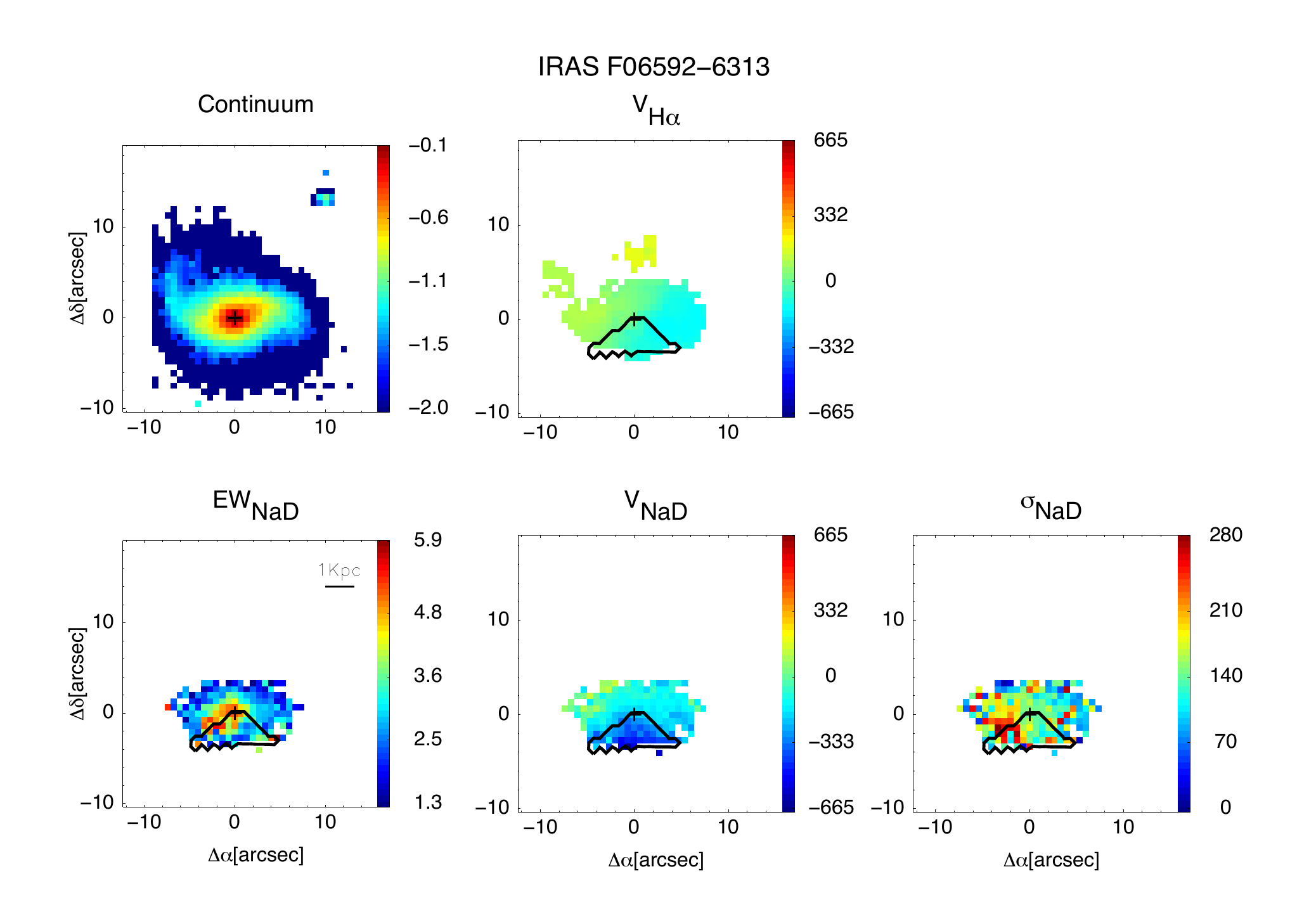}  
 \centering
\includegraphics[trim = -0.0cm 8.5cm .5cm 7.0cm, clip=true, width=.8\textwidth]{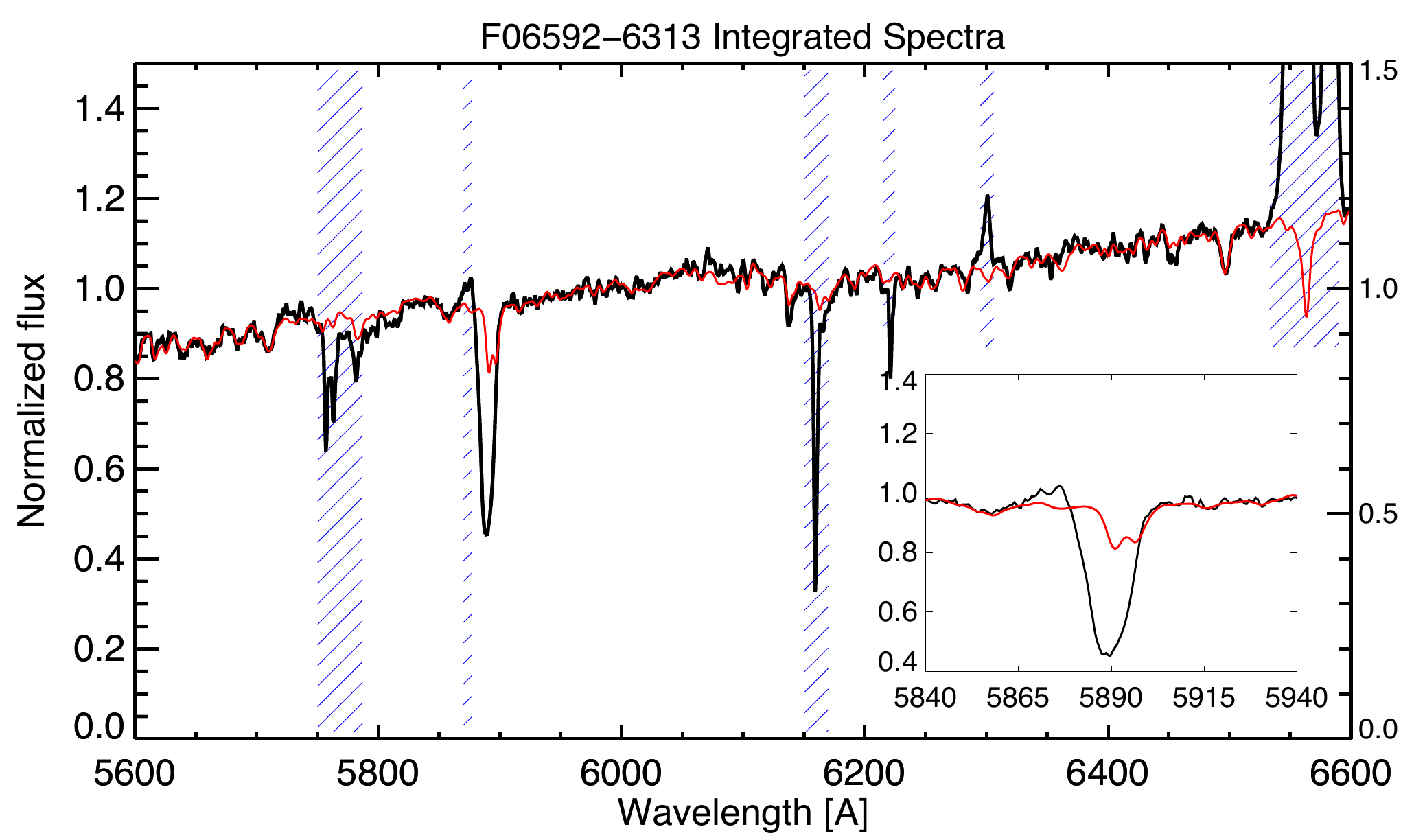}  
\caption{As Fig.~\ref{Panel_F01159} but for IRAS F06592-6313.}
 \label{Panel_F06592}            
\end{figure*}
\clearpage

     
\begin{figure*}
\centering
\includegraphics[trim = -0.0cm 0.2cm .5cm 13.0cm, clip=true, width=1.\textwidth]{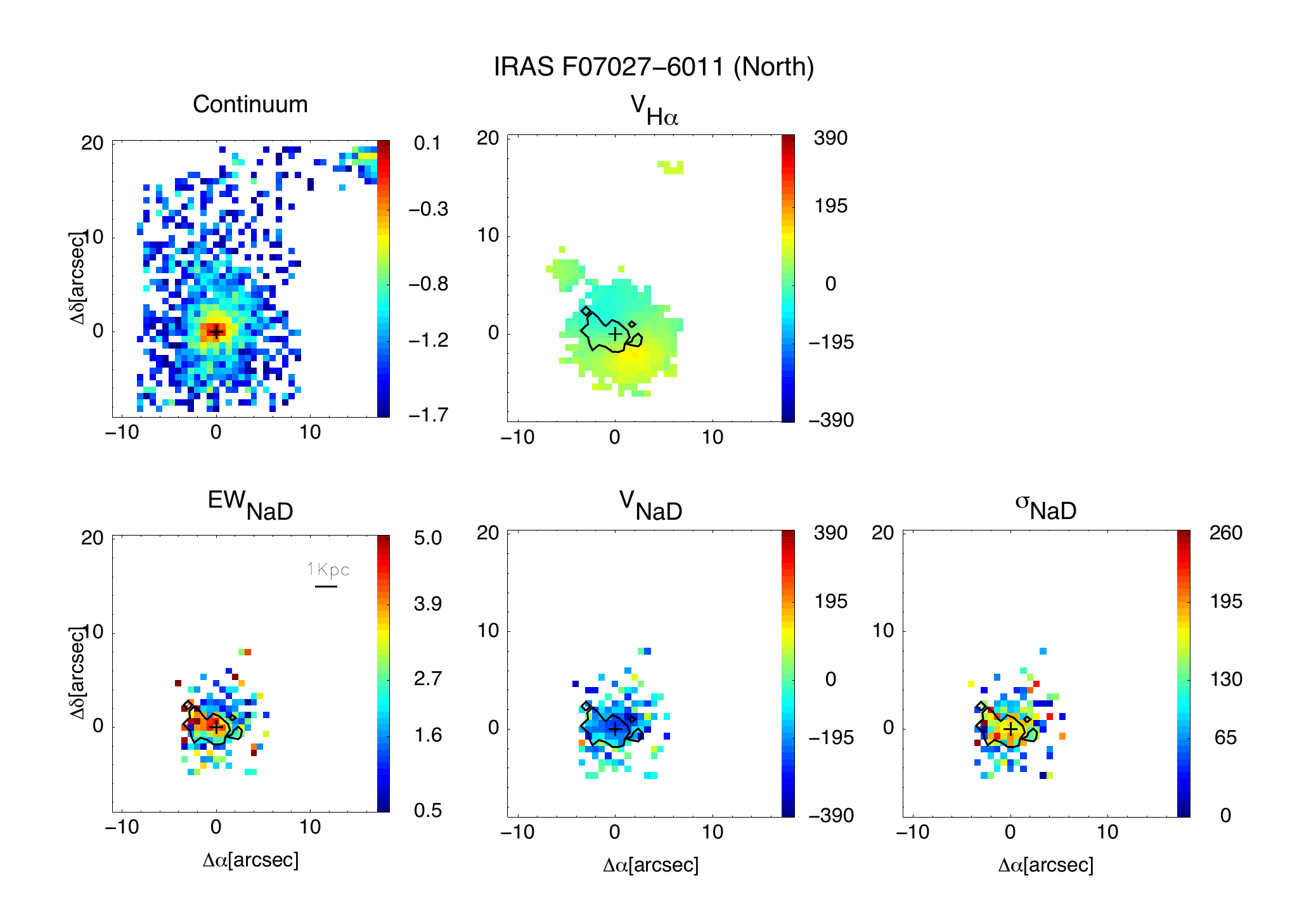}  
\centering
\includegraphics[trim = -0.0cm 8.5cm .5cm 7.0cm, clip=true, width=.7\textwidth]{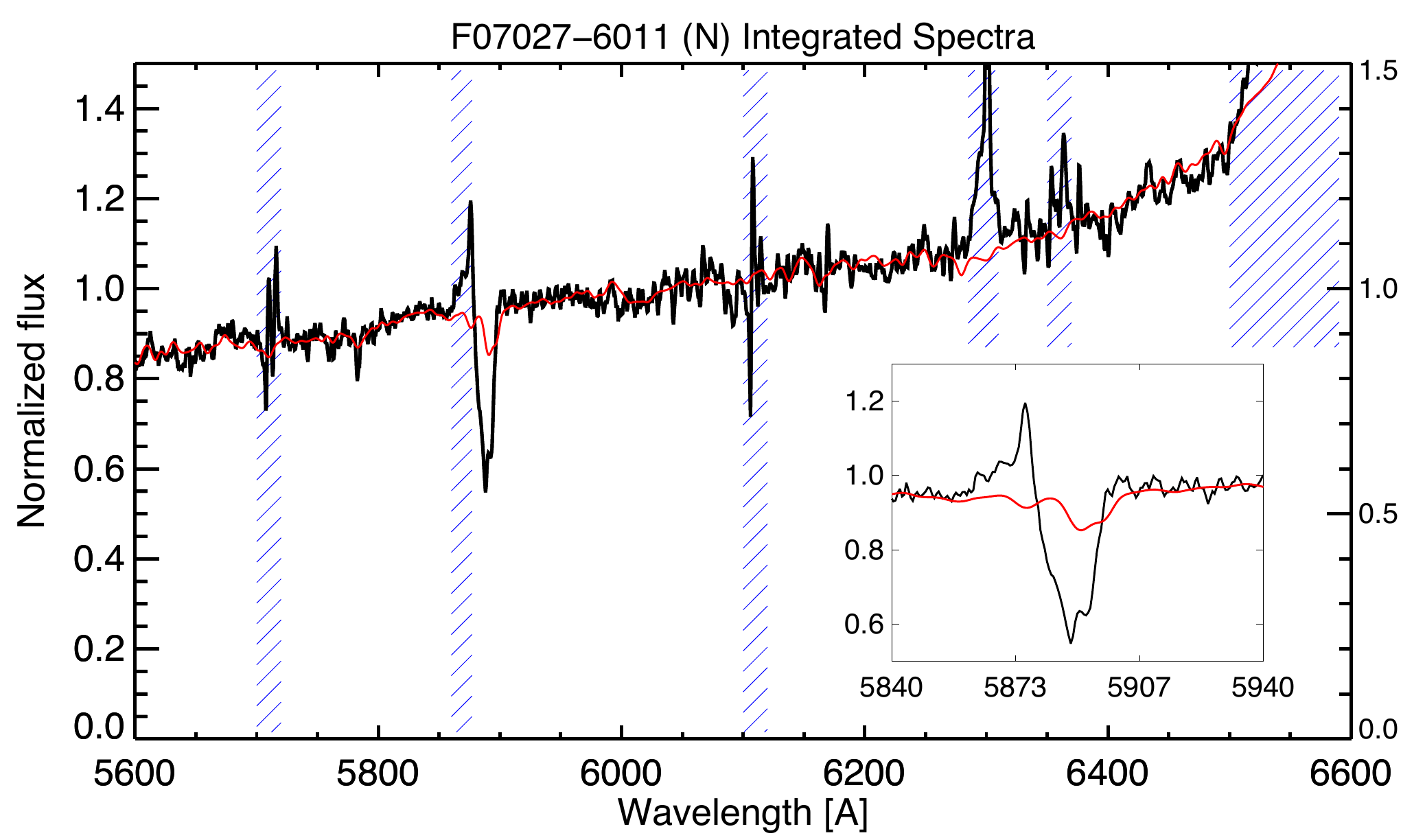}  
\caption{As Fig.~\ref{Panel_F01159} but for IRAS F07027-6011\,(N).}
\label{Panel_F07027N}            
\end{figure*}
\clearpage

       
\begin{figure*}
\centering
 \includegraphics[trim = -0.0cm .5cm .5cm 13.0cm, clip=true, width=.8\textwidth]{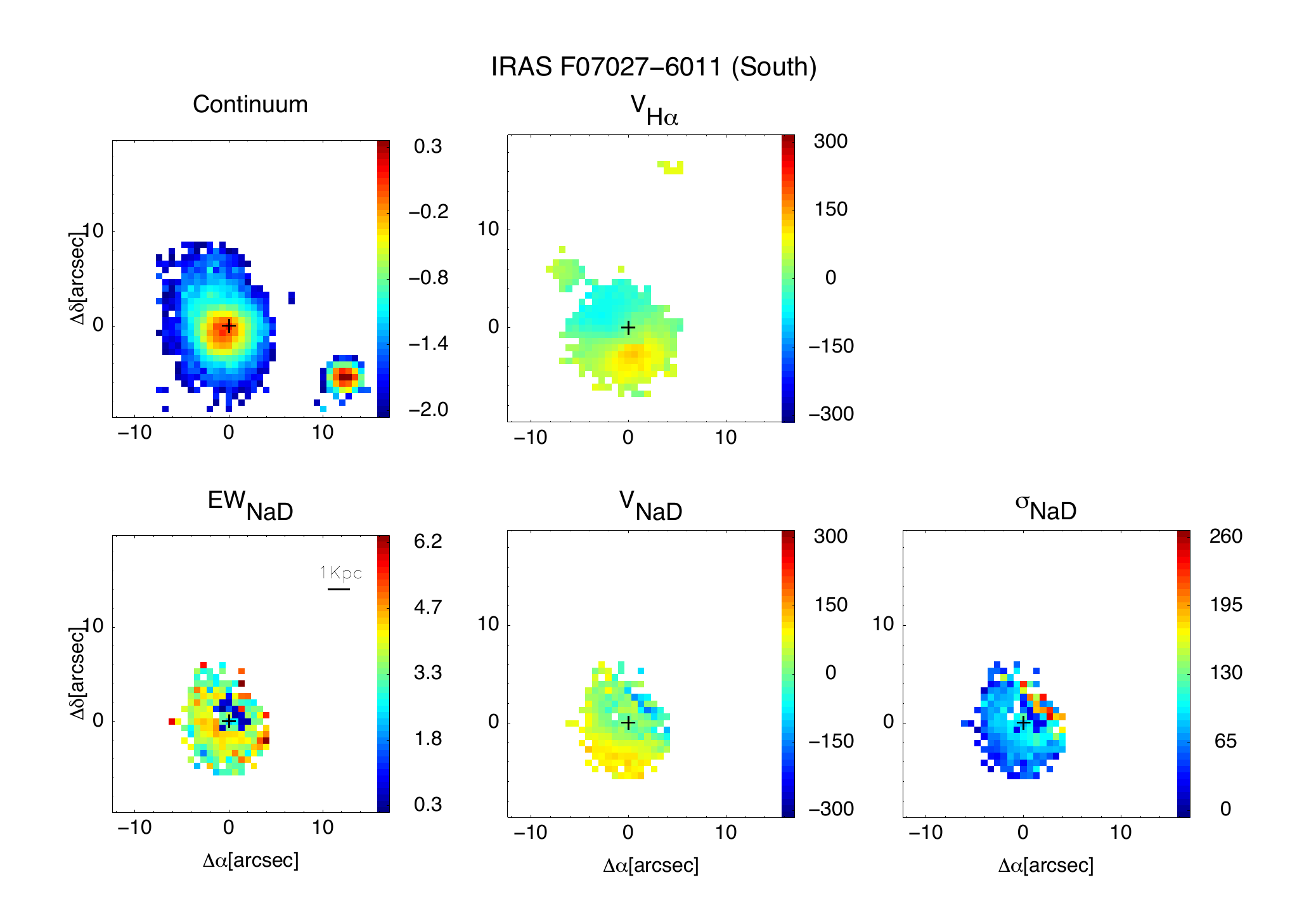} 
\centering
\includegraphics[trim = -0.0cm 0.5cm .5cm 19.5cm, clip=true, width=.8\textwidth]{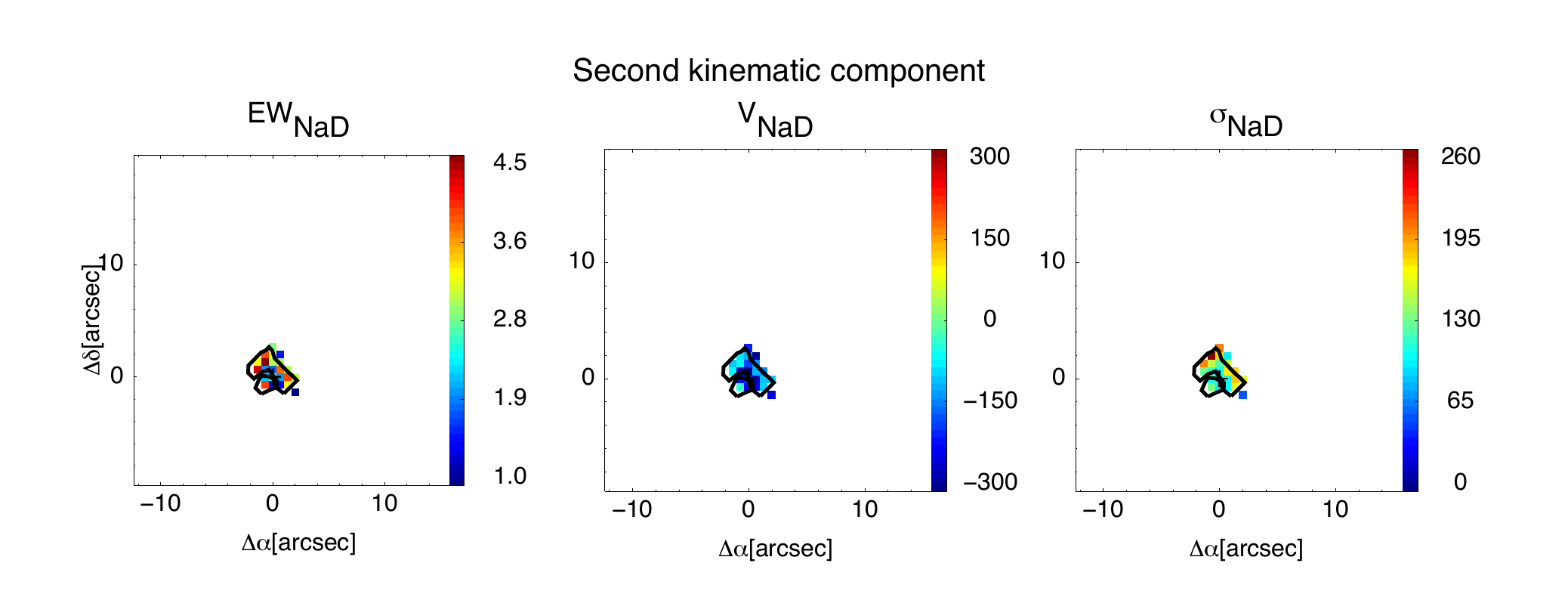}  
\centering
\includegraphics[trim = -0.0cm 8.5cm .5cm 7.0cm, clip=true, width=.7\textwidth]{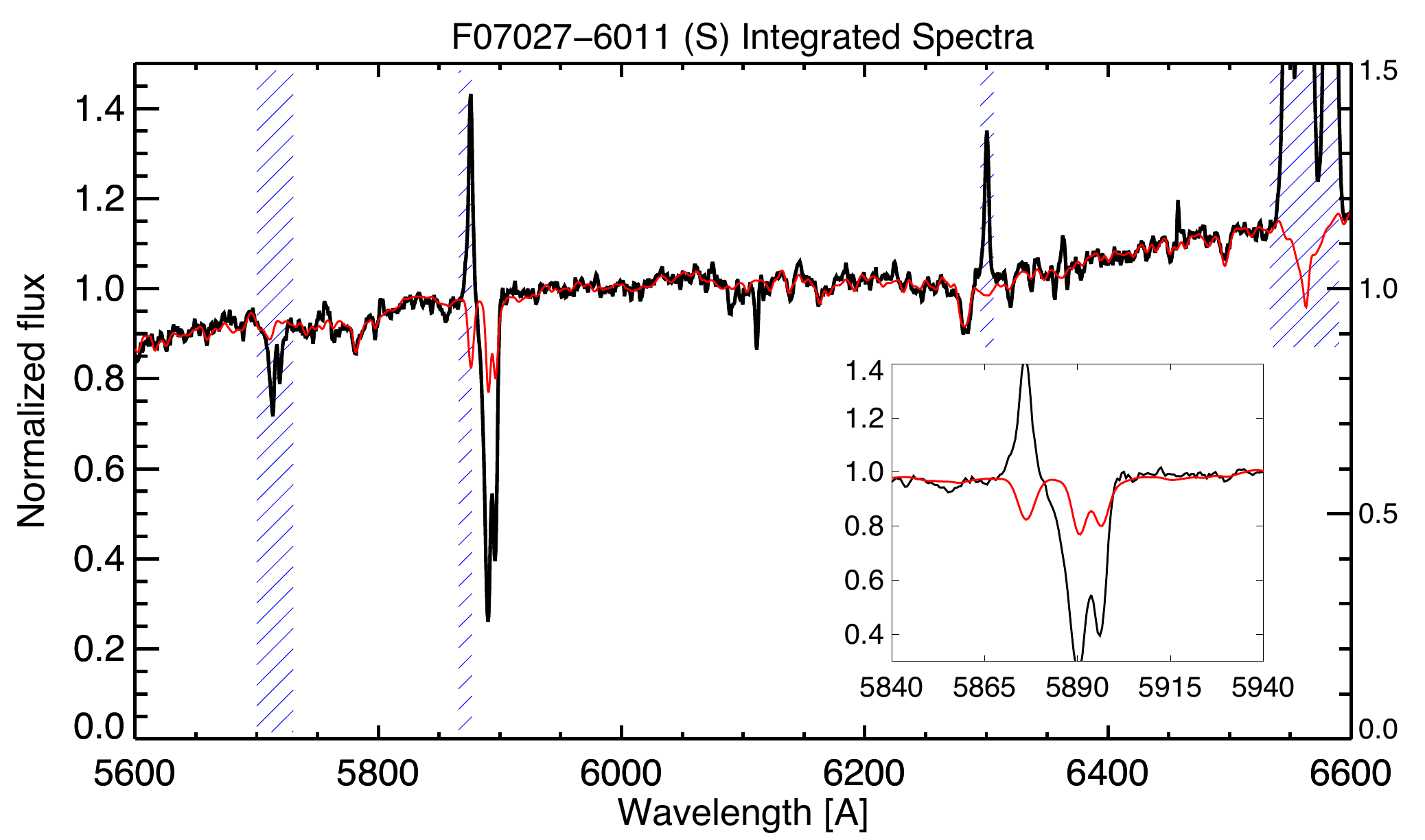}  
\caption{As Fig.~\ref{Panel_F01159} but for IRAS F07027-6011\,(S). Maps of the second kinematic component of NaD are also included (third row.)}
 \label{Panel_F07027S}           
\end{figure*}
\clearpage

            
\begin{figure*}
\centering
\includegraphics[trim = -0.0cm 0.2cm .5cm 13.0cm, clip=true, width=1.\textwidth]{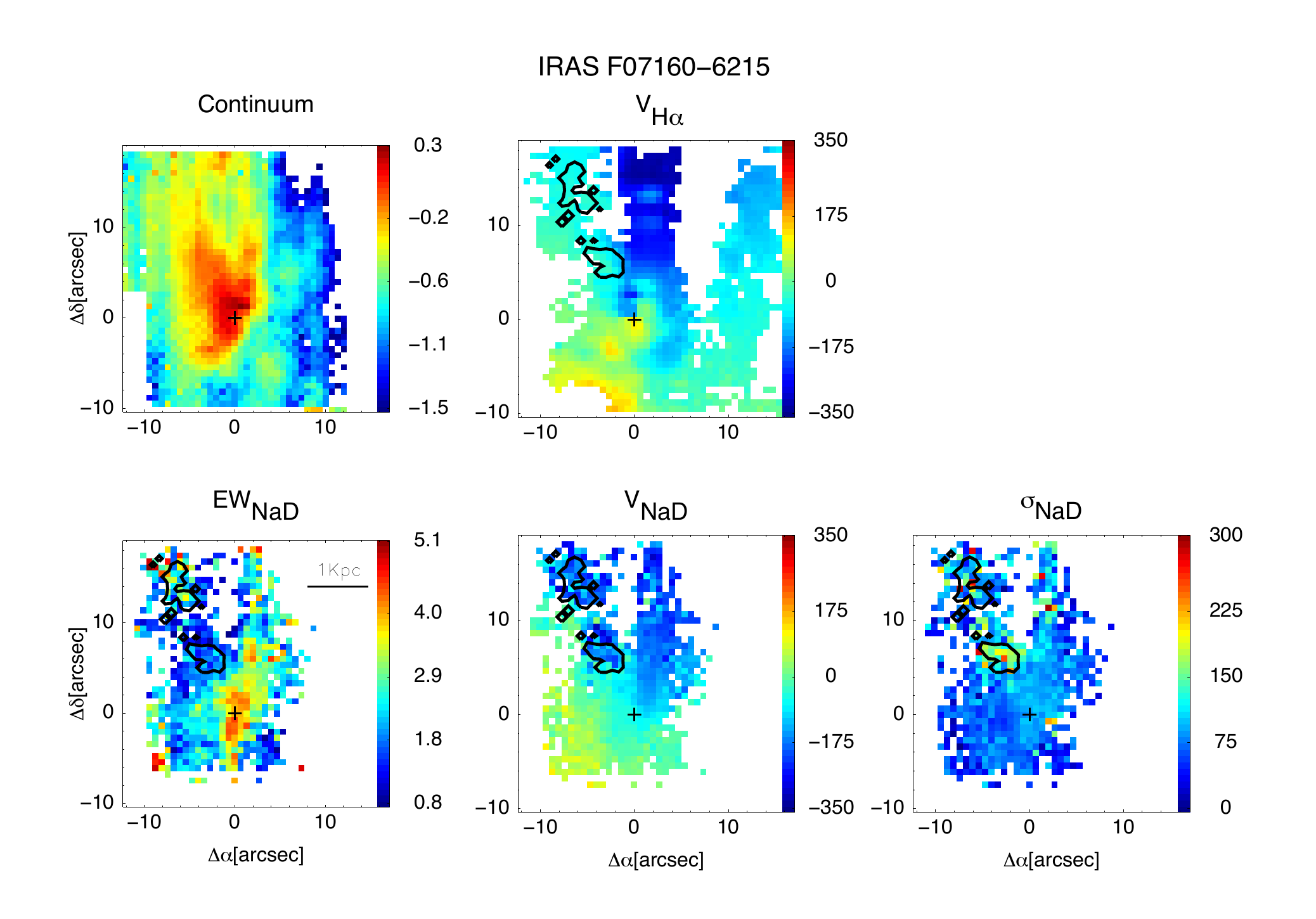}   
\centering
 \includegraphics[trim = -0.0cm 8.5cm .5cm 7.0cm, clip=true, width=.8\textwidth]{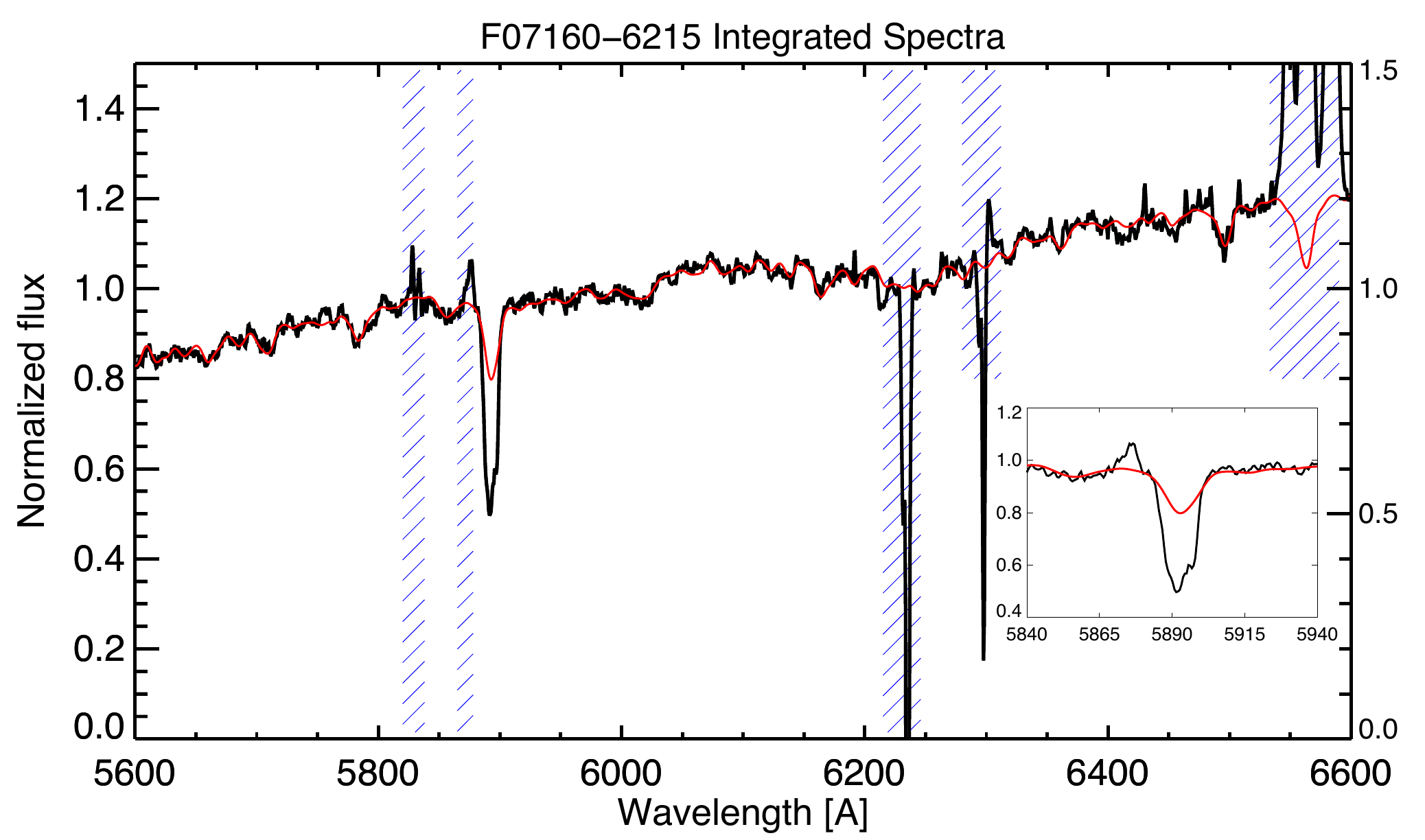}  
\caption{As Fig.~\ref{Panel_F01159} but for IRAS F7160-6215.}
 \label{Panel_F07160}            
\end{figure*}
\clearpage


\begin{figure*}
\centering
\includegraphics[trim = -0.0cm 0.2cm .5cm 13.0cm, clip=true, width=1.\textwidth]{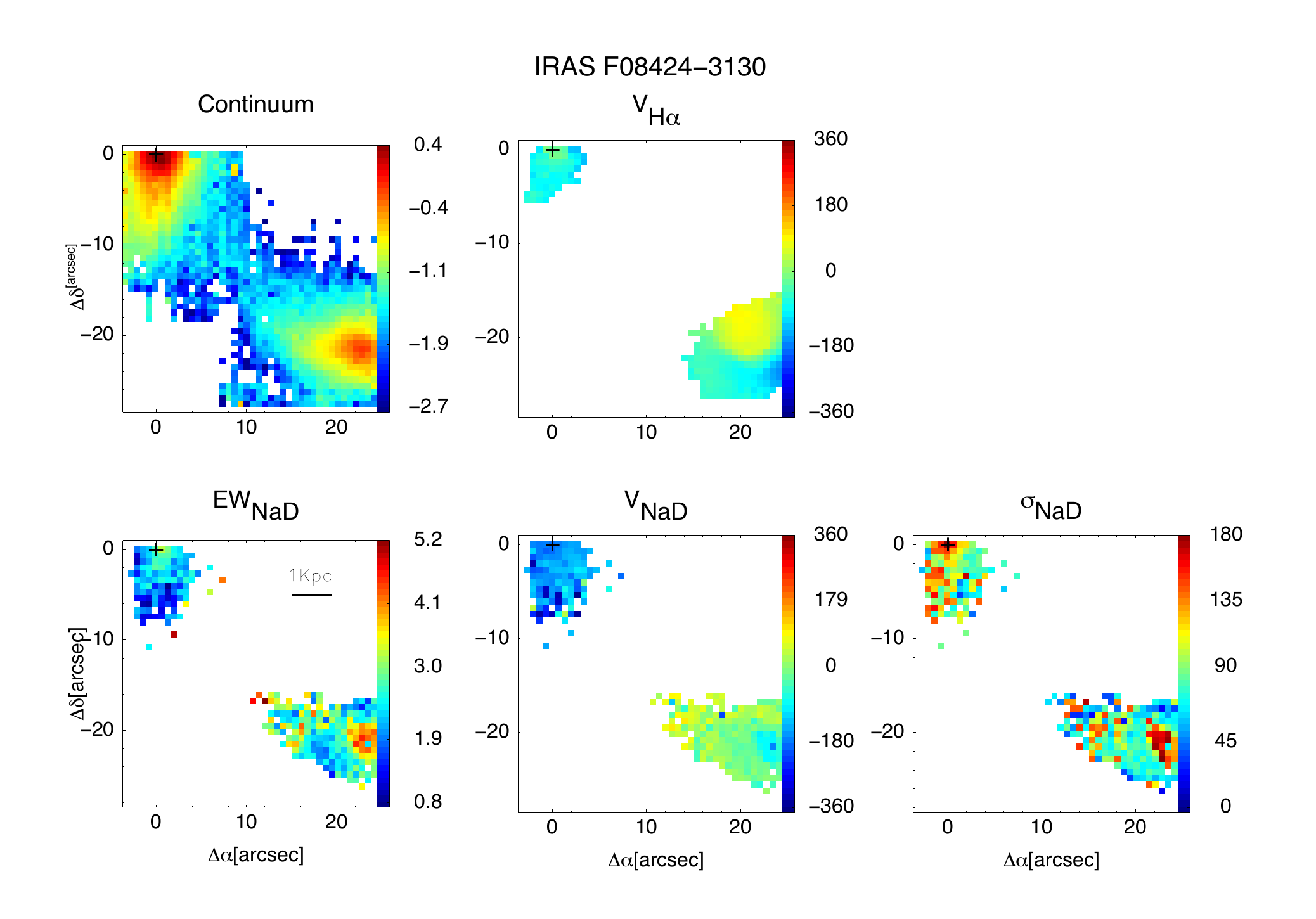}   
 \centering
\includegraphics[width=.95\textwidth]{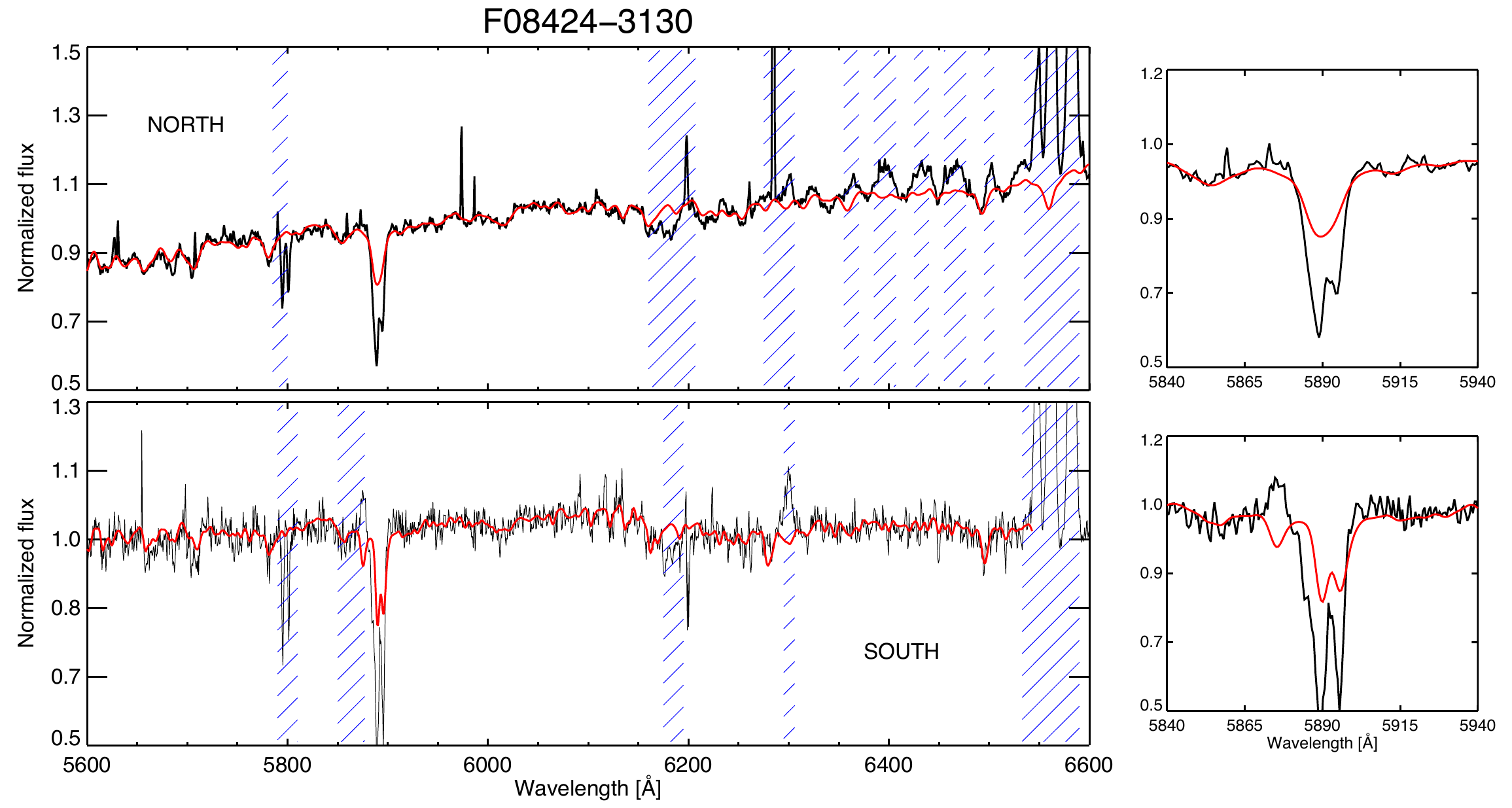}  
\caption{As Fig.~\ref{Panel_F01159} but for IRAS F08424-3130.}
 \label{Panel_F08424}            
\end{figure*}
\clearpage

                
\begin{figure*}
\centering
\vspace{4.em}   
\includegraphics[trim = -0.0cm 8.5cm .5cm 7.0cm, clip=true, width=.72\textwidth]{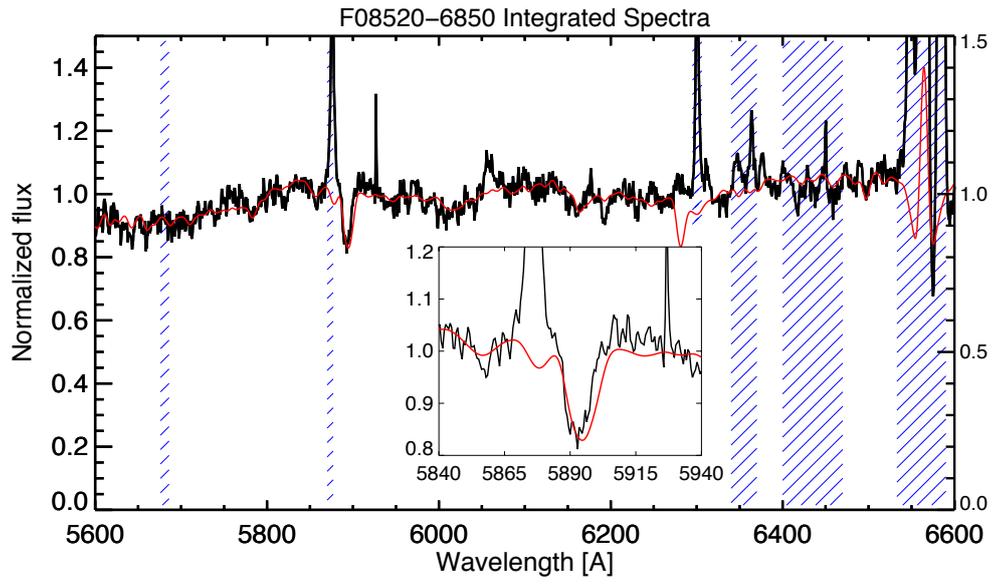} 
\caption{As in the lower panel of Fig.~\ref{Panel_F01159} but for IRAS 08520-6850.}
\end{figure*}


\begin{figure*}
\centering
\includegraphics[trim = -0.0cm 0.2cm .5cm 13.0cm, clip=true, width=1.\textwidth]{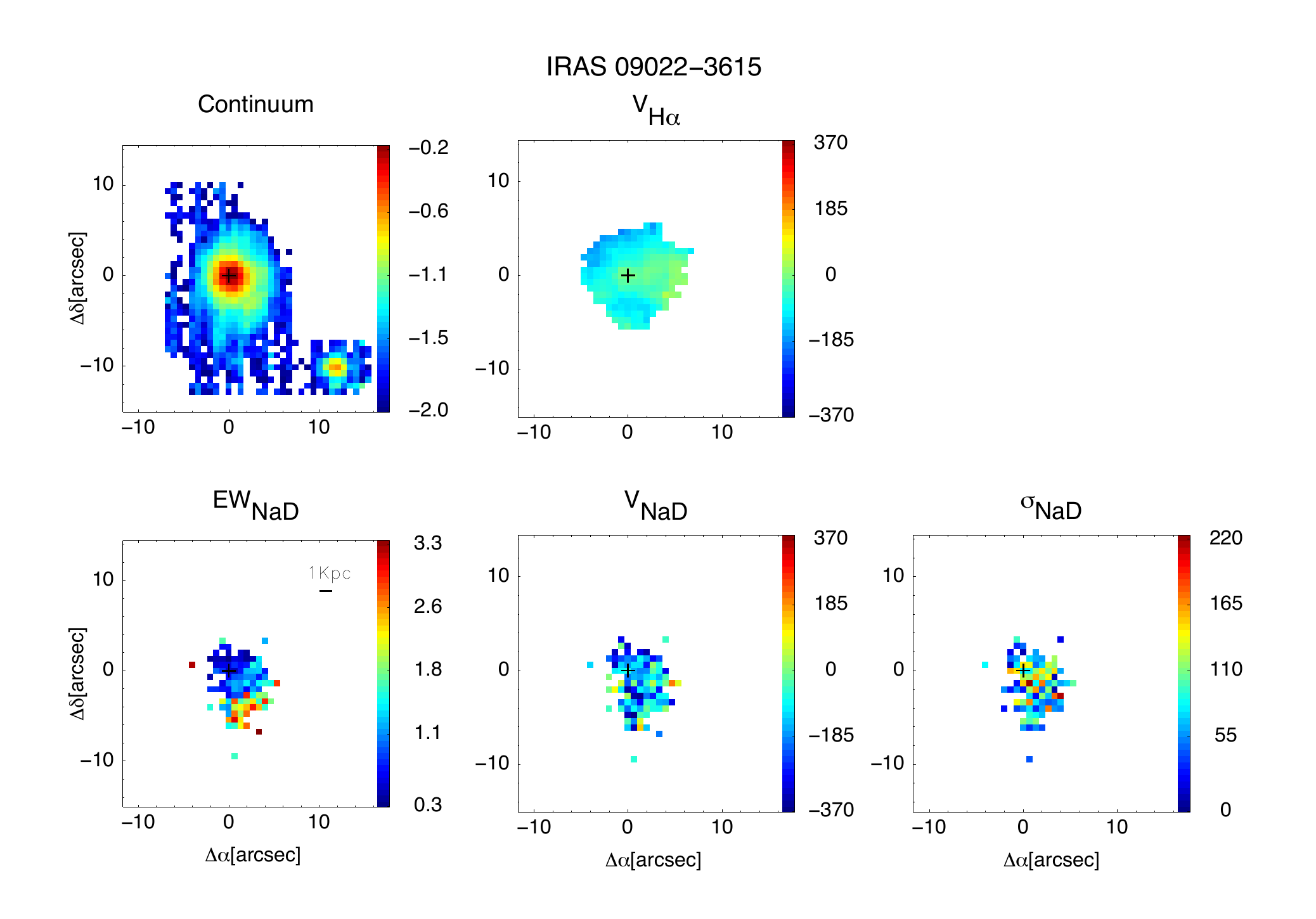}  
\centering
\includegraphics[trim = -0.0cm 8.5cm .5cm 7.0cm, clip=true, width=.8\textwidth]{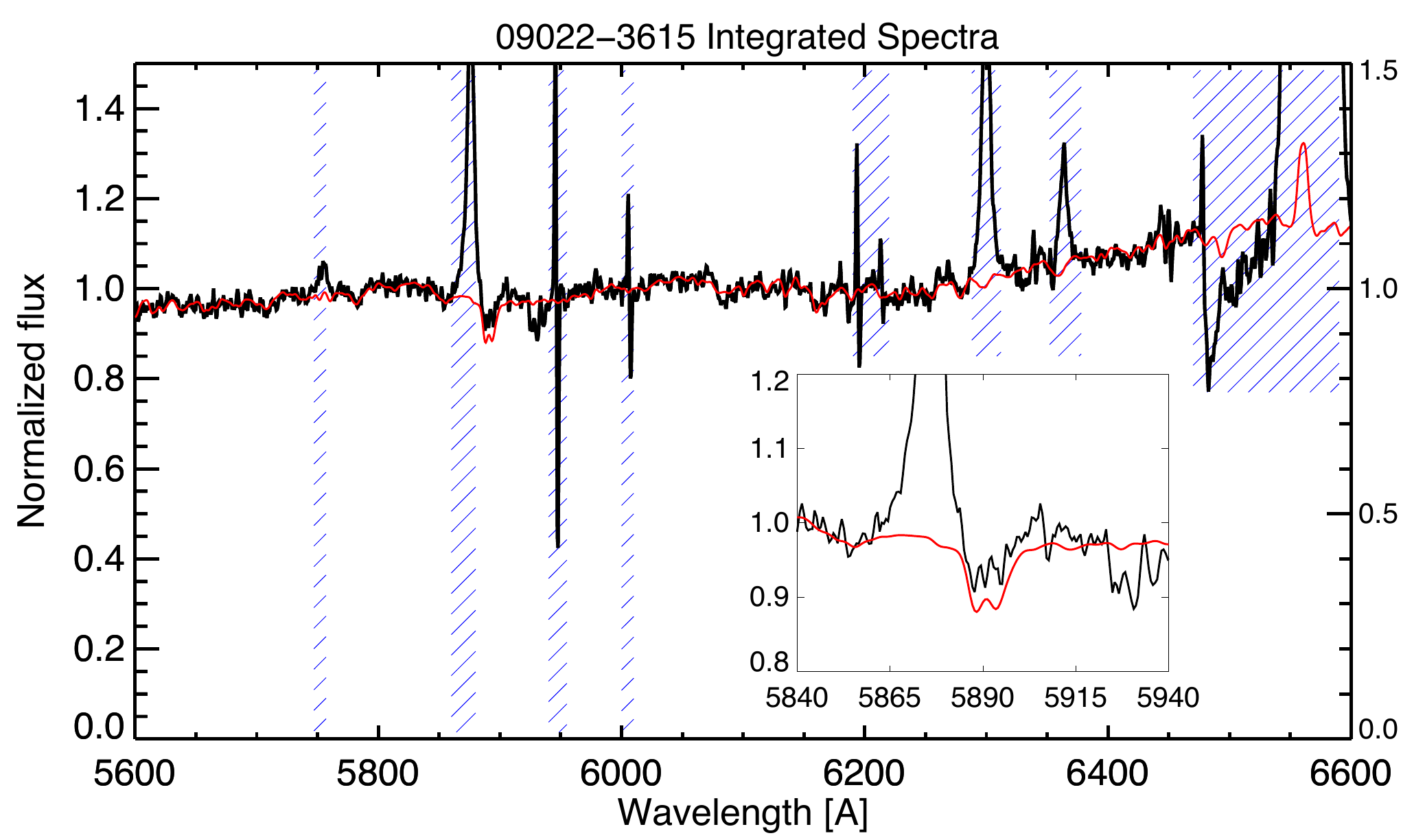}  
\caption{As Fig.~\ref{Panel_F01159} but for IRAS 09022-3615.}
\label{Panel_F9022}              
\end{figure*}
\clearpage
   

\begin{figure*}
\centering
\includegraphics[trim = -0.0cm 0.2cm .5cm 13.0cm, clip=true, width=1.\textwidth]{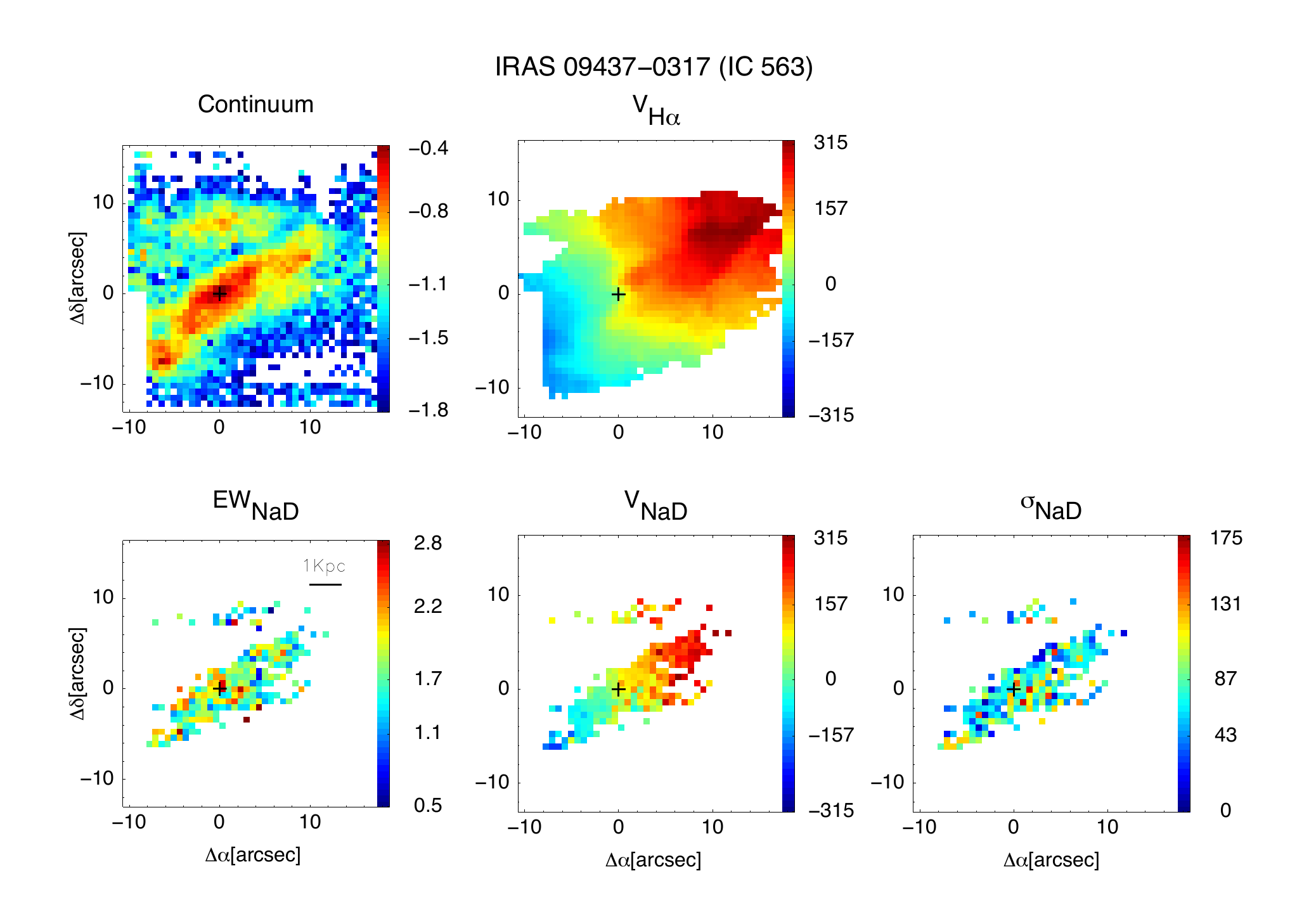}  
\centering
\includegraphics[trim = -0.0cm 8.5cm .5cm 7.0cm, clip=true, width=.8\textwidth]{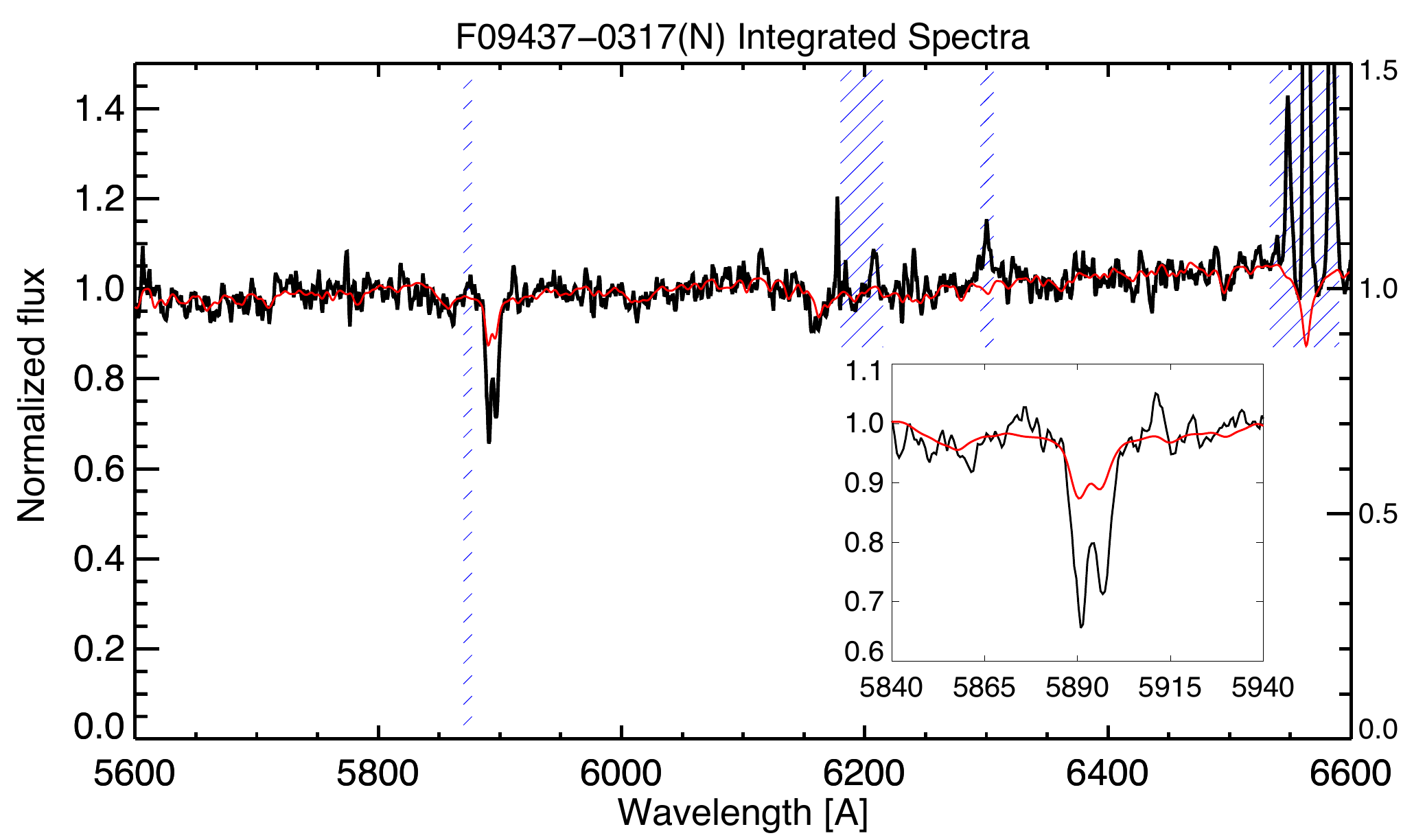}  
\caption{As Fig.~\ref{Panel_F01159} but for IRAS F09437-0317\,(N).}
\label{Panel_F09437}             
\end{figure*}
\clearpage


\begin{figure*}
\centering
 \includegraphics[trim = -0.0cm 0.2cm .5cm 13.0cm, clip=true, width=1.\textwidth]{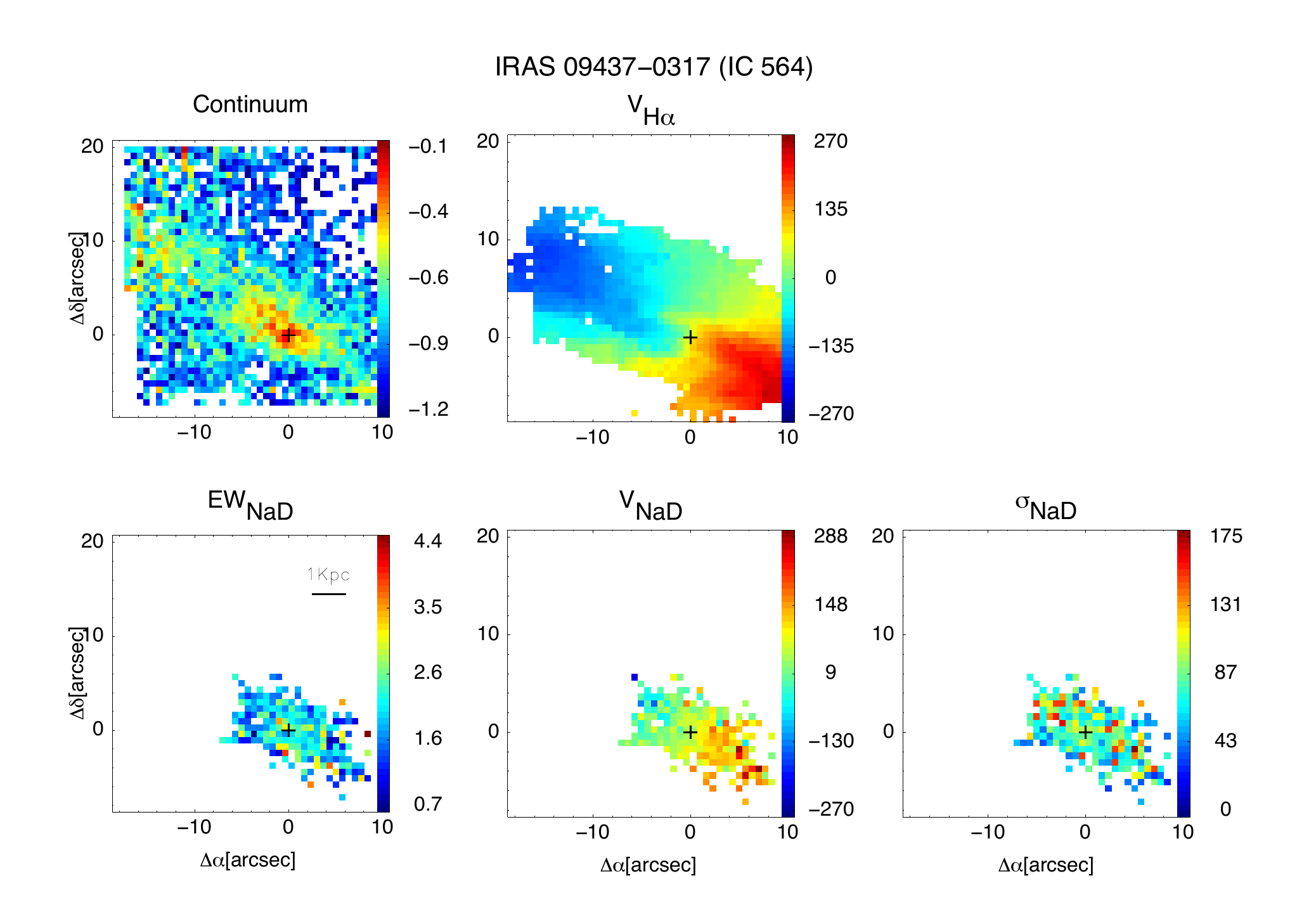}  
\centering
\includegraphics[trim = -0.0cm 8.5cm .5cm 7.0cm, clip=true, width=.8\textwidth]{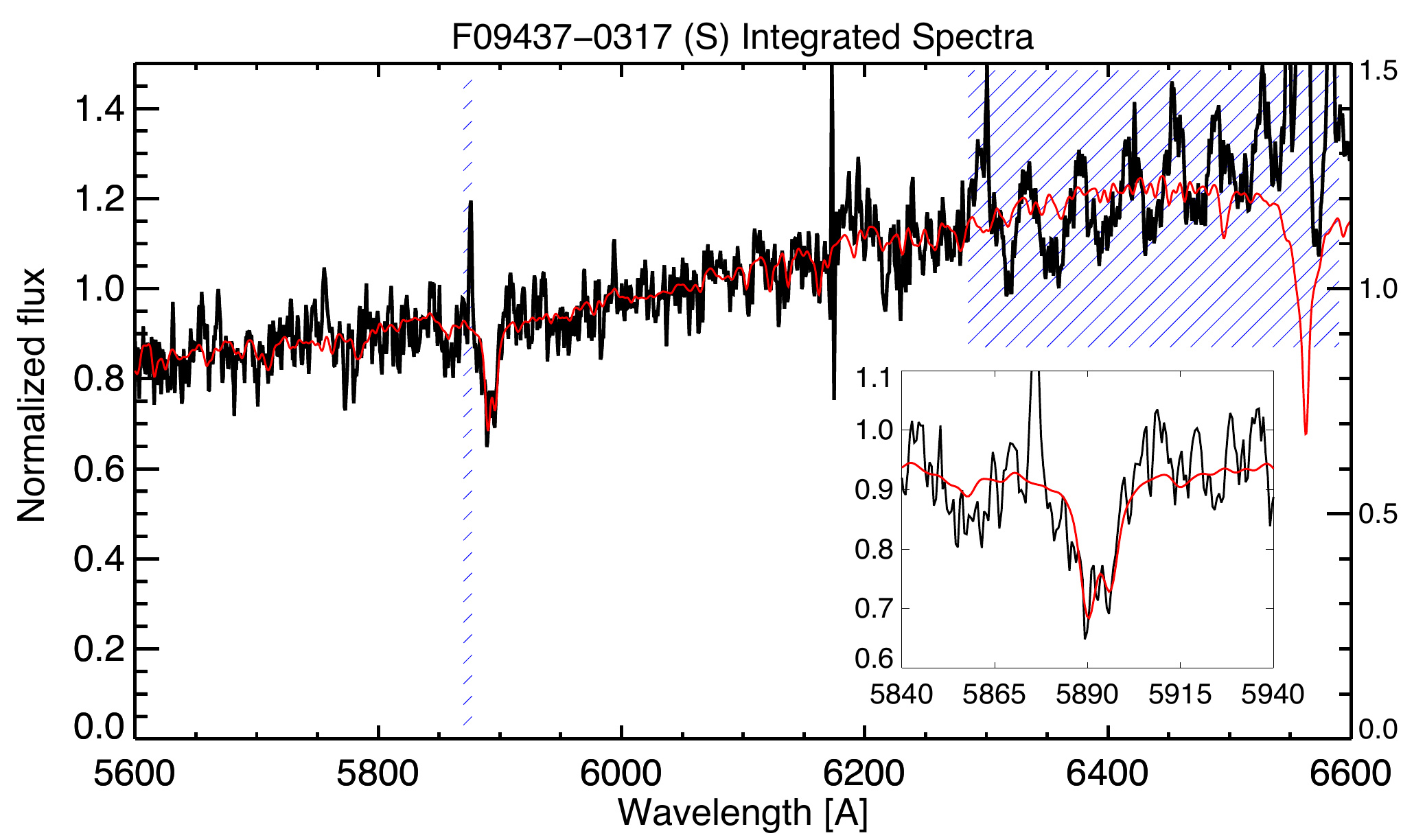}   
\caption{As Fig.~\ref{Panel_F01159} but for  IRAS F09437-0317.}
\label{Panel_F09437N1}           
\end{figure*}
\clearpage


\begin{figure*}
\centering
 \includegraphics[trim = -0.0cm 0.2cm .5cm 13.0cm, clip=true, width=1.\textwidth]{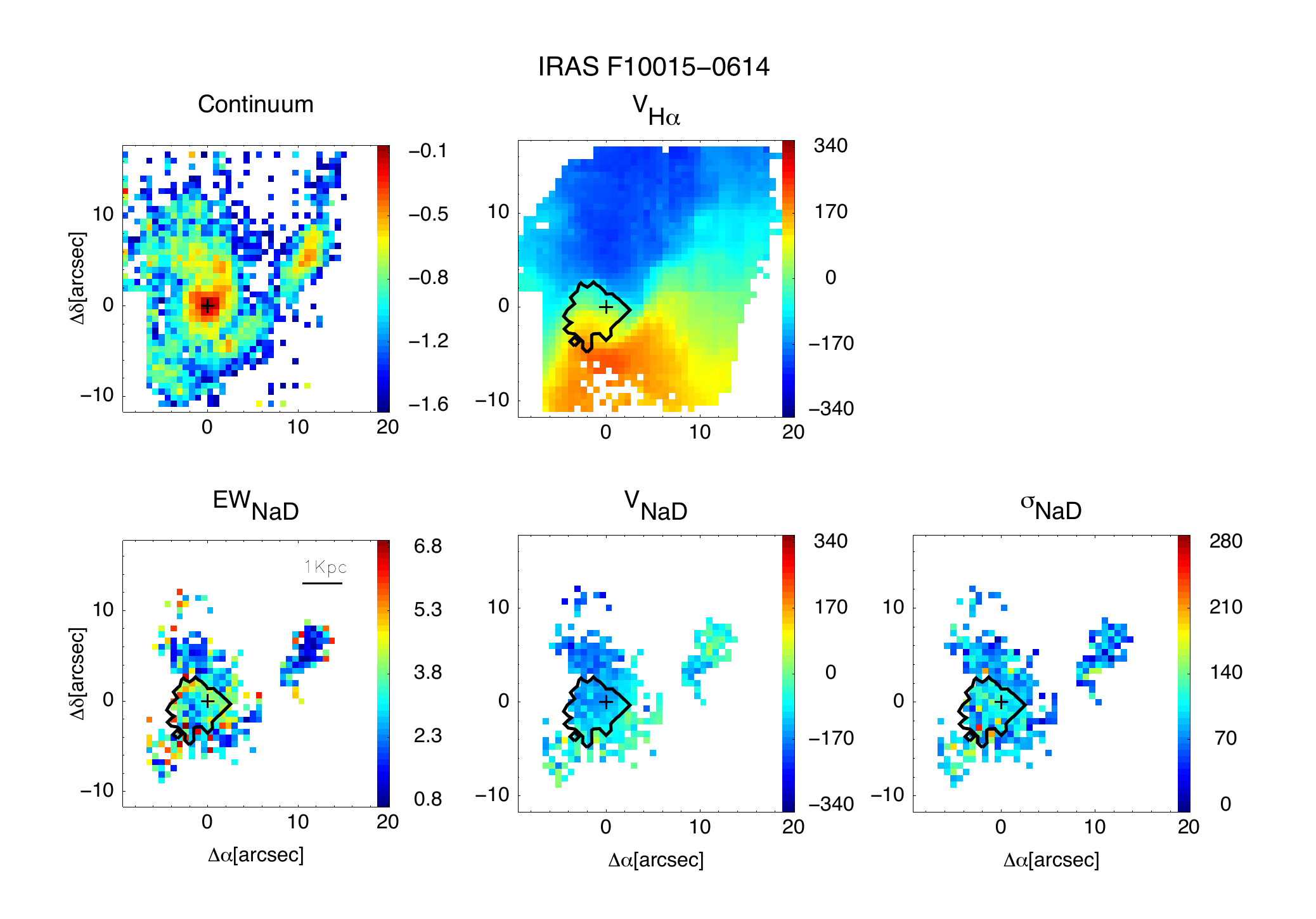}  
 \centering
 \includegraphics[trim = -0.0cm 8.5cm .5cm 7.0cm, clip=true, width=.8\textwidth]{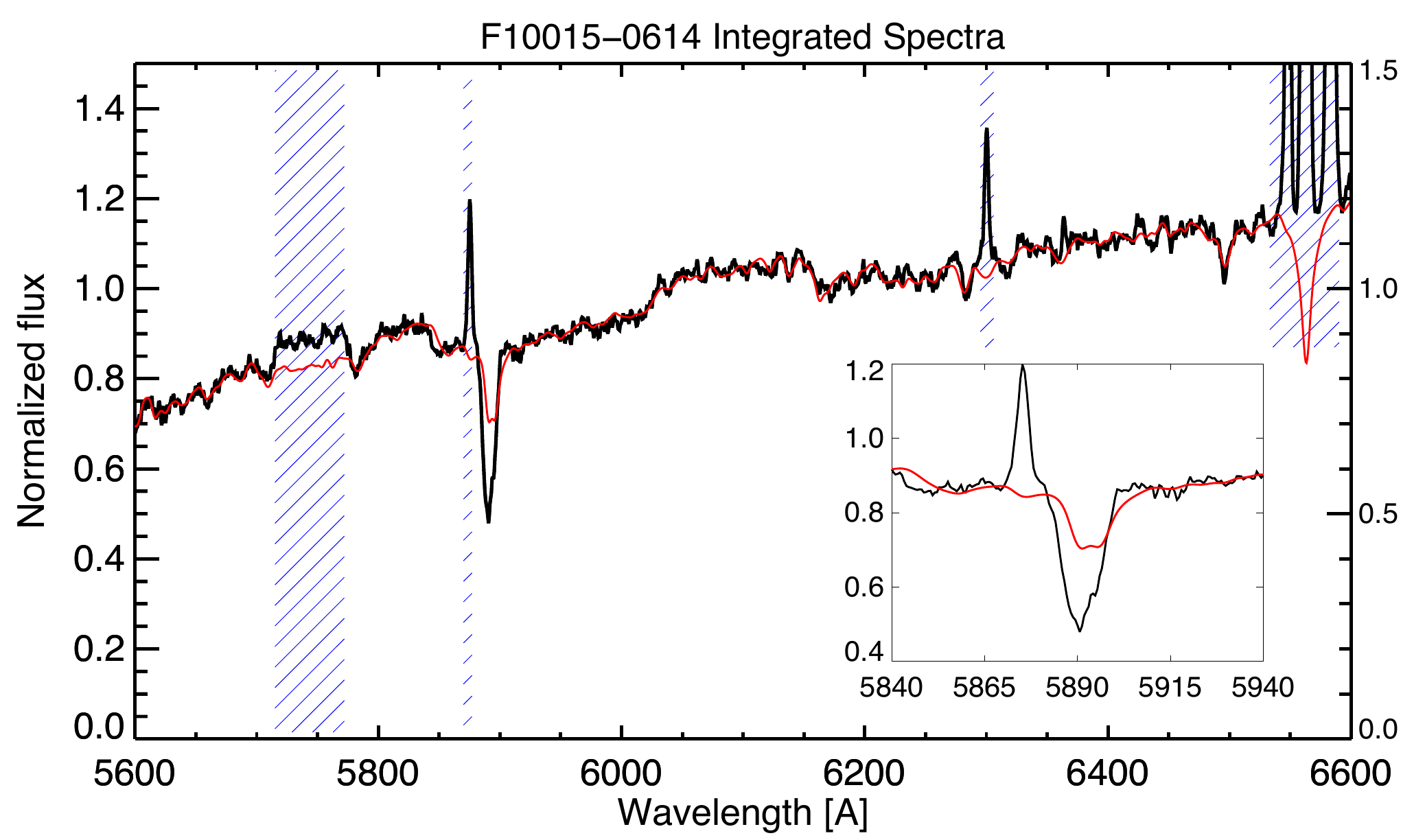}  
\caption{As Fig.~\ref{Panel_F01159} but for IRAS F10015-0614.}
 \label{Panel_F10015}            
\end{figure*}
\clearpage


\begin{figure*}
\centering
\includegraphics[trim = -0.0cm 0.2cm .5cm 13.0cm, clip=true, width=1.\textwidth]{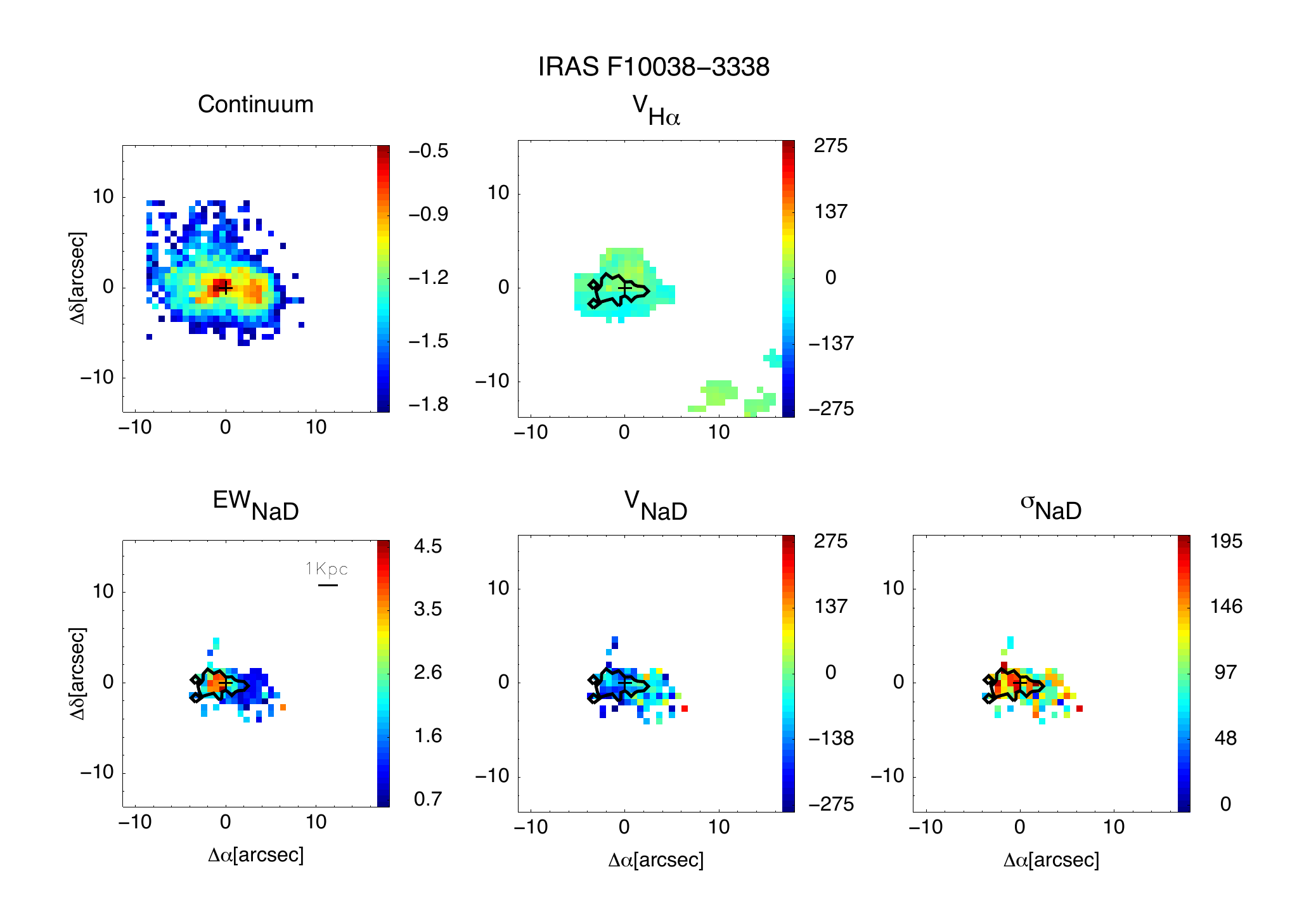}  
  \centering
\includegraphics[trim = -0.0cm 8.5cm .5cm 7.0cm, clip=true, width=.8\textwidth]{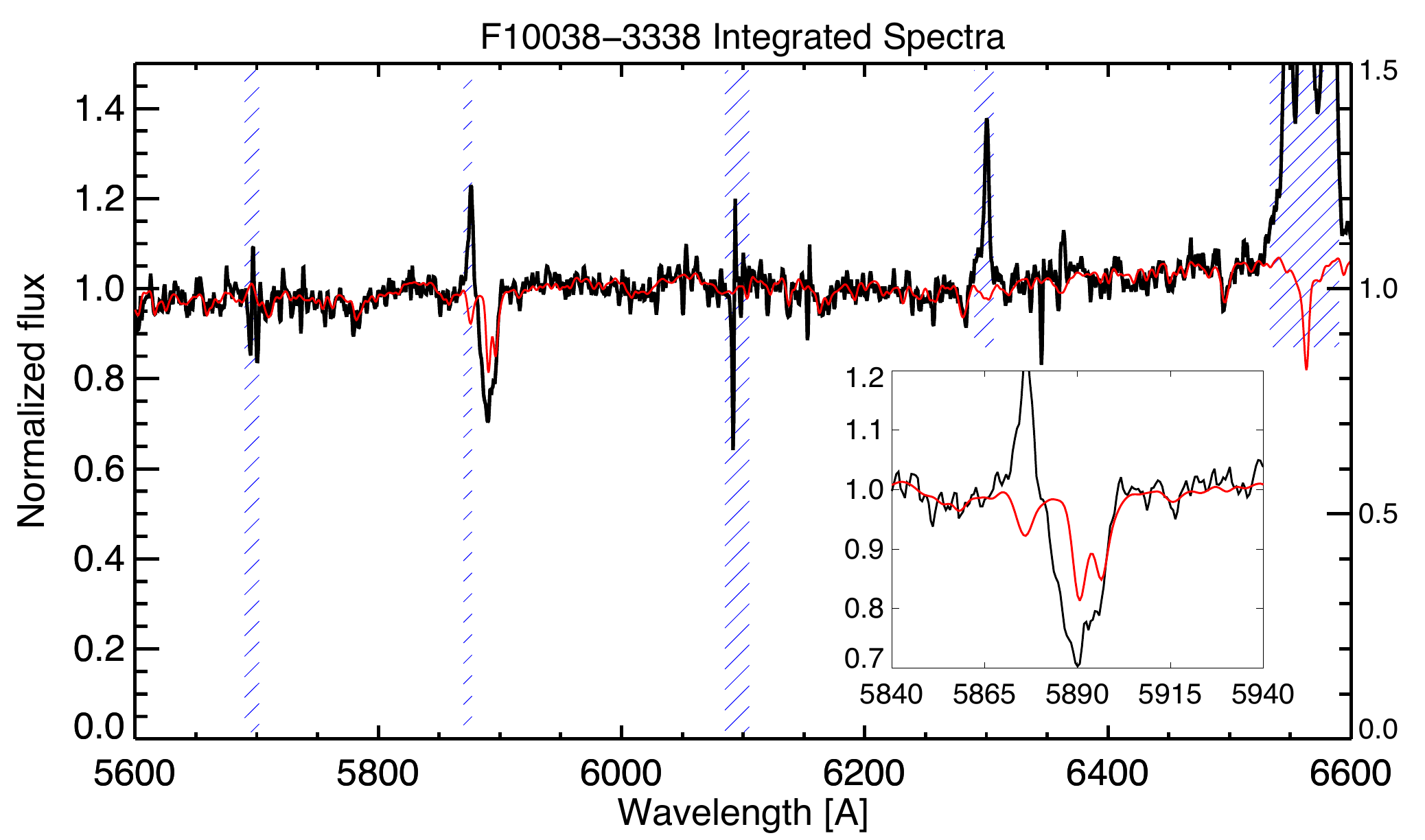}  
\caption{As Fig.~\ref{Panel_F01159} but for IRAS F10038-3338.}
\label{Panel_F10038}             
\end{figure*}
\clearpage
   

\begin{figure*}
\centering
\includegraphics[trim = -0.0cm 0.2cm .5cm 13.0cm, clip=true, width=1.\textwidth]{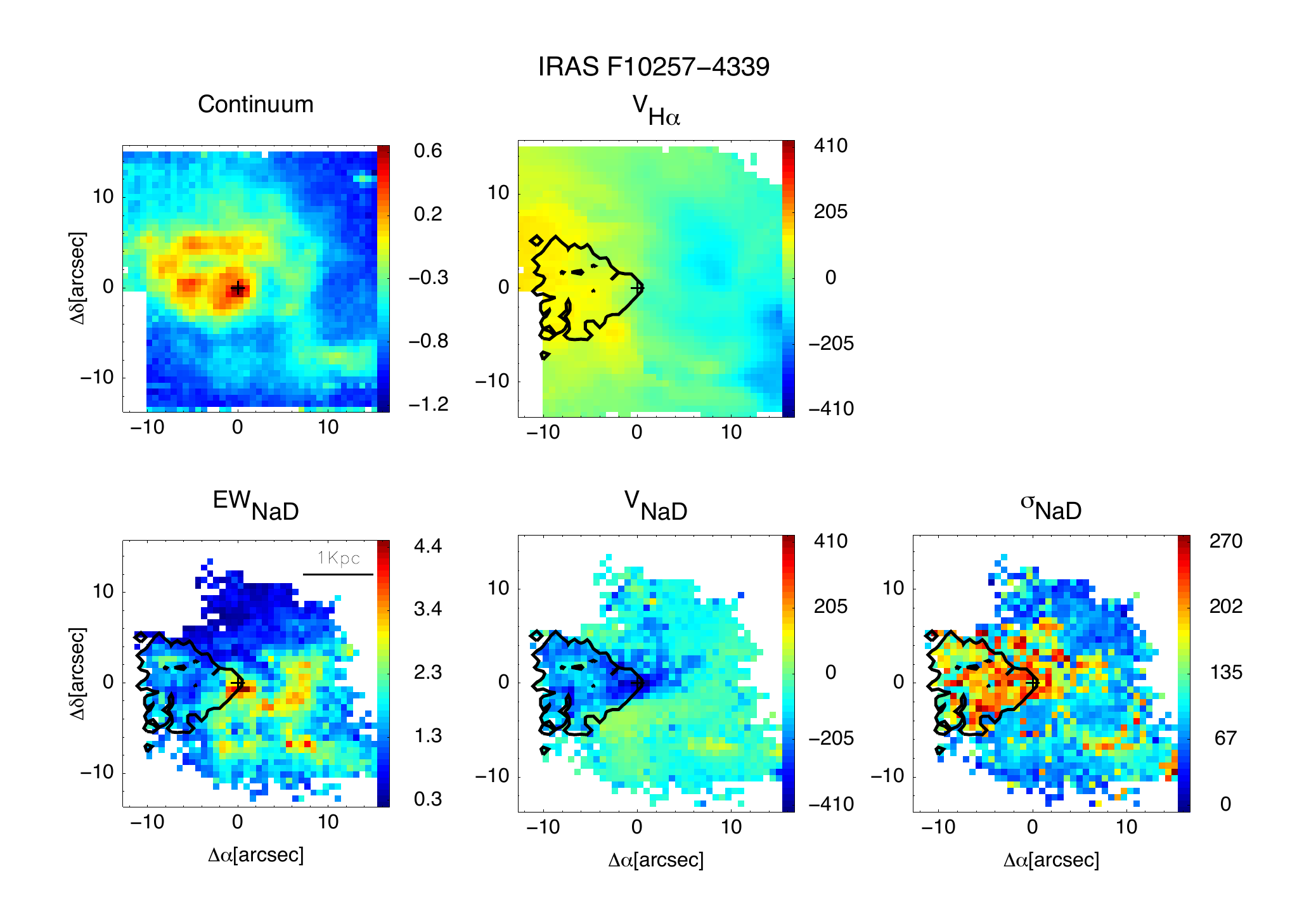}  %
\centering
\includegraphics[trim = -0.0cm 8.5cm .5cm 7.0cm, clip=true, width=.7\textwidth]{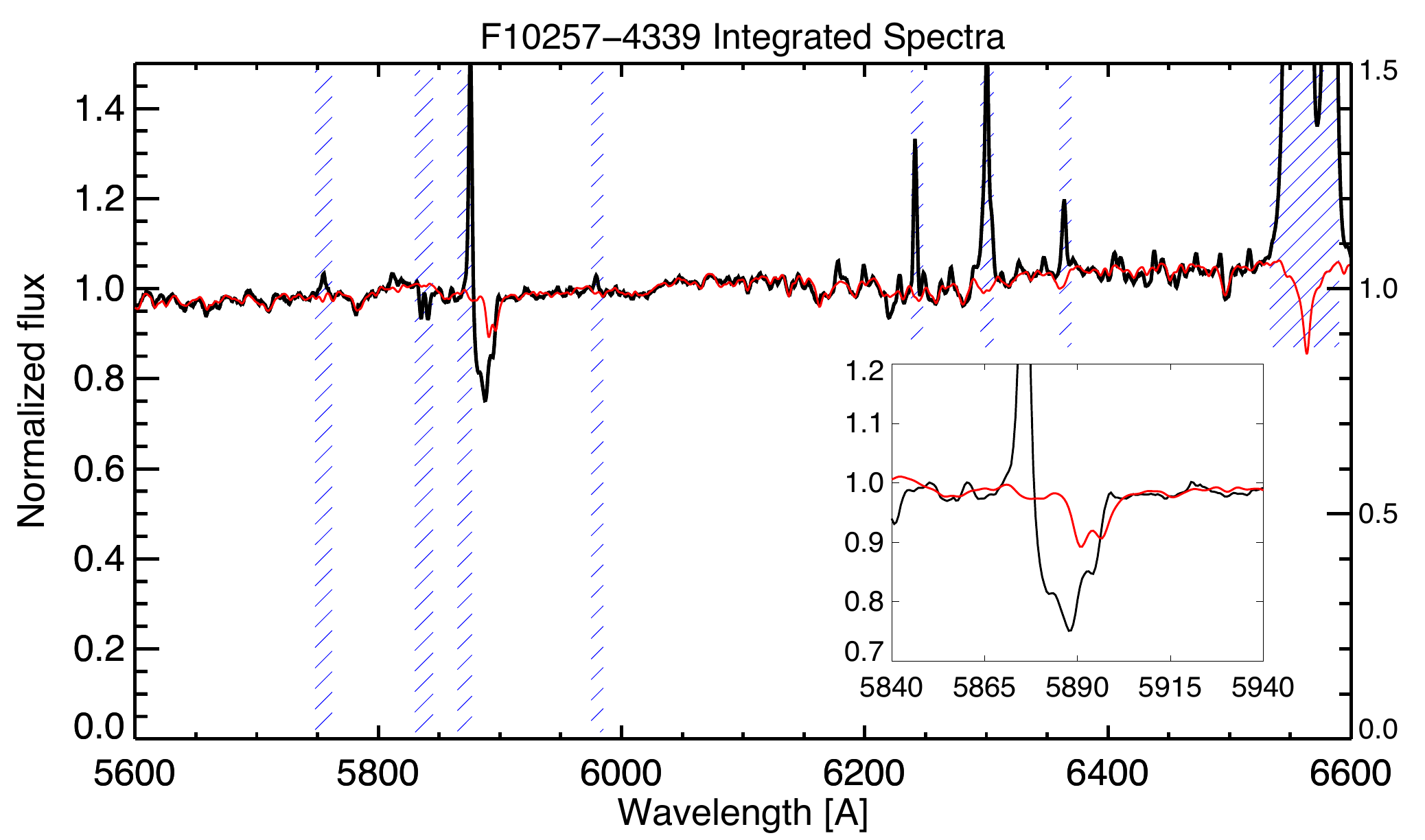}  
 \caption{As Fig.~\ref{Panel_F01159} but for IRAS F10257-4339.}
 \label{Panel_F10257}            
\end{figure*}
\clearpage


\begin{figure*}
\centering
\includegraphics[trim = -0.0cm 1.5cm .5cm 13.0cm, clip=true, width=.8\textwidth]{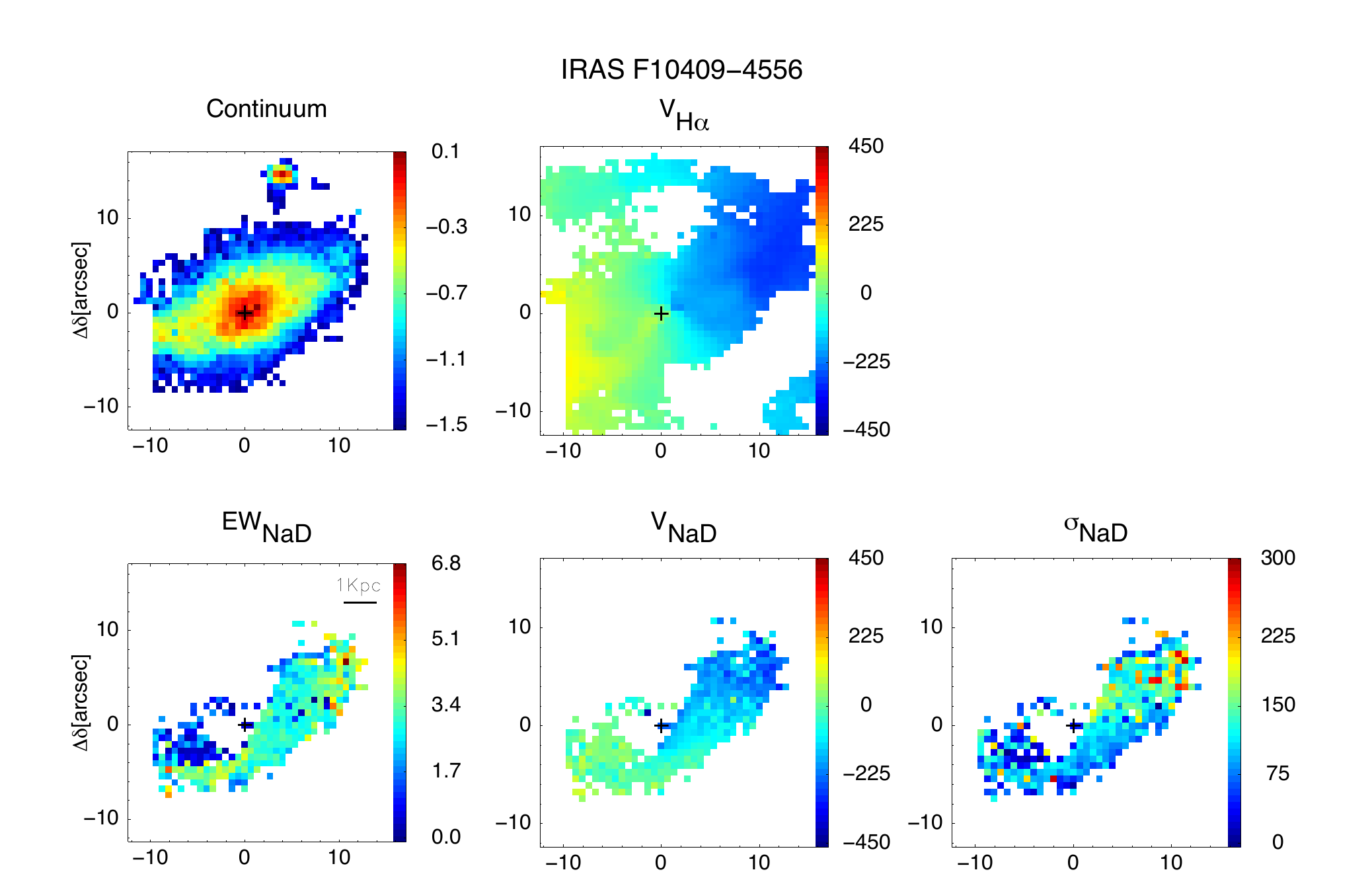} 
 \centering
 \includegraphics[trim = -0.0cm 0.5cm .5cm 19.5cm, clip=true, width=.8\textwidth]{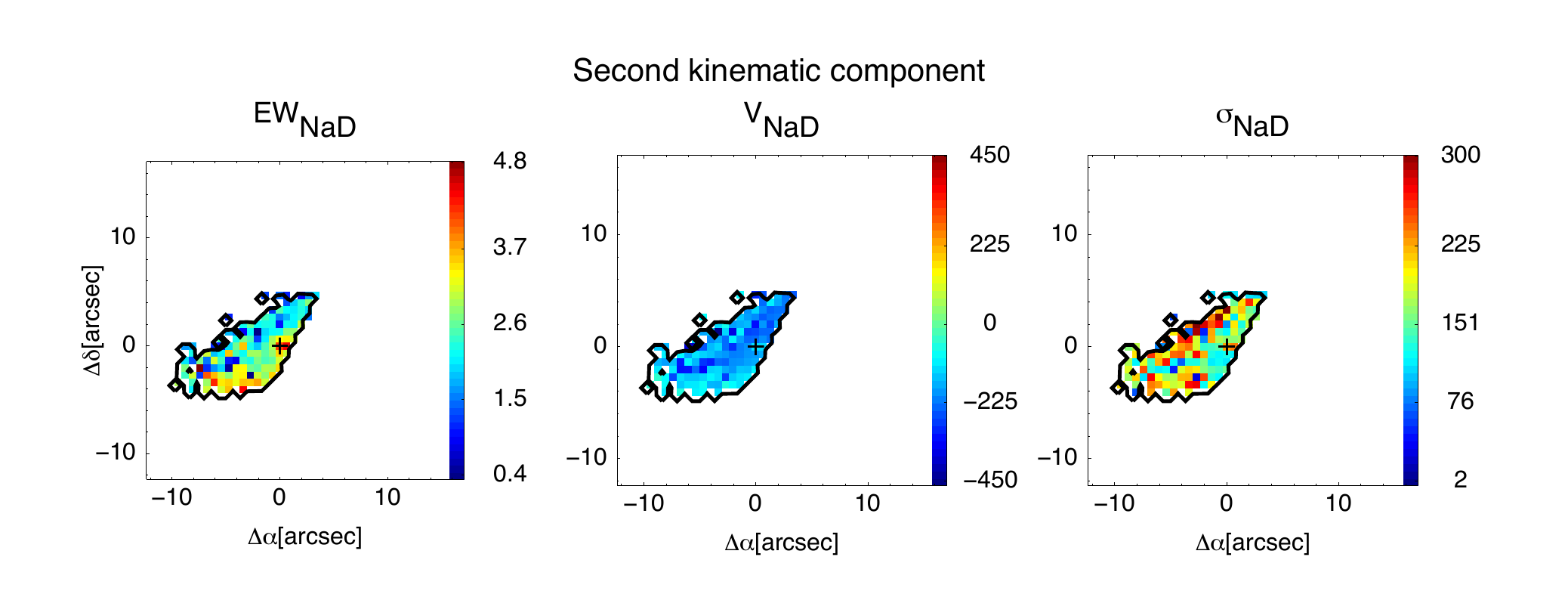}  
\centering
 \includegraphics[trim = -0.0cm 8.5cm .5cm 7.0cm, clip=true, width=.7\textwidth]{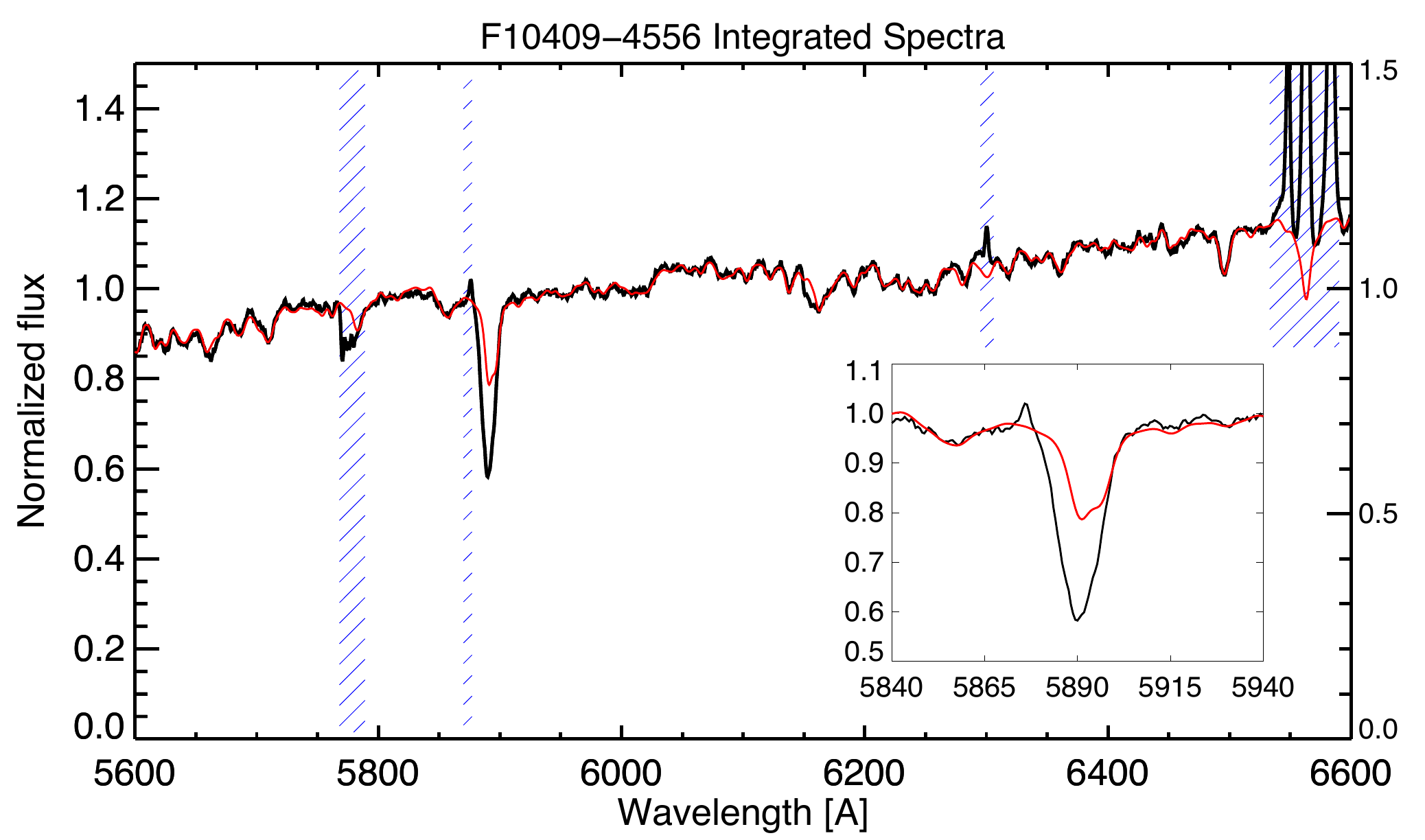}  
\caption{As Fig.~\ref{Panel_F07027S} but for IRAS F10409-4556.}
  \label{Panel_F10409}           
\end{figure*}
\clearpage

 
\begin{figure*}
\centering
\includegraphics[trim = -0.0cm 0.2cm .5cm 13.0cm, clip=true, width=1.\textwidth]{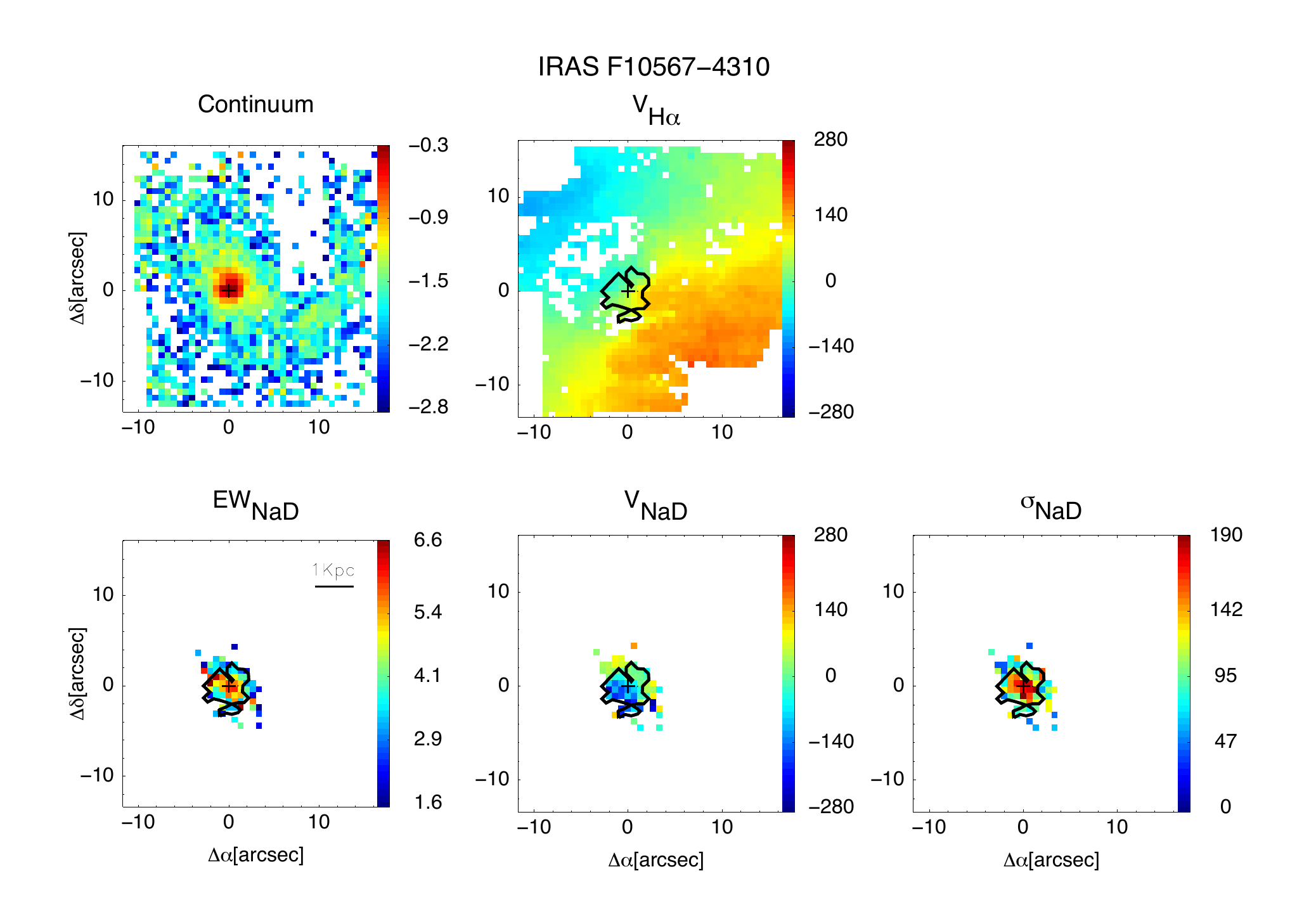}  
\centering
\includegraphics[trim = -0.0cm 8.5cm .5cm 7.0cm, clip=true, width=.8\textwidth]{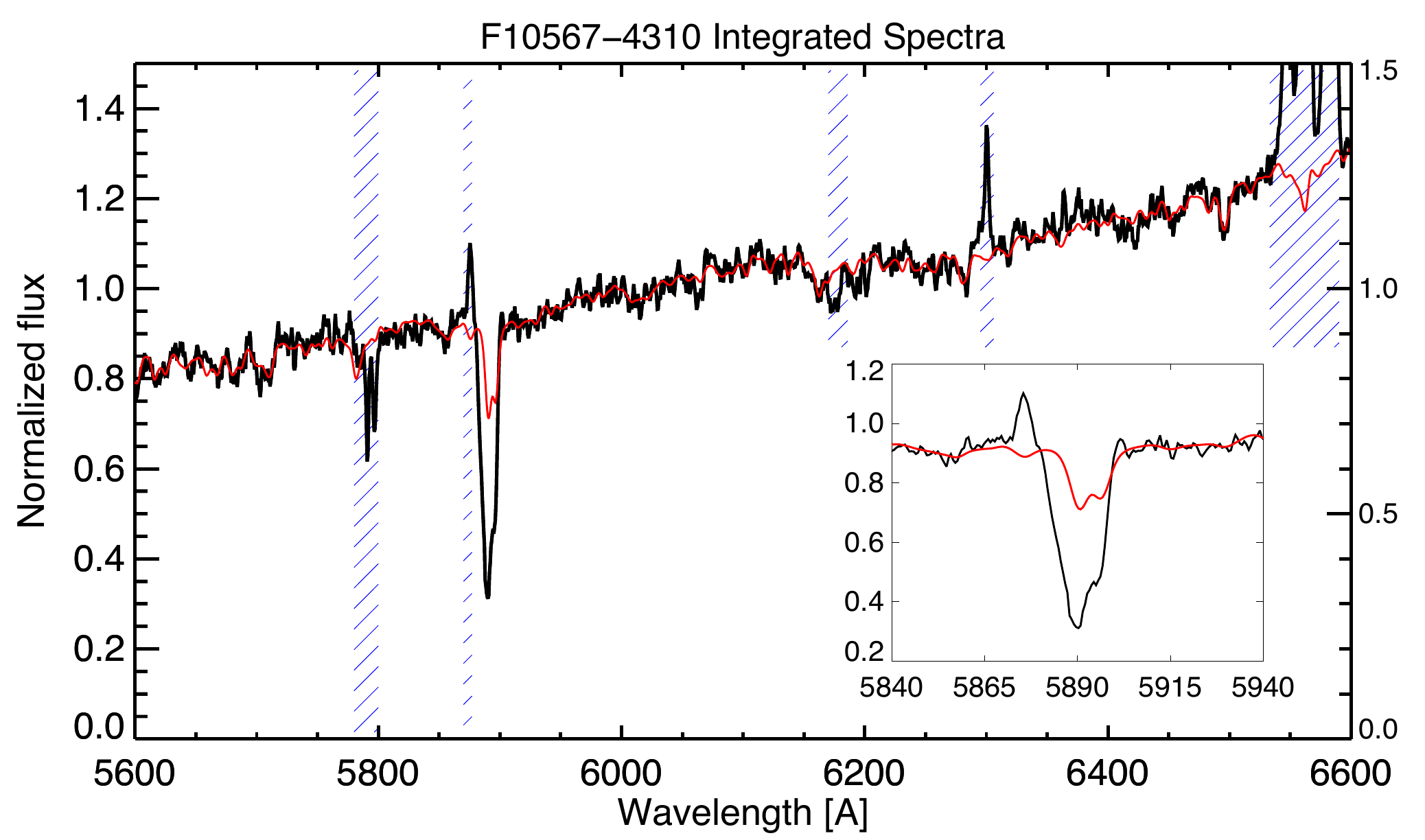}  
\caption{As Fig.~\ref{Panel_F01159} but for IRAS 10567-4310.}
\label{Panel_F10567}             
\end{figure*}
\clearpage


\begin{figure*}
\centering
 \includegraphics[trim = -0.0cm 0.2cm .5cm 13.0cm, clip=true, width=1.\textwidth]{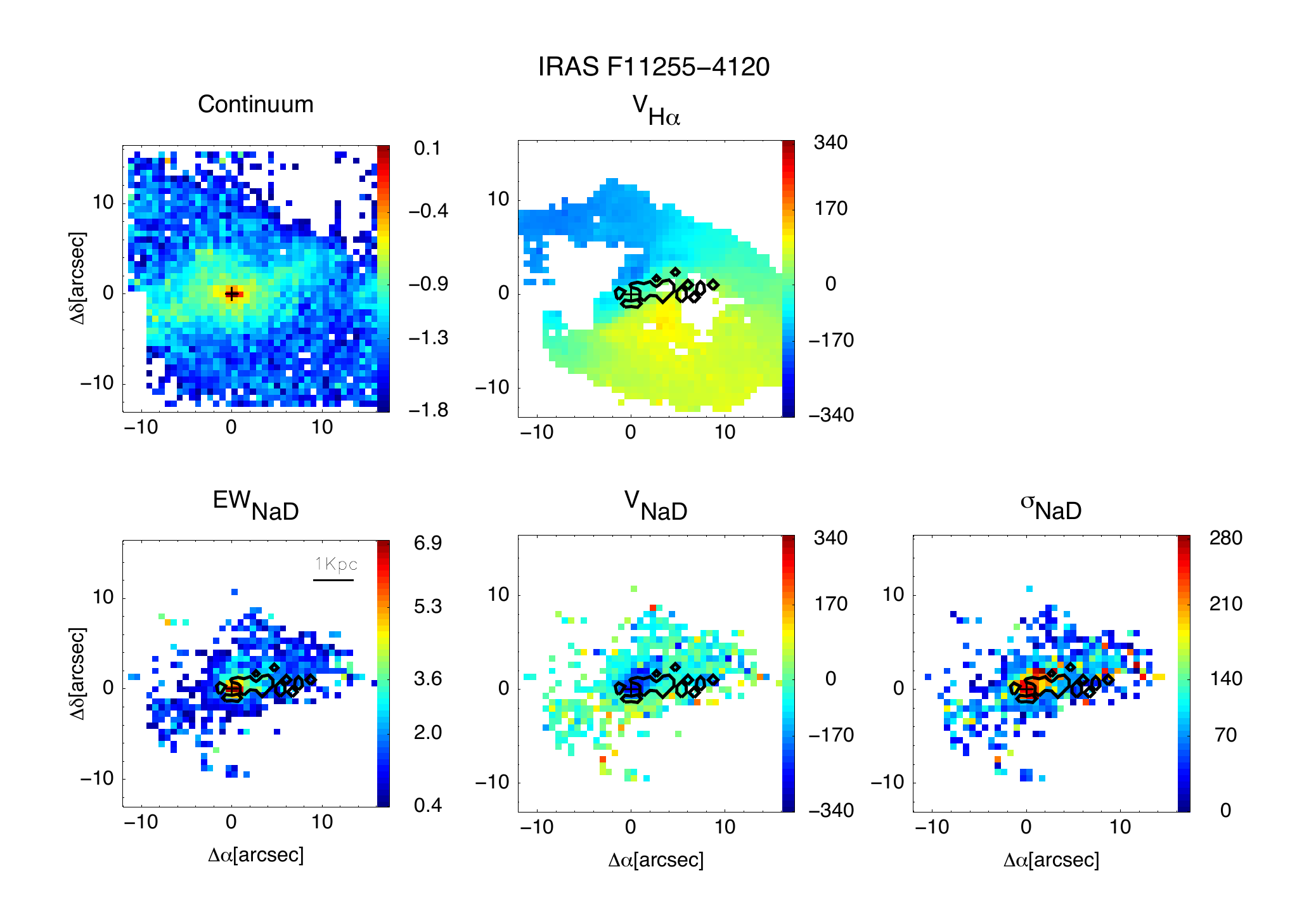}               
 \centering
 \includegraphics[trim = -0.0cm 8.5cm .5cm 7.0cm, clip=true, width=.8\textwidth]{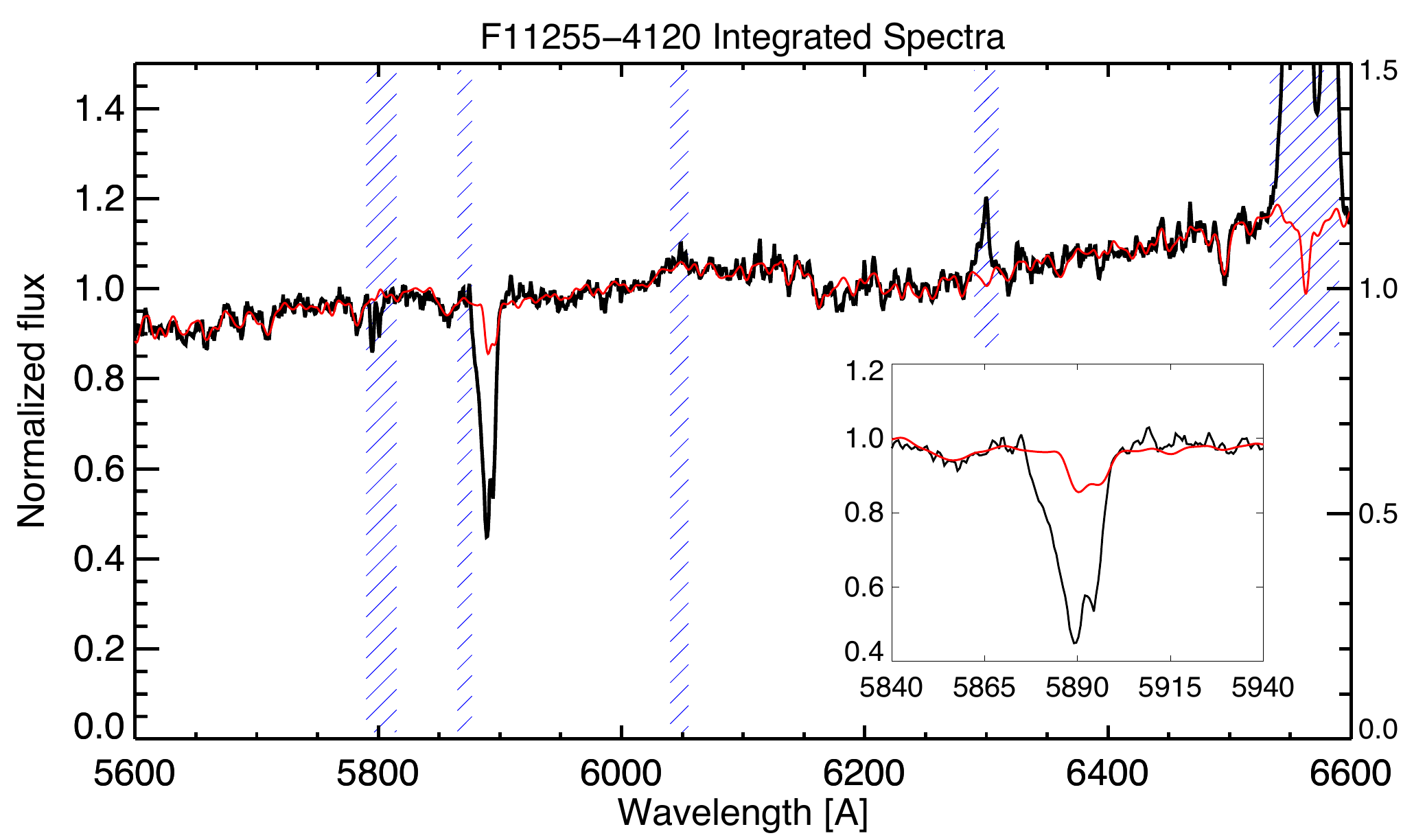}  
\caption{As Fig.~\ref{Panel_F01159} but for IRAS F11254-4120}
 \label{Panel_F11254}            
\end{figure*}
\clearpage

       
\begin{figure*}
\centering
\includegraphics[trim = -0.0cm 0.2cm .5cm 13.0cm, clip=true, width=1.\textwidth]{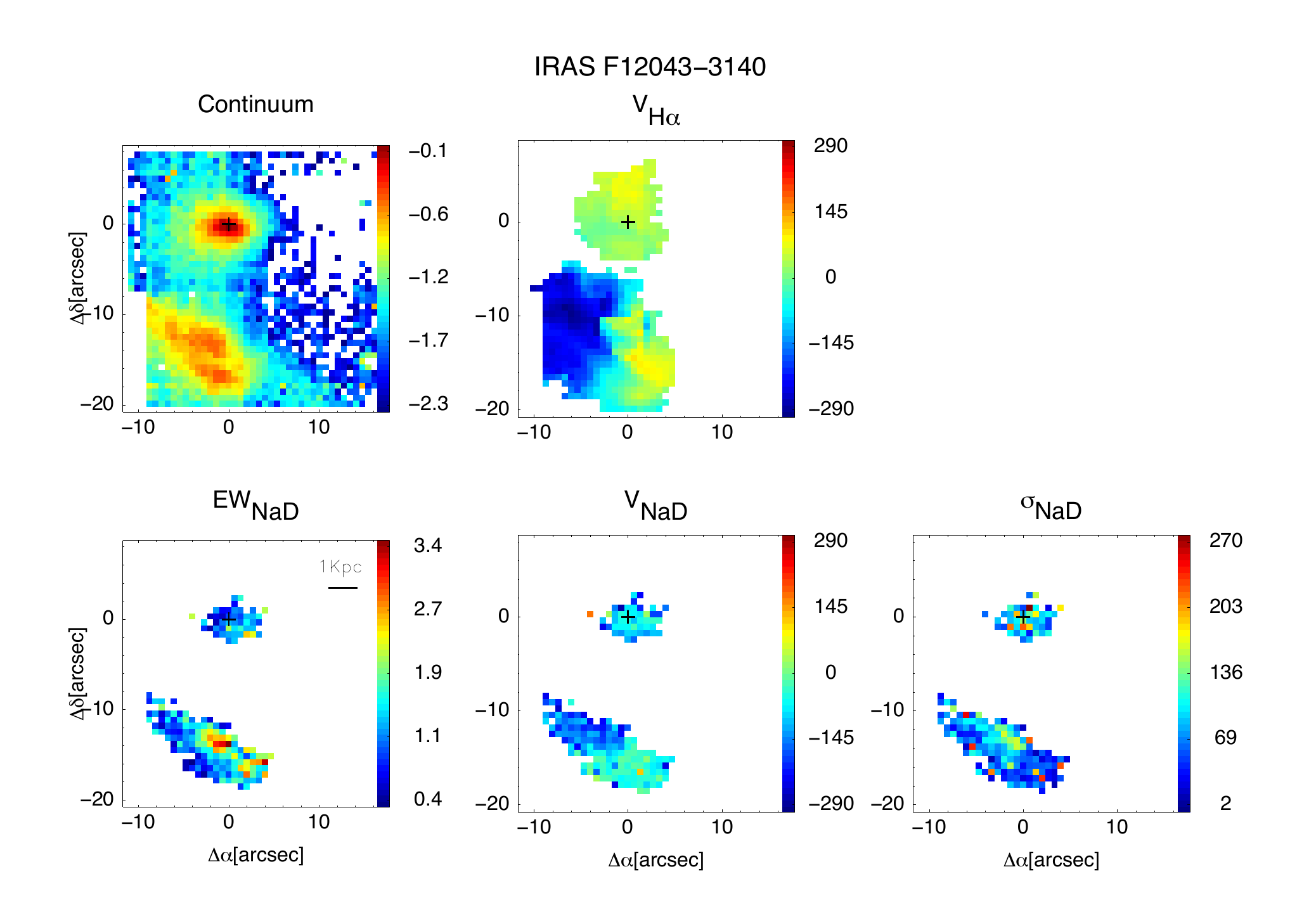}         
\centering                        
 \includegraphics[trim = 2.5cm 6.5cm 2.5cm 2.0cm, clip=true, width=.9\textwidth]{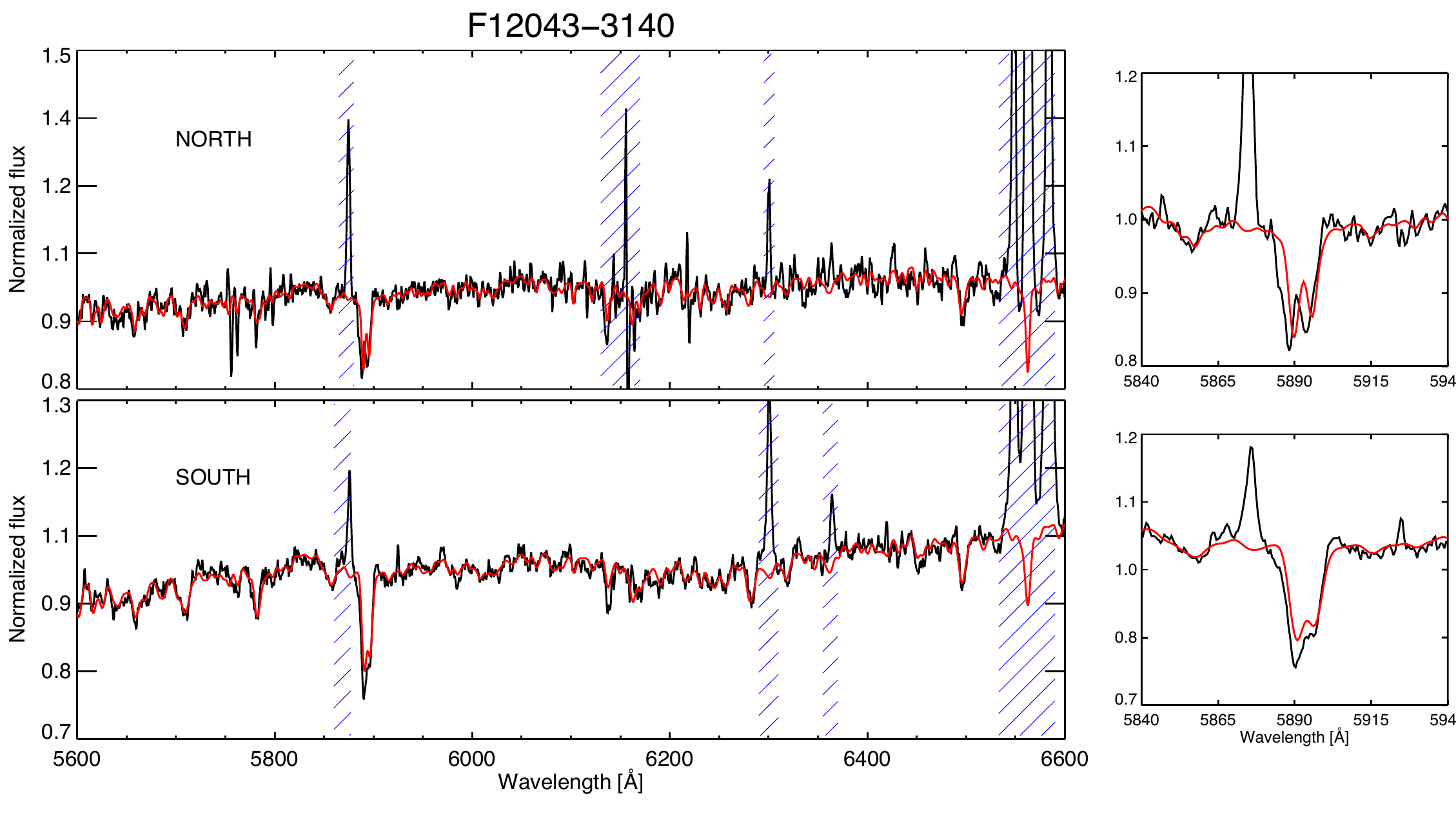}                  
 \caption{As Fig.~\ref{Panel_F01159} but for IRAS F12043-3140.}
\label{Panel_F12043}             
\end{figure*}
\clearpage


\begin{figure*}
\centering
 \includegraphics[trim = -0.0cm 0.2cm .5cm 13.0cm, clip=true, width=1.\textwidth]{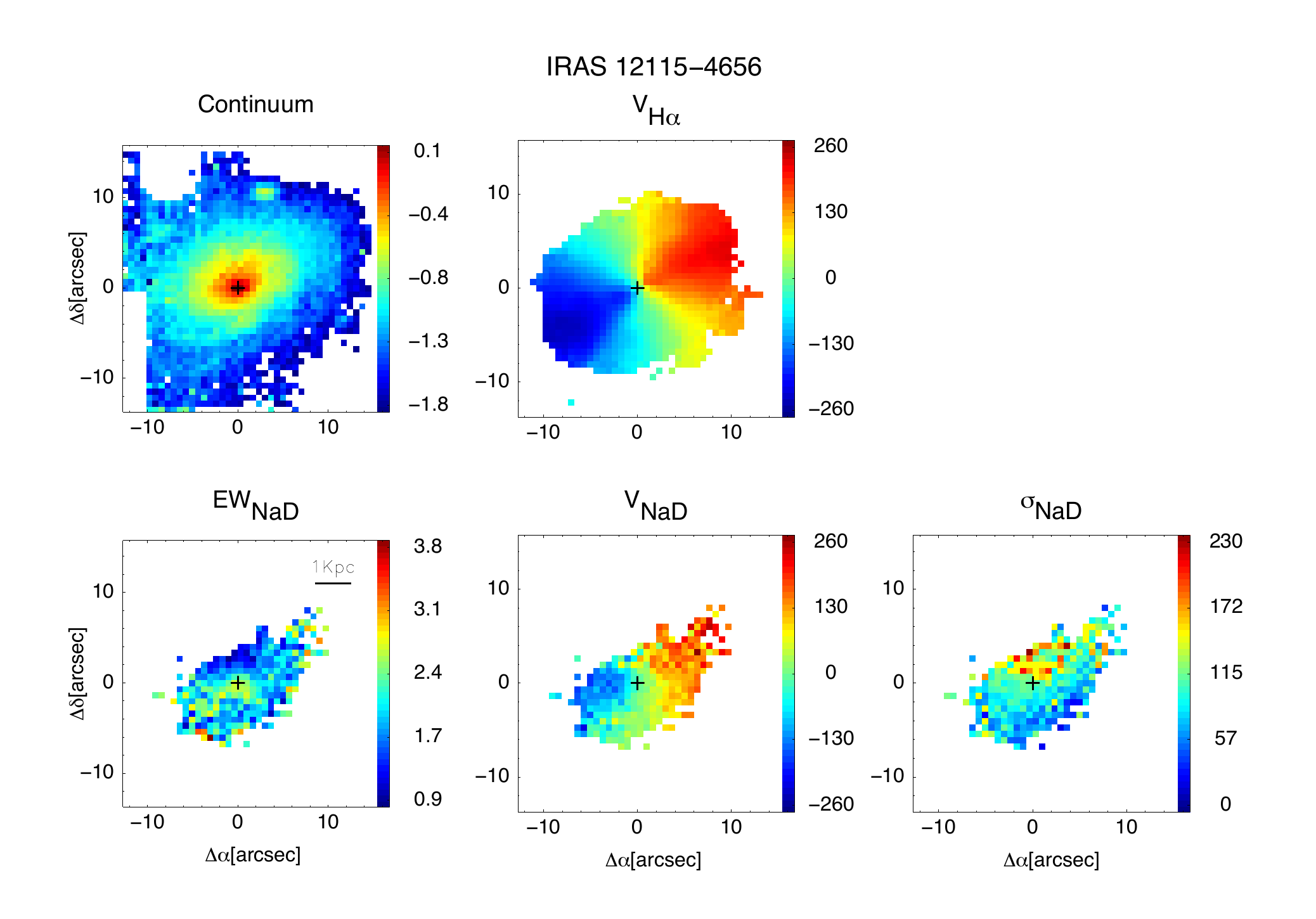}  
\centering
 \includegraphics[trim = -0.0cm 8.5cm .5cm 7.0cm, clip=true, width=.8\textwidth]{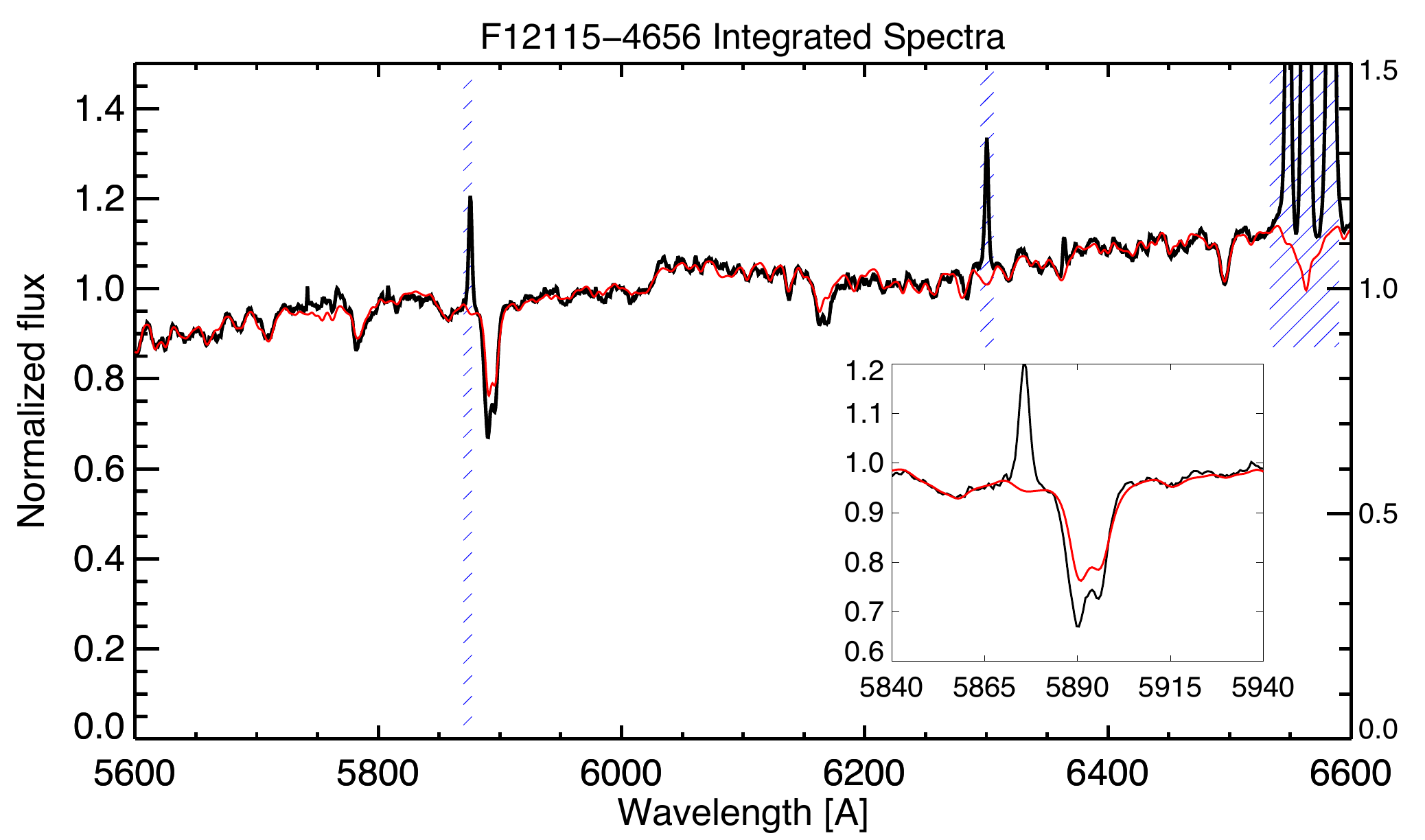}  
  \caption{As Fig.~\ref{Panel_F01159} but for IRAS F12115-4656.}
\label{Panel_F12115}             
\end{figure*}
\clearpage


\begin{figure*}
\centering
 \includegraphics[trim = -0.0cm .5cm .5cm 13.0cm, clip=true, width=.8\textwidth]{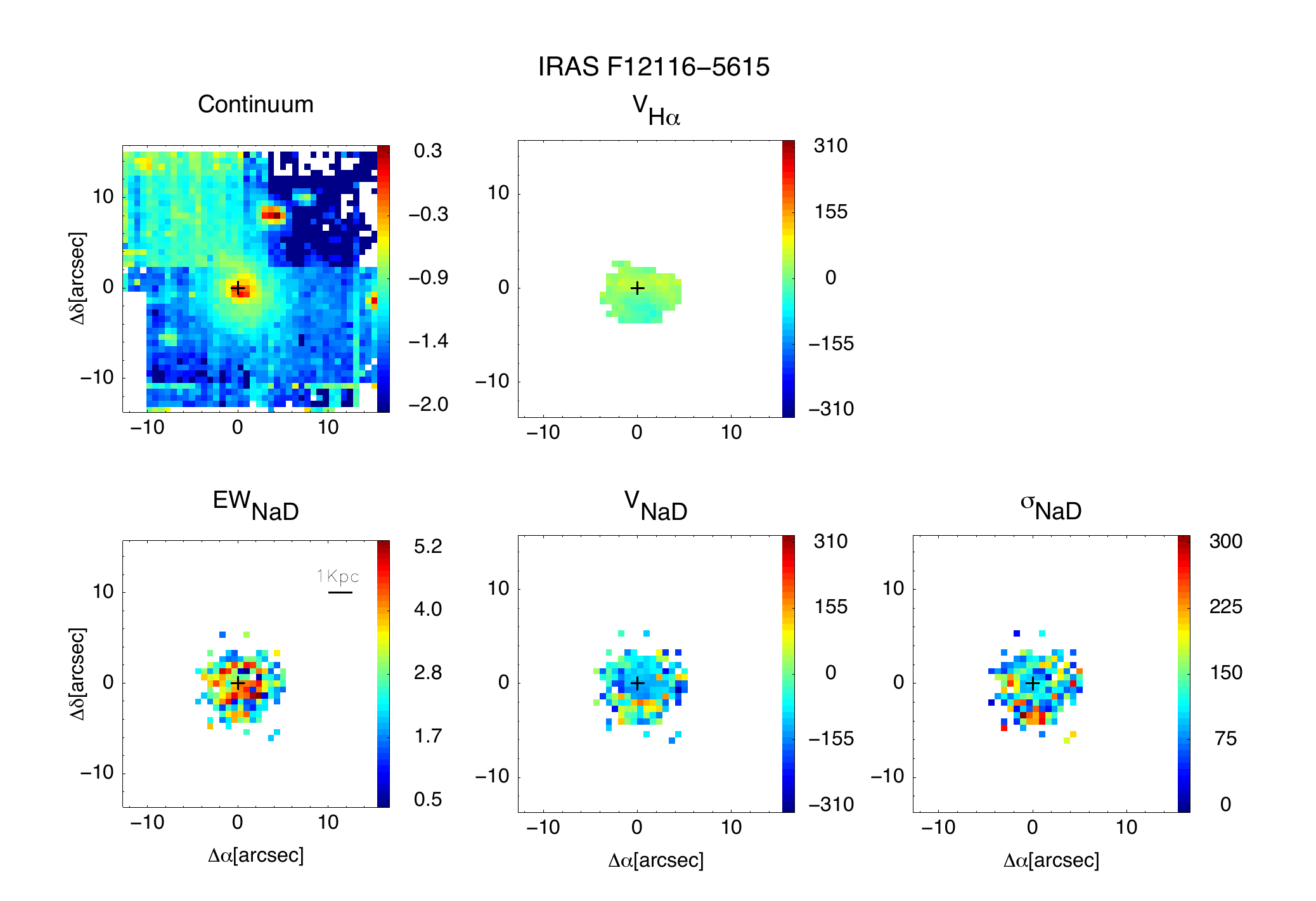} 
 \centering
 \includegraphics[trim = -0.0cm 0.5cm .5cm 19.5cm, clip=true, width=.8\textwidth]{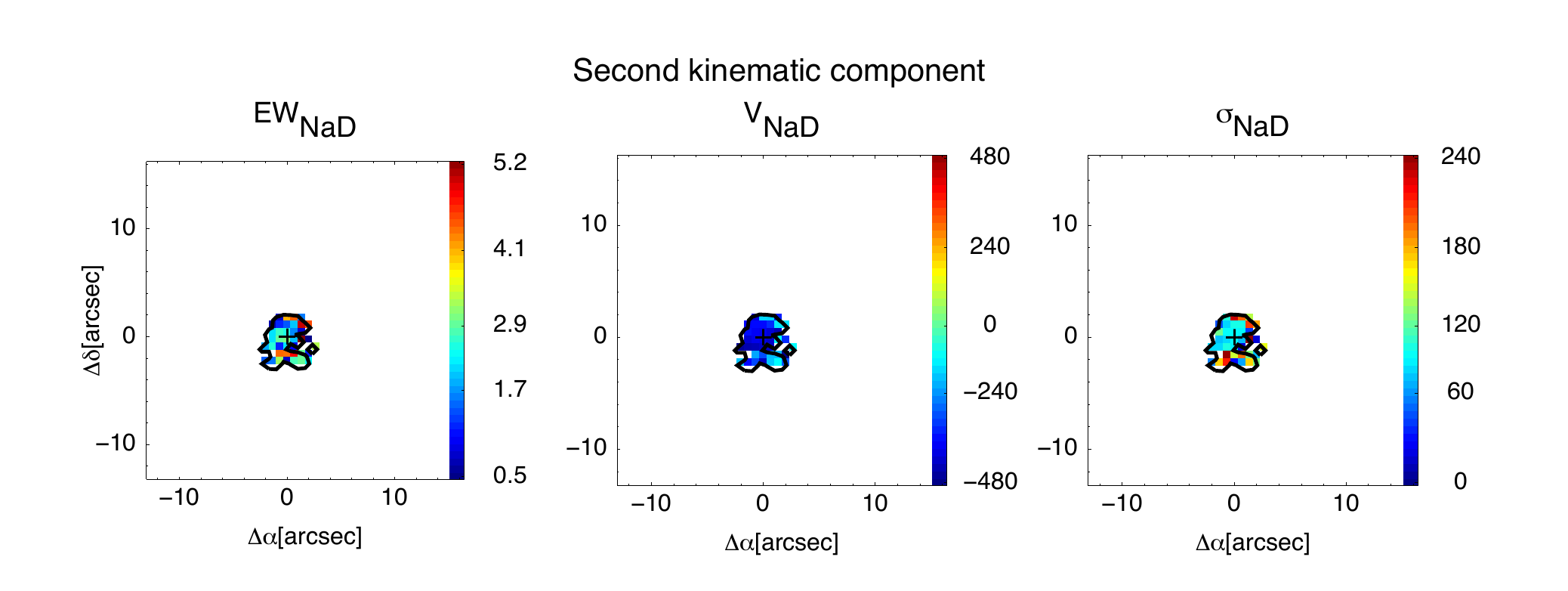}  
\centering
 \includegraphics[trim = -0.0cm 8.5cm .5cm 7.0cm, clip=true, width=.7\textwidth]{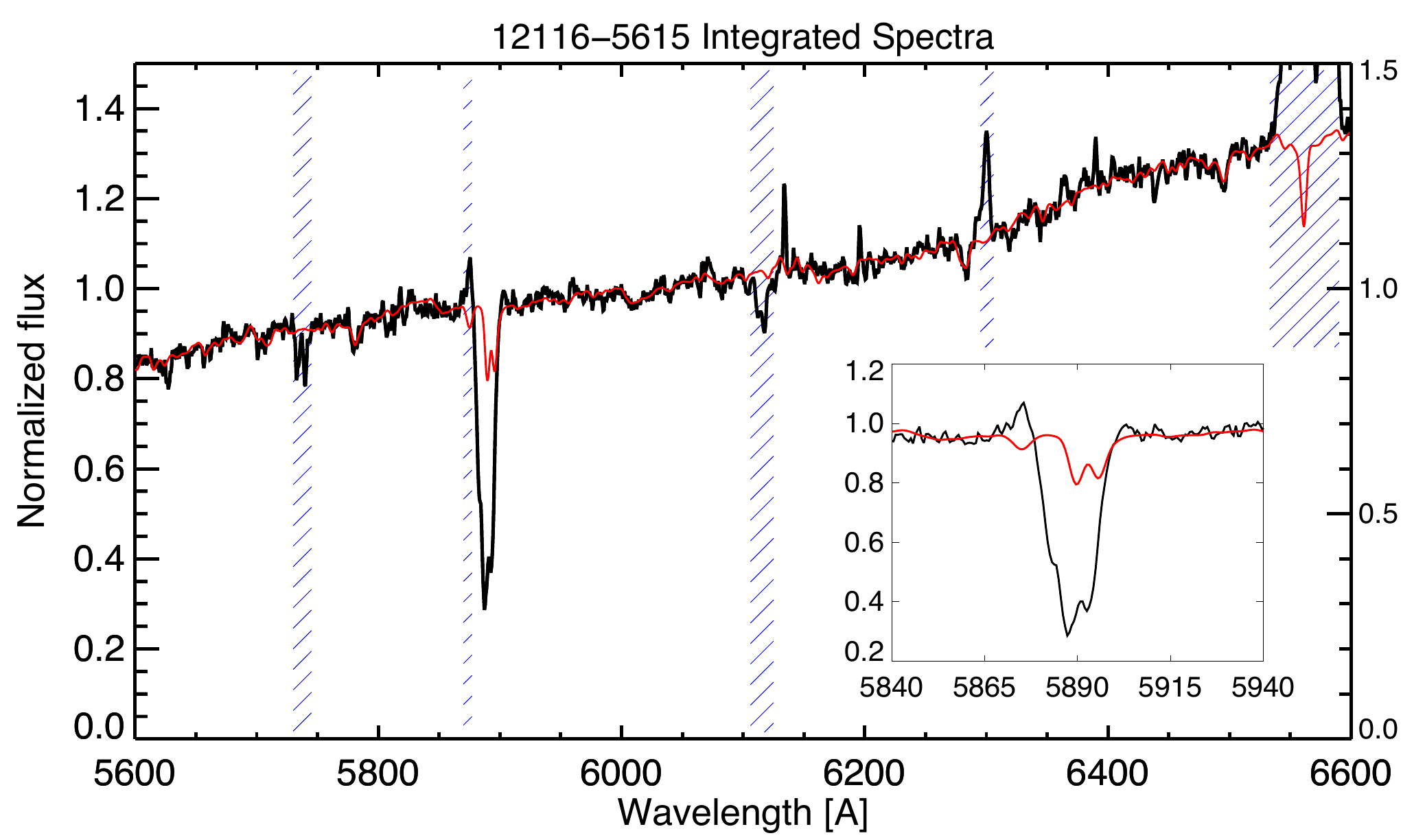}  
\caption{As Fig.~\ref{Panel_F07027S} but for IRAS F12116-2339.}
\label{Panel_F12116}             
\end{figure*}
\clearpage

          
\begin{figure*}
\centering
\vspace{4.em}   
   \includegraphics[trim = -0.0cm 8.5cm .5cm 7.0cm, clip=true, width=.75\textwidth]{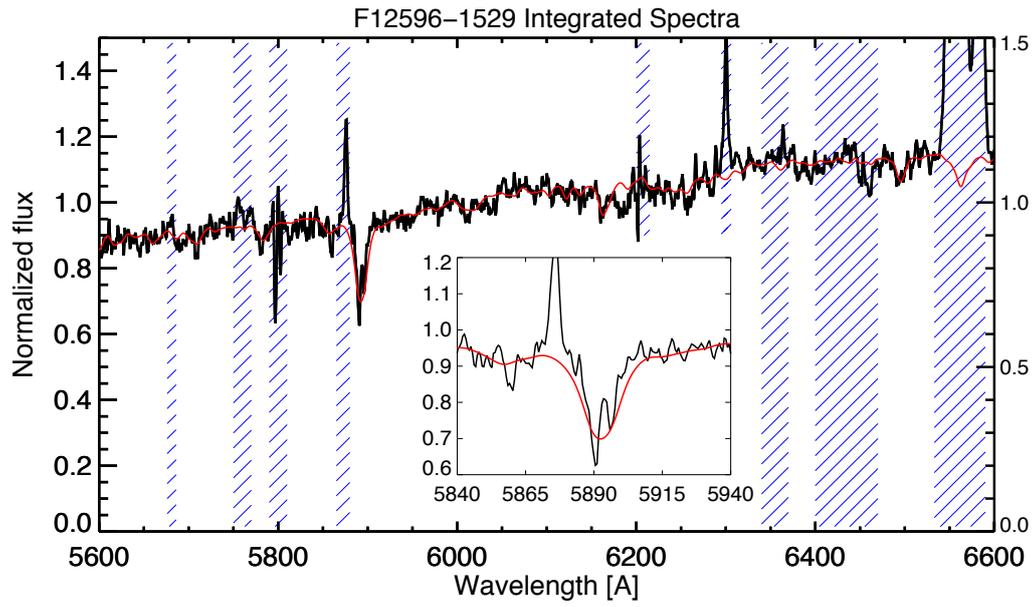}
  \caption{As in the lower panel of Fig.~\ref{Panel_F01159} but for 12596-1529 north and south (top to the bottom, respectively)}
  \label{Panel_12596}            
\end{figure*}
                

\begin{figure*}
\centering
 \includegraphics[trim = -0.0cm 0.2cm .5cm 13.0cm, clip=true, width=1.\textwidth]{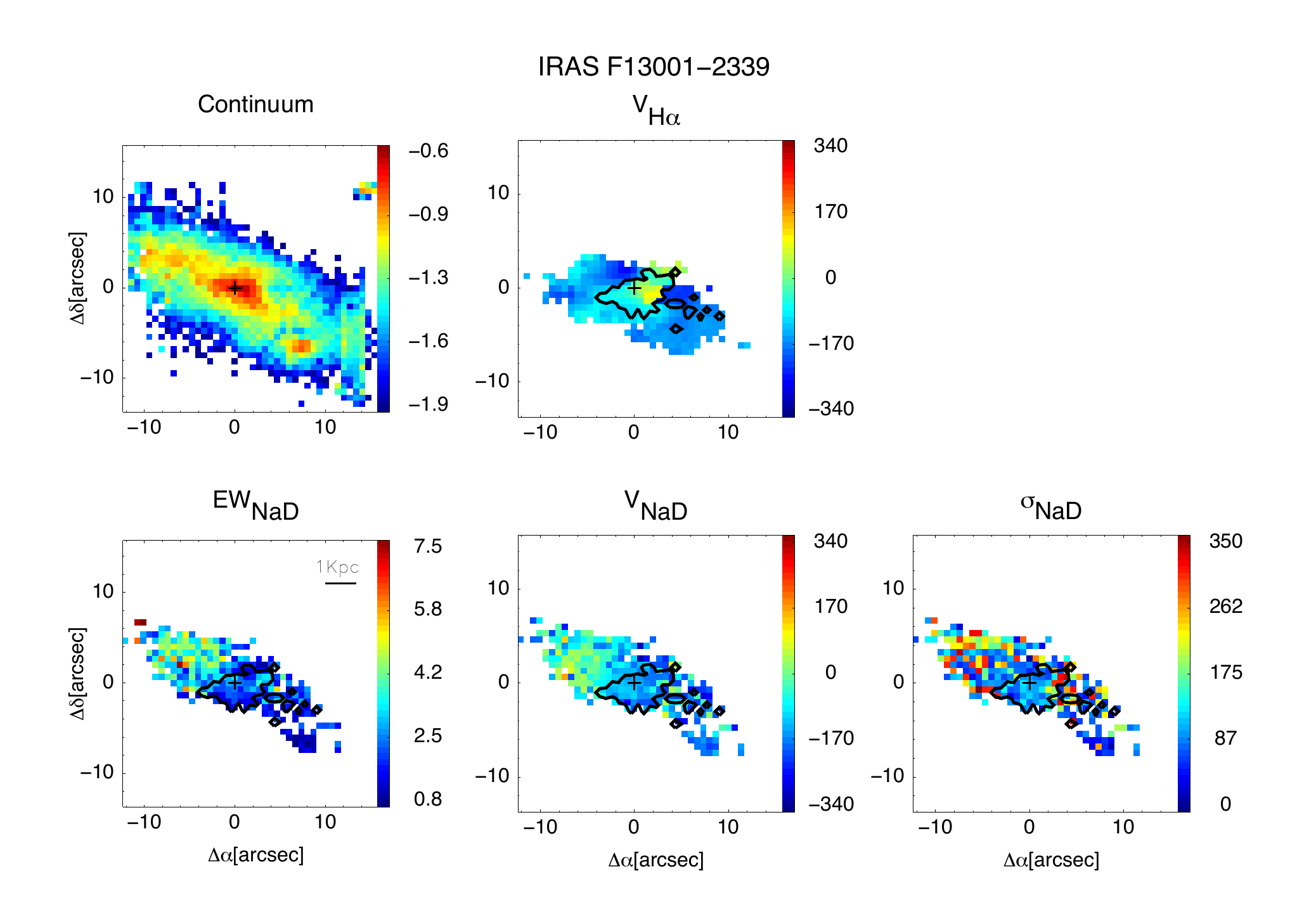}  
  \includegraphics[trim = -0.0cm 8.5cm .5cm 7.0cm, clip=true, width=.8\textwidth]{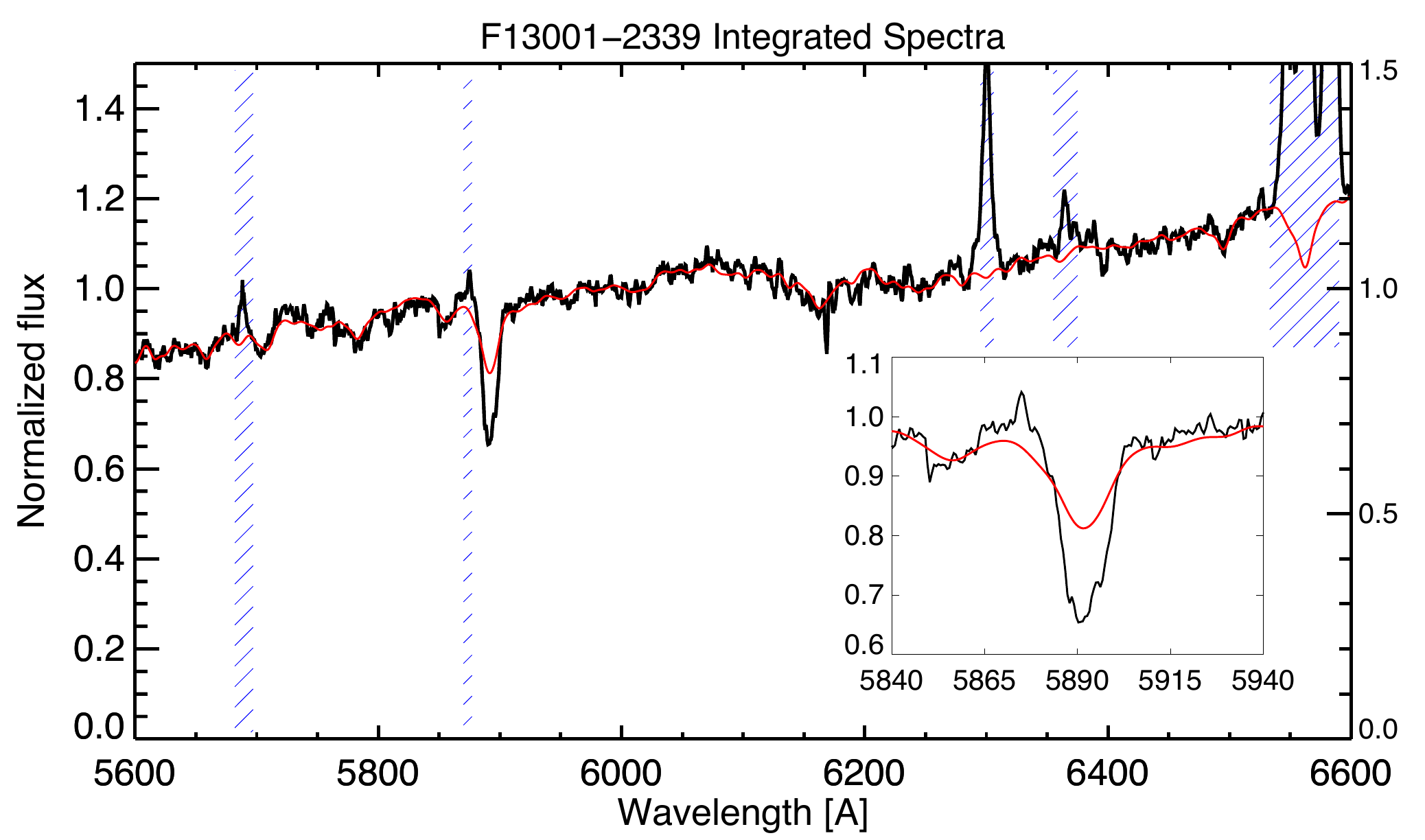}  
 \caption{As Fig.~\ref{Panel_F01159} but for IRAS-F13001-2339.}
 \label{Panel_F13001}            
\end{figure*}
\clearpage


\begin{figure*}
\centering
 \includegraphics[trim = -0.0cm 0.2cm .5cm 13.0cm, clip=true, width=1.\textwidth]{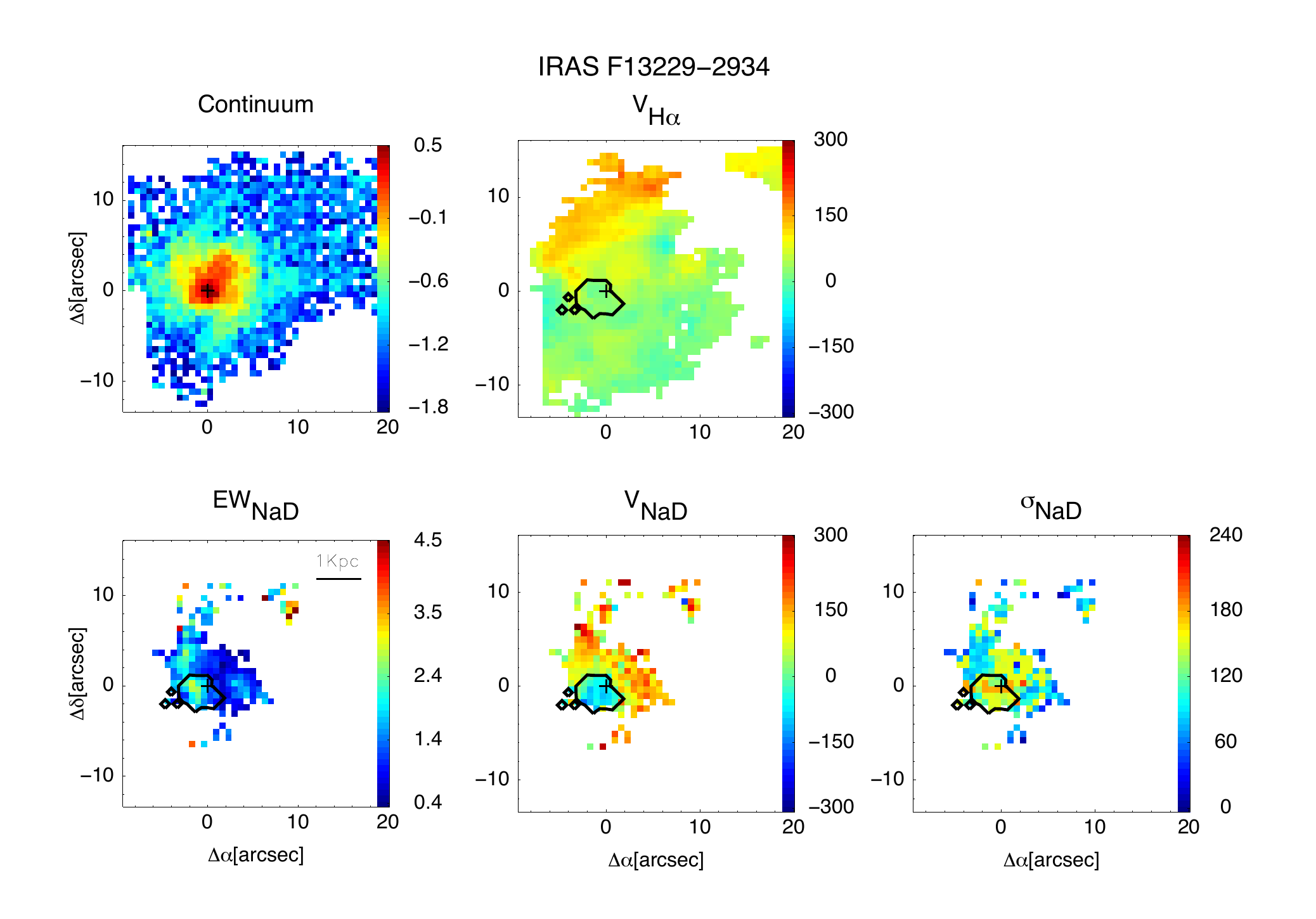}  
\centering
 \includegraphics[trim = -0.0cm 8.5cm .5cm 7.0cm, clip=true, width=.8\textwidth]{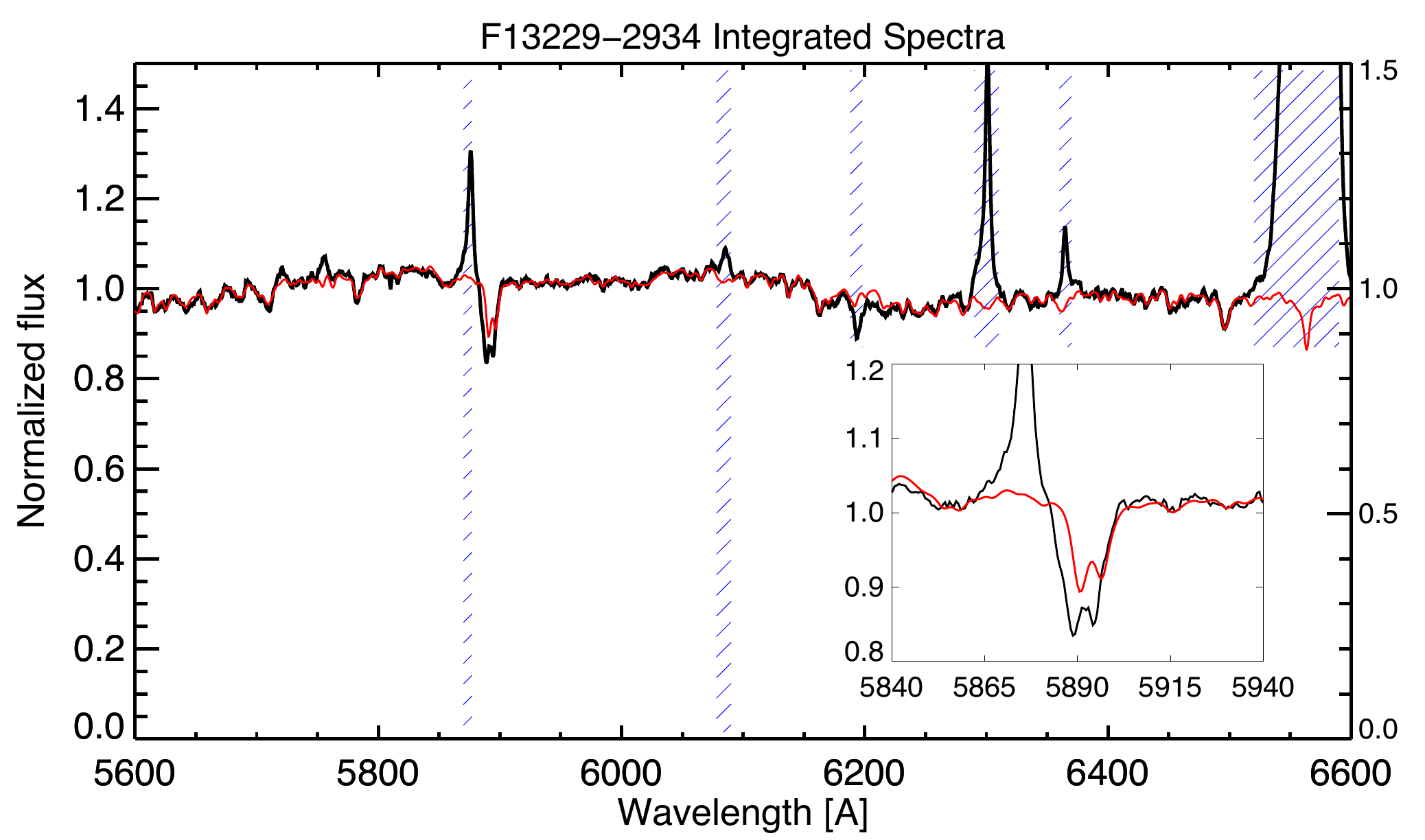}  
 \caption{As Fig.~\ref{Panel_F01159} but for IRAS F13229-2934.}
 \label{Panel_F13229}            
\end{figure*}
\clearpage

    
\begin{figure*}
\centering
 \includegraphics[trim = -0.0cm 0.2cm .5cm 13.0cm, clip=true, width=1.\textwidth]{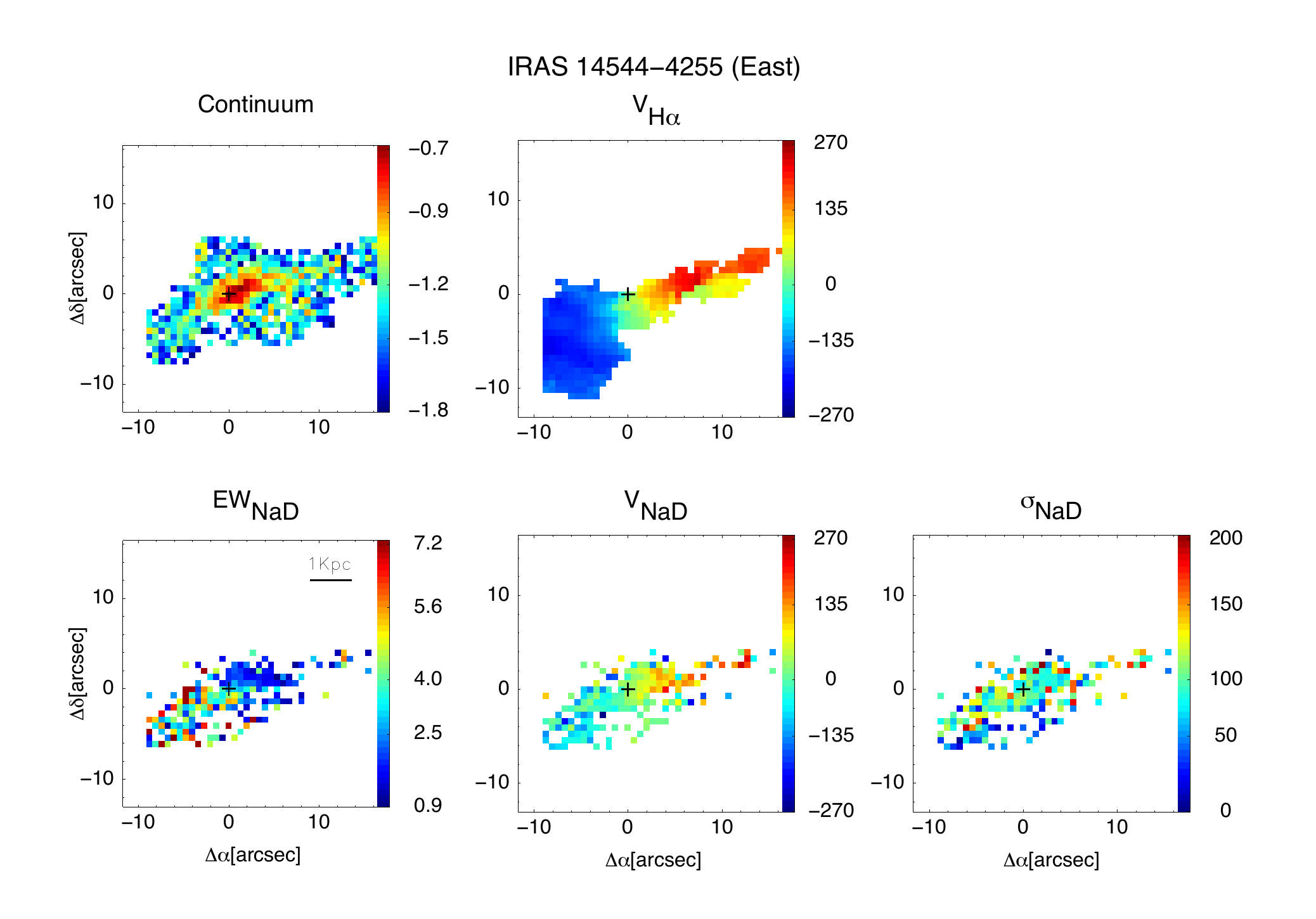}  
 \centering
 \includegraphics[trim = -0.0cm 8.5cm .5cm 7.0cm, clip=true, width=.8\textwidth]{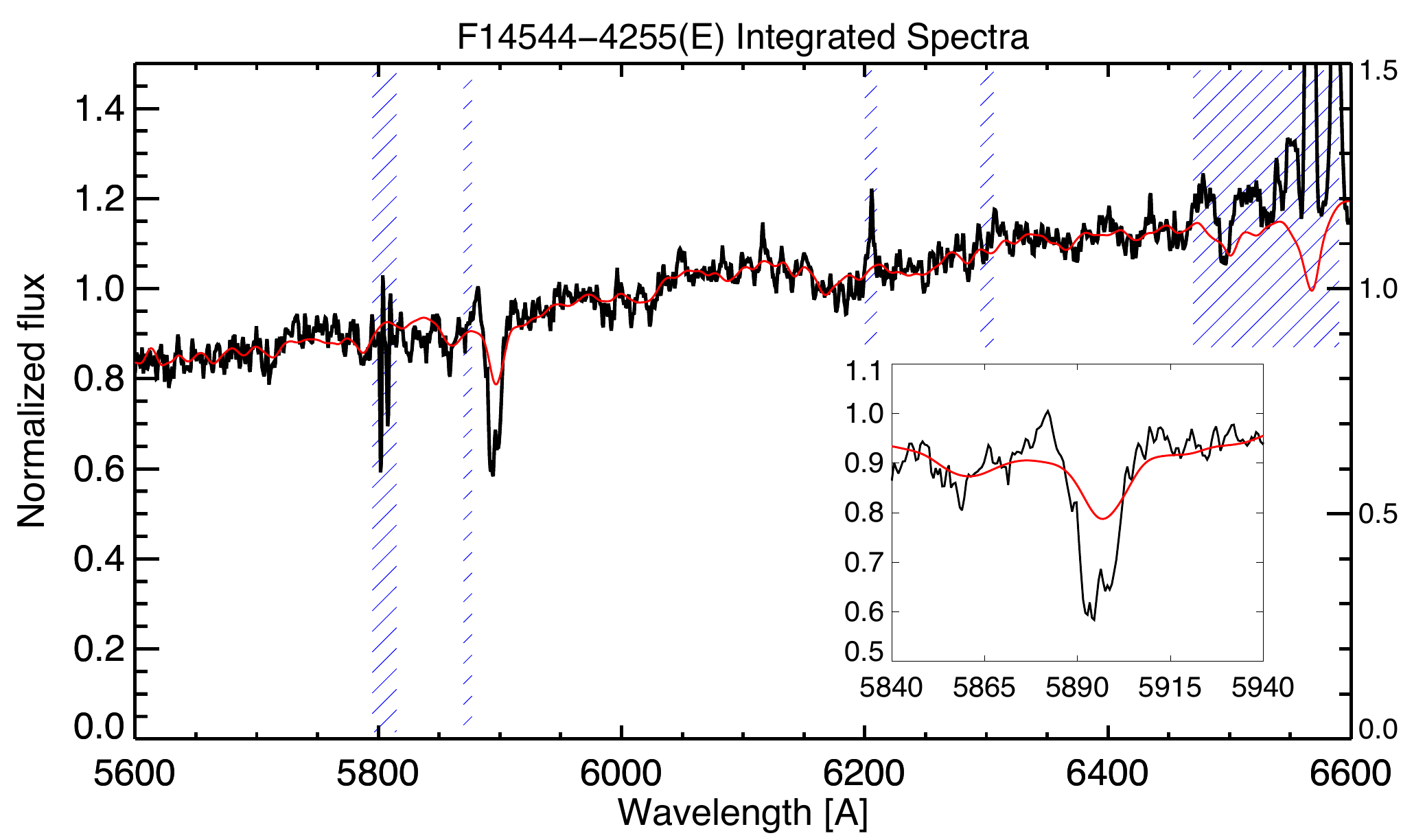}  
\caption{As Fig.~\ref{Panel_F01159} but for IRAS 14544-4255\,(E)}
 \label{Panel_F14544E}           
\end{figure*}
\clearpage

      
\begin{figure*}
\centering
\includegraphics[trim = -0.0cm 0.2cm .5cm 13.0cm, clip=true, width=1.\textwidth]{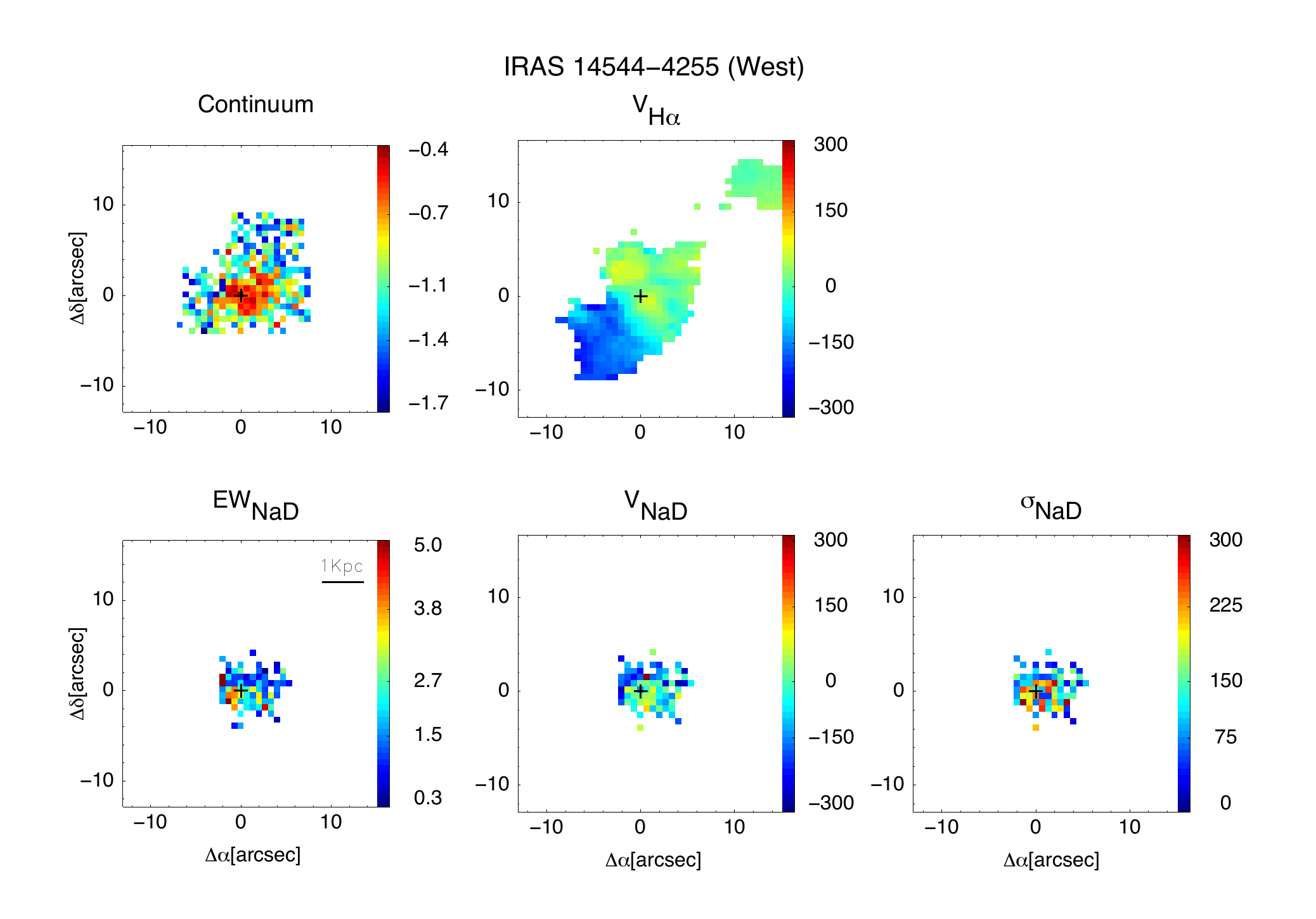}  
\centering
\includegraphics[trim = -0.0cm 8.5cm .5cm 7.0cm, clip=true, width=.8\textwidth]{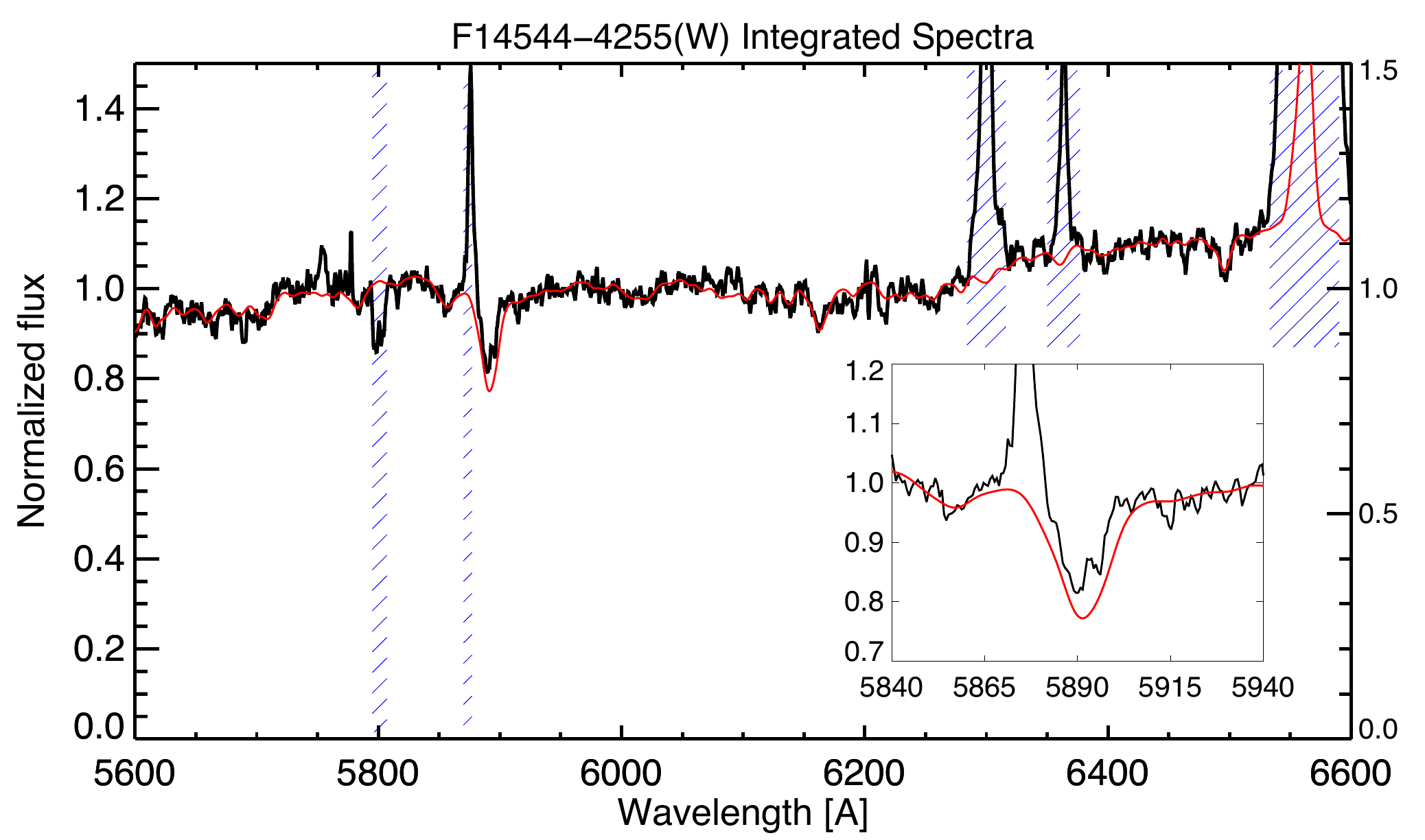}  
\caption{As Fig.~\ref{Panel_F01159} but for IRAS F14544-4255\,(W).}
  \label{Panel_F14544W}                  
\end{figure*}
\clearpage

 
\begin{figure*}
 \vspace{4.em}  
\centering
  \includegraphics[trim = -0.0cm 8.5cm .5cm 7.0cm, clip=true, width=.72\textwidth]{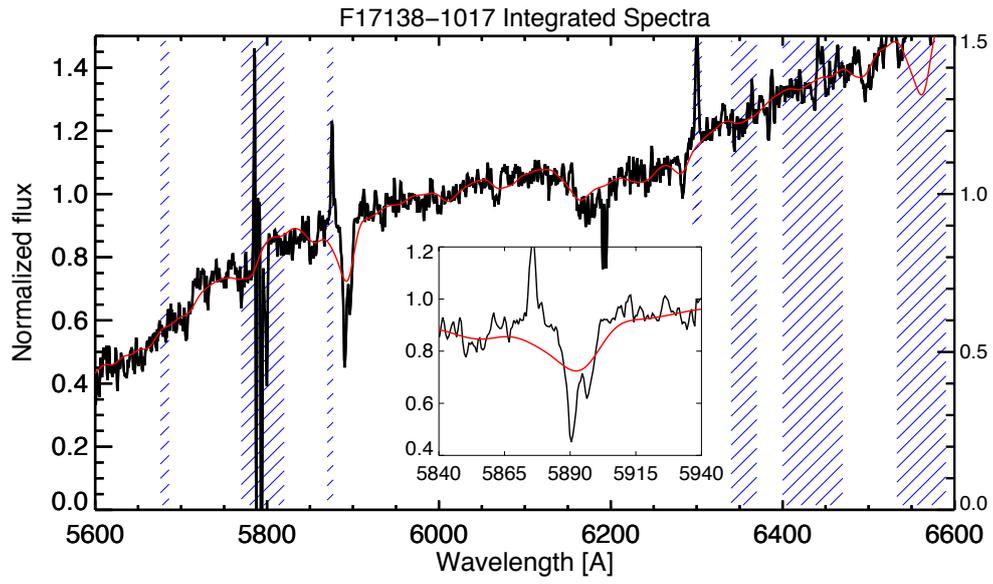}
   \caption{As in the lower panel of Fig.~\ref{Panel_F01159} but for IRAS 17138-1017}
\end{figure*}
\clearpage              
                
       
\begin{figure*}
\centering
\includegraphics[trim = -0.0cm 0.2cm .5cm 13.0cm, clip=true, width=1.\textwidth]{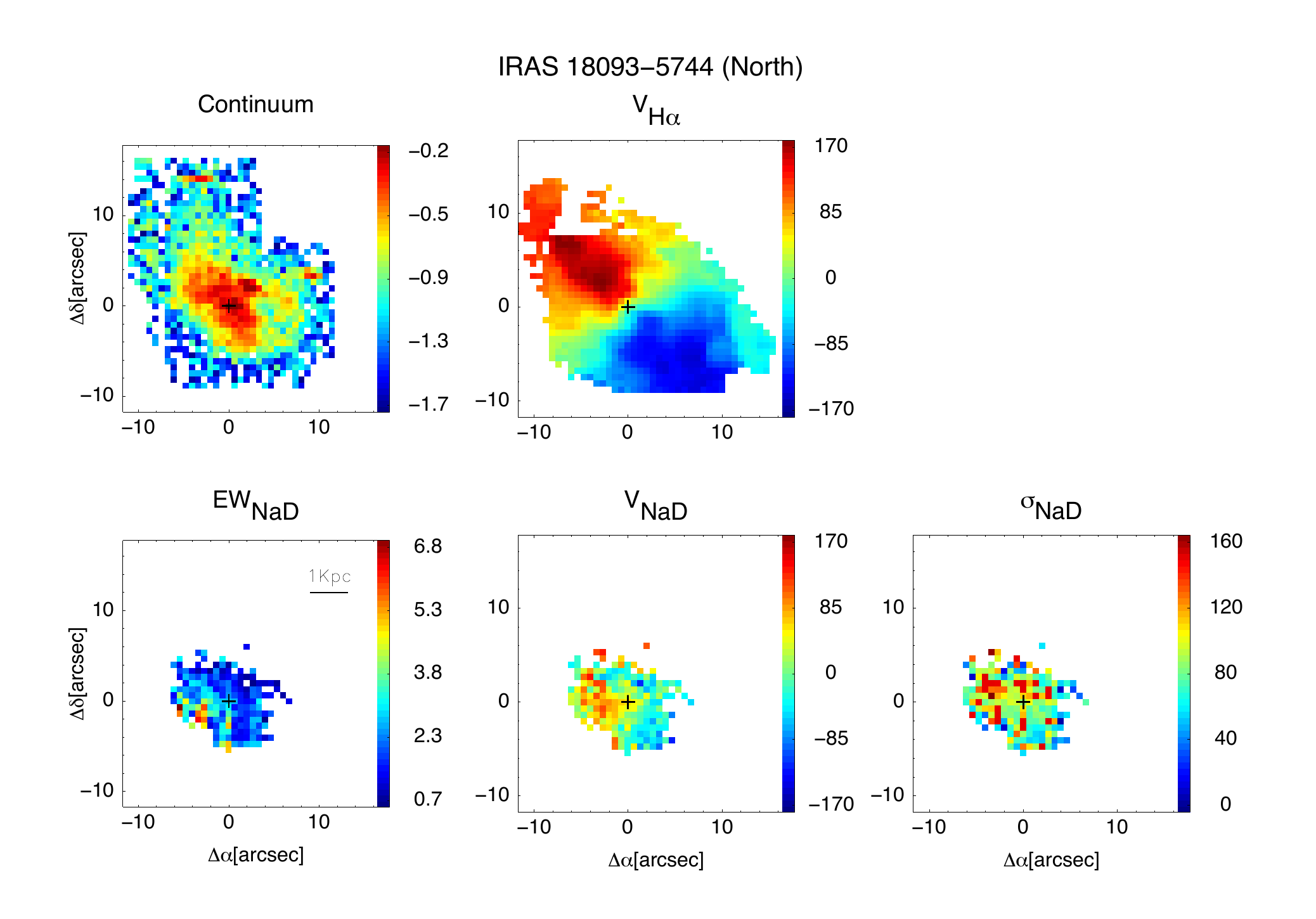}  
 \includegraphics[trim = -0.0cm 8.5cm .5cm 7.0cm, clip=true, width=.8\textwidth]{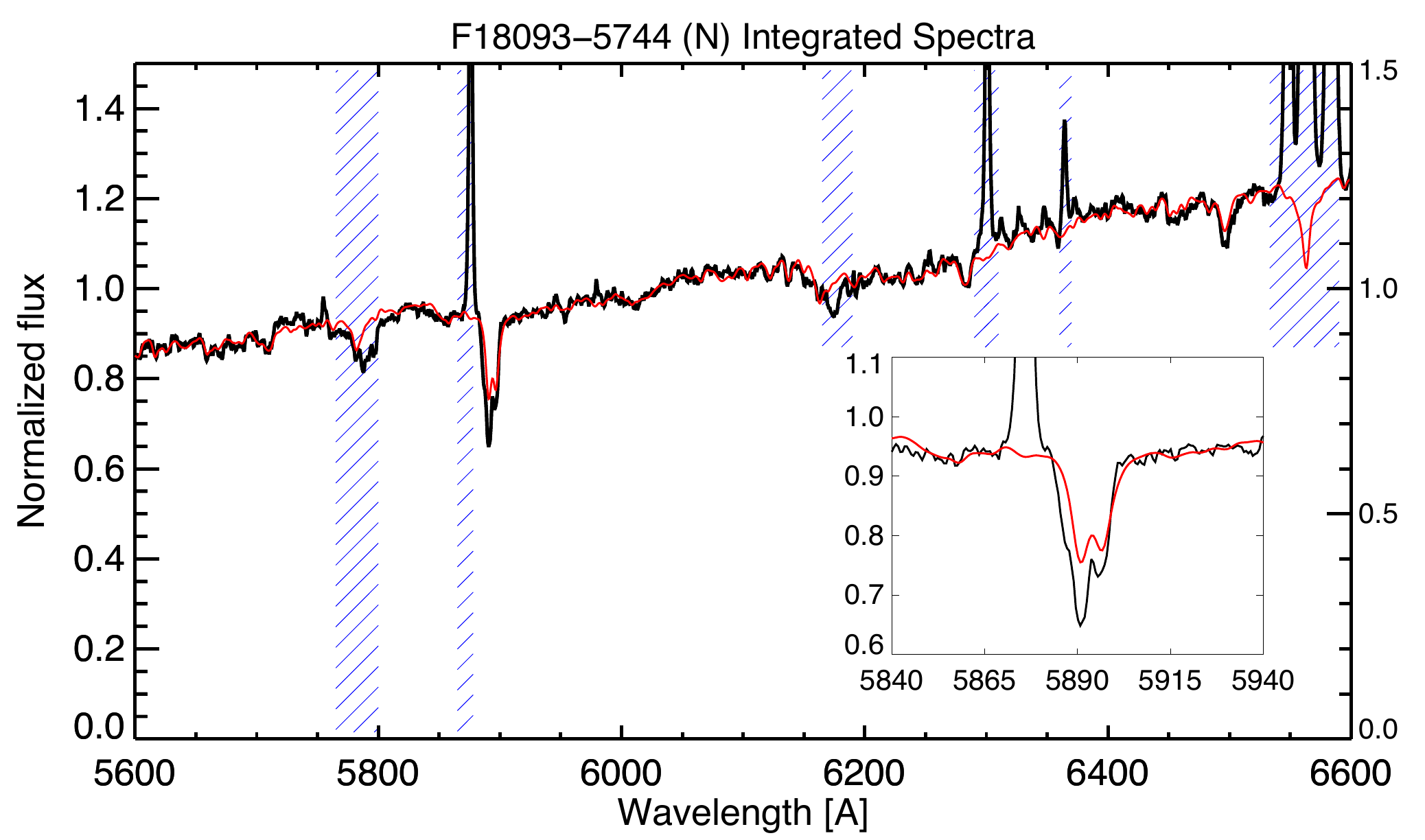}  
\caption{As Fig.~\ref{Panel_F01159} but for IRAS F18093-5744 (N).}
\label{Panel_F18093N}            
\end{figure*}
\clearpage


\begin{figure*}
\centering
\includegraphics[trim = -0.0cm 0.2cm .5cm 13.0cm, clip=true, width=1.\textwidth]{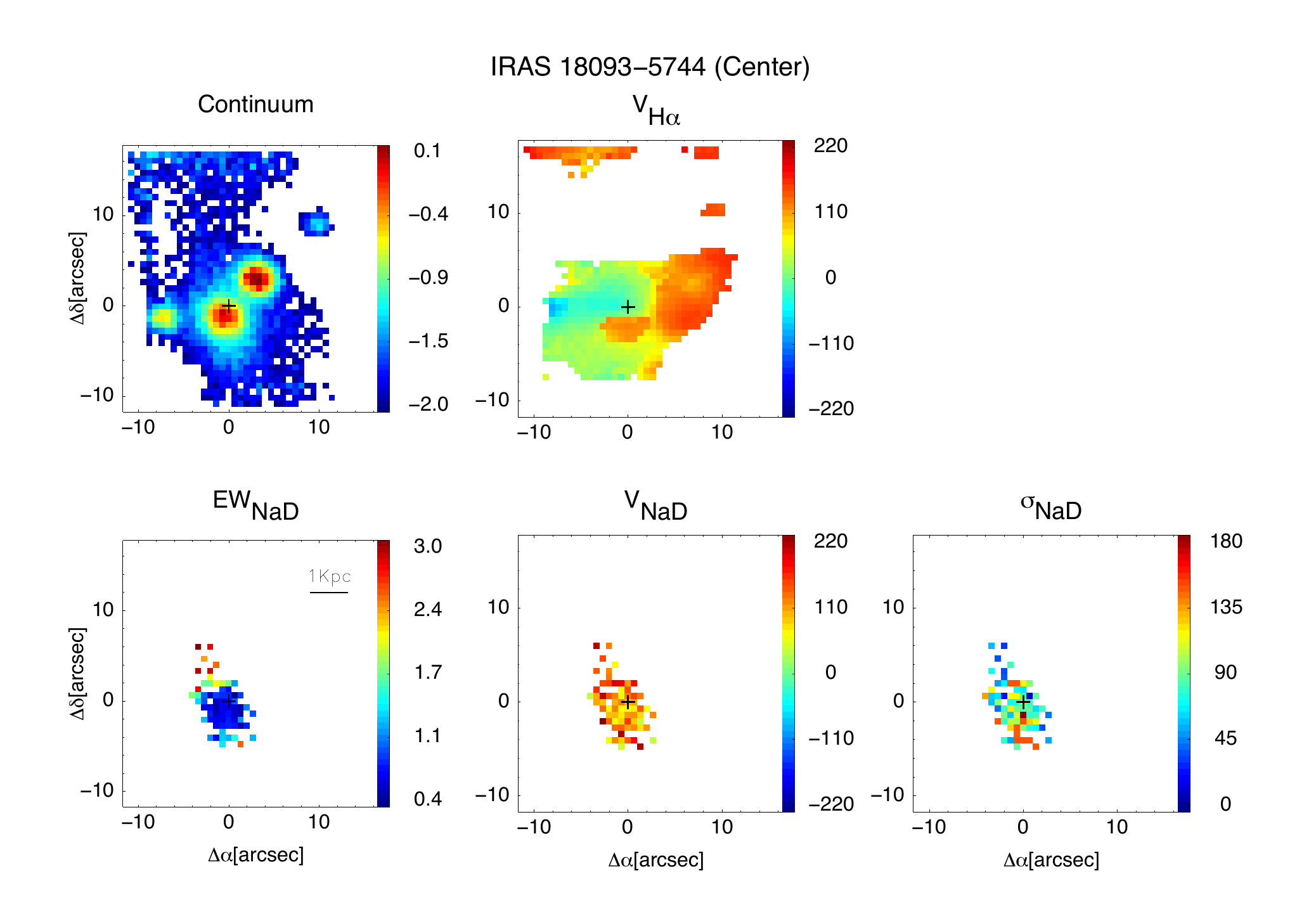}  
\centering
 \includegraphics[trim = -0.0cm 8.5cm .5cm 7.0cm, clip=true, width=.8\textwidth]{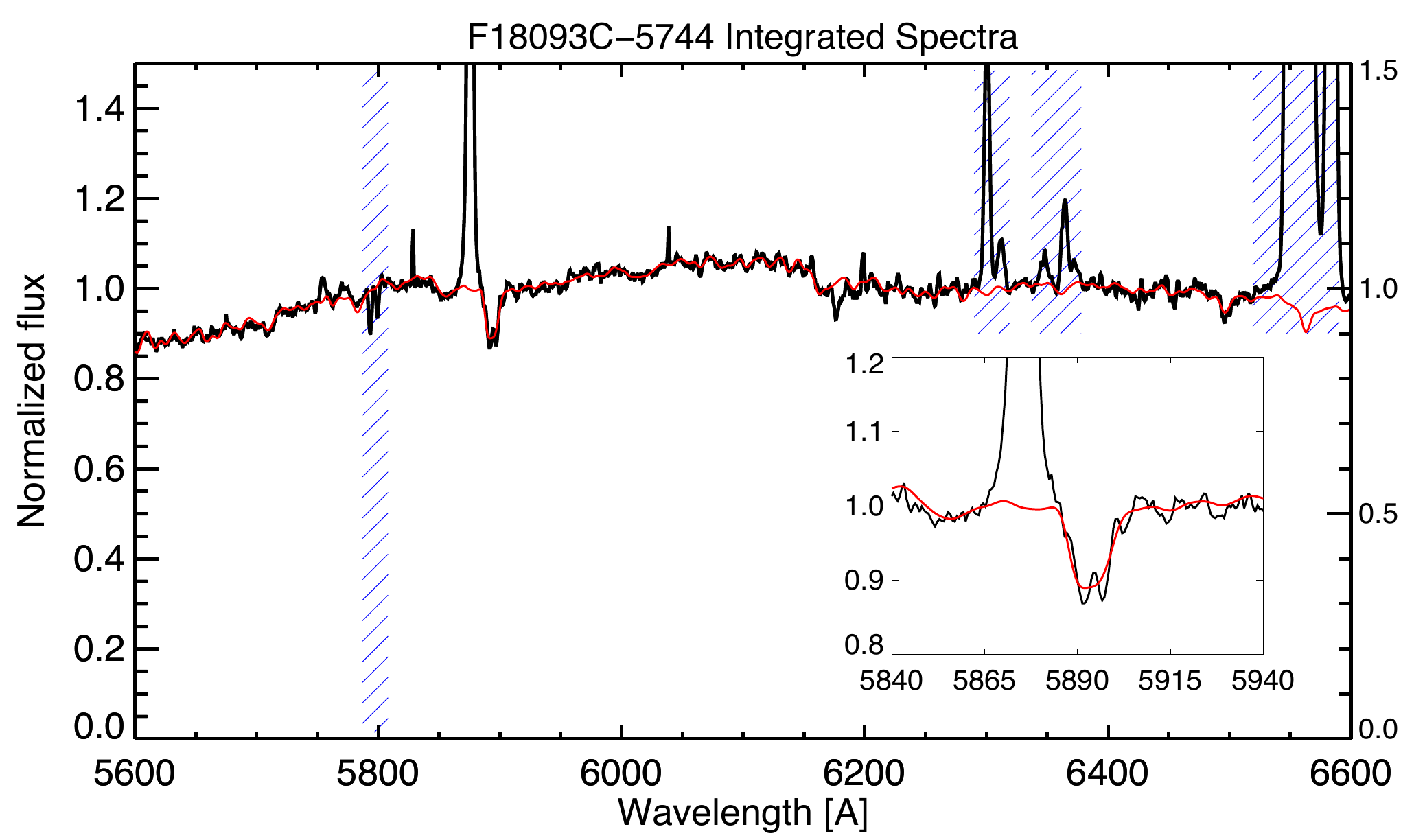}  
 \caption{As Fig.~\ref{Panel_F01159} but for IRAS 18093-5744\,(C).}
 \label{Panel_F18093C}           
\end{figure*}
\clearpage


\begin{figure*}
\centering
 \includegraphics[trim = -0.0cm 0.2cm .5cm 13.0cm, clip=true, width=1.\textwidth]{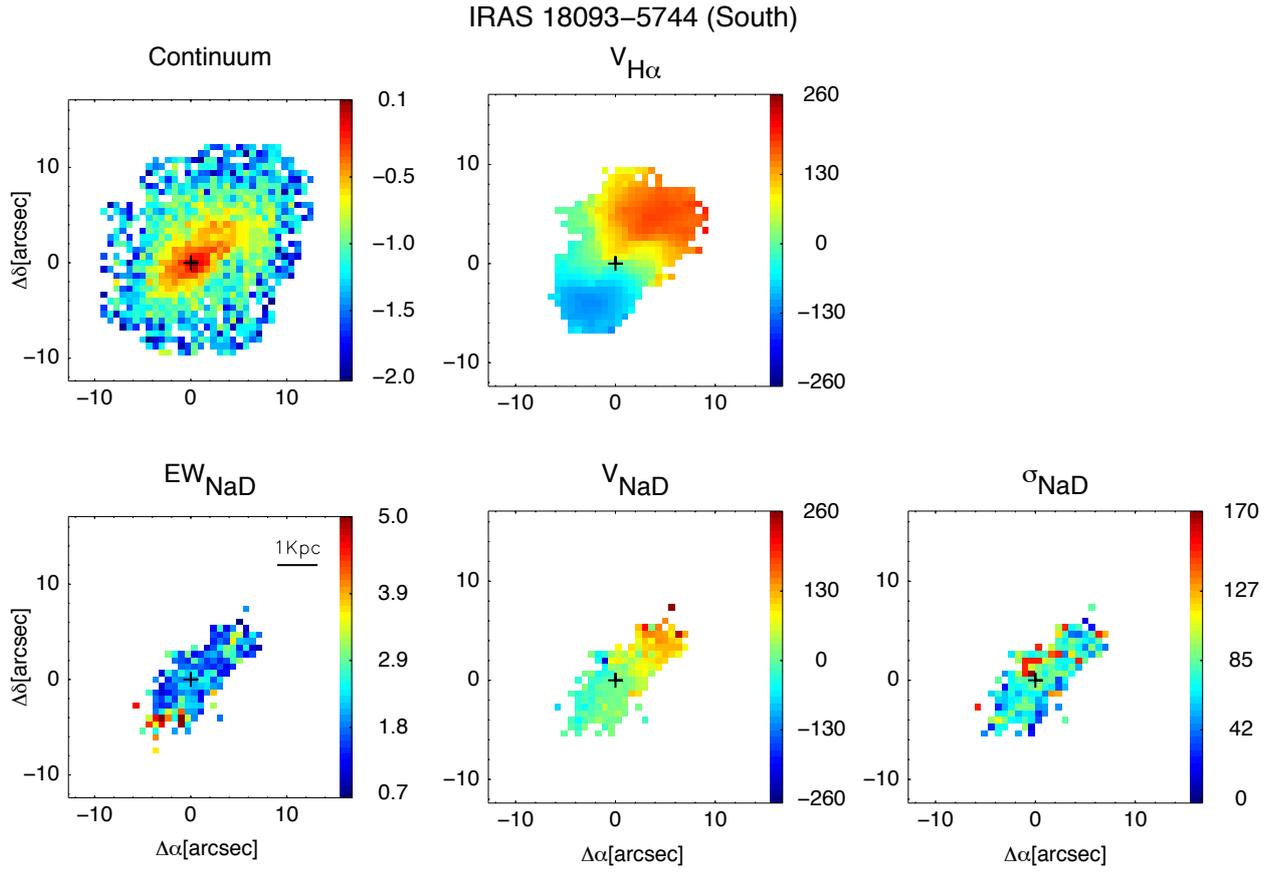}  
  \centering
 \includegraphics[trim = -0.0cm 8.5cm .5cm 7.0cm, clip=true, width=.8\textwidth]{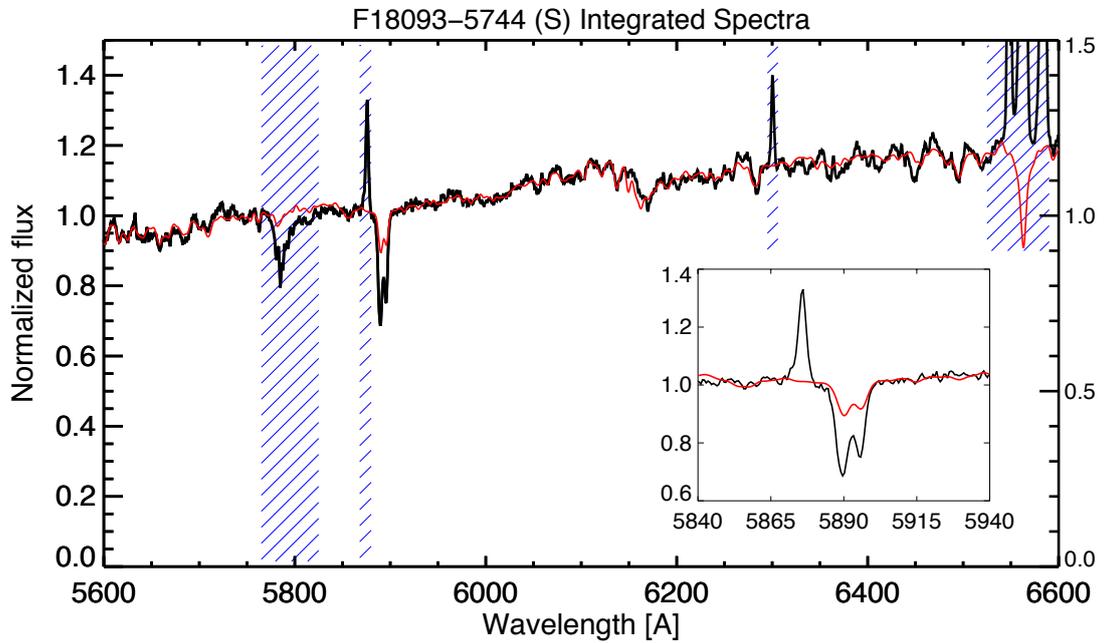}  
 \caption{As Fig.~\ref{Panel_F01159} but for IRAS 18093-5744 (S).}
 \label{Panel_F18093S}           
\end{figure*}
\clearpage

\begin{figure*}
\centering
 \includegraphics[trim = -0.0cm 0.2cm .5cm 13.0cm, clip=true, width=1.\textwidth]{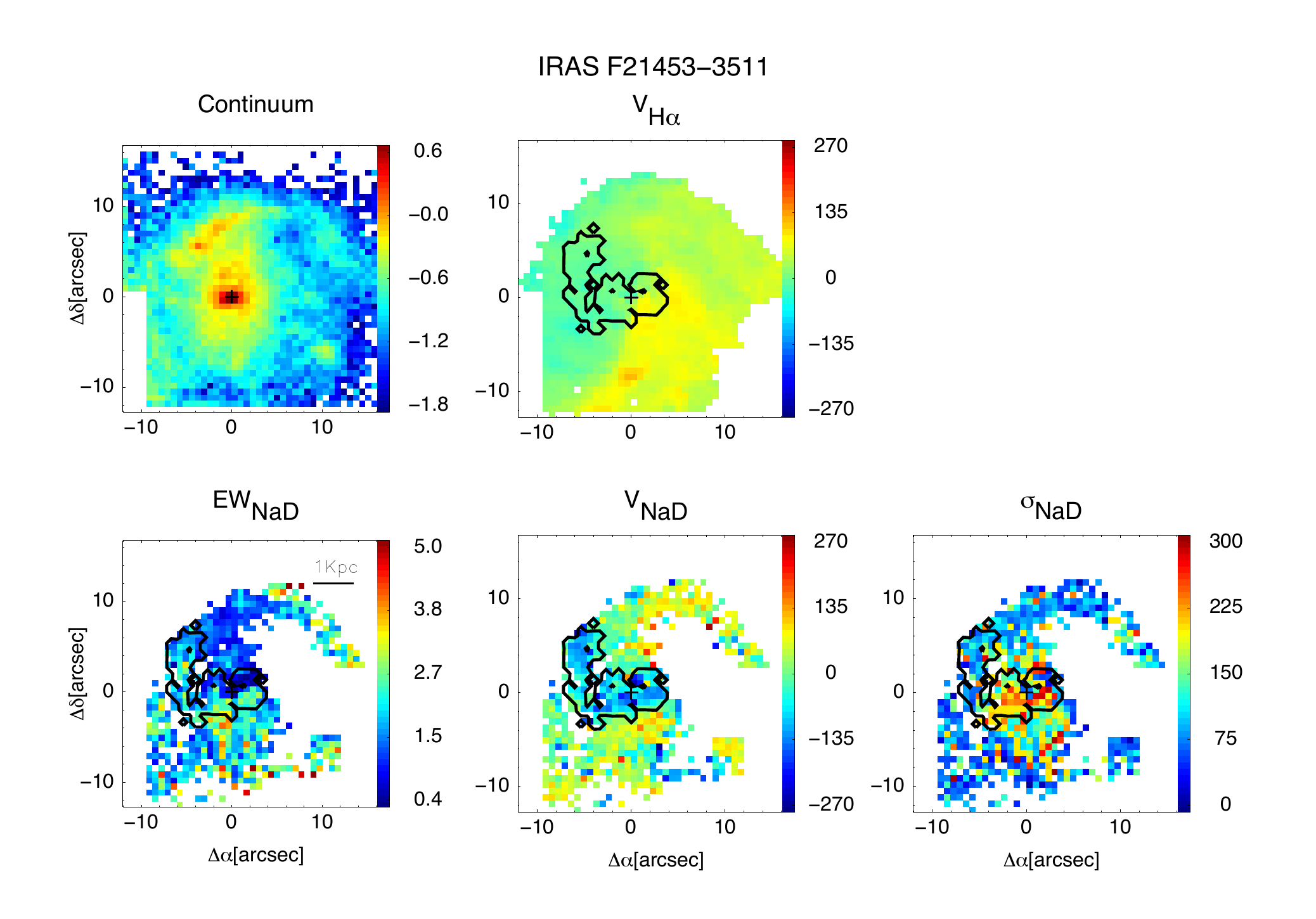}  
\centering
 \includegraphics[trim = -0.0cm 8.5cm .5cm 7.0cm, clip=true, width=.8\textwidth]{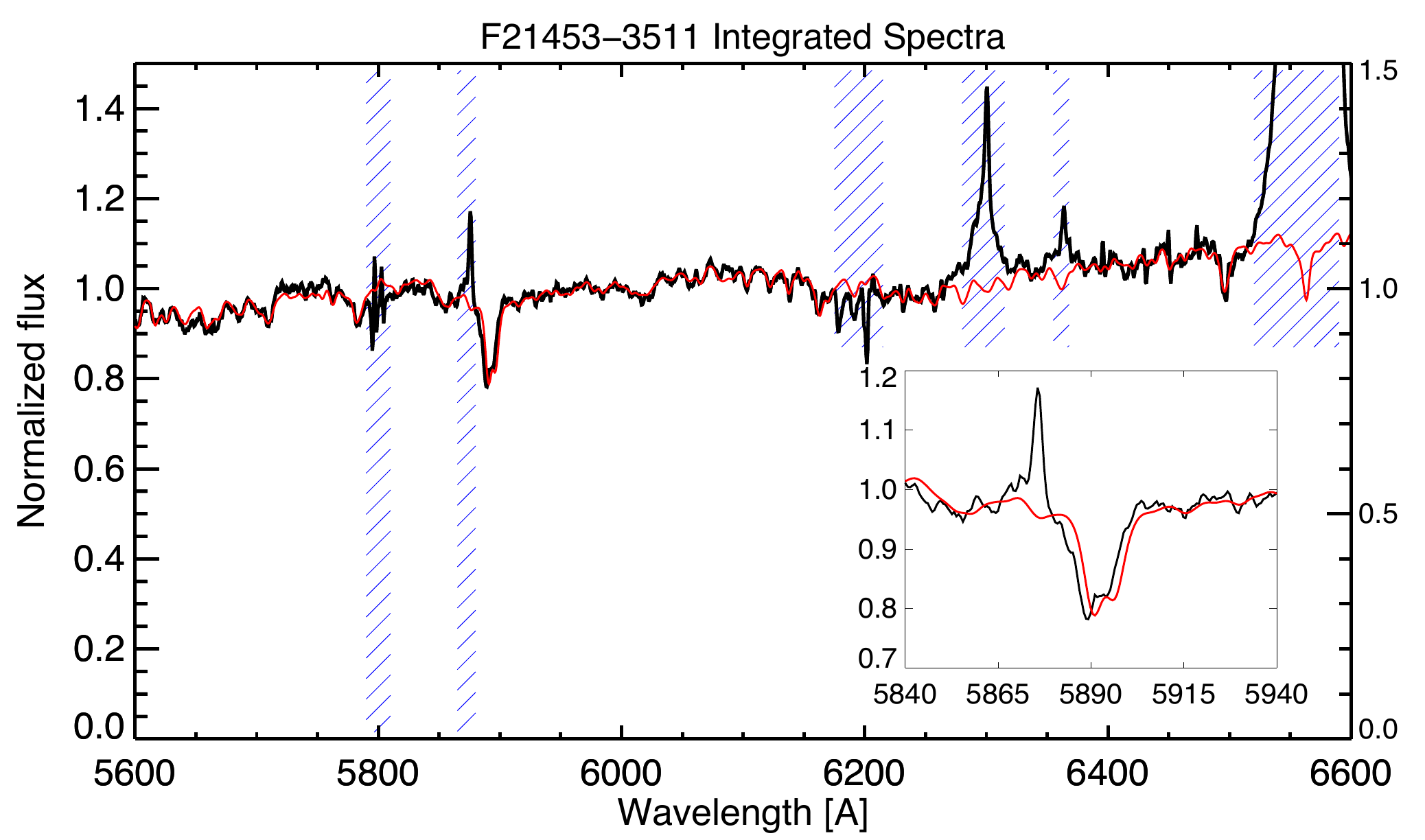}  
 \caption{As Fig.~\ref{Panel_F01159} but for IRAS F21453-3511.}
 \label{Panel_F21453}            
\end{figure*}
\clearpage


\begin{figure*}
\centering
  \includegraphics[trim = -0.0cm 0.2cm .5cm 13.0cm, clip=true, width=1.\textwidth]{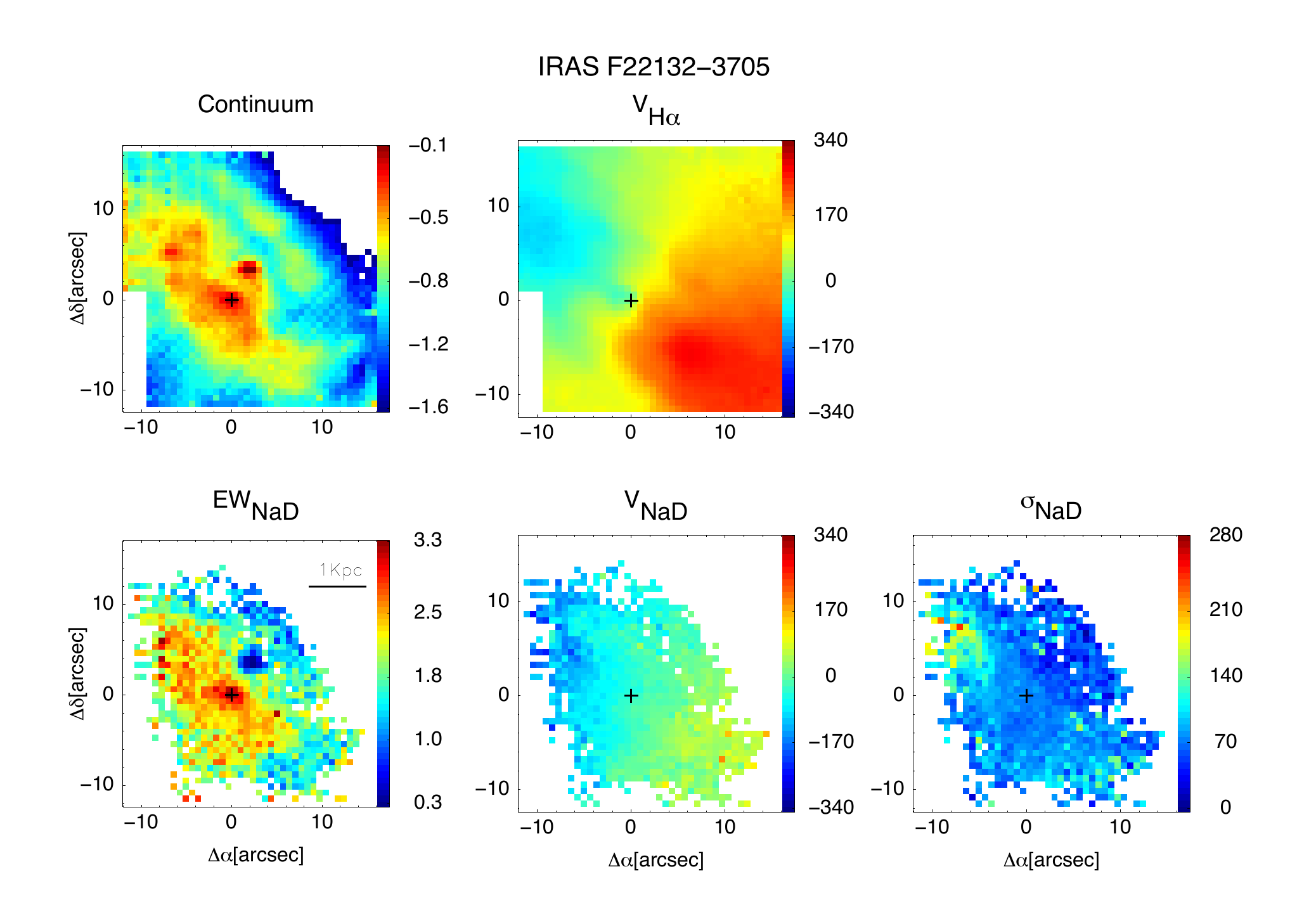}    
 \centering
 \includegraphics[trim = -0.0cm 8.5cm .5cm 7.0cm, clip=true, width=.8\textwidth]{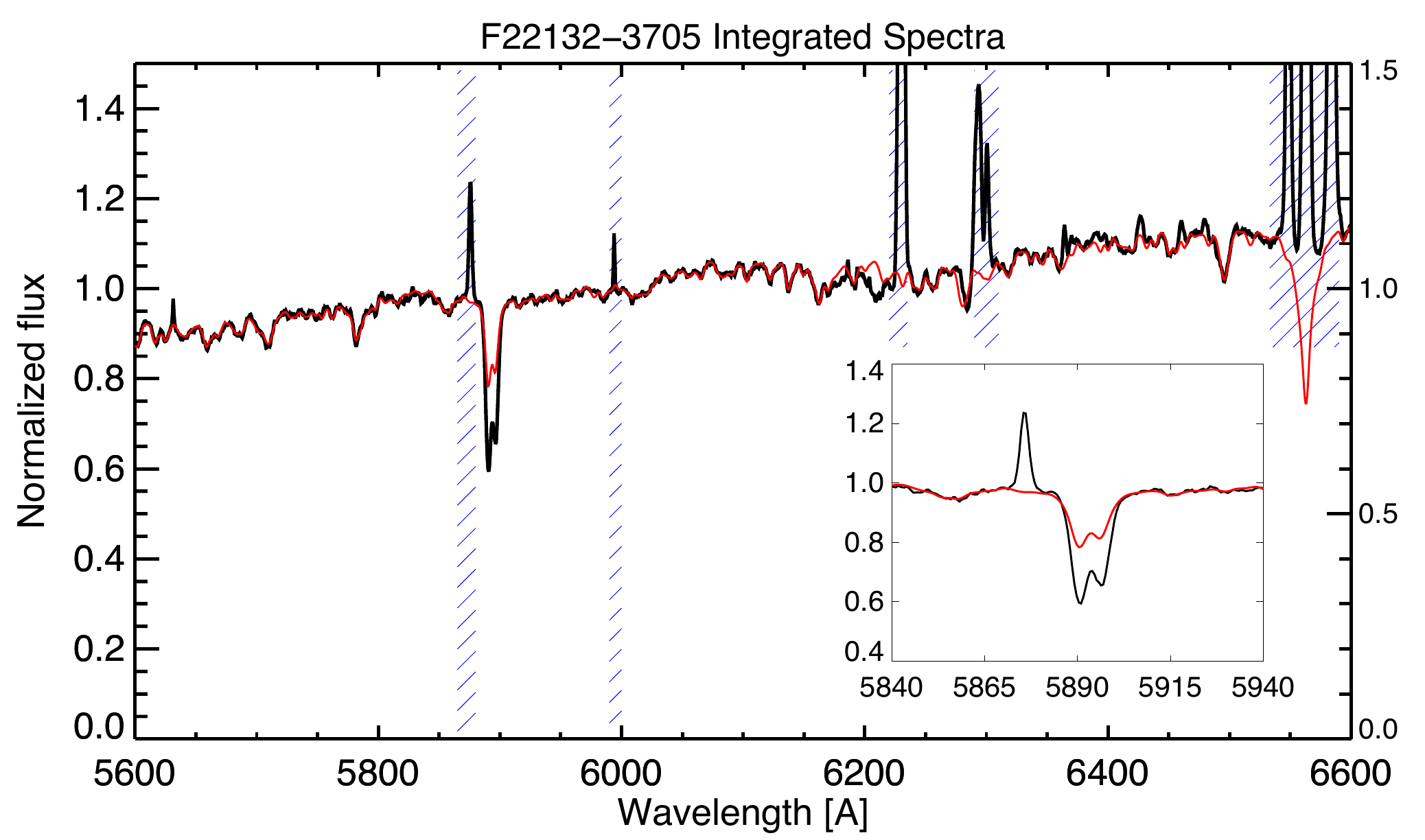}  
 \caption{As Fig.~\ref{Panel_F01159} but for IRAS F22132-3705.}
  \label{Panel_F22132}           
\end{figure*}
\clearpage

       \begin{figure*}
  \vspace{4.em} 
 \centering
   \includegraphics[trim = -0.0cm 8.5cm .5cm 7.0cm, clip=true, width=.75\textwidth]{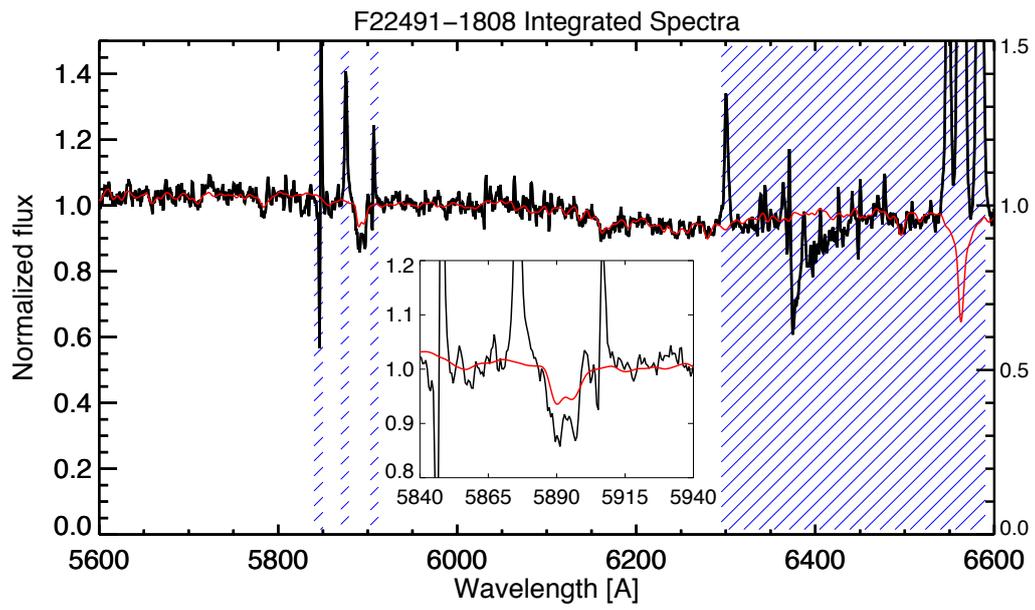}
  \caption{As in the lower panel of Fig.~\ref{Panel_F01159} but for IRAS 22491-1808.}
 \label{Panel_22491}             
\end{figure*}
\clearpage              
                 
\begin{figure*}
\centering
\includegraphics[trim = -0.0cm 0.2cm .5cm 13.0cm, clip=true, width=1.\textwidth]{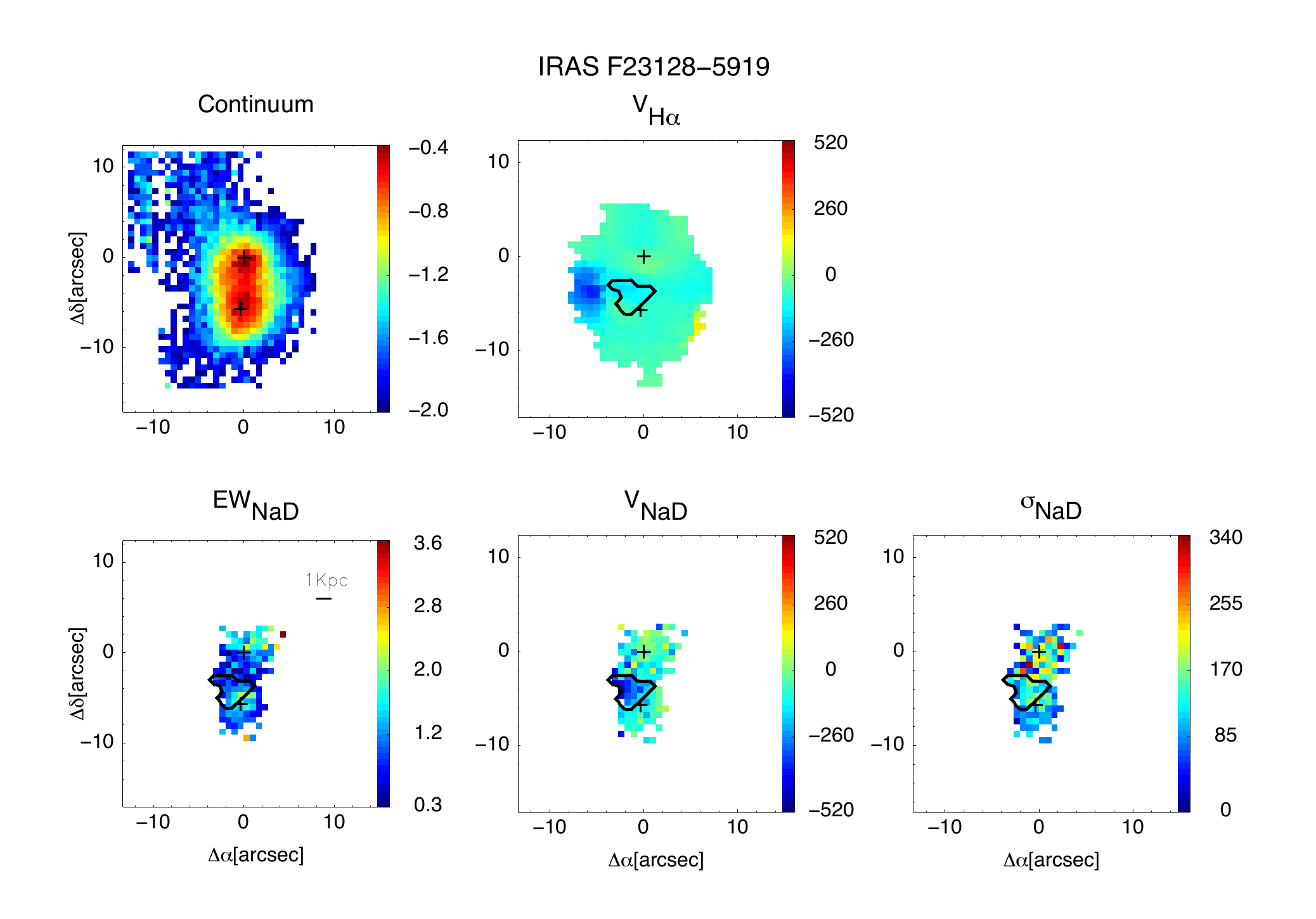}  
\includegraphics[trim = -0.0cm 5.5cm .5cm 2.0cm, clip=true, width=.95\textwidth]{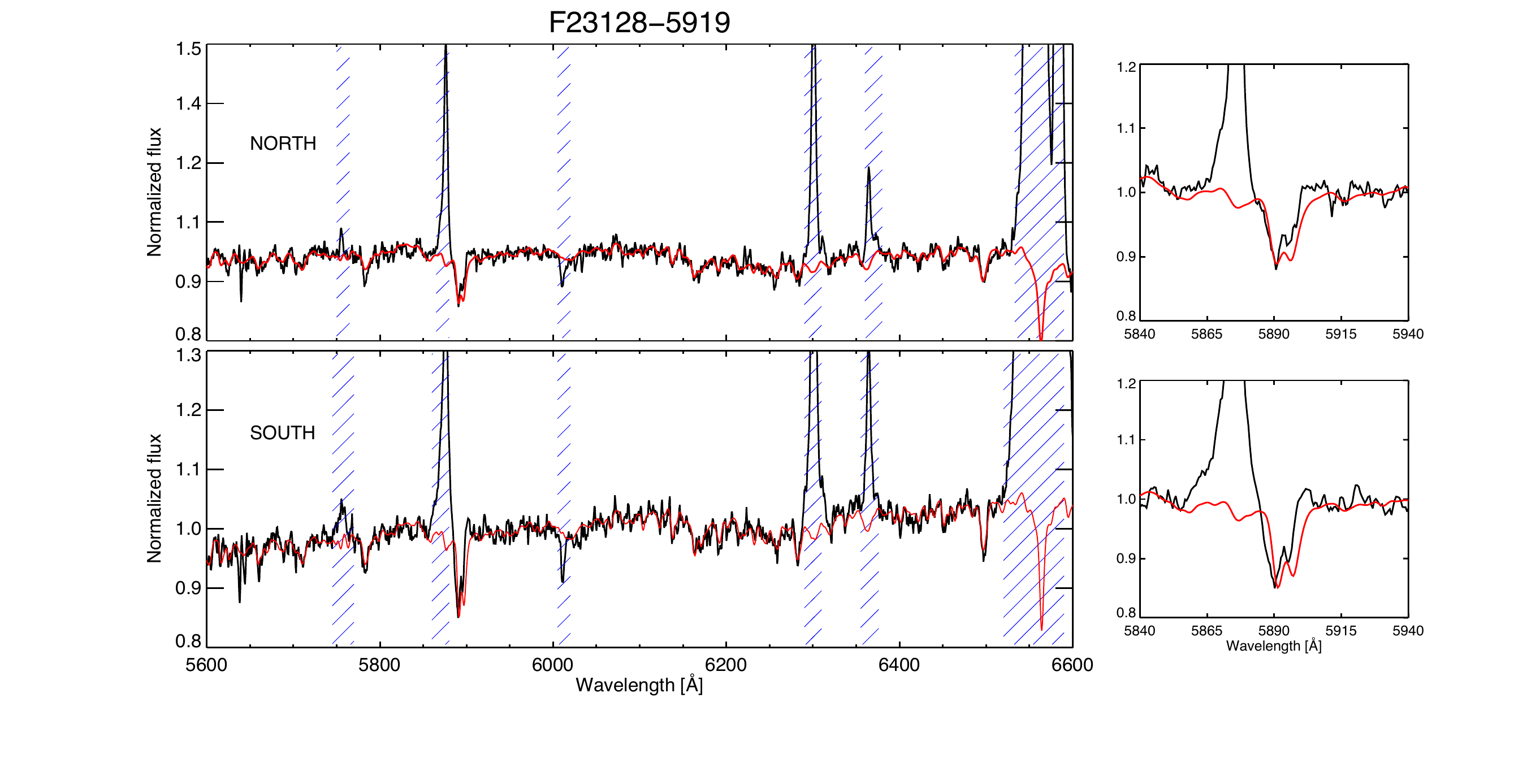}  
 \caption{As Fig.~\ref{Panel_F01159} but for IRAS F23128-5919. In this panel the nuclei of both galaxies are shown with a cross.}
 \label{Panel_F23128}            
\end{figure*}
\clearpage
                                      
                            
  \end{appendix}

\end{document}